# Magnetization Plateaus by the Field-Induced Partitioning of Spin Lattice

Myung-Hwan Whangbo[1],* Hyun-Joo Koo,[2] Reinhard K. Kremer,[3] and Alexander N. Vasiliev[4],*

[1] Department of Chemistry, North Carolina State University, Raleigh, NC 27695-8204, USA
[2] Department of Chemistry and Research Institute for Basic Sciences, Kyung Hee University, Seoul 02447, Republic of Korea
[3] Max Planck Institute for Solid State Research, Heisenberg Strasse 1, Stuttgart D-70569, Germany
[4] Department of Low Temperature Physics and Superconductivity, Lomonosov Moscow State University, Moscow 119991, Russia

whangbo@ncsu.edu
anvas2000@yahoo.com

**Abstract**

In this work we survey the crystal structures and the spin lattices of those magnets exhibiting plateaus in their magnetization vs. magnetic field ($M$ vs. $H$) curves in one or several regions of $H$. We lay out a conceptual picture describing the magnetization plateau phenomenon by probing the three questions: (a) why only certain magnets exhibit magnetization plateaus, (b) why there occur several different types of magnetization plateaus, and (c) what controls the widths of magnetization plateaus. Our work shows that the answers to these questions lie in how the magnets under field absorb Zeeman energy hence changing their magnetic structures. The magnetic structure of a magnet insulator is commonly described by a model Hamiltonian once its spin lattice is identified, which requires the determination of the nonnegligible spin exchanges between the magnetic ions. Our survey strongly suggests that, under magnetic field, the spin lattice of a magnet is partitioned into either antiferromagnetic (AFM) or ferrimagnetic fragments by breaking its weak magnetic bonds. By analyzing how these fragments are formed under magnetic field, we show that the answers to the three questions (a) – (c) emerge naturally, and that our supposition of the field-induced partitioning of a spin lattice into magnetic fragments is supported by the anisotropic magnetization plateaus of Ising magnets and by the highly anisotropic width of the 1/3-magnetization plateau in azurite.

**Table of Content**

# 1. Introduction

The properties of a magnet are primarily characterized by measuring thermodynamic quantities (e.g., magnetization $M$ and/or magnetic specific heat $C_{mag}$) as a function of temperature and external magnetic field. The values of $M(H, T)$ and $C_{mag}(H, T)$ for a given magnet depend on its magnetic energy spectrum and on how the individual states of this spectrum are thermally populated. The individual magnetic states differ in their magnetic properties, and their population at a given temperature is governed by the Boltzmann factor, so the thermodynamic quantity is a weighted average of the properties of various states with weights given by their Boltzmann factors at that particular temperature. As a function of temperature, the Boltzmann distributions of the individual states change. States with lower energy become exponentially more populated as the temperature is decreased. Thus, measuring the temperature dependence of the $M$ or $C_{mag}$ of a magnet is an indirect way of probing its magnetic energy spectrum. To confirm whether or not a magnet undergoes a long-range magnetic ordering as the temperature is lowered is often judged from the temperature dependence of its specific heat. The occurrence of such an ordering is signaled, e.g., by the presence of an anomaly in the $C_{mag}$ vs. $T$ curve, which reflects the loss of the magnetic entropy associated with the long-range magnetic ordering.

Often, to a very good approximation, the energy spectrum of a magnet can be described by the Heisenberg spin Hamiltonian $\hat{H}_{spin}$ (Eq. 1.1a), which is written as the sum of the pairwise spin exchange interactions between spin operators $\hat{S}_i$ and $\hat{S}_j$ located at the magnetic ion sites i and j, respectively.

$$\hat{H}_{spin} = \sum_{i>j} J_{ij}\, \hat{S}_i \cdot \hat{S}_j \qquad (1.1a)$$

Often, the spin operators $\hat{S}_i$ and $\hat{S}_j$ can be regarded as the classical spin vectors $\vec{S}_i$ and $\vec{S}_j$, respectively, and hence the spin Hamiltonian has the classical expression,

$$H_{spin} = \sum_{i>j} J_{ij}\, \vec{S}_i \cdot \vec{S}_j = \sum_{i>j} J_{ij}\, S_i S_j \cos\theta \qquad (1.1b)$$

where $\theta$ is the angle between the two spin vectors $\vec{S}_i$ and $\vec{S}_j$.

Here, we consider the simplest case of the symmetric spin exchange and omit contributions of antisymmetric exchange to the Hamiltonian or anisotropies in the exchange. Being a dot product, the extrema of $\vec{S}_i \cdot \vec{S}_j$ occur when the two spins are parallel and antiparallel to each other, respectively. Thus, with the spin Hamiltonian defined as in Eq. 1.1, the two spins prefer an AFM coupling if the spin exchange is positive ($J_{ij} > 0$), but a ferromagnetic (FM) coupling if it is negative ($J_{ij} < 0$). The "spin lattice" of a magnet refers to the repeat pattern of its spin exchange paths of nonnegligible strengths. If we consider such spin exchange paths as magnetic bonds, then the spin lattice of a magnet is the lattice of its magnetic bonds of various strengths. The spin lattice of a magnet is crucially important because it allows one to generate the energy spectrum relevant for the magnet using a model Hamiltonian $H_{spin}$ with a minimal number of spin exchange parameters.

Magnets including Heisenberg magnetic ions, with nonzero quantized spin moments in all directions, are described by the Heisenberg spin Hamiltonian, for which $\vec{S}_i \cdot \vec{S}_j = S_{ix}S_{jx} + S_{iy}S_{jy} + S_{iz}S_{jz}$. The magnets of uniaxial (i.e., Ising) magnetic ions, which possess nonzero spin moments only in one direction (by convention, the z-direction) so that $\vec{S}_i \cdot \vec{S}_j = S_{iz}S_{jz}$, are described by the Ising spin Hamiltonian

$$H_{Ising} = \sum_{i>j} J_{ij}\, S_{iz}S_{jz} \qquad (1.1c)$$

Transition-metal magnetic ions M in oxide magnets ions form $MO_n$ (typically, n = 3 – 6) polyhedra. The Ising magnetism is found for a magnetic ion M when the d-states of its $MO_n$ polyhedron has an unevenly-occupied degenerate d-state (in the non-spin-polarized, one-electron picture of electronic structure description).[1,2] Such a magnetic ion is susceptible to a Jahn-Teller distortion, which tends to lift, though weakly, the degeneracy responsible for the Ising magnetism.[3] Thus, true Ising magnets are rather rare.

The dependence of magnetization $M$ on external magnetic field $\mu_0 H$ is usually measured at the lowest possible temperature to minimize the contributions of magnetic excited states lying close to the magnetic ground state through the Boltzmann averaging. As a function of the magnetic field $\mu_0 H$, the magnetization of a paramagnet is well described by a Brillouin function, which increases steadily from zero to the magnetic saturation $M_{sat}$. On increasing the magnetic field $\mu_0 H$, the magnetization of an antiferromagnet exhibits spin flop and spin flip transitions (see below) while that of a ferromagnet rapidly reaches the saturation, depending on anisotropy and dipolar energies.

For some magnets among the wide variety of magnetic materials, their magnetization versus magnetic field ($M$ vs. $H$) curves exhibit plateaus at rational fractions $f = m/n$, where $m$ and $n$ are integers with $m < n$ (most commonly, $m = 1$), of their saturation magnetization $M_{sat}$. This phenomenon occurs not only in magnets undergoing a long-range magnetic order at low temperatures but also in low-dimensional or spin-frustrated magnets that do not undergo a magnetic order down to the lowest temperatures. In discussing the $M$ vs. $H$ curves observed for such magnets, it is convenient to distinguish magnets with and without uniaxial (i.e., Ising) anisotropy.[1,2] The idealized $M$ vs. $H$ curves observed for non-Ising magnets are illustrated in **Fig. 1.1a-c**, and those for Ising magnets in **Fig. 1.1d-f**. Consider first the $M$ vs. $H$ behaviors of non-Ising magnets. On increasing the magnetic field from zero, a gradual increase of $M$ from zero to $(m/n)M_{sat}$ precedes before reaching the $m/n$-magnetization plateau at $M = (m/n)M_{sat}$ (**Fig. 1.1a**), the $(m/n)$-magnetization plateau occurs as soon as the field increases from zero (**Fig. 1.1b**), or the zero-magnetization plateau at $M = 0$ precedes until the field reaches a value from which a gradual increase of $M$ to $(m/n)M_{sat}$ begins (**Fig. 1.1c**). For an Ising magnet, the spin moment is nonzero only along one specific direction in space. An Ising magnet exhibits a highly anisotropic $M$ vs. $H$ behavior; its $M$ vs. $H$ curve exhibits a "step-like" feature when the applied field is parallel to the direction of the spin moment (**Fig. 1.1d,e**), i.e., the easy axis direction, but the magnetization does not change with field showing no magnetization plateau when the applied field is perpendicular to the spin moment direction (**Fig. 1.1f**). Experimentally, a very slight linear increase with field is observed, but this is often due to a minute misalignment of the crystal with respect to the external magnetic field.

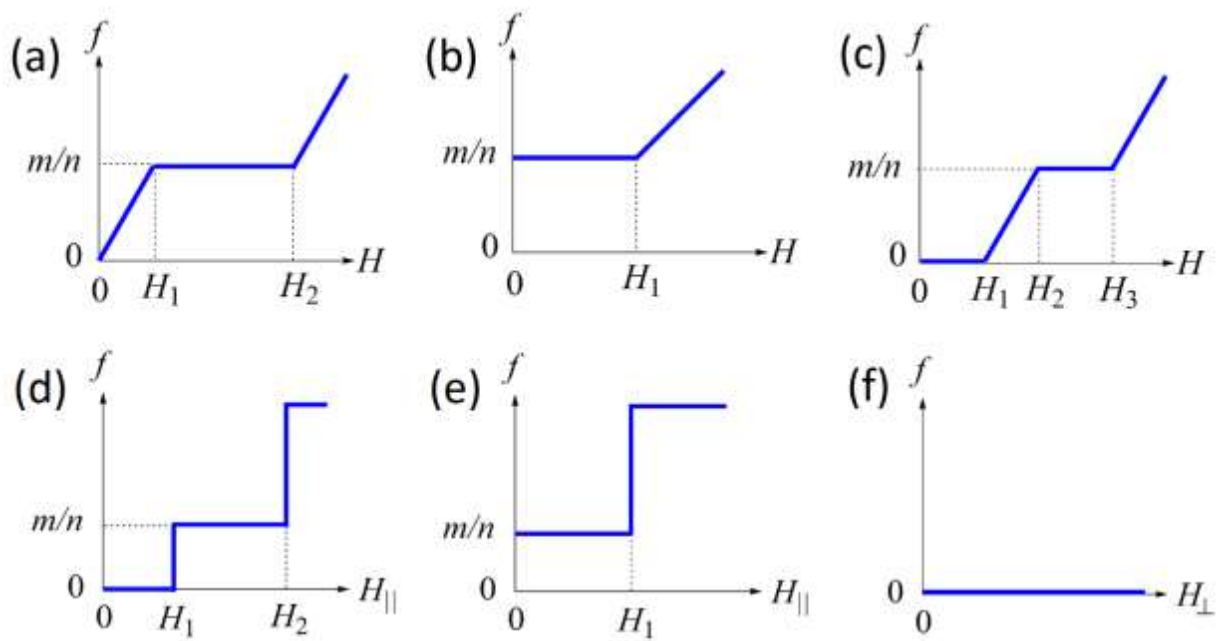

**Fig. 1.1.** Idealized magnetization versus magnetic field ($M$ vs. $H$) curves for two types of magnets exhibiting magnetization plateaus, where the magnetization is given as the fraction $f$ of the saturation magnetization $M_{sat}$. (a – c) $M$ vs. $H$ curves expected for magnets with isotropic magnetism, and (d – f) those expected for magnets with Ising magnetism. On increasing the magnetic field from 0, a gradual increase of $M$ from 0 to $(m/n)M_{sat}$ precedes the $m/n$-magnetization plateau in (a), the magnetization plateau occurs immediately at $(m/n)M_{sat}$ in (b), and a 0-magnetization plateau occurs at $M = 0$ until the field reaches a value from which a gradual increase of $M$ to $(m/n)M_{sat}$ starts in (c). $M$ vs. $H$ curves for Ising magnets exhibit step-like features when the field parallel to the direction of the spin moment as illustrated in (d) and (e), but the magnetization does not change with field when the field is perpendicular to the direction of the spin moment as depicted in (f).

Magnetic plateaus have been found in a large variety of magnets. Their spin lattices can be one-dimensional, two-dimensional (2D) or three-dimensional (3D), their ground state can be AFM or ferrimagnetic in the absence of external magnetic field, their spin lattices may or may not be spin frustrated, and their structures can be extended or discrete. For magnets of high symmetry, a number of theoretical studies examined their magnetization plateaus from the viewpoint of their magnetic energy spectra generated by model spin Hamiltonians.[4,5] So far, however, there has been no systematic study aimed at providing a conceptual picture for the magnetization plateau phenomenon. The primary objective of our survey is to come up with a conceptual framework useful for chemists, materials scientists and experimental physicists in organizing and thinking about magnetization plateaus. Therefore, we pursue the qualitative answers to the three questions (a) – (c) by analyzing not only the structural chemistry associated with the magnetic ions but also the relative magnitudes and the signs of the spin exchange interactions between them. Our study strongly suggests that the spin lattice of a magnet exhibiting a magnetic plateau is partitioned into ferrimagnetic or AFM fragments by breaking the weakest magnetic bonds one at a time by absorbing Zeeman energy provided by an external magnetic field. The $M$ vs. $H$ curve of a magnet is divided into two different regions; the regions where a magnet does not absorb Zeeman energy so nonzero ($m/n > 0$) magnetic plateaus occur and the regions where the magnet absorbs Zeeman energy so no magnetization plateau, except for the zero ($m/n = 0$) magnetization plateau, occurs.

Our survey is organized as follows: In Section 2, we analyze why the spin lattice of a certain magnet is partitioned into smaller magnetic fragments and what types of magnetization plateaus are possible. Section 3 describes the magnetization plateaus of magnets whose spin lattices are partitioned into AFM fragments (with an even number of spin sites) under field, and in Section 4 those of magnets whose spin lattices are partitioned into ferrimagnetic fragments (with an odd number of spin sites) under field. The magnetization plateaus of magnets possessing kagomé and trigonal layers are discussed in Section 5, and those of magnets with complex magnetic fragments in Section 6. Finally, our conclusions are summarized in Section 7.

We note that this work is not a comprehensive review on magnetization plateaus, but a survey on studies of magnetization plateaus that enabled us to put forward the concept that a magnet under field absorbs Zeeman energy by breaking its weak magnetic bonds. The associated partitioning of its spin lattice into magnetic fragments gives rise to magnetization plateaus. For magnets with low-symmetry crystal structures possessing a large number of magnetic ions per unit cell, describing their magnetic structures quantitatively using a model spin Hamiltonian is practically impossible. This difficulty has led us to search for a qualitative description of such magnets on the basis of their spin exchanges (i.e., magnetic bonds), because they can readily determined by employing density functional theory (DFT) calculations. Our studies on numerous such magnets over the past two decades revealed that the spin exchanges obtained from DFT calculations are quite accurate in their relative magnitudes and are therefore reliable in finding which magnetic bonds are weak and hence will be broken preferentially under field. This realization led us to the concept of the field-induced partitioning of a spin lattice into magnetic fragments, initially from our own studies on magnets exhibiting magnetic plateaus. We then checked whether this concept is applicable to other magnets for which magnetization plateaus were reported. When the spin exchanges of these magnets are not available, we determined them by performing DFT calculations as summarized in the supporting information. This survey is the outcome of these efforts. The choice of our references is not comprehensive as expected for a review article but is rather confined to those central to our supposition of the field-induced partitioning of a spin lattice into magnetic fragment.

## 2. Field-induced partitioning of spin lattices
### 2.1. Zeeman energy and magnetic bonds

Two spins of an AFM exchange path tend to align antiparallel to each other, so it requires energy to force them to be ferromagnetically aligned. At very low temperatures where magnetization measurements are usually carried out, the energy needed for such a conversion in a spin exchange path of a magnet is supplied by Zeeman energy, $E_Z$. For a magnetic ion with spin moment $\vec{\mu}_S = -g\mu_B \vec{S}$ under a magnetic field $\mu_0 \vec{H}$, the Zeeman energy is given by

$$E_Z = g\mu_0\mu_B \vec{H} \cdot \vec{S}, \qquad (2.1)$$

which is positive and negative if $\vec{H}$ and $\vec{S}$ are parallel and antiparallel, respectively. As the magnetic field is gradually increased from zero, the conversion from antiparallel to parallel spin alignment occurs initially in weak AFM exchange paths, i.e., those with small exchange J are converted first. For the convenience of our discussion, an AFM exchange path will be termed "a magnetic bond" if the two spins of the path are antiferromagnetically coupled. Likewise, an AFM exchange path may be termed "a broken magnetic bond" or "a magnetic antibond" if the two spins of the path are forced to be ferromagnetically coupled. Thus, an AFM magnetic bond can be broken by the Zeeman energy $E_Z$ (**Fig. 2.1a**). Similarly, an FM exchange path may be termed "a magnetic bond" if the two spins of the path are ferromagnetically aligned, but "a broken magnetic bond or a "magnetic antibond" if the two spins of the path are antiferromagnetically aligned. Thus, an FM magnetic bond can be broken by Zeeman energy (**Fig. 2.1b**). For convenience of our discussion, we will represent the up-spin and down-spin at a magnetic ion site by unshaded and shaded circles, respectively (**Fig. 2.1**). [Here, the up-spin and down-spin are parallel and antiparallel to the direction of an external magnetic field (taken to be the z-direction by convention), respectively. In the absence of an external field, what matters is that the up-spin and down-spin are antiparallel to each other, regardless of their absolute directions in space.] Breaking an AFM magnetic bond increases the spin moment (**Fig. 2.1a**), while breaking an FM bond can either decrease or increase the spin moment (**Fig. 2.1b**) because an FM coupling in a FM magnetic bond can be represented by the (↑↑) or (↓↓) spin arrangement. It is the FM bond breaking from (↓↓) to (↑↓), not from (↑↑) to (↑↓), that is relevant for our discussion of magnetization because the total magnetic moment should not decrease with field (see below for further discussion). Summarizing, magnetic bonds should be referred to as either AFM magnetic bonds or FM magnetic bonds. However, for most magnets showing magnetization plateaus, one deals with breaking AFM magnetic bonds, and it is very rare to find magnets whose magnetization requires the breaking of FM magnetic bonds. Thus, in the following, we use the term "magnetic bonds" to describe AFM bonds, unless stated otherwise.

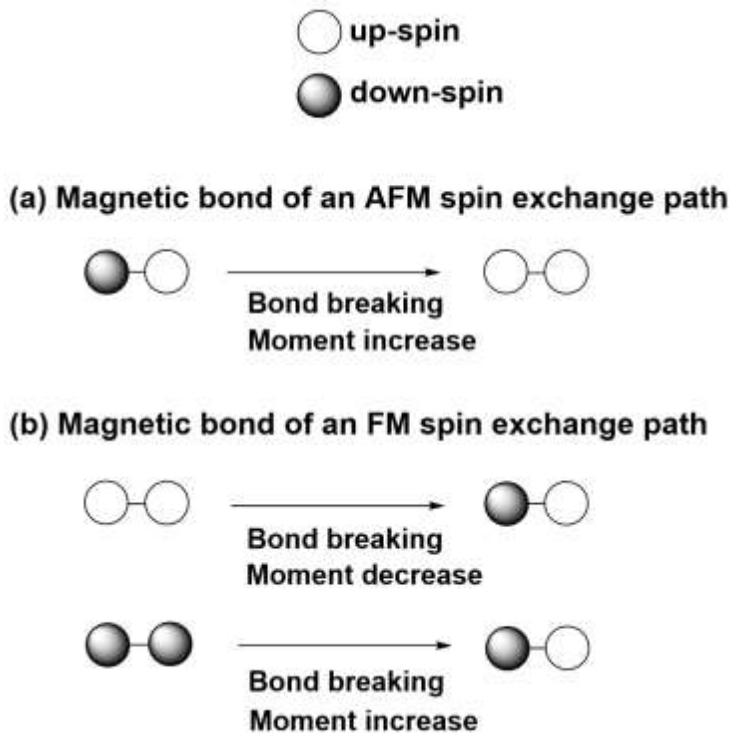

**Fig. 2.1.** Conventions and terminologies employed in discussing the magnetization behaviors of various magnets.

To avoid a possible confusion in using the terminology, the broken or unbroken magnetic bond, it is necessary to distinguish between the eigenstates and the broken-symmetry states. For all practical evaluations of spin exchanges for any magnet, broken symmetry states are used instead of the eigenstates simply because the latter are very difficult to determine.[6] For example, consider a spin dimer made up of two $S = 1/2$ magnetic ions representing, for example, the molecular $Cu_2Cl_6^{2-}$ ion of edge-sharing $CuCl_4$ square planes (see Section 3.2.2). This dimer can be described by the singlet and triplet states $|S\rangle$ and $|T\rangle$, which are the eigenstates of the dimer (**Fig. 2.2a**). Then, the energy difference between the two states is the spin exchange J, i.e., $E_T - E_S = J$ (**Fig. 2.2b**). If the dimer is described by the broken-symmetry states ↑↑ and ↑↓, then the energy difference between the two states is given by $E_{↑↑} - E_{↑↓} = J/2$ (**Fig. 2.2c**). In the following, by breaking an AFM J bond in an extended magnet, we mean the conversion from the AFM coupling ↑↓ to the FM coupling ↑↑. In the case of an isolated dimer, the breaking the AFM J bond means the excitation from the singlet to the triplet state.

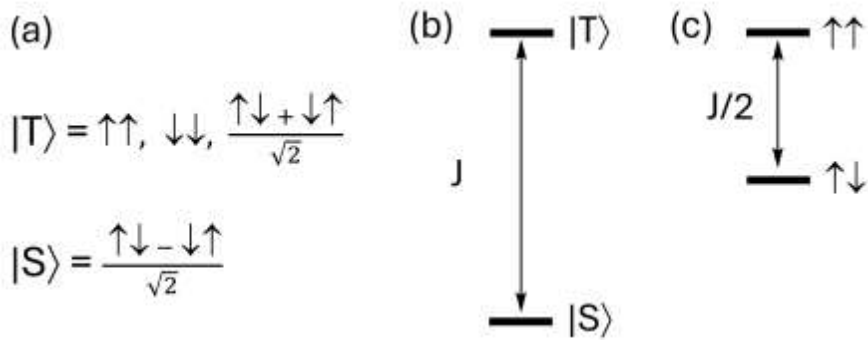

**Fig. 2.2**. (a) Expressions of the singlet and triplet states of an isolated spin dimer made up of two $S = 1/2$ magnetic ions. (b) Single and triplet states of a spin dimer in which the singlet state is lower in energy than the triplet state. (c) Broken symmetry states of a spin dimer in which the AFM coupling is more stable than the FM coupling.

## 2.2. Causes for magnetization plateaus

In the following we put forward the supposition that, during the magnetization process, the spin lattice of a magnet becomes partitioned into either AFM or ferrimagnetic fragments by breaking its weak magnetic bonds. As an example, consider how a 0-magnetization plateau arises by considering a chain in which AFM dimers described by spin exchange $J_1$ are antiferromagnetically coupled in the tail-to-tail bridging pattern to make $J_2$ bonds between adjacent dimers such that $J_1 > J_2$. An AFM chain made up of alternating $J_1$ and $J_2$ bonds is presented in **Fig. 2.3a**. Since $J_2$ is weaker than $J_1$, the $J_2$ bond will be broken 'successively' (**Fig. 2.3b,c**) as the magnetic field is increased from 0 until all $J_2$ bonds are broken (**Fig. 2.3d**). In **Fig. 2.3b–d**, the red ellipses are used to indicate that the (↓↑) dimers, resulting from the (↑↓) dimers of **Fig. 2.3a**, break the inter-dimer bonds $J_2$. The energy needed for breaking a $J_2$ bond is supplied by Zeeman energy, but the magnetization remains at zero while the field increases until all $J_2$ boinds are broken because the spins stay paired in the $J_1$ bonds. This leads to a 0-magnetization plateau (e.g., **Fig. 1.1c**).

As the magnetic field increases further, the $J_1$ bonds become broken one at a time as depicted in **Fig. 2.3e-g**, where the green ellipses are used to indicate the broken dimers, (↑↑), resulting from the (↓↑) dimers of **Fig. 2.3d**. In principle, a broken dimer can be equally well represented by the configuration (↓↓). However, throughout our discussion, a broken dimer will be represented by (↑↑), because the spin moments of a magnet under magnetic field should not decrease with field and because we use the convention that up-spin and down-spin have the positive and negative moments, respectively. Since each $J_1$ bond breaking creates unpaired up-spins, the magnetization $M$ increases with the field; $M = M_{sat}/4$, if one out of four $J_1$ bonds is broken (**Fig. 2.3e**), $M = M_{sat}/3$ if one out of three $J_1$ bonds is broken (**Fig. 2.3f**), and $M = M_{sat}/2$ if one out of two $J_1$ bonds is broken (**Fig. 2.3g**).

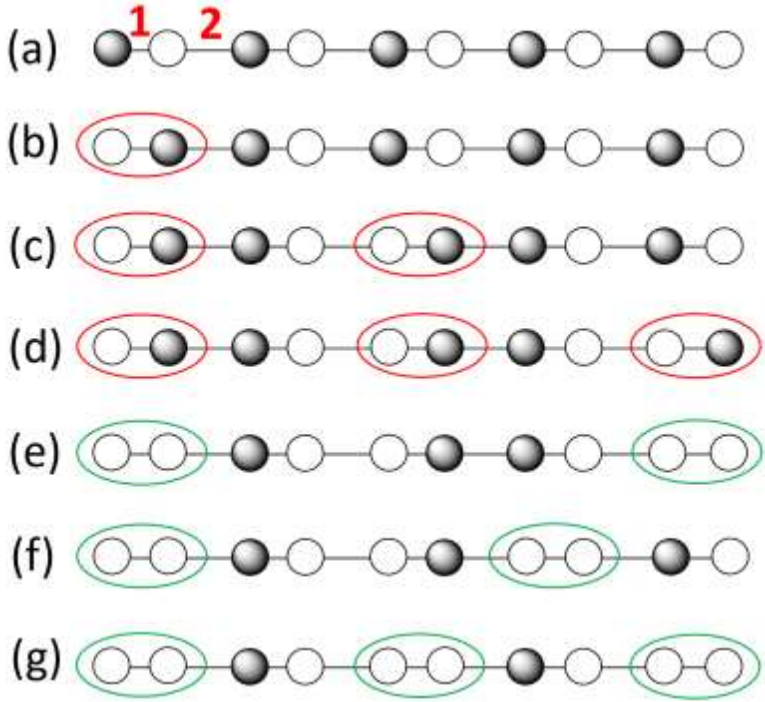

**Fig. 2.3.** Effect of the external magnetic field on the magnetic structure of an AFM chain made up of AFM dimers, where labels 1 and 2 refer to the spin exchanges $J_1$ and $J_2$, respectively. It is assumed that the intra-dimer exchange $J_1$ is stronger than the inter-dimer-exchange $J_2$: (a) Ground state in the absence of the external magnetic field. (b, c) Breaking of the $J_2$ bonds one at a time with increasing field. (d) State in which all $J_2$ bonds are broken. (e) $M = M_{sat}/4$ state that results when one out of four $J_1$ bonds is broken. (f) $M = M_{sat}/3$ state that results when one out of three $J_1$ bonds is broken. (g) $M = M_{sat}/2$ state that results when one out of two $J_1$ bonds is broken.

In general, when the spin lattice of a magnet becomes partitioned into identical AFM fragments by breaking its weak magnetic bonds interconnecting them, the magnet acquires a zero-magnetization given by the AFM fragment regardless of how many inter-fragment magnetic bonds there are. If there exist a large number of different arrangements between the AFM fragments, which differ only in the number of their inter-fragment magnetic bonds, then the magnetization of a magnet remains unchanged while the inter-fragment magnetic bonds are being broken by increasing the magnetic field. The 0-magnetization plateau of the AFM chain made up of AFM dimers discussed above is an example. To increase the magnetization beyond this level, a weak magnetic bond of each AFM fragment needs to be broken.

Suppose that the spin lattice of a magnet becomes partitioned into identical ferrimagnetic fragments when the field increases by breaking its weak inter-fragment magnetic bonds. Then the magnetization increases gradually with field until all inter-fragment bonds are broken so that all ferrimagnetic fragments become ferromagnetically coupled, leading to a certain level of magnetization given by the ferrimagnetic fragments. To increase the magnetization beyond this level, it is necessary to break a weak magnetic bond of each ferrimagnetic fragment. If this magnetic bond is strong, the bond breaking will not happen unless the field reaches a high enough value, hence leading to a magnetization plateau.

### 2.3. Magnetic bonding pattern affecting the nature of magnetization plateaus

Consider now an AFM chain in which ferrimagnetic linear trimers with the (↑↓↑) configuration are antiferromagnetically coupled in a tail-to-tail pattern (**Fig. 2.4a**) so that the (↑↓↑) and (↓↑↓) trimers alternate. We assume that the intra-trimer $J_1$ bond is stronger than the inter-trimer bond $J_2$. Then, as the field increases from 0, the $J_2$ bonds become broken one at a time (**Fig. 2.4b**) until all $J_2$ bonds are broken (**Fig. 2.4c**) by converting each (↓↑↓) trimer to a (↑↓↑) trimer, as indicated by the red ellipses. Since each trimer constitutes a ferrimagnetic unit, the magnetization increases with the number of broken $J_2$ bonds until it reaches $M_{sat}/3$ where all $J_2$ bonds are broken. When the magnetic field is further raised, the bonds to break are the two $J_1$ bonds in each (↑↓↑) trimer as indicated by the green ellipse in **Fig. 2.4d**. The magnetic bonds $J_1$ are strong, and two $J_1$ bonds of a trimer should be broken simultaneously. Therefore, until the field reaches a high enough value, they are not broken hence leading to no increase of the magnetization. The magnetization plot of **Fig. 1.1a** is a characteristic feature for an AFM chain in which ferrimagnetic fragments are antiferromagnetically coupled in a tail-to-tail manner.

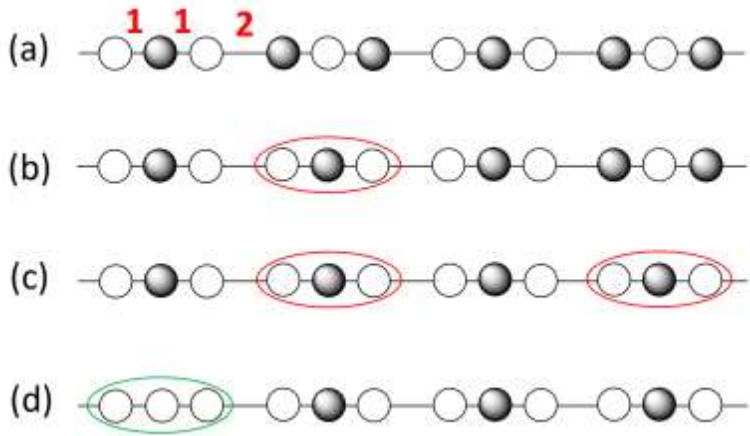

**Fig. 2.4.** Effect of the external magnetic field on the magnetic structure of an AFM chain made up of linear AFM trimers in a tail-to-tail bridging pattern, where labels 1 and 2 refer to the spin exchanges $J_1$ and $J_2$, respectively. It is assumed that the intra-trimer exchange $J_1$ is stronger than the inter-trimer-exchange $J_2$: (a) Ground state in the absence of the external magnetic field. (b) Breaking of the $J_2$ bonds one at a time with increasing field. (c) State in which all $J_2$ bonds are broken, leading to the ferrimagnetic state with $M = M_{sat}/3$. (d) Breaking of two $J_1$ bonds of a linear trimer, enhancing the magnetization toward $M = M_{sat}$.

When the ferrimagnetic linear trimers are combined in a head-to-tail bridging pattern, the resulting chain is a ferrimagnetic chain (**Fig. 2.5a**), with magnetization $M = M_{sat}/3$. Under magnetic field, a $J_2$ bond of this ferrimagnetic chain cannot be broken because, if broken, the resulting ferrimagnetic trimer will have the (↓↑↓) configuration (indicated by the red ellipse in **Fig. 2.5b**) and hence will reduce the overall moment of the chain. Therefore, the only way of absorbing magnetic energy is to break the two $J_1$ bonds of a ferrimagnetic trimer successively (indicated by the green ellipses in **Fig. 2.5c,d**) hence increasing the magnetization toward $M_{sat}$. The $J_1$ bond is strong and a simultaneous breaking of two $J_1$ bonds requires high energy, so this will not occur until the applied field is high enough. Thus, the 1/3-magnetization plateau will be wide. The magnetization curve of the ferrimagnetic chain is well described by **Fig. 1.1b**.

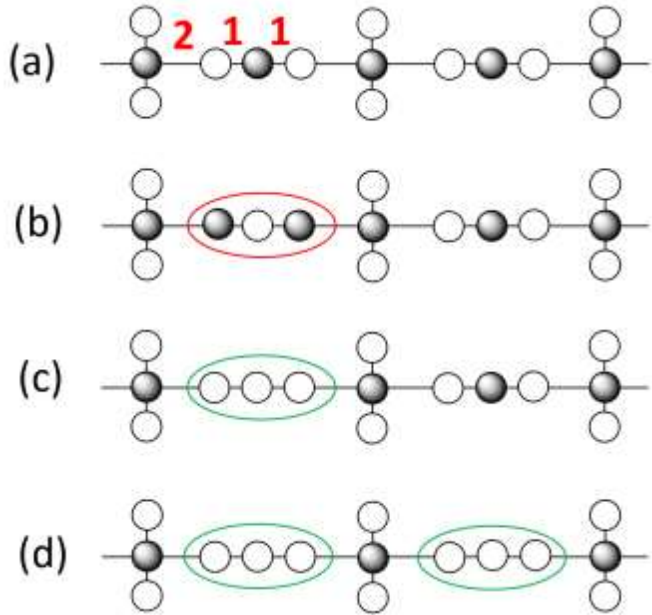

**Fig. 2.5.** Effect of the external magnetic field on the magnetic structure of a ferrimagnetic chain made up of linear AFM trimers in a head-to-tail bridging pattern, where labels 1 and 2 refer to the spin exchanges $J_1$ and $J_2$, respectively. It is assumed that the intra-trimer exchange $J_1$ is stronger than the inter-trimer-exchange $J_2$: (a) Ferrimagnetic ground state in the absence of the external magnetic field. (b) Breaking one $J_2$ bond leads to one trimer in the (↓↑↓) configuration, which reduces the overall moment of the chain. Hence breaking a $J_2$ bond will not take place since the moment of a magnet cannot decrease under field. (c, d) Breaking of two $J_1$ bonds of a linear trimer, enhancing the magnetization toward $M = M_{sat}$.

## 2.4. Field-induced ferrimagnetic fragments in spin-frustrated lattices

As described above, field-induced partitioning of a spin lattice into magnetic fragments lies at the heart of the magnetization plateau phenomenon. In understanding this field-induced partitioning, it is crucial to identify the weak magnetic bonds of a given spin lattice that can be readily broken by moderate magnetic fields. The absorption of Zeeman energy by a magnet is a consequence of the Le Chatelier's principle. There are cases when it is not immediately obvious how a spin lattice under field will be partitioned into magnetic fragments when the spin lattice is defined by a few spin exchanges of comparable magnitude is spin-frustrated, e.g., trigonal, kagomé and diamond chain spin lattices. For such cases as well, Le Chatelier's principle enables us to put forward the supposition that a spin lattice is partitioned into small ferrimagnetic fragments of nonzero spin $\vec{S}$ such that these fragments fill the spin lattice without overlapping between them.

This partitioning reduces the extent of spin frustration by absorbing Zeeman energy and hence breaking the inter-fragment bonds and making the ferrimagnetic fragment absorb Zeeman energy further when the field is raised. For example, consider a magnet consisting of diamond chains made up of two AFM spin exchange $J_1$ and $J_2$ with $J_1 > J_2$ (**Fig. 2.6a**). The weaker magnetic bonds $J_2$ form a continuous chain, but the 1/3-magnetization phenomenon observed for such a magnet can be readily understood by supposing that the diamond chain undergoes a field-induced partitioning into triangular ferrimagnetic fragments as depicted by shaded triangles in **Fig. 2.6b,c**.

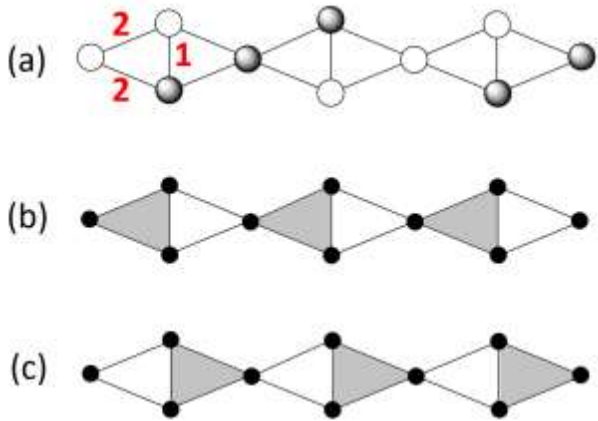

**Fig. 2.6.** (a) Diamond chain defined by two magnetic bonds $J_1$ and $J_2$, where $J_1 > J_2$. (b, c) Two possible ways of fragmenting a diamond chain into non-overlapping magnetic triangles.

Each triangular fragment can have six possible spin arrangements (**Fig. 2.7**). In the absence of applied magnetic field, the (↑↑↑) and (↓↓↓) trimers are less stable than the other four ferrimagnetic trimers, and the (↑↓↓) and (↓↑↓) trimers with net negative moment are as stable as the (↓↑↑) and (↑↓↑) trimers with net positive moment. Applying a magnetic field stabilizes the latter but destabilizes the former. Likewise, applied field stabilizes the (↑↑↑) trimer while destabilizing the (↓↓↓) trimer. Thus, for the discussion of the 1/3-magnetization plateau of the diamond chains, only the ferrimagnetic (↑↓↑) or (↓↑↑) trimers are relevant.

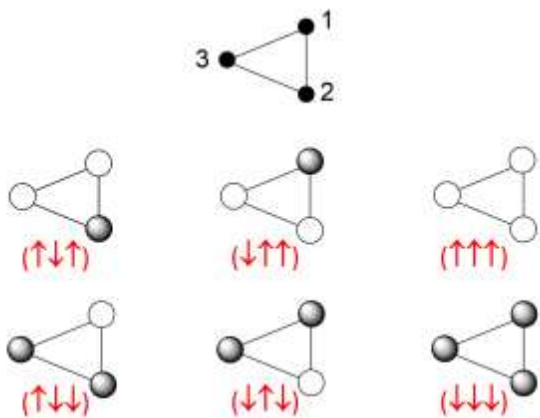

**Fig. 2.7.** Spin arrangements possible for a triangular fragment. In each case, the arrows from the left to right indicate the spins at the magnetic sites 1, 2 and 3, respectively.

Let us consider theoretical and experimental justifications for our supposition that a spin-frustrated spin lattice will undergo a field-induced partitioning into small ferrimagnetic fragments of nonzero spin $\vec{S}$. If a magnet can produce such ferrimagnetic fragments, the energy of each ferrimagnetic fragment is raised by Zeeman energy, $E_Z = g\mu_0\mu_B HS$, which can be used for the magnet to break the magnetic bonds necessary for the partitioning. However, if a magnet cannot interact with magnetic field, such fragmentation cannot occur so that the magnet cannot develop any magnetization plateau. An experimental test for these predictions is provided by magnetization studies on Ising magnets. The spins of an Ising magnet are nonzero in one direction in space (say, the ∥z direction) but zero in all directions perpendicular to this direction (i.e., ⊥z). For field $\vec{H}_\parallel$ along the ∥z direction, the Zeeman energy is nonzero ($\vec{H}_\parallel \cdot \vec{S} \neq 0$). For field $\vec{H}_\perp$ along the ⊥z direction, however, the Zeeman energy is zero ($\vec{H}_\perp \cdot \vec{S} = 0$). (Here, we have assumed that the anisotropy energy that forced the spins to align either parallel or antiparallel to z is much larger than the Zeeman energy.) Therefore, an Ising magnet can have a magnetization plateau when the field is along the ∥z direction but cannot if the field is along the ⊥z direction (**Fig. 1.1d-f**). The step-like feature of the magnetization curves found for Ising magnets under field $\vec{H}_\parallel$ indicates that a large number of ferrimagnetic fragments develop simultaneously, because spin flipping is necessary for magnetic bond breaking.

## 2.5. Spin-lattice interactions

In our discussion of magnetic plateaus so far, the field-induced partitioning of a spin lattice into ferrimagnetic or antiferromagnetic fragments is discussed without considering spin-lattice interactions. The magnetic fragments broken off differ in their surroundings from the spin lattice (e.g., **Fig. 2.4a-c**) and hence would require some relaxation in their atomic/electronic structures through magnetoelastic coupling. The concomitant change in the spin-lattice interaction can therefore affect the stability and nature of a magnetization plateau. A magnetization plateau generates a certain pattern of up-spin and down-spin arrangement and hence the associated spin-lattice interactions. When the spin-lattice interaction, associated with a certain magnetization plateau, leads to an energy-lowering relaxation, this magnetization plateau will arise. Otherwise, it will not be observed. It goes without saying that the spin-lattice interaction will be important for magnets of compact and high-symmetry atomic structure, because the pattern of up-spin and down-spin arrangement should be compatible with the symmetry of the underlying spin lattice.

Magnets of spinel structure such as $CdCr_2O_4$ and $LiGaCr_4O_8$ have a pyrochlore spin lattice (see **Section 3.1.2**). The ideal spinel structure is cubic. Studies on $CdCr_2O_4$[7] and $LiGaCr_4O_8$[8,9] reveal that their observed magnetization plateaus, namely, the 1/2-magnetization plateaus, are strongly stabilized by the magnetoelastic (i.e., spin-lattice) coupling. Another high-symmetry magnet showing the importance of the spin-lattice interaction is the quasi-2D tetragonal magnet $SrCu_2(BO_3)_2$. Early magnetization measurements on single-crystal samples of $SrCu_2(BO_3)_2$ identified 1/8-, 1/4- and 1/3-plateaus of $M_{sat}$ (**Fig. 3.2b**).[10] More recent experiments with fields up to ~140 T[11,12] revealed that the transition into the regions of the 1/2-magnetization plateau is accompanied by strong magnetoelastic effects.[13] In stabilizing the magnetization plateaus of other magnets with less rigid and low symmetry atomic structure, the magnetoelastic coupling would also be important although elaborate studies such as carried for the spinel magnets and $SrCu_2(BO_3)_2$ are mostly not available yet. Thus, in what follows, we will discuss the field-induced fragmentation of a spin lattice based solely on the interaction of the external magnetic field with the spin arrangement in the spin lattice.

## 2.6. Different magnetization behaviors of Heisenberg and Ising magnets

Heisenberg and Ising magnets change the direction of their magnetic moments as the external magnetic field increases, and exhibit a slightly different dependence on the magnetic field. Though often described by an idealized Heisenberg spin Hamiltonian, such a magnet exhibits weak magnetic anisotropy if the orbital moment quenching of its magnetic ions is incomplete. In a similar manner, an ideal Ising magnet described by an Ising Hamiltonian is difficult to find because the associated Jahn-Teller distortion can lift, though weakly, the degeneracy of the d-state responsible for the Ising magnetism.[3] Therefore, by Heisenberg and Ising magnets, we mean those whose magnetic properties are well approximated by Heisenberg and Ising spin Hamiltonians, respectively.

The plateau formation at fractional values of the saturation magnetization $M_{sat}$ compete and coexist with spin-flop transitions in Heisenberg antiferromagnets and metamagnetic transitions in Ising antiferromagnets. In the absence of a magnetization plateau, the magnetization processes of Heisenberg antiferromagnets can be described as illustrated in **Fig. 2.8**, and those of Ising antiferromagnets as illustrated in **Fig. 2.9**. Most Heisenberg magnets possess weak magnetic anisotropy, so describing them with a Heisenberg spin Hamiltonian is an approximation. Similarly, the $MO_n$ polyhedra of the magnetic ions M in most oxide Ising magnets are weakly distorted due to the Jahn-Teller instability, it is an approximation to describe Ising magnets using an Ising spin Hamiltonian. The majority of magnetic measurements are conducted mainly on polycrystalline samples. For measurements on such samples, it is often difficult to distinguish whether a magnetic transition involved is a spin-flop or a metamagnetic type. However, it is generally observed that the $M(H)$ curve is concave for spin-flop transitions, but convex for metamagnetic transitions.

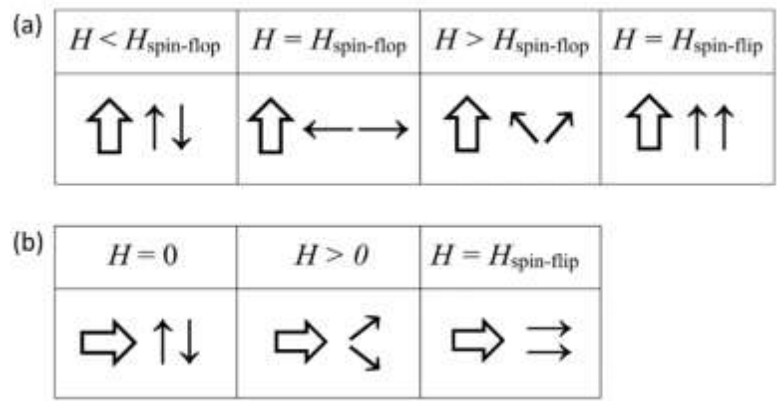

**Fig. 2.8.** Change in the spin moment orientations in Heisenberg antiferromagnets as a function of the magnetic field strength when the magnetic field is (a) parallel and (b) perpendicular to the direction of the spin moment. The thick white arrows represent the magnetic field direction, while thin black arrows represent the directions of the spin moments.

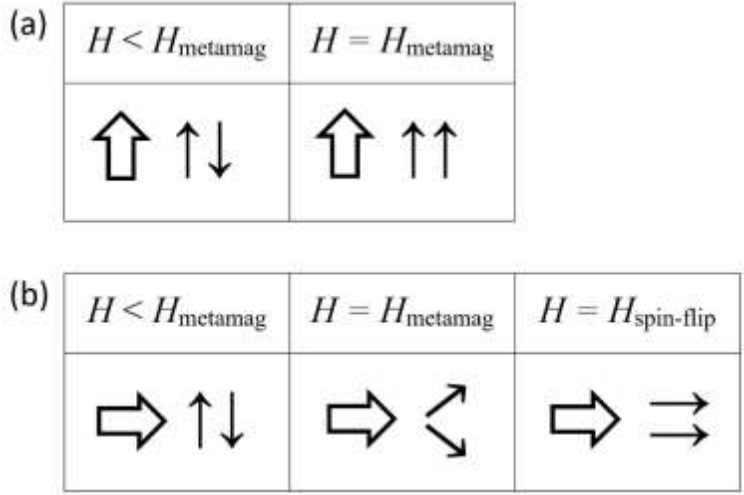

**Fig. 2.9.** Change in the spin moment orientations in Ising antiferromagnets as a function of the magnetic field strength when the magnetic field is (a) parallel and (b) perpendicular to the direction of the spin moments. The thick white arrows represent the magnetic field direction, while thin black arrows represent the directions of the spin moments.

### A. Spin flop and spin flip processes of antiferromagnets

From the viewpoint of phenomenological description, Heisenberg and Ising magnets differ in the strength of the single-ion magnetic anisotropy D with respect to that of the spin exchange J. The positions of the spin-flop and metamagnetic transitions in the magnetization curves are determined by J and D. General aspects of field-induced transitions were described by Néel,[14] detailed discussions of spin-flop transitions by Morosov and Sigov,[15] and those of metamagnetic transitions by Strujewski and Giordano.[16] Consider an antiferromagnet with single-ion anisotropy aligning the magnetic moments along a preferred crystal direction commonly called 'easy axis', which is generally defined as the z-direction. For such a magnet, its behaviors in a field directed either along or perpendicular to the easy magnetization axis are important to analyze. First, let us consider a Heisenberg-type antiferromagnet with a small magnetic anisotropy D compared with the exchange J, which has the magnetic moments of the two magnetic sublattices, $M_1$ and $M_2$ (namely, the up-spin and down-spin sublattices). Consider, for example, $Fe_2O(SeO_3)_2$ [17] in which the Fe atoms, Fe1, Fe2 and Fe3, lead to three spin exchange paths $J_1$, $J_2$ and $J_3$ (for the nearest-neighbor Fe1-Fe2, Fe2-Fe3 and Fe3-Fe3 paths, respectively). These paths form 2D nets parallel to the ab-plane (**Fig. 2.10a**) which are stacked along the c-direction. The Fe1, Fe2 and Fe3 atoms exist as $Fe^{3+}$ ($d^5$, S =5/2) ions, so their magnetic anisotropy is weak (i.e., small D). The neutron diffraction studies [17] reveal that, in all these spin exchange paths, the spin moments parallel to the b-axis are antiferromagnetically coupled. The magnetic susceptibility of this magnet (**Fig. 2.10b**) shows that, for an external magnetic field along the easy axis, $\mu_0 H \| b$, the transition to the AFM state at the Néel temperature $T_N$ manifests itself as a sharp decrease in the magnetic susceptibility $\chi_\|$. For an external field perpendicular to the easy axis, $\mu_0 H \perp b$, however, the magnetic susceptibility $\chi_\perp$ at $T < T_N$ remains almost unchanged. At low temperatures, $\chi_\| < \chi_\perp$. Thus, upon reaching a certain critical field, the difference in the energy of the magnetic moments $M_1$ and $M_2$ oriented either parallel or perpendicular to an external magnetic field reaches a critical value

$$\Delta E = -\frac{1}{2}(\chi_\perp - \chi_\|)\mu_0 H^2_{spin-flop}, \qquad (2.2)$$

at which there is a 90° rotation of the magnetic moments $M_1$ and $M_2$ to the direction perpendicular to the magnetic field. Taking into account the fact that these two moments are related by the exchange interaction J, the field of the spin-flop transition is determined by the expression

$$\mu_0 H_{spin-flop} = (2DJ)^{1/2}. \tag{2.3}$$

At this field and at temperatures small compared to the Néel temperature $T_N$, there is a sharp jump in the magnetization $M_\|$, after which the magnetization monotonically increases up to the saturation magnetization

$$M_{saturation} = |M_1| + |M_2|, \tag{2.4}$$

which is reached in the field of the spin-flip transition

$$\mu_0 H_{spin-flip}^{Heisenberg} = J. \tag{2.5}$$

In an external magnetic field $\mu_0 H \perp b$, the magnetization $M_\perp$ increases monotonically, reaching the saturation magnetization in the same field $\mu_0 H_{spin-flip}$ (we neglect here the anisotropy of the g factor). The field dependence of the magnetization of the easy-axis antiferromagnet $Fe_2O(SeO_3)_2$ is shown in **Fig. 2.10c**.

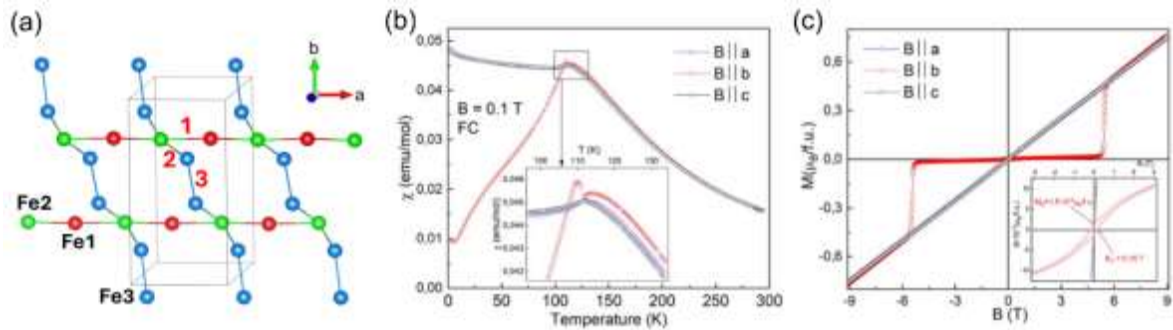

**Fig. 2.10**. (a) Arrangement of the three spin exchanfe paths $J_1$, $J_2$ and $J_3$ forming 2D nets parallel to the ab-plane [17], where the labels 1, 2 and 3 refer to $J_1$, $J_2$ and $J_3$, respectively. (b) Temperature dependence of the magnetic susceptibility measured under the magnetic field of 0.1 T along the a, b, and c axes in $Fe_2O(SeO_3)_2$ [17]. The inset shows a zoomed-in view around the magnetic transition. (c) Magnetization curve at 2 K. The inset shows a zoomed-in view for the magnetization at low fields [17].

We now turn to an Ising-type antiferromagnet with a magnetic anisotropy D comparable in strength to or exceeding the spin exchange J. When the magnetic field is directed along the magnetic moments of the two sublattices $M_1$ and $M_2$ (i.e., along the b axis), the magnetic susceptibility $\chi_\|$ of an Ising antiferromagnet is similar to that observed for a Heisenberg antiferromagnet, as shown in **Fig. 2.11a** for the francisite-type compound.[18] When an external magnetic field $\mu_0 H \| b$ reaches the critical value

$$\mu_0 H_{metamagnetic} = J, \tag{2.6}$$

one of the sublattices (either $M_1$ or $M_2$) will reverse its moment direction by 180° producing a sharp jump, as shown in **Fig. 2.11b**.[19] In this case, $\mu_0 H_{metamagnetic}$ is equivalent to $\mu_0 H_{spin\text{-}flip}$ and corresponds to $M_{saturation}$. In a magnetic field $\mu_0 H \perp b$, both $M_1$ and $M_2$ moments continuously rotate to the direction of the external magnetic field reaching the saturation magnetization $M_{saturation}$ at

$$\mu_0 H^{Ising}_{spin-flip} = J + D. \tag{2.7}$$

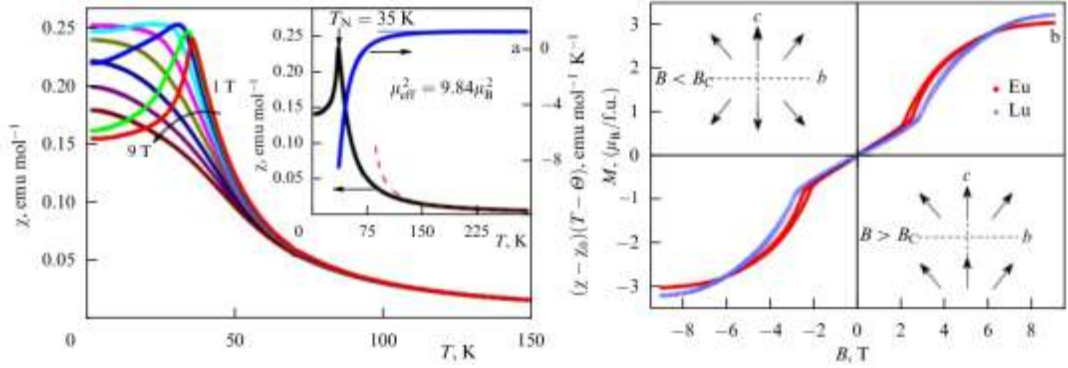

**Fig. 2.11.** (a) Magnetic susceptibility of $Cu_3Y(SeO_3)_2O_2Cl$ at various probe magnetic fields between 1 - 9 T[18] measured with field $\mu_0 H \| b$. The inset shows the susceptibility measured at 0.1 T, and the Curie constant C refers to $(\chi - \chi_0)(T - \Theta)$. The dashed line (see the inset) represents the Curie-Weiss law. (Reproduced with permission from reference 18.) (b) Metamagnetic phase transition in $Cu_3Eu(SeO_3)_2O_2Cl$ and $Cu_3Lu(SeO_3)_2O_2Cl$ under field $\mu_0 H \| b$. Inset: Schematic representations of the $Cu^{2+}$ spin moments in weak and strong magnetic fields.[19] (Reproduced with permission from reference 19.)

**B. Magnetization plateaus**

The magnetization behaviors of Heisenberg and Ising magnets differ in their $M$ vs. $H$ curves preceding a magnetization plateau (**Fig. 1**); the magnetization exhibits a linear increase with field for a Heisenberg magnet but does not depend on field for an Ising magnet. We briefly comment on why this difference comes about. In a Heisenberg exchange coupled system, the exchange energy depends only on the relative orientation of the participating spin moments. To a first approximation, the effective spin Hamiltonian for an Ising magnet contains the z-components of the spins only. The only options of the crystal field anisotropy are to favor either a parallel or an antiparallel alignment of the spin moments along the easy axis. Therefore, an external magnetic field not only competes with the spin exchange but also with the anisotropy. Consequently, the magnetic response of an Ising magnet to an external magnetic field depends sensitively on the alignment of the magnetic field with respect to the easy axis as well as their relative magnitudes. Spin-flop transitions with sudden jumps of the magnetization from very low to large values are a phenomenon connected to the presence of crystal field anisotropy. Thus, the magnetization curve of an Ising magnet deviates somewhat from the step-like features (**Fig. 1.1d,e**) when the magnetic field is parallel to the easy axis, and from the flat line (**Fig. 1.1f**) when the magnetic field is perpendicular to the easy axis.

**2.7. Quantitative evaluations of spin exchange interactions**

In understanding the magnetic properties of a magnet, it is essential to know the strengths of its magnetic bonds, i.e., the values of its spin exchanges $J_{ij}$. The parameters $J_{ij}$ specify the spin Hamiltonian (Eq. 1.1) and determine the energy spectrum for a given magnet. These days, it is almost routine to evaluate the spin exchanges of a magnet composed of transition-metal magnetic ions by using the energy-mapping analysis[6,20,21] based on density functional theory (DFT) calculations. The quantitative values of the spin exchanges are determined by mapping the energy spectrum of a magnet generated by the spin Hamiltonian onto that obtained for a set of broken-symmetry states of the magnet by spin polarized DFT calculations.

The spin exchanges of some magnets discussed in our survey have not been determined before. To gain better insight into these magnets we determined them by carrying out the energy-mapping analyses. For our DFT calculations, we employ the frozen core projector augmented plane wave (PAW) method[22] encoded in the Vienna ab Initio Simulation Packages (VASP),[23] and the PBE exchange-correlation functional.[24] To take into consideration of the electron correlation associated with transition-metal magnetic ions, we perform DFT+U calculations[25] with an effective on-site repulsion $U_{eff} = U - J = 3$ and 4 eV on the magnetic ions to ensure that all broken-symmetry states employed for a magnet are magnetic insulating. For a certain magnet, the effect of spin-orbit coupling (SOC) was tested by doing DFT+U+SOC calculations.[26] Unless stated otherwise, the values of the calculated spin exchanges (in K), which are included in a figure describing each magnet, are those determined by DFT+U or DFT+U+SOC calculations with $U_{eff}$. Other details of our calculations are reported in the supporting information (SI).

## 3. Magnets of AFM fragments
### 3.1. Spin clusters with even number of spin sites
#### 3.1.1. Orthogonal spin dimers in $SrCu_2(BO_3)_2$

In the quasi-2D tetragonal compound $SrCu_2(BO_3)_2$, the $CuBO_3$ layers alternate with Sr layers along the *c* axis. In each $CuBO_3$ layer, planar $Cu_2O_6$ dimers of two edge-sharing $CuO_4$ units are interconnected by $BO_3$ triangles (**Fig. 3.1a**). The two dominant spin exchanges of this layer are the intradimer exchange $J_1$ of the Cu-O-Cu type and the interdimer exchange $J_2$ of the Cu-O…O-Cu type.[6] In each $CuBO_3$ layer, the $(Cu^{2+})_2$ dimer ions have an orthogonal arrangement such that the $J_1$ and $J_2$ exchange paths are interconnected by a "head-to-tail" bridging pattern (**Fig. 3.1b**). The values of $J_1$ and $J_2$, estimated to be 84.7 and 53.4 K ($J_2/J_1 = 0.63$), respectively, by LDA+U calculations,[27] are very close to the experimental values.[28] This orthogonal arrangement of the dimers represents a realization of the so-called Shastry-Sutherland spin lattice.[29] The magnetic susceptibility of $SrCu_2(BO_3)_2$ has a sharp peak at about 20 K. Once the contribution of magnetic defects is removed, the susceptibility drops to zero indicating a nonzero spin gap Δ (**Fig. 3.2a**).[30] At low temperatures, early magnetization measurements on single-crystal samples of $SrCu_2(BO_3)_2$ identified 1/8-, 1/4- and 1/3-plateaus of $M_{sat}$ (**Fig. 3.2b**).[10] More plateaus (in particular, 2/5 and 1/2) were found in recent experiments with fields up to ~140 T.[11,12] In the plateau regions, the triplet dimers [namely, the $(Cu^{2+})_2$ dimers with broken $J_1$ bonds] crystallize into magnetic superstructures, so the transitions into the plateau regions are accompanied by substantial magnetoelastic effects. The latter are detected by changes in the magnetostrictive length and volume and a drastic decrease in the sound velocity.[13] To theoretically model the 1/2-magnetization plateau, the interlayer spin exchange needs to be taken into account.[13]

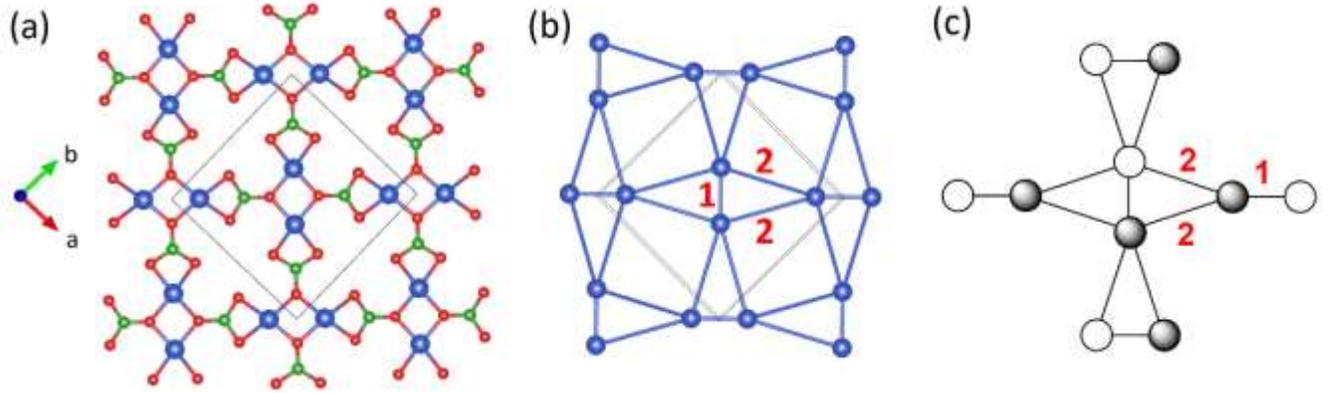

**Fig. 3.1.** (a) A projection view of one CuBO$_3$ layer of SrCu$_2$(BO$_3$)$_2$ along the $c$ direction, where the blue, green and red circles represent the Cu, B and O atoms, respectively. (b) The spin lattice of a CuBO$_3$ layer showing an orthogonal arrangement of (Cu$^{2+}$)$_2$ dimers, where the labels 1 and 2 refer to the spin exchanges J$_1$ and J$_2$, respectively. (c) Arrangement of the intradimer bonds J$_1$ in the Shastry-Sutherland spin lattice leading to a J$_2$ bond and a broken J$_2$ bond between adjacent dimers.

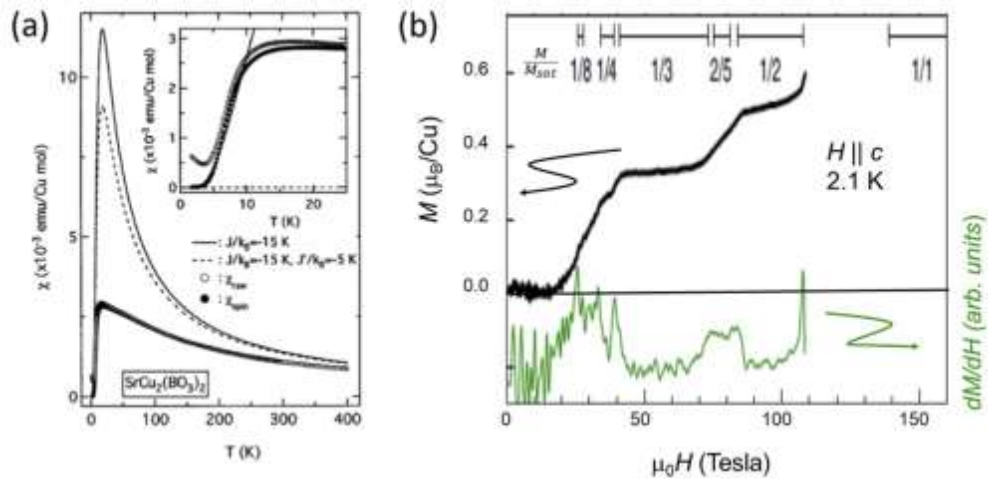

**Fig. 3.2.** (a) Temperature dependence of the magnetic susceptibility measured for SrCu$_2$(BO$_3$)$_2$ powder. The solid and dashed lines show the theoretical approximations. The inset enlarges low-temperature data.[30] (Reproduced with permission from reference 30.) (b) Field dependence of the magnetization measured for SrCu$_2$(BO$_3$)$_2$ single crystal. Adapted from Matsuda et al.[12] (Adapted with permission from reference 12.)

The essential aspect of the Shastry-Sutherland spin lattice is illustrated in **Fig. 3.1c**. In the ground state, each J$_1$ bond is surrounded by an equal number of unbroken and broken J$_2$ bonds. Thus, the contribution of the J$_2$ bonds to the total energy vanishes. The magnetization of SrCu$_2$B$_2$O$_6$ is zero between 0 and ~20 T (**Fig. 3.2b**), because it requires the breaking of J$_1$ bonds to increase magnetization and because all different arrangements of the J$_2$ and broken J$_2$ bonds are identical in energy as long as the J$_1$ bonds remain unbroken. The 1/n-magnetization plateau of SrCu$_2$B$_2$O$_6$ requires that one out of every n dimers have a broken J$_1$ bond, because the resulting spin configuration (↑↓)$_{n-1}$(↑↑) has two net spins out of every 2n spins hence leading to the 1/n-plateau.

The magnetic superstructure describing the 1/$n$-plateau is determined by how a (↑↑) dimer is arranged with respect to every ($n$-1) (↑↓) dimers. **Fig. 3.2b** reveals that the widths of the magnetization plateaus are not uniform, i.e., they decrease in the order, 1/3- > 1/2- > 2/5-, 1/4- > 1/8-plateau. This variation would be caused by their spin-lattice interactions and hence the associated energy lowering, because they will be different due to the difference in their magnetic superstructures. Several theoretical studies examined the magnetic textures and superstructures of the $CuBO_3$ layer predicting many more plateaus,[10] which were mostly detected in magnetization, magnetostriction, magnetocaloric effect, and nuclear magnetic resonance measurements.

### 3.1.2. Spin tetrahedra in spinel $CdCr_2O_4$

$CdCr_2O_4$ is a spinel-type compound based on $CrO_6$ octahedra containing $Cr^{3+}$ ($S = 3/2$) ions. It is convenient to describe the structure of this compound in terms of the $Cr_4O_{16}$ cluster (**Fig. 3.3a**), which is made up of four $CrO_6$ octahedra by sharing their edges to form a $Cr_4O_4$ distorted cube containing a $Cr_4$ tetrahedron. The spinel structure of $CdCr_2O_4$ is obtained by corner-sharing the $Cr_4O_{16}$ clusters, which is accompanied by the corner-sharing $Cr_4$ tetrahedra such that each $Cr_4$ tetrahedron shares a corner with four $Cr_4$ tetrahedra in a tetrahedral arrangement (**Fig. 3.3b**). Thus, the resulting arrangement of the $Cr^{3+}$ ions is a pyrochlore spin lattice, which is highly spin-frustrated.

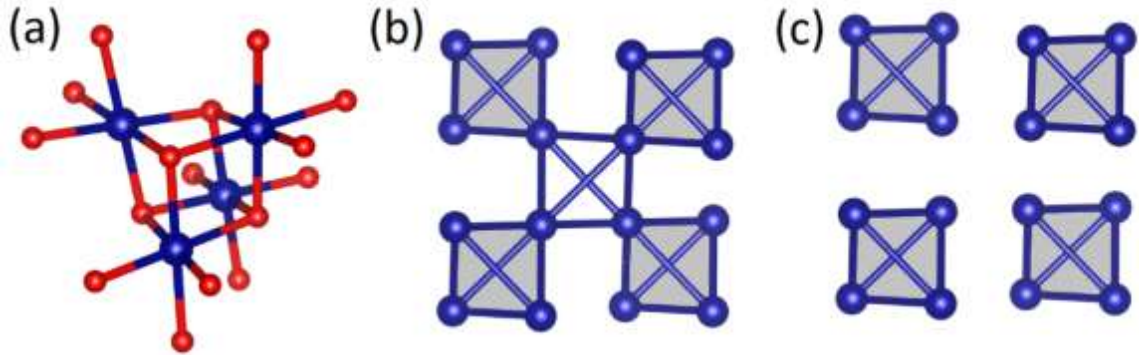

**Fig. 3.3.** (a) $Cr_4O_{16}$ cluster in $CdCr_2O_4$ made up of three $CrO_6$ octahedra by edge-sharing. (b) Pyrochlore spin lattice of $Cr^{3+}$ ions in which each $(Cr^{3+})_4$ tetrahedron is corner-shared with four $(Cr^{3+})_4$ tetrahedra in a tetrahedral manner. (c) Generating isolated $(Cr^{3+})_4$ tetrahedra that fill the pyrochlore lattice without overlapping between them. In (c), the shaded $(Cr^{3+})_4$ tetrahedra represent those that become isolated when the inter-tetrahedra interactions are neglected.

$CdCr_2O_4$ undergoes a long-range AFM ordering and a tetragonal lattice distortion at $T_N$ = 7.8 K.[31] By neutron diffraction measurements, the ground state spin configuration was found to be a helical spin structure.[32] The magnetization curve determined at low temperatures reveals a gradual increase with field as the field increases from zero, which is followed by a sharp transition into a 1/2-plateau at $M_{sat}/2 = 1.5$ $\mu_B/Cr^{3+}$ (**Fig. 3.4a**). To account for the nature of the observed $M$ vs. $H$ curve, we treat the pyrochlore arrangement of $Cr^{3+}$ ions as composed of isolated $(Cr^{3+})_4$ tetrahedra by neglecting the interactions between them, as shown in **Fig. 3.3c**. In the zero-magnetization state, each tetrahedron has the (2↑2↓) configuration (**Fig. 3.4b**), which has four $J_1$ and two broken $J_1$ bonds. When a tetrahedron has the (3↑1↓) configuration (**Fig. 3.4b**), which has three $J_1$ and three broken $J_1$ bonds, the magnetization increases. The 1/2-magnetization, $M = M_{sat}/2$, is reached when all isolated tetrahedra have the (3↑1↓) configuration. The energy change required for each $(Cr^{3+})_4$ tetrahedron to undergo the $(2\uparrow 2\downarrow)$ to $(3\uparrow 1\downarrow)$ transition is to break one $J_1$ bond per tetrahedron. The sharp jump in the magnetization of $CdCr_2O_4$ takes place at about 29 T. At 1.8 K, $M(H)$ increases linearly with $H$ until ~28 T, where $M \approx 0.75$ $\mu_B$, namely, when half the $(Cr^{3+})_4$ tetrahedra have the $(3\uparrow 1\downarrow)$ configuration. On increasing the field beyond this point, the $J_1$ bond breaking occurs simultaneously everywhere such that all $(Cr^{3+})_4$ tetrahedra have the $(3\uparrow 1\downarrow)$ configuration. As the number of broken $J_1$ bonds increases, the $J_1$ breaking at one tetrahedron becomes correlated with those at other places, due to the inter-cluster tetrahedra, which were neglected in our discussion. Above 28 T, $CdCr_2O_4$ shows a flat magnetization. For the magnetization to increase beyond $M_{sat}/2$, a tetrahedron must undergo the configuration change from $(3\uparrow 1\downarrow)$ to $(4\uparrow 0\downarrow)$. The $(4\uparrow 0\downarrow)$ configuration has no $J_1$ bond (**Fig. 3.4**), while $(3\uparrow 1\downarrow)$ has three $J_1$ bonds, so the $(3\uparrow 1\downarrow)$ to $(4\uparrow 0\downarrow)$ transition requires to break three $J_1$ bonds per tetrahedron. That is, this transition requires more energy than does the $(2\uparrow 2\downarrow)$ to $(3\uparrow 1\downarrow)$ transition (i.e., one $J_1$ bond per tetrahedron). This explains why the magnetization of $CdCr_2O_4$ is flat above 28 T, and the 1/2-magnetization plateau belongs to the intra-fragment mechanism.

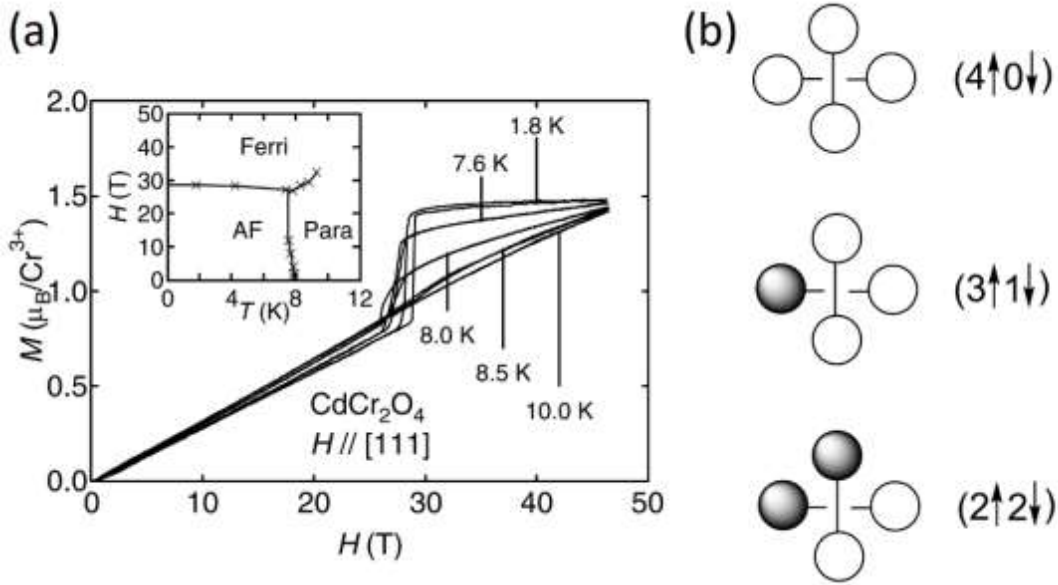

**Fig. 3.4.** (a) Field dependence of the magnetization $M(H)$ observed for a $CdCr_2O_4$ single crystal at various temperatures.[31] (Reproduced with permission from reference 31.) (b) Three spin configurations of a $(Cr^{3+})_4$ tetramer, where the tetramer is shown in terms of two dimers.

A spinel magnet with the ideal pyrochlore spin lattice would not undergo a 3D magnetic long-range order due to the severe spin frustration. However, most spinel magnets undergo a 3D magnetic long-range order because the degeneracy of their ground state is lifted by a structural distortion. In a sense, this distortion can be considered as a 3D spin-Peierls transition. A theoretical analysis of $CdCr_2O_4$[7] showed that its 1/2-magnetization plateau is stabilized by the magnetoeleastic coupling and this plateau is robust. The pyrochlore spin lattice of the spinel magnet $LiGaCr_4O_8$ differs from that found for $CdCr_2O_4$ in that it consists of small and large tetrahedral $(Cr^{3+})_4$ clusters, which alternate by corner-sharing.[8] The magnetization and magnetostriction studies of $LiGaCr_4O_8$ under magnetic fields of up to 600 T[9] show that it exhibits a two-step coupled magnetic and structural phase transition between 150 T and 200 T, followed by a robust 1/2-magnetization plateau up to ~420 T, and that the intermediate-field phase is stabilized by the strong spin-lattice coupling. This phase can be considered as a tetrahedron-based superstructure with a 3D periodic array of (3↑1↓) and canted (2↑2↓) configurations.

### 3.1.3. Spin hexamers in pyroxene CoGeO$_3$ and anisotropic magnetization plateau

CoGeO$_3$ consists of two nonequivalent Co atoms, Co1 and Co2, forming Co1O$_6$ and Co2O$_6$ octahedra. By sharing their edges these octahedra form zigzag ribbon chains parallel to the *bc*-plane, as shown in **Fig. 3.5a**. The 3D structure of CoGeO$_3$ is obtained from these zigzag ribbon chains when their oxygen atoms are shared with GeO$_4$ tetrahedra.[33] The magnetic properties of CoGeO$_3$ present a novel feature.[34] The 1/3-magnetization plateau of CoGeO$_3$ is uniaxially anisotropic, that is, CoGeO$_3$ exhibits a pronounced 1/3-plateau when measured with field applied along the *c*-direction but does not show any magnetization plateau when measured with field perpendicular to the *c*-direction (**Fig. 3.6a**). This observation provides an experimental support for our supposition that field-induced partitioning of a spin lattice into ferrimagnetic fragments is essential for magnetization plateaus. To probe the cause for the anisotropic character of the 1/3-magnetization plateau in CoGeO$_3$ mentioned above, we evaluate the four spin exchanges $J_1 – J_4$ defined in **Fig. 3.5b** by DFT+U calculations. The intra-chain exchanges $J_1 – J_3$ are of the Co-O-Co type exchange, while the inter-chain exchange $J_4$ is of the Co-O…O-Co type. Results of these calculations are summarized in **Fig. 3.5c** (see Section S1 of the SI).

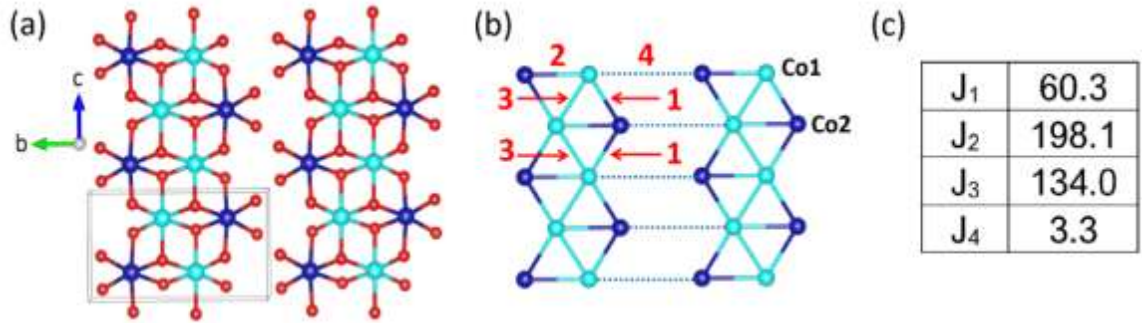

**Fig. 3.5.** (a) Zigzag ribbon chains of CoGeO$_3$ parallel to the *bc*-plane. (b) Arrangement of Co1$^{2+}$ and Co2$^{2+}$ ions in the ribbon chains with the four spin exchange paths $J_1 – J_4$ represented by the labels 1 – 4, respectively. (c) Values of $J_1 – J_4$ (in K) determined by DFT+U calculations.

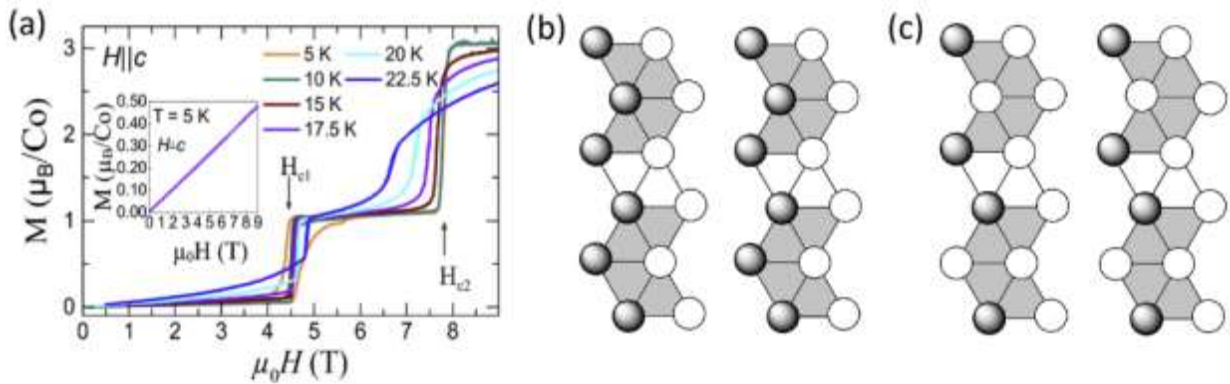

**Fig. 3.6.** (a) Magnetization $M(H)$ of CoGeO$_3$ for $H\|c$. (Data obtained by ramping up to 9 T and down to 0 T after virgin zero-field cooling from 200 K down.) Pronounced 1/3 magnetization plateaus can be seen. The magnetization for $H\perp c$ obtained at 5 K (inset) is linear in $H$ and unsaturated up to 9 T.[34] (b) Spin arrangement of the AFM ground state, where the intrachain spin arrangement is dictated by two strong AFM spin exchanges J$_2$ and J$_4$. (c) Spin arrangement of the ferrimagnetic state representing the 1/3-magnetization plateau. In (b) and (c), the "six-spin" units are shaded.

The spin exchanges J$_1$ – J$_4$ are all AFM, and the interchain exchange J$_4$ is considerably weaker than the intrachain exchanges J$_1$ – J$_3$. Since J$_1$ is considerably weaker than J$_2$ and J$_3$, the magnetic ground state for the layer of the double chains has the AFM spin arrangement as depicted in **Fig. 3.6b**, where each double chain is made up of J$_2$ and J$_3$ magnetic bonds as well as broken J$_1$ bonds. The smallest fragment that can generate a ferrimagnetic fragment of 1/3-magnetization is the hexamer composed of three J$_2$ bonds with (3↑3↓) spin configuration (indicated by shading in **Fig. 3.6b**). The ferrimagnetic fragment of (4↑2↓) configuration is generated when one of the three J$_2$ bonds is broken, ultimately leading to the ferrimagnetic state (**Fig. 3.6c**) when every hexamer has the (4↑2↓) spin configuration. The conversion from a (3↑3↓) to a (4↑2↓) is facilitated because the breaking a J$_2$ bond is accompanied by the formation of two J$_1$ bonds.

CoGeO$_3$ exhibits uniaxial (i.e., Ising) magnetism with the spin moments oriented along the $c$-direction.[33] According to the selection rules governing the preferred spin orientations of magnetic ions,[35] either Co1$^{2+}$ or Co2$^{2+}$ or both ions of CoGeO$_3$ prefer to have their spins oriented along the $c$-direction. Consider an ideal axially-compressed CoO$_6$ octahedron containing Co$^{2+}$ (d$^7$, S = 3/2) ion with the short Co-O bonds oriented along the z-axis, as depicted in **Fig. 3.7a**. The t$_{2g}$-state of such an octahedron is split into the degenerate (xz, yz) state lying above the xy state, assuming that the axially-compressed octahedron has an ideal shape with four-fold rotational symmetry. With two d-electrons to occupy the down-spin d-states, the split t$_{2g}$ states become occupied as depicted in **Fig. 3.7a**. Thus, between the highest-occupied and the lowest-unoccupied d-states, the minimum difference in their magnetic quantum numbers, $|\Delta L_z|$, is zero so that the preferred spin orientation is parallel to the z-axis.[35] Of the Co1O$_6$ and Co2O$_6$ octahedra of CoGeO$_3$, only the Co2O$_6$ octahedra have a structure close to an axial-compression [namely, Co2-O$_{ax}$ = 1.994 (×2), and Co2-O$_{eq}$ = 2.118 (×2), 2.278 (×2) Å]. Since the Co2O$_6$ octahedra have no four-fold rotational symmetry, their xz and yz states are not degenerate, but they still lie above the xy state due to the strong axial compression. The latter guarantees $|\Delta L_z|$ = 0 hence predicting that the spins of the Co2O$_6$ octahedra are oriented along the short Co2-O bonds, i.e., along the $c$-direction. The Co1$^{2+}$ ions adopt the spin orientation of the Co2$^{2+}$ ions to maximize their spin exchanges (J$_2$ and J$_4$) with the Co2$^{2+}$ ions. This explains why CoGeO$_3$ exhibits uniaxial magnetism with spin moment along the $c$-direction.

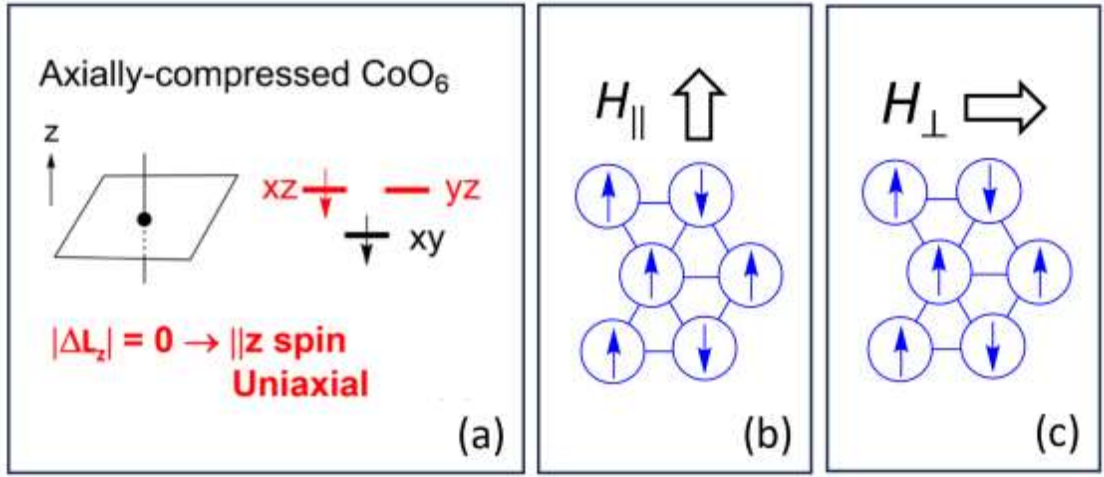

**Fig. 3.7.** (a) Split down-spin $t_{2g}$ states of an ideally axially-compressed $CoO_6$ octahedron. With one electron to fill the degenerate (xz, yz) state, so the smallest difference in the $L_z$ values of the highest-occupied and the lowest-unoccupied d-states of such an octahedron is zero, namely, $|\Delta L_z| = 0$. (b, c) Magnetic field $H\|c$ and $H\perp c$ acting on the (4↑2↓) ferrimagnetic fragment with spins oriented along the c-direction.

Based on the uniaxial magnetism of $CoGeO_3$, one can understand why it shows a 1/3-magnetization plateau only when the field is along the *c*-direction. As already discussed in Section 2, the Zeeman energy $E_Z$ between the spin moment $\vec{\mu}_s = -g\mu_B \vec{S}$ and magnetic field $\mu_0 \vec{H}$ is given by $E_Z = g\mu_0\mu_B \vec{H} \cdot \vec{S}$ (Eq. 2.1). Therefore, when the spin moment and magnetic field are parallel to each other (**Fig. 3.7b**), $E_Z > 0$ so that the formation of a ferrimagnetic fragment is energetically favored. However, $E_Z = 0$ if the magnetic field is perpendicular to the spin moment. In such a case, the energy needed to break magnetic bonds and hence form ferrimagnetic fragments is not available. This explains why the 1/3-magnetization plateau of $CoGeO_3$ has a uniaxial character. In addition, this finding is in support of our suggestion that, for a magnet to exhibit magnetic plateaus, its spin lattice should undergo a field-induced partitioning into ferrimagnetic clusters. It should be noted that the magnetization curves of $CoGeO_3$ (**Fig. 3.6a**) have a "step-like" feature, because the uniaxial magnetism favors a spin flip mechanism for magnetization.

### 3.2. Bose-Einstein condensates

In a certain magnet composed of discrete units possessing two magnetic ions, such "dimers" have an S = 0 ground state, and the interactions between adjacent dimers are weak so that the first excited state of each dimer, which has S > 0, lies close to the S = 0 ground state. In such a case, the magnetic states of the magnet are well approximated by those of its dimer. The $|S, S_z\rangle = |S, -S\rangle$ substate of the excited state is lowered in energy under magnetic field $\mu_0 H$. When $\mu_0 H$ exceeds a certain value, $\mu_0 H_c$, the $|S, -S\rangle$ substate becomes lower in energy than the ground state $|0, 0\rangle$, the magnetic ground state of each dimer becomes an S > 0 state. Magnets showing such a behavior, known as Bose-Einstein condensates, have been reviewed by Zapf et al.[36] It should be noted that S = 0 dimers can be discrete molecular units such as $Cu_2Cl_6^{2-}$ anions containing two magnetic ions or dimers composed of two monomers such as $(MnO_4^{3-})_2$ (see below). In both cases, the inter-dimer spin exchange is weaker than the intra-dimer exchanges. In this section, we discuss the magnetization phenomena observed in two Bose-Einstein condensates.

### 3.2.1. 0- and 1/2-plateaus of $Ba_3Mn_2O_8$

The trigonal compound $Ba_3Mn_2O_8$[37] is composed of $MnO_4$ tetrahedra containing $Mn^{5+}$ (S = 1) ions. Every two tetrahedra combine to form a dimer unit $(MnO_4)_2$ such that one Mn-O bond of each $MnO_4$ is parallel to the *c*-axis (hereafter the Mn-O$_\parallel$ bond). The two Mn-O$_\parallel$ bonds of each dimer are pointed in opposite directions (**Fig. 3.8a**), and these dimers form trigonal layers. Adjacent layers are shifted from each other such that each dimer of one layer is pointed to the center of three dimers of the two adjacent layers (**Fig. 3.8a**). Consequently, every $(Mn^{5+})_2$ dimer ion of one layer is surrounded by six $(Mn^{5+})_2$ dimer ions (**Fig. 3.8b**). The spin exchanges of $Ba_3Mn_2O_8$ are dominated by the intradimer exchange $J_0$ and the interdimer exchange $J_1$ (**Fig. 3.8c**). ($J_0$ = 15.2 K and $J_1$ = 1.4 K according to our DFT+U calculations, see Section S2 of the SI).

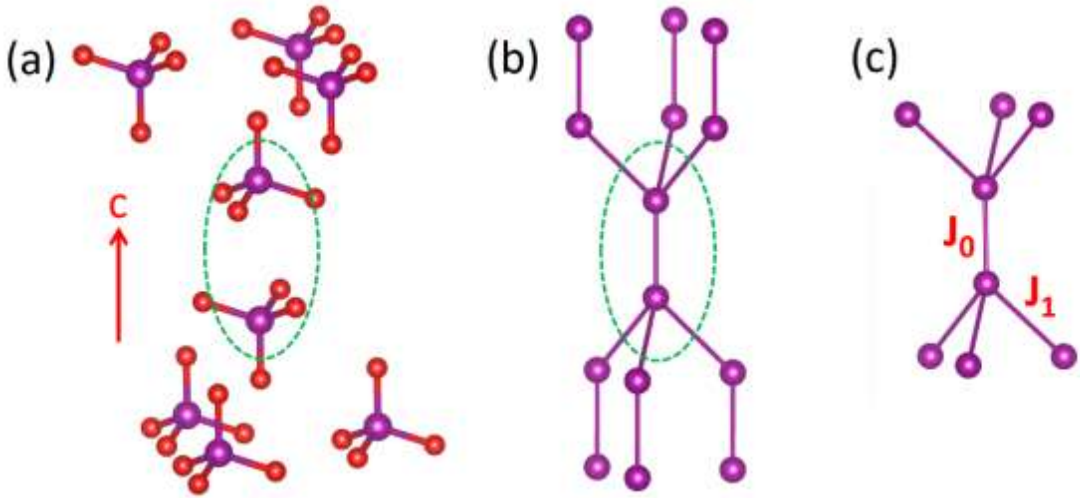

**Fig. 3.8.** (a) Arrangement of the $MnO_4$ tetrahedra in $Ba_3Mn_2O_8$, around one dimer unit $(MnO_4)_2$ indicated by a green ellipse. (b) Simplified view of the arrangement of six $(Mn^{5+})_2$ dimer ions surrounding one $(Mn^{5+})_2$ dimer ion indicated by a green ellipse. (c) Definitions of the intradimer exchange $J_0$ and the interdimer exchange $J_1$.

The allowed spin states of each $(Mn^{5+})_2$ dimer ion are singlet, triplet and quintuplet since $Mn^{5+}$ is an S = 1 ion, apart from usually small zero-field splitting of these multiplets. The temperature-dependence of the magnetic susceptibility $\chi$ measured for $Ba_3Mn_2O_8$ is shown in **Fig. 3.9a**,[38] which evidences that $Ba_3Mn_2O_8$ is in a singlet ground state with spin gap $\Delta$ = 11.2 K, in which all $(Mn^{5+})_2$ dimer ions are in the singlet state. At low temperatures, the magnetization plateaus are observed at $M$ = 0 zero and $M_{sat}/2$ = 2 $\mu_B$ per formula unit (**Fig. 3.9b**).[38] We now examine how these plateaus are related to the breaking of the $J_1$ and $J_0$ bonds. The most stable and least stable arrangements of $J_1$ bonds around a $J_0$ bond are shown in **Fig. 3.10a**, and those around a broken $J_0$ bond in **Fig. 3.10b**. There are many other arrangements of the $J_1$ and broken $J_1$ bonds whose stabilities lie in between these two extremes. In general, the arrangement becomes more stable if it has more $J_1$ bonds but becomes less stable if it has more broken $J_1$ bonds. As the field increases from 0 to $H_{c1}$, each $J_1$ bond begins to break without breaking the $J_0$ bonds. Thus, $M$ = 0 between 0 and $H_{c1}$. As the field increases from $H_{c1}$, the $J_0$ bond breaking proceeds, hence increasing

*M*. Two dimers with one $J_0$ and one broken $J_0$ bond have the (3↑1↓) configuration. When half the $J_0$ bonds are broken, the $M = M_{sat}/2$ point at $H_{c2}$ is reached. The 1/2-plateau between $H_{c2}$ and $H_{c3}$ means that there are more $J_1$ bonds than broken $J_1$ bonds at $H_{c2}$, while the opposite is the case at $H_{c3}$. That is, magnetic energy is absorbed without increasing magnetization from $M_{sat}/2$. Since $J_1$ is a weak magnetic bond, the width of the 1/2-magnetization plateau is narrow. When the field is stronger than $H_{c3}$, more $J_0$ bonds begin to break, increasing the magnetization.

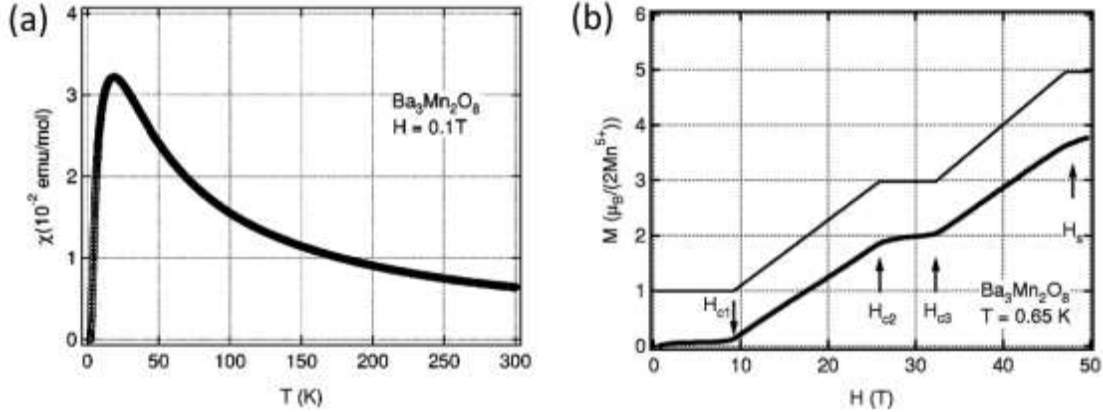

**Fig. 3.9.** (a) Temperature dependence of magnetic susceptibility in $Ba_3Mn_2O_8$ powder at 0.1 T. (b) Field dependence of magnetization at 1.4 K.[38] (Reproduced with permission from reference 38.)

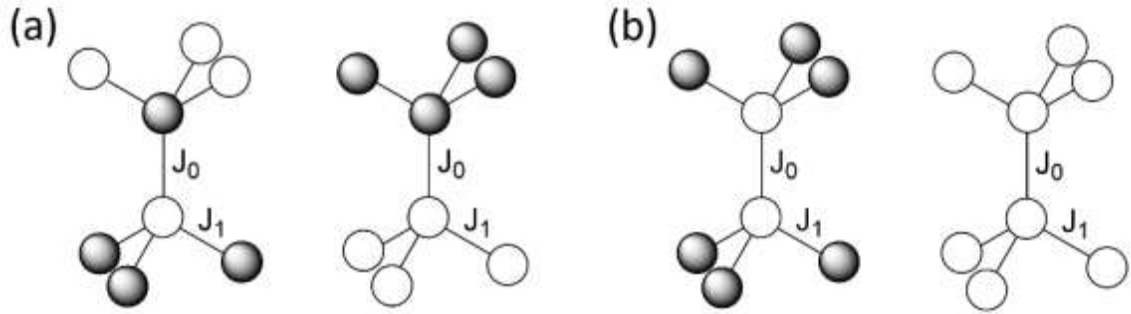

**Fig. 3.10.** (a) Two arrangements of $J_1$ bonds around a $J_0$ bond in $Ba_3Mn_2O_8$. (b) Two arrangements of $J_1$ bonds around a broken $J_0$ bond in $Ba_3Mn_2O_8$.

### 3.2.2. Gapped and gapless ground states of ACuCl₃ (A = K, Tl, NH₄)
**A. Singlet to triplet excitations under magnetic field**

The molecular magnets $ACuCl_3$ (A = K, Tl, NH₄)[39-41] consist of planar $Cu_2Cl_6^{2-}$ anions, which are made up of two $CuCl_4$ square planes containing $Cu^{2+}$ ($S = 1/2$) ions by edge-sharing (**Fig. 3.11a**). Thus, each $Cu_2Cl_6^{2-}$ anion contains a spin dimer $(Cu^{2+})_2$. The ground spin state for such a dimer can be either singlet ($\Delta_{ST} > 0$, 'singlet dimer', **Fig. 3.11b**) or triplet ($\Delta_{ST} < 0$, 'triplet dimer' **Fig. 3.11c**). When such a spin dimer is exposed to a magnetic field $\mu_0H$, the triplet state $|S, S_z\rangle$ ($S = 1$, $S_z = -1, 0, 1$) is split while the singlet state $|S, S_z\rangle$ ($S = 0$, $S_z = 0$) remains unaffected. For a singlet spin dimer ($\Delta_{ST} > 0$) under magnetic field, the triplet state becomes more stable than the singlet state if the field is greater than a critical value $\mu_0H_c$ (**Fig. 3.11d**), so that every spin dimer occupies the $S_z = -1$ state, and the system undergoes a Bose-Einstein condensation. Likewise, for a triplet spin dimer ($\Delta_{ST} < 0$) under magnetic field, the singlet state becomes more stable than the triplet state if the field is higher than a critical value (**Fig. 3.11e**). For a singlet dimer below $\mu_0 H_c$, there are three possible spin-flip transitions from the singlet to the triplet under magnetic field, namely, $|0, 0\rangle \rightarrow |1, S_z\rangle$ ($S_z$ = -1, 0, 1), and the energy difference between the two states can be accessed, e.g., by inelastic neutron spectroscopy techniques. The energy difference immediately provides the magnitude of the spin exchange J in the spin dimer. For such transitions to be observed by inelastic neutron scattering measurements, the singlet state $|0, 0\rangle$ should be thermally populated and should be more populated than the triplet state(s) into which the transition occurs. This is the case for a singlet dimer because, for field lower than $\mu_0 H_c$, the $|0, 0\rangle$ state is the lowest-lying in energy than any of the three triplet branches (**Fig. 3.11d**). For a triplet dimer ($\Delta_{ST}$ < 0), the $|0, 0\rangle$ state can be thermally more populated than one branch of the triplet, i.e., the $|1, +1\rangle$ state, only when the field is substantially greater than $\mu_0 H_c$ (**Fig. 3.11e**). Under this condition, the $|0, 0\rangle \rightarrow |1, +1\rangle$ transition can take place in a triplet dimer.

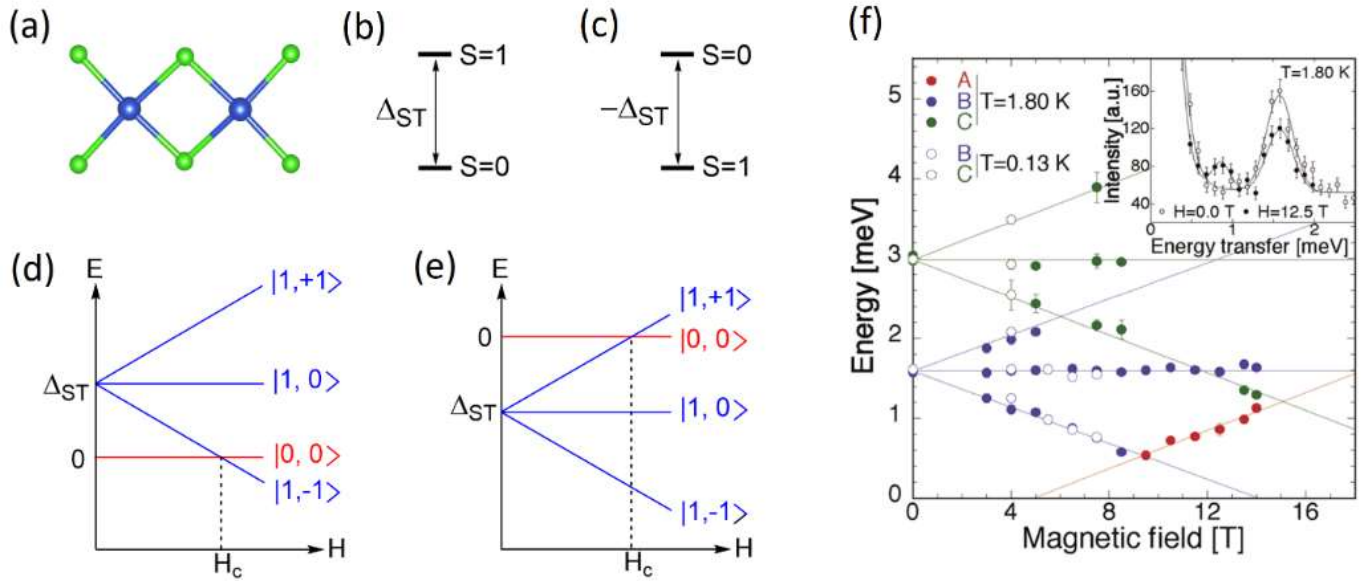

**Fig. 3.11**. (a) Planar $Cu_2Cl_6^{2-}$ anion of ACuCl$_3$. (b) Singlet dimer with $\Delta E$ > 0. (c) Triplet dimer with $\Delta E$ < 0. (d) Splitting of a singlet spin dimer as a function of $\mu_0 H_c$. (e) Splitting of a triplet spin dimer as a function of $\mu_0 H_c$. (f) Singlet to triplet excitation energies of NH$_4$CuCl$_3$ measured by neutron scattering experiments.[42] (Reproduced with permission from reference 42.)

NH$_4$CuCl$_3$ is known to consist of three different spin dimers, termed A, B and C, in the 1:2:1 ratio. Results of inelastic neutron scattering experiments carried out for NH$_4$CuCl$_3$ at 0.13 and 1.80 K are summarized in **Fig. 3.11f**,[42] which shows the three branches of the triplet state for the dimers B and C at both temperatures (1.80 and 0.13K). This finding proves that B and C are singlet dimers with $\Delta E_{ST}$ values of ~1.6 and ~3.0 meV, respectively. For dimer A, however, only one branch has been found, which becomes visible only above ~8.5 T and only at 1.80 K, but apparently was not observed for the measurement at 0.13 K. A singlet-triplet splitting of 0.5 meV ($\approx$ 5.8 K) with the spin triplet lower than the singlet ($\Delta_{ST}$ < 0) derived by extrapolating the observed $|1,+1\rangle$ branch to zero field is consistent with this finding. The small energy difference between the triplet and singlet implies that any excitations with energies below ~0.5 meV, could have been masked under the elastic peak or accidentally coincide with excitations for dimers B and C (see the inset in Fig. 4 of ref. 30). These results suggest that the $(Cu^{2+})_2$ dimer A, though weak, is coupled by ferromagnetic spin exchange, i.e., it is a triplet dimer. Further temperature dependent inelastic neutron scattering investigations are necessary to verify this conclusion. We found that the magnetization curve observed for NH$_4$CuCl$_3$ at 0.5 K using the $H\|a$ field is reasonably well reproduced by assuming that dimers A, B and C are all singlet dimers with the intradimer spin exchanges of 2.7(1), 11.3 (1) and 19.3 (2) K, respectively (see **Fig. S1** in Section S3 of the SI). This suggests that dimer A is a singlet dimer but is inconsistent with the inelastic neutron scattering study described above. With dimer A as a triplet dimer, it is straightforward to understand the gapless excitation in the magnetization measurements of NH$_4$CuCl$_3$ (see below) because its triplet dimers A have a nonzero spin moment even in the absence of field. Though isostructural with NH$_4$CuCl$_3$ as far as the atom positions of the heavier atoms are concerned, KCuCl$_3$ and TlCuCl$_3$ consist of only one kind of singlet spin dimers. The excitation energy gaps measured for KCuCl$_3$ and TlCuCl$_3$ are 2.6 and 0.7 meV, respectively.[43,44]

**B. Different magnetization behaviors of ACuCl$_3$ (A = K, Tl, NH$_4$)**

Magnetization processes of KCuCl$_3$, TlCuCl$_3$ and NH$_4$CuCl$_3$ have been investigated up to 39 T at low temperatures (**Fig. 3.12**). Both KCuCl$_3$ and TlCuCl$_3$ exhibit a 0-magnetization plateau, and this plateau has a much wider width for KCuCl$_3$ (**Fig. 3.12a** and **3.12b**). The transition from a singlet ground state to a magnetic excited state occurs when the field is greater than ~6 and ~20 T for TlCuCl$_3$ and KCuCl$_3$, respectively. These critical fields $\mu_0H_c$ are consistent with the excitation energy gaps of 0.7 and 2.6 meV observed for TlCuCl$_3$ and KCuCl$_3$, respectively.[43,44] Except for the 0-magnetization plateau, the magnetization of TlCuCl$_3$ and KCuCl$_3$ increases continuously with field showing no more plateau. The magnetization of NH$_4$CuCl$_3$ shows a very different behavior. As the field increases from zero, the magnetization reveals gapless excitations toward a 1/4-plateau, which is followed by a 3/4-plateau before reaching full saturation $M_{sat}$ (**Fig. 3.12c**).[45]

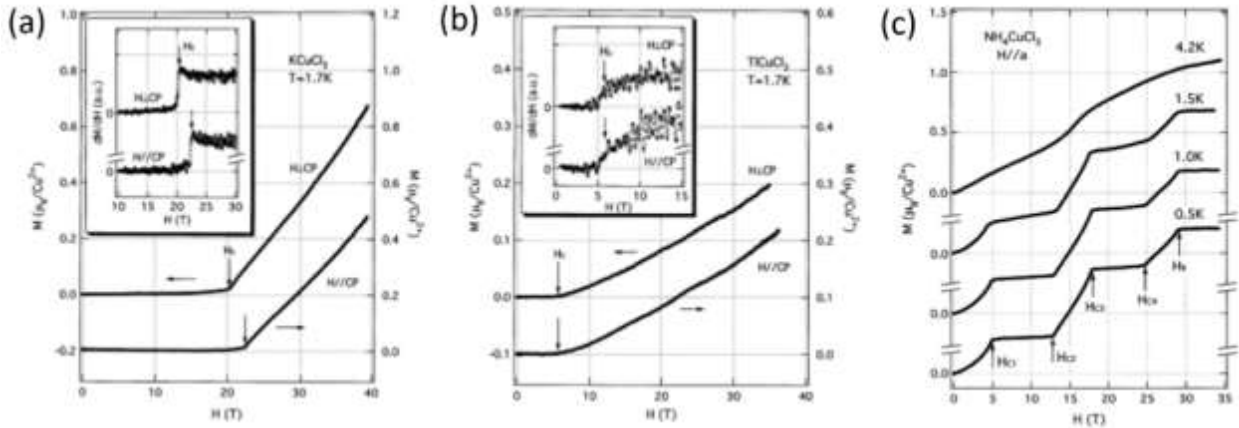

**Fig. 3.12.** Field dependence of the magnetization $M$ in (a) KCuCl$_3$, (b) TlCuCl$_3$ and (c) NH$_4$CuCl$_3$.[45] (Reproduced with permission from reference 45.)

The magnetization behaviors of KCuCl$_3$ and TlCuCl$_3$ can be readily understood by considering how the magnetic bonds of their spin dimers are broken under magnetic field. When the field is zero, both KCuCl$_3$ and TlCuCl$_3$ have zero moment because each spin dimer has the singlet configuration, $(\uparrow\downarrow - \downarrow\uparrow)/\sqrt{2}$. The magnetization of KCuCl$_3$ and TlCuCl$_3$ can increase from zero only if dimers start to break their magnetic bonds, one at a time, to assume the triplet configuration (↑↑). The critical field $\mu_0H_c$ needed to break each dimer magnetic bond is much higher for KCuCl$_3$ than for TlCuCl$_3$ (~20 vs. ~6 T) because the singlet-triplet energy difference $\Delta_{ST}$ is much larger for KCuCl$_3$ than for TlCuCl$_3$ (2.0 vs. 0.7 meV). This explains why the 0-magnetization plateau is much wider for KCuCl$_3$. With increasing the field, the singlet to the triplet magnetic bond breaking will continue until all magnetic bonds are broken, namely, until the saturation magnetization is reached.

The magnetization behaviors of NH$_4$CuCl$_3$, though apparently more complex, can be similarly explained by noting that spin dimers A, B and C constitute 25 %, 50 % and 25 % of all the dimers, and that dimers A are triplet dimers while dimers B and C are singlet dimers, and the singlet-triplet energy gap is greater for C than for B (3.0 vs.1.6 meV).[42] When $\mu_0H = 0$, triplet dimers A should exist half in the (↑↑) configuration and half in the (↓↓) configuration. As $\mu_0H$ increases from 0, dimers A with (↓↓) configuration will switch their configuration to (↑↑), successively, until all dimers A attain the (↑↑) configuration at $\mu_0H_{c1}$, where $M = M_{sat}/4$ because all A dimers have the (↑↑) configuration while the dimers B and C are in the (↑↓) configuration and because 25 % of the dimers are dimers of type A. The 1/4-plateau continues until $\mu_0H_{c2}$. When the field is greater than $\mu_0H_{c2}$, the magnetic bonds of dimers B start to break, one at a time, until all dimers B break their bonds at $\mu_0H_{c3}$, where $M = 3M_{sat}/4$ because all dimers A and B have the (↑↑) configuration while the dimers C have the (↑↓) configuration and because dimers A and B with (↑↑) configuration represent 75 % of the total dimers. The 3/4-plateau continues until $\mu_0H_{c4}$. When the field is greater than $\mu_0H_{c4}$, the magnetic bonds of dimers C start to break, successively, until all dimers C break their bonds at $\mu_0H_{c5}$, and the saturation magnetization is finally reached.

## C. Crystal structures of ACuCl$_3$ (A = K, Tl, NH$_4$)

As discussed above, the spin dimers of KCuCl$_3$ are very different from those of TlCuCl$_3$ in the singlet-to-triplet excitation energies. However, the Cu$_2$Cl$_6^{2-}$ ions of KCuCl$_3$ are very similar in crystal structure to those of TlCuCl$_3$.[35,36] Neutron scattering measurements reveal the existence of three different spin dimers in NH$_4$CuCl$_3$,[42] but the neutron diffraction studies to determine the crystal structure[46] carried out for ND$_4$CuCl$_3$ at various temperatures show that there is only one kind of Cu$_2$Cl$_6^{2-}$ anions in ND$_4$CuCl$_3$. These apparently puzzling observations imply that the spin dimers used in interpreting the experimental magnetic data are the effective spin dimers which are affected by the interactions between dimers and by those with the cations A$^+$ (A = K, Tl, NH$_4$). Therefore, it is necessary to examine the crystal structures of ACuCl$_3$ (A = K, Tl, NH$_4$) in more detail with focus on why the magnetic behavior of NH$_4$CuCl$_3$ differs from those of KCuCl$_3$ and TlCuCl$_3$.

In ACuCl$_3$ the Cu$_2$Cl$_6^{2-}$ anions form stacks along the *a*-direction (**Fig. 3.13a**). The spin exchanges describing the interactions within each stack are the intradimer exchange J$_1$ and the two interdimer exchanges J$_a$ and J$'_a$. An important interdimer exchange between adjacent stacks of Cu$_2$Cl$_6^{2-}$ anions is J$_2$ (**Fig. 3.13b**). The spin exchanges J$_1$ and J$_2$ are contained in a layer of Cu$_2$Cl$_6^{2-}$ anions and A$^+$ cations, which is parallel to the *ad*-plane, where the repeat vector **d** is defined as **d** = **a** + **c**/2 (**Fig. 3.13b**). In this layer each Cu$_2$Cl$_6^{2-}$ anion is surrounded by six A$^+$ cations, and the adjacent J$_1$-J$_2$-J$_1$-J$_2$ alternating chains interact by the interdimer exchanges J$_3$ and J$_4$.

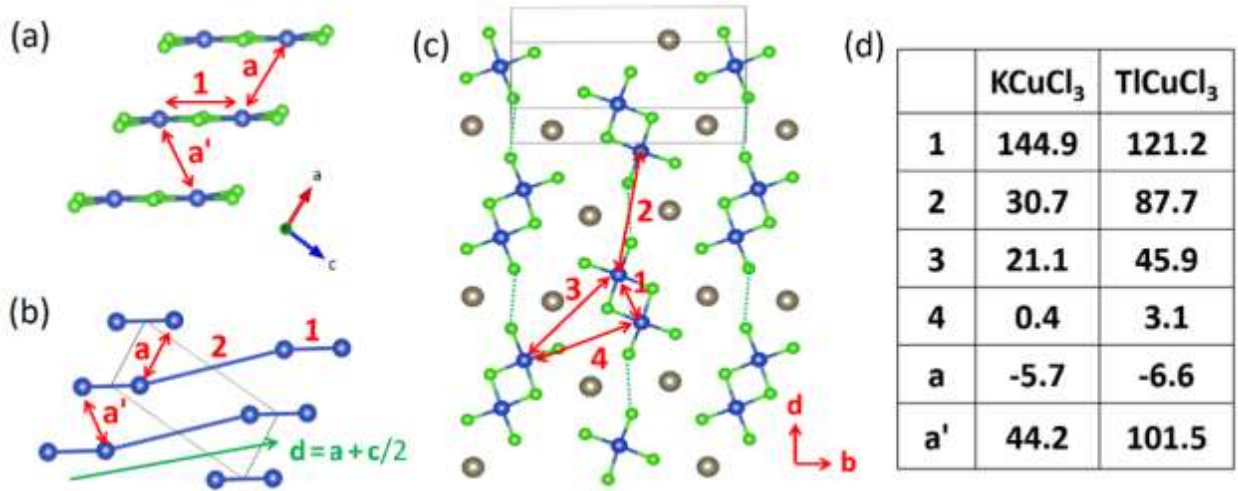

**Fig. 3.13.** (a) A stack of $Cu_2Cl_6^{2-}$ anions along the *a*-direction. (b) Two adjacent stackes of $Cu_2Cl_6^{2-}$ anions viewed along the *b*-direction. (c) Arrangement of the $Cu_2Cl_6^{2-}$ anions and the $A^+$ cations in a layer parallel to the *ad*-plane. (d) Spin exchanges calculated for $KCuCl_3$ and $TlCuCl_3$ by DFT+U calculations (see text).

The crystal structure of $NH_4CuCl_3$ is slightly more complex than those of $KCuCl_3$ and $TlCuCl_3$ due to the orientation of each $NH_4^+$ cation. The crystal structures of $ND_4CuCl_3$ including the D atom positions were determined by neutron diffraction at various temperature.[46] In these structure determinations, the presence of three different dimers A, B and C were not taken into consideration (see below for further discussion).

**D. Interdimer exchanges of $KCuCl_3$ and $TlCuCl_3$**

The values (in K) of the spin exchanges defined in **Fig. 3.13a-c**, evaluated by using the energy-mapping analysis based on DFT+U calculations (see Sections S4 and S5 of the SI), are presented in **Fig. 3.13d**. The intradimer exchange $J_1$ is stronger than the interdimer exchanges in both $KCuCl_3$ and $TlCuCl_3$, and the interdimer spin exchanges are substantially stronger for $TlCuCl_3$ than for $KCuCl_3$. These findings are consistent with the results of the neutron scattering study of Matsumoto et al.[43] (The intradimer exchange $J_1$ is slightly smaller for $KCuCl_3$ in their study, while the opposite is the case in our calculations.) Thus, the excitation energy is substantially smaller for $TlCuCl_3$ than for $KCuCl_3$ essentially because the interdimer spin exchanges are substantially stronger for $TlCuCl_3$ as found previously.[43]

To find why the interdimer exchanges are stronger for $TlCuCl_3$ than for $KCuCl_3$, we consider the spin exchanges $J_2$ and $J'_a$ as representative examples. The $x^2-y^2$ magnetic orbital of each $Cu^{2+}$ ion lies in the $CuCl_4$ square plane. Thus, the two $Cu^{2+}$ ions of a spin exchange path are represented by two $CuCl_4$ square planes, a $(CuCl_4)_2$ dimer for short. The $(CuCl_4)_2$ dimers of the exchange paths $J_2$ and $J_a'$ make short contacts with $A^+$ cations, as depicted in **Fig. 3.14a** and **3.14b**, respectively. In each $(CuCl_4)_2$ dimer, the $x^2-y^2$ magnetic orbitals of two $Cu^{2+}$ ions form in-phase and out-of-phase combinations (see **Fig. S2**, Section S3 of the SI), which we represent by the labels (+) and (-), respectively. The frontier orbitals of $K^+$ and $Tl^+$ that can interact with the (+) and (-) d-states are K 4s, Tl 6s and Tl 6p orbitals (**Fig. 3.14c**). By symmetry, the K 4s orbital interacts with the (+) state, so the (+) level is lowered in energy (**Fig. 3.14c**). The Tl 6s orbital interacts with the (+) state, which raises the (+) level, but the Tl 6p orbital interacts with the (-) state, which lowers the (-) level (**Fig. 3.14c**). Such interactions occur at every Cl…$A^+$…Cl bridge each $(CuCl_4)_2$ dimer makes with the surrounding A$^+$ cations. Consequently, the energy gap between the (+) and (−) d-states for the interdimer spin exchanges is larger for TlCuCl$_3$ than for KCuCl$_3$. The intra-stack exchange J$_a'$ is calculated to be slightly stronger than the strongest inter-stack exchange J$_2$ in both KCuCl$_3$ and TlCuCl$_3$ (**Fig. 3.13d**). This reflects that the J$_a'$ path has four A$^+$ cations making the Cl…A$^+$…Cl bridges (**Fig. 3.14b**), while the J$_2$ path has only two such bridges (**Fig. 3.14a**).

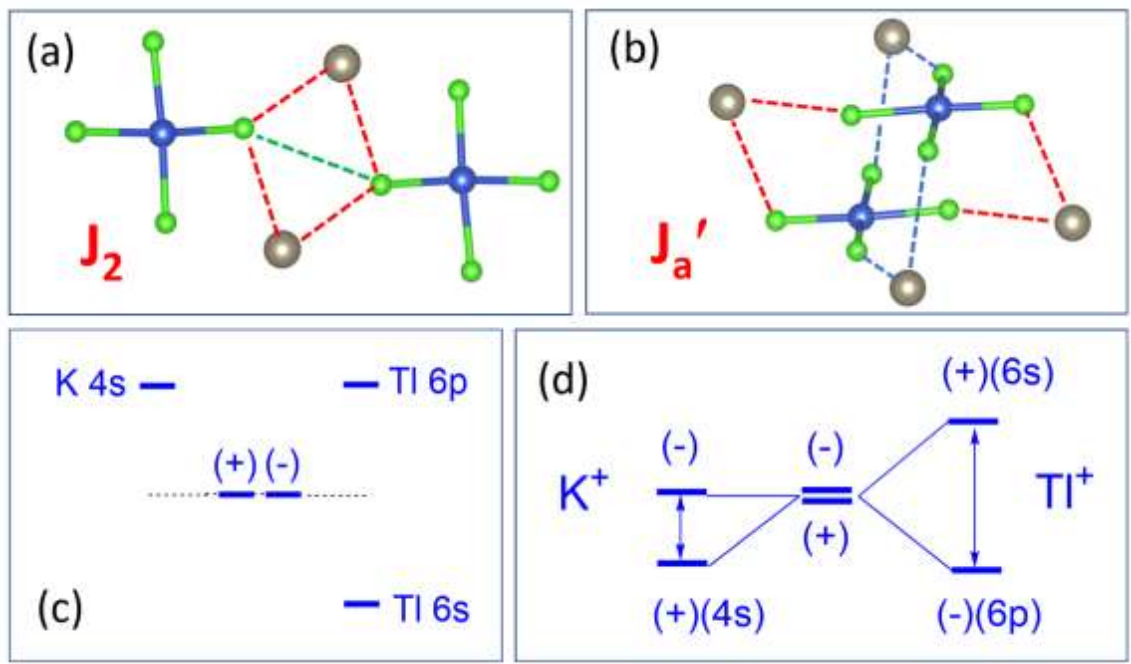

**Fig. 3.14.** (a, b) Arrangements of the cations A$^+$ around the (CuCl$_4$)$_2$ dimers constituting the inter-dimer spin exchange paths J$_2$ and J$_a'$. (c) The orbitals of the cations K$^+$ and Tl$^+$ that can interact with the d-states, (+) and (−), states of the (CuCl$_4$)$_2$ dimers. (d) A schematic diagram showing how the d-states, (+) and (−), of the (CuCl$_4$)$_2$ dimer are affected by the K 4s orbital in KCuCl$_3$ (left), and by the Tl 6s and Tl 6p orbitals in TlCuCl$_3$ (right).

**E. Intradimer exchange of NH$_4$CuCl$_3$**

Let us now examine how the spin exchanges of NH$_4$CuCl$_3$ depend on the orientations of the NH$_4^+$ cations with respect to the Cu$_2$Cl$_6^{2-}$ anions they surround (**Fig. 3.13c**). Each N-H bond of a NH$_4^+$ cation has a σ*$_{N-H}$ orbital, which is highly anisotropic in shape because it is oriented along the N-H bond. Based on the crystal structure determined by X-ray diffraction,[41] and assuming that the rotational mobility of the NH$_4^+$ cations ceases at low temperatures, we construct three model orientations, termed YY, NY and NN, of the two NH$_4^+$ cations that bridge the either side of the Cl…Cl contact in every J$_2$ exchange path (**Fig. 3.15a**). Under the constraint that two NH bonds of each NH$_4^+$ group are coplanar with the Cl…Cl contact and the other NH$_2$ group bisects the Cl…Cl contact, only three different NH$_4^+$ arrangements are possible; both NH$_4^+$ cations make N-H…Cl hydrogen bonds with the Cl…Cl contact in the YY arrangement, only one NH$_4^+$ cation does in the NY arrangement, and no NH$_4^+$ cation does so in the NN arrangement (**Fig. 3.15b**). The orientations of six NH$_4^+$ cations surrounding each Cu$_2$Cl$_6^{2-}$ ion in the YY, NY and NN arrangements (see **Fig. S3**, Section S3 of the SI).

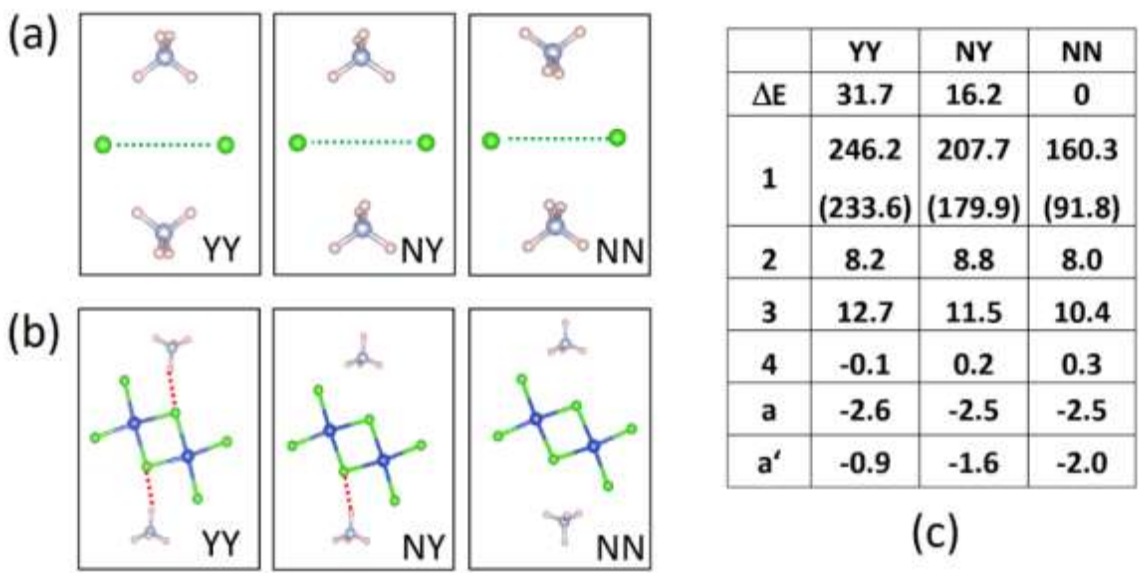

**Fig. 3.15.** (a) The YY, NY and NN arrangements of two $NH_4^+$ cations around the Cl…Cl contact of each $J_2$ exchange path in $NH_4CuCl_3$. (b) Short N-H…Cl contacts to the mid Cl atoms of the $Cu_2Cl_6^{2-}$ anion in the YY, NY and NN arrangements of two $NH_4^+$ cations. (c) The relative energies $\Delta E$ (in meV/f.u.) of $NH_4CuCl_3$ with the YY, NY and NN arrangements, and values of their intradimer exchanges (in K), where the labels 1 – 4, a and a′ refer to the spin exchanges $J_1 - J_4$, $J_a$ and $J_a'$, respectively. The $J_1$ values in the parentheses were obtained using the optimized $NH_4CuCl_3$ structures.

The relative energies of these three structures and the values of their intradimer spin exchange $J_1$ are summarized in **Fig. 3.15c**, from which we note the following: (a) The YY, NY and NN arrangements of $NH_4CuCl_3$ have considerably different relative stabilities, with the stability increasing in the order, YY < NY < NN. (b) The intradimer exchange $J_1$ of $NH_4CuCl_3$ depends strongly on the $NH_4^+$ orientations, with its value increasing in the order, NN < NY < YY. (c) The interdimer exchanges, $J_2$ and $J_a'$ of $NH_4CuCl_3$ are much weaker than those of $KCuCl_3$ and $TlCuCl_3$. This reflects the fact that the $\sigma^*_{N-H}$ orbitals of $NH_4^+$ are strongly contracted compared with the K 4s and the Tl 6s/6p.

Since there are two equivalent ways of having the NY arrangement, the statistical probabilities for the YY, NY and NN arrangements are 1:2:1. The observations (a) and (b) are consistent with the experimental observation suggesting that $NH_4CuCl_3$ consists of three different $Cu_2Cl_6^{2-}$ anions in the 1:2:1 ratio.[42] The $Cu_2Cl_6^{2-}$ anions in the YY, NY and NN structures are surrounded by $NH_4^+$ cations with different orientations (**Fig. S3**, Section S3 of the SI), and hence will undergo different local relaxations further changing the values of their spin exchanges. To test this hypothesis, we optimized the YY, NY and NN structures by relaxing only the Cu and Cl positions and then calculate the spin exchanges for the resulting structures (**Fig. 3.15c**), to find a reduction of $J_1$ by 0.05, 13 and 43 % for the YY, NY and NN structures, respectively. These reductions reflect that the mid Cl atom of the $Cu_2Cl_6^{2-}$ anion with (without) the short N-H…Cl contact moves away from (toward) the N atom thereby increasing (decreasing) the ∠Cu-Cl-Cu angle of the $Cu_2Cl_6^{2-}$ anion [namely, 96.28° (×2) for the YY, 95.43° and 96.27° for the NY, and

95.35° (×2) for the NN structure]. The effect of the structure relaxation on other spin exchanges is weak (see Sections S6 – S8 of the SI).

### F. Consequence of the interaction between $NH_4^+$ and $Cu_2Cl_6^{2-}$ in $NH_4CuCl_3$

It is of interest to find why the intradimer exchange $J_1$ of $NH_4CuCl_3$ depends so sensitively on the $NH_4^+$ orientations. The two d-states of a $Cu_2Cl_6^{2-}$ anion, termed the [+] and [-] states in **Fig. 3.16a**, are the in-phase and out-of-phase combinations of the two $x^2-y^2$ magnetic orbitals. With respect to the long axis of the $Cu_2Cl_6^{2-}$ ion, the two p-orbitals at each bridging Cl atom (hereafter, the mid-Cl atom) in **Fig. 3.16a** are hybridized to become a perpendicular p-orbital ($p_\perp$) in the [+] state, but a parallel p-orbital ($p_\parallel$) in the [-] state. The $p_\perp$ orbital is spatially more extended out toward the surrounding cations $NH_4^+$ than is the $p_\parallel$ orbital and is hence more effective in the cation-anion interactions in the short $NH_4^+$…Cl contacts. The observation (b) reflects that the energy lowering by the ($p_\perp$-$\sigma^*_{N-H}$) interaction occurs in one and two places in the NY and YY structures, respectively (**Fig. 3.16b**). Thus, the energy gap between the [+] and [-] states of a $Cu_2Cl_6^{2-}$ anion interacting with the surrounding $NH_4^+$ cations increases in the order, NN < NY < YY, as depicted in **Fig. 3.16c**.

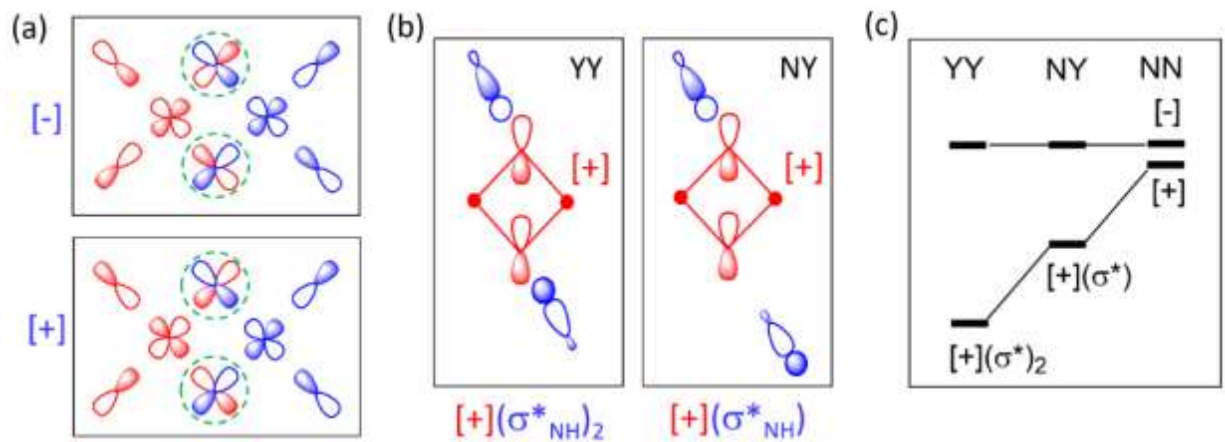

**Fig. 3.16.** In-phase, [+], and (b) the out-of-phase, [-], combinations of the two $x^2-y^2$ magnetic orbitals describing the intradimer exchange path $J_1$ of a $Cu_2Cl_6^{2-}$ anion. The two magnetic orbitals are given in red and blue colors to ease of distinction. The two p-orbitals at each bridging Cl atoms (encircled by a dashed green circle) in the [+] state become a $p_\perp$ orbital perpendicular to the Cu…Cu axis, and those in the [-] state a $p_\parallel$ orbital parallel to the Cu…Cu axis. (b) Sigma bonding ($p_\perp$-$\sigma^*_{N-H}$) interaction(s) in the YY and NY structures of $NH_4CuCl_3$ that the $\sigma^*_{NH}$ orbital of $NH_4^+$ makes with the $p_\perp$ orbital(s) in the [+] d-state of $Cu_2Cl_6^{2-}$ ion. (c) Lowering of the [+] level by the ($p_\perp$-$\sigma^*_{N-H}$) interaction(s) in the YY and NY structures of $NH_4CuCl_3$.

We now examine an important implication of the observation made in **Fig. 3.16c**. In general, the spin exchange J of a spin dimer made up of two S = 1/2 ions is written as $J = J_F + J_{AF}$.[6,47] If the spin sites at i and j are represented by magnetic orbitals $\phi_i$ and $\phi_j$, respectively, the FM component $J_F$ (< 0) increases in magnitude with the overlap density $\rho_{ij} = \phi_i\phi_j$, and the AFM component $J_{AF}$ (> 0) with the magnitude of the overlap integral $S_{ij} = \langle\phi_i|\phi_j\rangle$. The interaction between $\phi_i$ and $\phi_j$ leads to the energy split $(\Delta e)_{ij}$ between them, which is related to $S_{ij}$ as $(\Delta e)_{ij} \propto (S_{ij})^2$. Therefore, the overall spin exchange J can be FM if $(\Delta e)_{ij}$ is small. **Fig. 3.16c** shows that the energy gap between the [+] and [-] d-states of $NH_4CuCl_3$ decreases in the order, YY > NY > NN. If the energy split $(\Delta e)_{ij}$ becomes smaller, then the associated spin exchange can become FM, hence the associated dimer becoming a triplet dimer. It is most likely that the three spin dimers A, B and C of $NH_4CuCl_3$ as experimentally observed might be assigned to the $Cu_2Cl_6^{2-}$ anions surrounded with the NN, NY and YY orientations of the $NH_4^+$ cations, respectively. This is a consequence that a given $NH_4CuCl_3$ sample does not have a uniform orientation of the $NH_4^+$ cations. It rather consists of regions possessing mainly YY, NY and NN orientations of the $NH_4^+$ cations.

## 4. Magnets of ferrimagnetic fragments
### 4.1. Linear trimers and chains
#### 4.1.1. Isolated linear trimers in $Mn_3(PO_4)_2$

Manganese diphosphates $Mn_3(PO_4)_2$ are found in several different phases, namely, α, β', and γ,[48] which undergo a long-range AFM order at $T_N$ = 21.9, 12.3, and 13.3 K, respectively. The 3D crystal structures of these phases consist of corner- and edge-sharing $MnO_5$ and $MnO_6$ polyhedra, which are further bridged by $PO_4$ tetrahedra. In γ-$Mn_3(PO_4)_2$, each $Mn2O_6$ octahedron corner-shares with two $Mn1O_5$ trigonal bipyramids to form a Mn1-Mn2-Mn1 linear trimer (**Fig. 4.1a**), and these trimers are edge-shared either in a head-to-tail (**Fig. 4.1b**) or tail-to-tail (**Fig. 4.1c**) fashion. The magnetization curves of these phases show spin-flop-like features at low magnetic field, but a 1/3-magnetization plateau is found only for the α- and γ-$Mn_3(PO_4)_2$ modifications. As shown in **Fig. 4.1d**, the 1/3-plateau of the γ-phase is very wide.

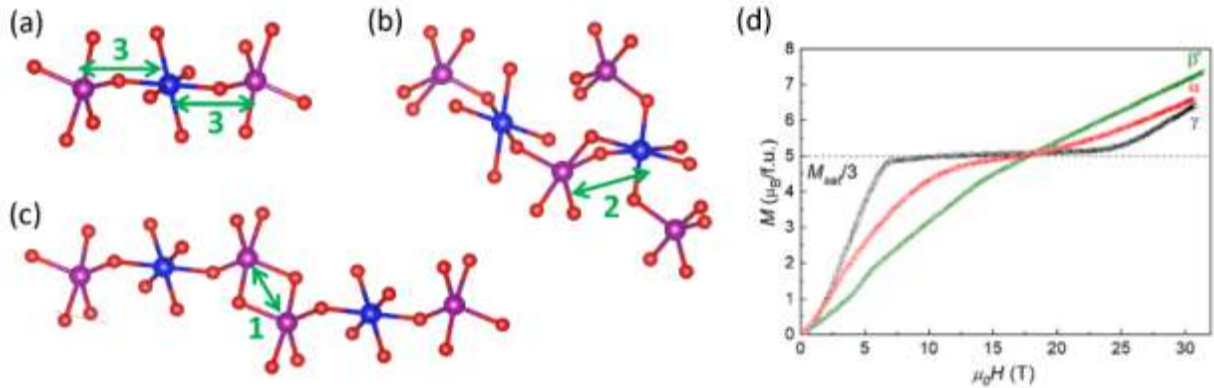

**Fig. 4.1.** (a) Linear Mn1-Mn2-Mn1 trimer in γ-$Mn_3(PO_4)_2$, which results when a $Mn2O_6$ octahedron corner-shares with two $Mn1O_5$ trigonal bipyramids. (b) Two Mn1-Mn2-Mn1 trimers edge-sharing in a head-to-tail fashion. (c) Two Mn1-Mn2-Mn1 trimers edge-sharing in a tail-to-tail fashion. The labels 1, 2 and 3 refer to the spin exchange paths $J_1$, $J_2$ and $J_3$, respectively. (d) Field dependence of the magnetization in α-, β'- and γ-phases of $Mn_3(PO_4)_2$ at 2 K.[48] (Reproduced with permission from reference 48.)

A simplified view of the layer that linear Mn1-Mn2-Mn1 trimers form by a head-to-tail bridging is presented in **Fig. 4.2a**. The spin lattice of this layer is defined by the intra-trimer exchange $J_3$ and the inter-trimer exchange $J_2$. Such layers make a 3D structure by a tail-to-tail bridging between the trimers lying in adjacent layers, which leads to the inter-layer exchange $J_1$ (**Fig. 4.2b**). The spin exchanges $J_1$, $J_2$ and $J_3$ of γ-$Mn_3(PO_4)_2$ (**Fig. 4.1a**) are all AFM and are estimated to be 1.7, 4.7 and 10.5 K, respectively.[48] Each layer defined by the exchanges $J_3$ and $J_2$ are ferrimagnetic because each linear trimer is ferrimagnetic due to the strong AFM exchange $J_3$, and because the head-to-tail coupling between two ferrimagnetic trimers does not cancel their moments (**Fig. 4.2c**). Such ferrimagnetic layers are coupled antiferromagnetically via the exchange $J_1$ to form an AFM magnetic ground state responsible for the AFM ordering.

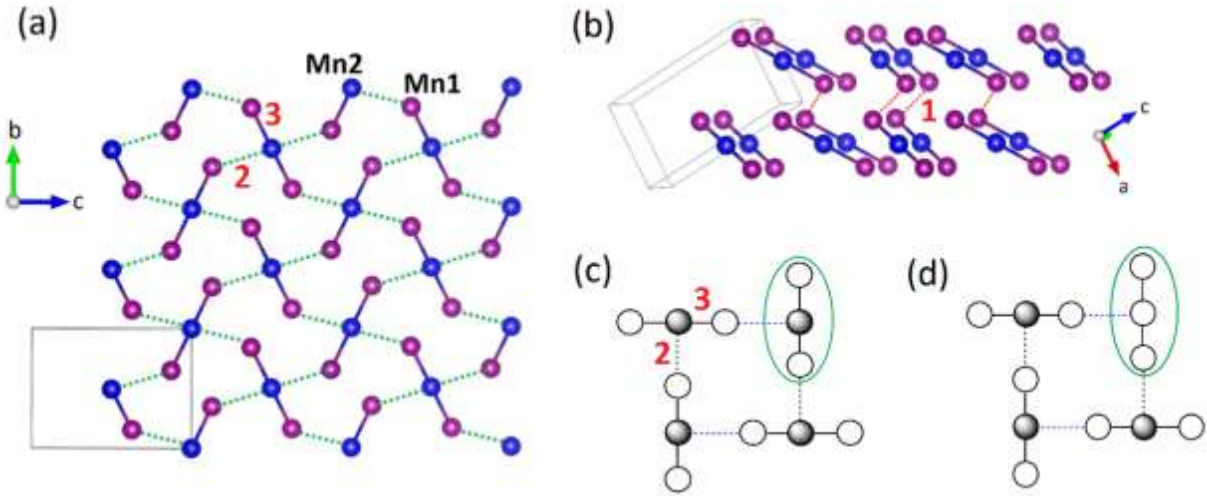

**Fig. 4.2.** (a) Layer of Mn1-Mn2-Mn1 linear trimers in γ-Mn$_3$(PO$_4$), parallel to the *bc*-plane, formed by a head-to-tail bridging. (b) Tail-to-tail bridging between the trimers lying in adjacent layers. (c) Ferrimagnetic state of a layer defined by the spin exchange $J_3$ and $J_2$ due to the head-to-tail coupling between the linear trimers. A green ellipse indicates a trimer that will undergo a field-induced $J_3$ bond breaking. In (a – c), the labels 1 – 3 refer to the spin exchanges $J_1 – J_3$, respectively. (d) Breaking of an inter-trimer bond $J_2$ as a consequence of breaking the two $J_3$ bonds of a linear trimer.

The gradual increase in the magnetization $M$ of γ-Mn$_3$(PO$_4$)$_2$ with increasing $\mu_0 H$ in the region of 0 – 7.5 T mirrors the breaking of the inter-layer $J_1$ bonds, leading to the ferrimagnetic layers. This field-induced ferrimagnetic state at 7.5 T explains the 1/3-plateau. For each ferrimagnetic layer to go beyond the 1/3-plateau, it is necessary to break two $J_3$ bonds of a trimer, which is accompanied by the breaking of a $J_2$ bond (**Fig. 4.2d**). The wide plateau between 7.5 – 23.5 T reflects the difficulty of simultaneously breaking one $J_2$ and two $J_3$ bonds when a linear AFM trimer becomes FM.

### 4.1.2. Bent trimers in Cu$_3$(P$_2$O$_6$OH)$_2$

The building blocks of Cu$_3$(P$_2$O$_6$OH)$_2$ are Cu2O$_6$ octahedra and Cu1O$_5$ trigonal bipyramids, as found in γ-Mn$_3$(PO$_4$)$_2$. However, each Cu2O$_6$ octahedron edge-shares with two Cu1O$_5$ trigonal bipyramids in Cu$_3$(P$_2$O$_6$OH)$_2$ (**Fig. 4.3a**)[49] to form linear Cu1-Cu2-Cu1 trimers, in contrast to the corner-sharing found in γ- Mn$_3$(PO$_4$)$_2$ (**Fig. 4.1a**). Cu$_3$(P$_2$O$_6$OH)$_2$ exhibits a 1/3-magnetization plateau above 12 T (**Fig. 4.4a**),[50] which was initially interpreted by supposing that its spin lattice is a $J_1$-$J_2$-$J_2$ chain made up of ferrimagnetic linear Cu1-Cu2-Cu1 trimers (**Fig. 4.3b**). However, this model is not consistent with the spin exchanges of Cu$_3$(P$_2$O$_6$OH)$_2$ evaluated by DFT+U calculations;[51] the latter found that the exchange $J_2$ is practically zero, and that Cu$_3$(P$_2$O$_6$OH)$_2$ has a 2D spin lattice made up of three spin exchanges $J_1$, $J_3$ and $J_6$ (479, 69 and 90 K, respectively) as shown in **Fig. 4.3c**.

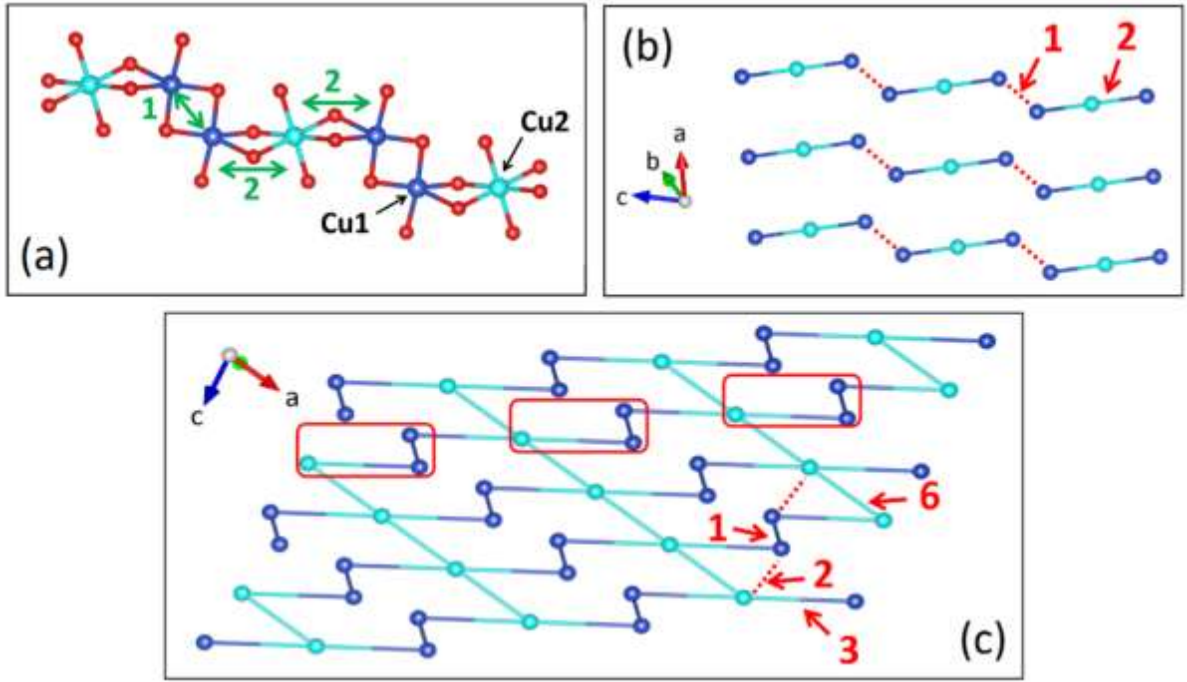

**Fig. 4.3.** (a) Chain of edge-sharing Cu1O$_5$ trigonal bipyramids and Cu2O$_6$ octahedra in Cu$_3$(P$_2$O$_6$OH)$_2$. (b) A spin lattice composed of ferrimagnetic linear Cu1-Cu2-Cu1 trimers making chains by a tail-to-tail coupling. (c) A spin lattice of ferrimagnetic bent Cu1-Cu1-Cu2 trimers (e.g., those enclosed in red rectangles) making chains by a tail-to-tail coupling, and such chains make a 2D net by a tail-to-tail coupling. The labels 1, 2, 3 and 6 refer to the spin exchanges $J_1$, $J_2$, $J_3$ and $J_6$, respectively.

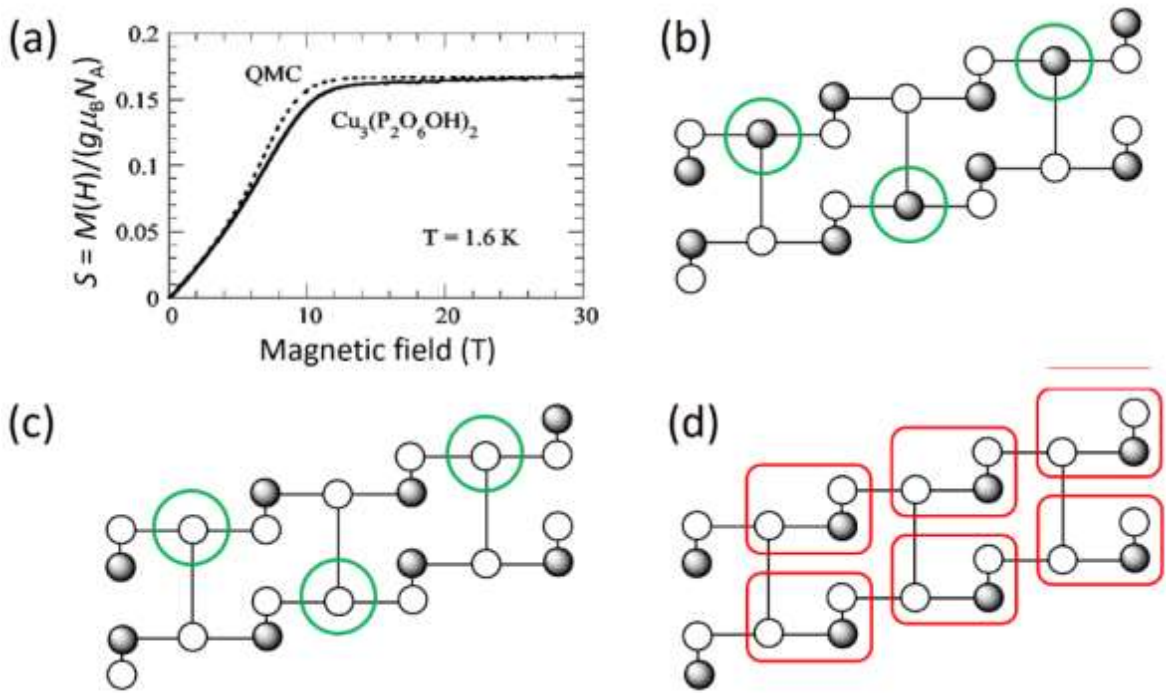

**Fig.4.4.** (a) Field dependence of magnetization of $Cu_3(P_2O_6OH)_2$ at 1.6 K and its Quantum Monte Carlo (QMC) simulation.[50] (Reproduced with permission from reference 50.) (b) AFM arrangement of ferrimagnetic bent trimers via the tail-to-tail coupling along the $J_3$ and $J_6$ exchange paths. The down-spin at each site encircled with a green circle becomes up-spin to increase the moment under magnetization. (c) A ferrimagnetic state resulting from the down-spin to up-spin conversion at each circled down-spin site in (b). (d) A ferrimagnetic state, equivalent to the one shown in (c), composed of ferrimagnetic bent trimers.

The three AFM exchanges $J_1$, $J_3$ and $J_6$ lead to an AFM spin arrangement in the 2D lattice (**Fig. 4.4b**), where half the Cu2 sites have up-spins, and the remaining half down-spins (i.e., those in green circles). Since $J_3$ and $J_6$ are considerably weaker than $J_1$, the increase in $M$ with $\mu_0 H$ is achieved by breaking these magnetic bonds, i.e., by flipping the down-spin to up-spin at the Cu2 sites, one at a time. This spin flipping simultaneously breaks one $J_6$ and two $J_3$ bonds. When all down spins at the Cu2 sites are flipped, a ferrimagnetic configuration with $M = M_{sat}/3$ is reached (**Fig. 4.4c**). Note that this spin arrangement is equivalent in energy to another ferrimagnetic spin arrangement shown in **Fig. 4.4d**. Either ferrimagnetic arrangement can be decomposed into bent ferrimagnetic trimers, as illustrated in **Fig 4.4d**. The plateau above 12 T is wide because the $J_1$ bond is strong and because a spin flip from (↑↓↑) to (↑↑↑) in each ferrimagnetic trimer, which must occur to increase the magnetization beyond $M = M_{sat}/3$, simultaneously breaks one $J_1$ and one $J_3$ bond.

### 4.1.3 Heisenberg chains in volborthite $Cu_3V_2O_7(OH)_2·2(H_2O)$

Volborthite, $Cu_3V_2O_7(OH)_2·2H_2O$, has a layered crystal structure, in which the layers of composition $Cu_3O_6(OH)_2$ parallel to the *ab*-plane are pillared by pyrovanadate $V_2O_7$ groups, and crystal water molecules occupy the voids between the layers. The $Cu^{2+}$ ions in each $Cu_3O_6(OH)_2$ layer have a kagomé-like arrangement (**Fig. 4.5a**). Below room temperature, volborthite undergoes two structural phase transitions, one at ~292 K from a *C*2/*c* phase to a *I*2/*a* phase, and the other at ~155 K from the *I*2/*a* phase to a *P*2₁/*c* phase.[52] The latter structural phase transition generates two kagomé layers slightly different in structure. Below 1.5 K, volborthite exhibits magnetic order, indicated by two anomalies in the magnetic specific heat.[53]

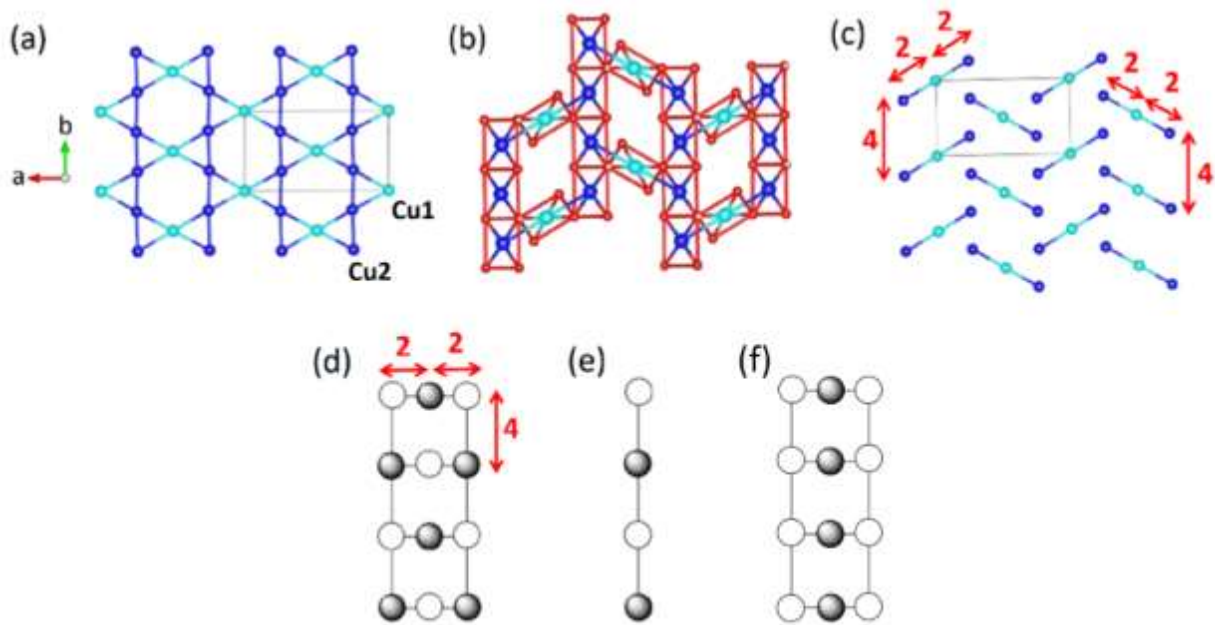

**Fig. 4.5.** (a) Arrangement of the $Cu^{2+}$ ions in a $Cu_3O_6(OH)_2$ layer of volborthite. (b) Arrangement of the $CuO_4$ square planes containing the $x^2$-$y^2$ magnetic orbitals in a $Cu_3O_6(OH)_2$ layer of volborthite. (c) Arrangement of Cu2-Cu1-Cu2 linear trimers in a $Cu_3O_6(OH)_2$ layer of volborthite. (d) AFM state of a two-leg spin ladder with rungs of ferrimagnetic linear trimers defined by $J_2$ and legs defined by $J_4$. In (c, d) the labels 2 and 4 refer to $J_2$ and $J_4$, respectively. (e) Effective S=1/2 AUH chain representing the two-leg spin ladder of (d) at low temperature, where thermal excitations within each rung are absent. (f) Ferrimagnetic state of a two-leg spin ladder with rungs of ferrimagnetic linear trimers.

Volborthite had been regarded as a kagomé spin lattice system.[54] However, according to a recent study,[53] it is not a kagomé spin lattice but an S=1/2 AFM uniform Heisenberg (AUH) chain that describes the magnetic properties of volborthite at low temperatures. This observation reflects the fact that the spin lattice of a magnet does not necessarily follow the geometrical pattern of its magnetic ion arrangement but is determined by that of strongly interacting spin exchange paths between the magnetic ions.[55] The $CuO_6$ octahedra of volborthite accommodating the $Cu^{2+}$ ions are axially elongated, so their $x^2$-$y^2$ magnetic orbitals lie in their $CuO_4$ square planes perpendicular to the elongated Cu-O bonds. The arrangement of these $CuO_4$ planes in volborthite, depicted in **Fig. 4.5b**, is highly anisotropic forming the Cu2-Cu1-Cu2 linear trimers bridged by Cu2-O-Cu1 linkages. Within each $Cu_3O_6(OH)_2$ layer, the Cu2-Cu1-Cu2 trimers are arranged as in **Fig. 4.5c**. The spin exchanges of volborthite determined by DFT+U calculations show[53] that the strongest AFM spin exchange, $J_2$ (550 and 582 K for the two different layers), makes each Cu2-Cu1-Cu2 linear trimer ferrimagnetic, and these ferrimagnetic trimers are coupled antiferromagnetically by the next strongest spin exchange $J_4$ (78 K for both layers) to form two-leg spin ladders. All other spin exchanges are negligibly small, and adjacent spin ladders are entangled in their legs (**Fig. 4.5c**). In essence, the kagomé-like arrangement of $Cu^{2+}$ ions in $Cu_3V_2O_7(OH)_2 \cdot 2H_2O$ gives rise to weakly interacting two-leg spin ladders with Cu2-Cu1-Cu2 trimers as rungs, which have an AFM spin arrangement as depicted in **Fig. 4.5d**.

At low temperatures where thermal excitations within each trimer rung are absent, each rung acts as an effective S=1/2 species due to a strong AFM coupling between adjacent $Cu^{2+}$ sites, so that each two-leg spin ladder should behave as an effective S=1/2 AUH chain (**Fig. 4.5e**).[53] Indeed, the magnetic susceptibility of volborthite at low temperatures (below 75 K) is very well described by an S=1/2 AUH chain model to find the nearest-neighbor spin exchange $J_C$ = 27.8(5) K, as shown in **Fig. 4.6a**.[53] On lowering the temperature, the susceptibility shows a broad maximum and converges to a nonzero value as the temperature approaches zero, a characteristic feature expected for an S=1/2 AUH chain. Volborthite exhibits an extremely wide 1/3 magnetization plateau above 28 T continuing over 74 T at 1.4 K (**Fig. 4.6b**).[56] Before reaching the value of $M = M_{sat}/3$, the magnetization increases with field because each linear (↓↑↓) trimer is converted to a linear (↑↓↑) trimer, breaking four $J_4$ bonds, eventually reaching the ferrimagnetic state (**Fig. 4.5f**) at ~28 T. A further increase in magnetic field does not increase magnetization leading to the 1/3-plateau because it requires breaking two $J_2$ bonds to convert a (↑↓↑) rung to a (↑↑↑) rung and because $J_2$ bond is very strong. There is a theoretical study on the magnetization plateau of a two-leg spin ladder.[57] In our analysis of the magnetization plateau of volborthite obtained at 1.4 K,[55] we employ the S=1/2 AUH chain model (**Fig. 4.6c**) because each (↑↓↑) rung acts as an effective S=1/2 entity at 1.4 K. As shown in **Fig. 4.6c**, the experimental magnetization data are very well described by the S=1/2 AUH chain model (**Fig. 4.6c**) using the nearest-neighbor spin exchange $J_C$ of 27.5 K,[53] just as are the magnetic susceptibility data below 75 K.

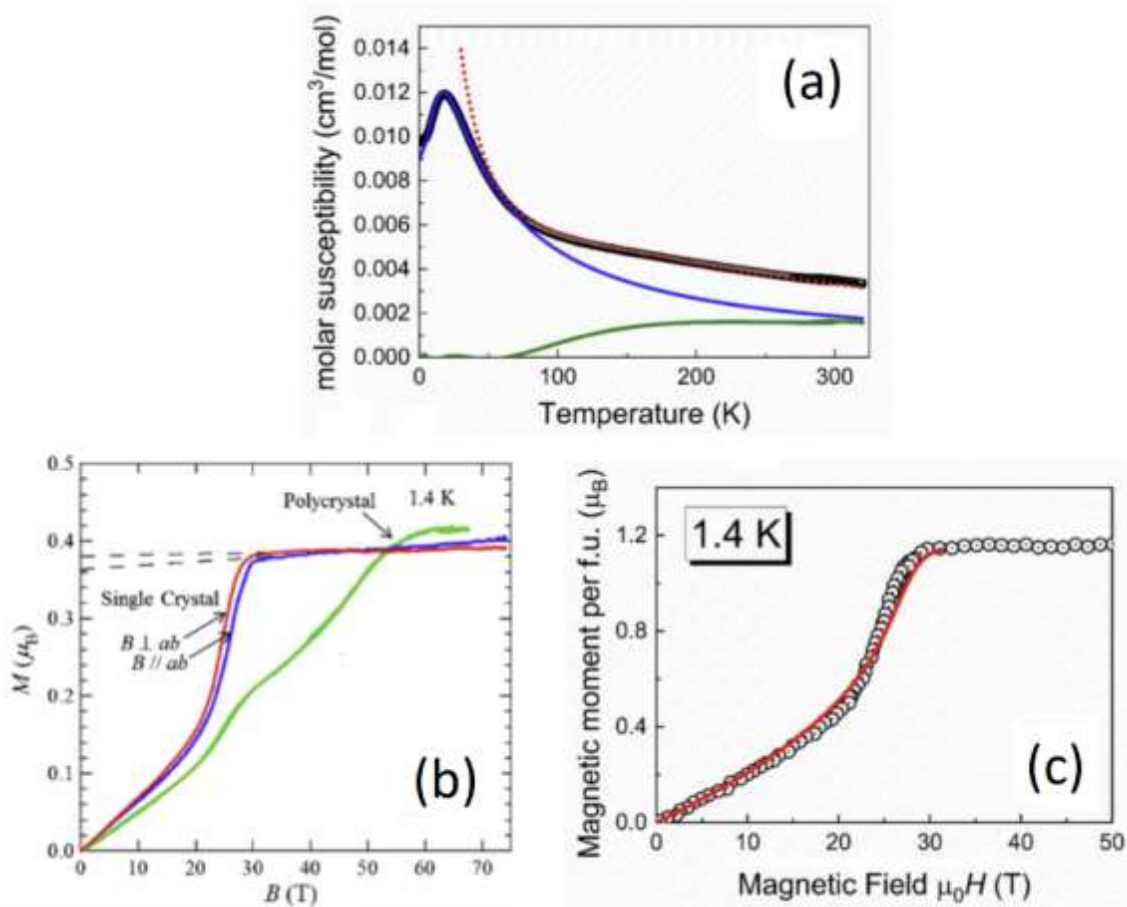

**Fig. 4.6.** (a) Magnetic susceptibility of volborthite for one formula unit (i.e., comprising three Cu atoms) of volborthite (black circles) with probe field applied along the crystallographic b axis fitted

(for T < 75 K) to the theoretical prediction for an S=1/2 AUH chain (see the solid blue curve).[53] The difference between the two is displayed as a solid green line. The experimental susceptibilities for 75 K ≤ T ≤ 320 K are well fitted by the susceptibility of a linear spin S = 1/2 trimer with spin exchange of 197 K (red dotted curve). (b) Field dependence of the magnetization (per Cu) measured for single crystal and polycrystalline samples of $Cu_3V_2O_7(OH)_2·2H_2O$ at 1.4 K.[55] (Reproduced with permission from reference 55.) (c) Field dependence of the magnetization (per three Cu) measured for volborthite at 1.4 K (taken from Ishikawa et al.[55]) compared with quantum Monte Carlo calculations for an S=1/2 AUH chain with $J_C$ = 27.5 K (solid red line).[53]

### 4.1.4. Head-to-tail coupling of bent trimers and anisotropic 1/3-plateau in $Cs_2Cu_3(SeO_3)_4·2(H_2O)$

$Cs_2Cu_3(SeO_3)_4·2H_2O$ consists of two nonequivalent Cu atoms, Cu1 and Cu2 in the 1:2 ratio, each forming $Cu1O_4$ and $Cu2O_4$ square planes, respectively. The 3D framework of $Cs_2Cu_3(SeO_3)_4·2H_2O$ is formed by the corner-sharing of $Cu1O_4$ and $Cu2O_4$ square planes.[58] As depicted in **Fig. 4.7a,b**, each $Cu1O_4$ square plane is corner-shared with four $Cu2O_4$ square planes such that the four Cu2 atoms around a Cu1 atom make a $Cu1(Cu2)_4$ tetrahedron (**Fig 4.7c**). Condensing such $Cu1(Cu2)_4$ tetrahedra by sharing their Cu2 corners leads to the 3D network of $Cu^{2+}$ ions of $Cs_2Cu_3(SeO_3)_4·2H_2O$, which can be described as resulting from the fusing of chair-shape hexagonal rings (**Fig. 4.7d**).

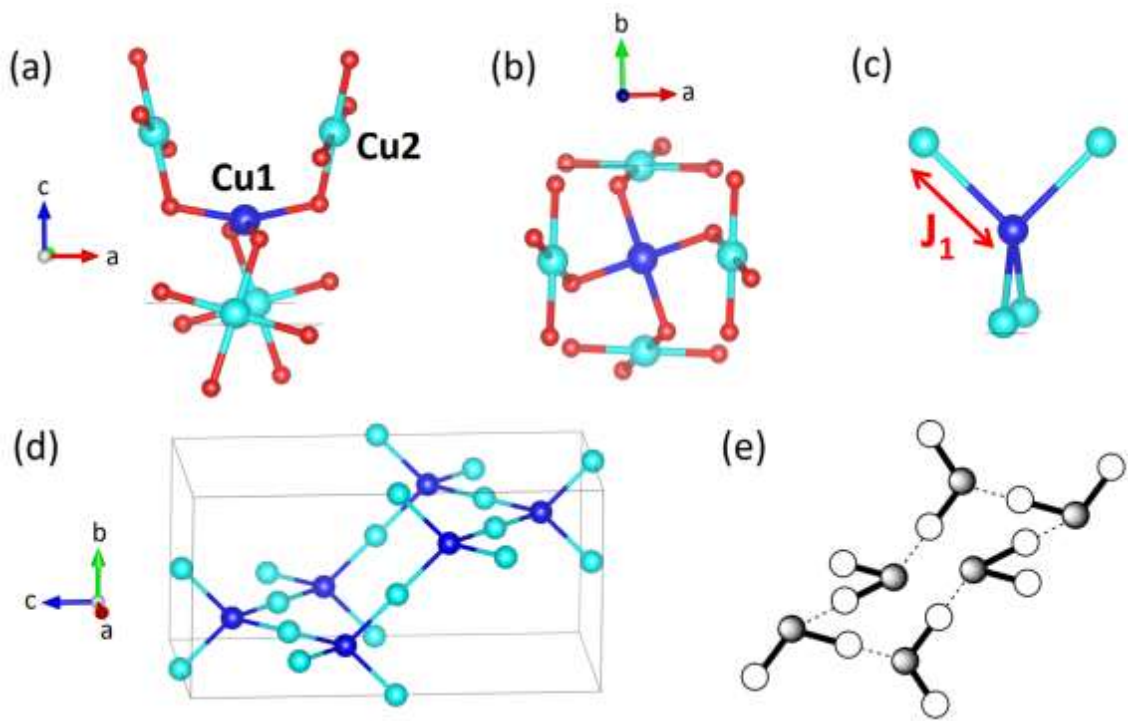

**Fig. 4.7.** (a, b) Four $Cu2O_4$ square planes sharing their oxygen corners with a $Cu1O_4$ square plane in $Cs_2Cu_3(SeO_3)_4(H_2O)_2$ viewed approximately along the b-direction in (a) and along the c-direction in (b). (c) $Cu1(Cu2)_4$ tetrahedron associated with the five $CuO_4$ planes in (a). (d) Chair-form hexagonal ring made up of $Cu1(Cu2)_4$ tetrahedra by sharing their Cu2 corners. (e) Head-to-tail coupling of the bent ferrimagnetic Cu2-Cu1-Cu2 units with (↑↓↑) spin configuration leading to the ferrimagnetic state of $Cs_2Cu_3(SeO_3)_4(H_2O)_2$.

Cs$_2$Cu$_3$(SeO$_3$)$_4$(H$_2$O)$_2$ is a ferrimagnet ordering at $T_C$ = 20 K with residual magnetization at about $M_{sat}$/3 (**Fig. 4.8a**). In general, ferrimagnetism occurs when ferrimagnetic fragments are combined antiferromagnetically in a head-to-tail bridging pattern so that the magnetic moment of each ferrimagnetic fragment is not quenched. Such ferrimagnetic units in Cs$_2$Cu$_3$(SeO$_3$)$_4$(H$_2$O)$_2$ should consist of one Cu1 and two Cu2 atoms, given that the Cu1 and Cu2 atoms occur in the 1:2 ratio. DFT+U calculations[58] show that the nearest-neighbor exchange J$_1$ (**Fig. 4.7c**) is strong (256 K) but other exchanges are negligibly weak. This makes all nearest-neighbor Cu1…Cu2 linkages antiferromagnetically coupled, so the ferrimagnetic fragments needed to explain the ferrimagnetism of Cs$_2$Cu$_3$(SeO$_3$)$_4$(H$_2$O)$_2$ are the bent Cu2-Cu1-Cu2 units with (↑↓↑) spin configuration.

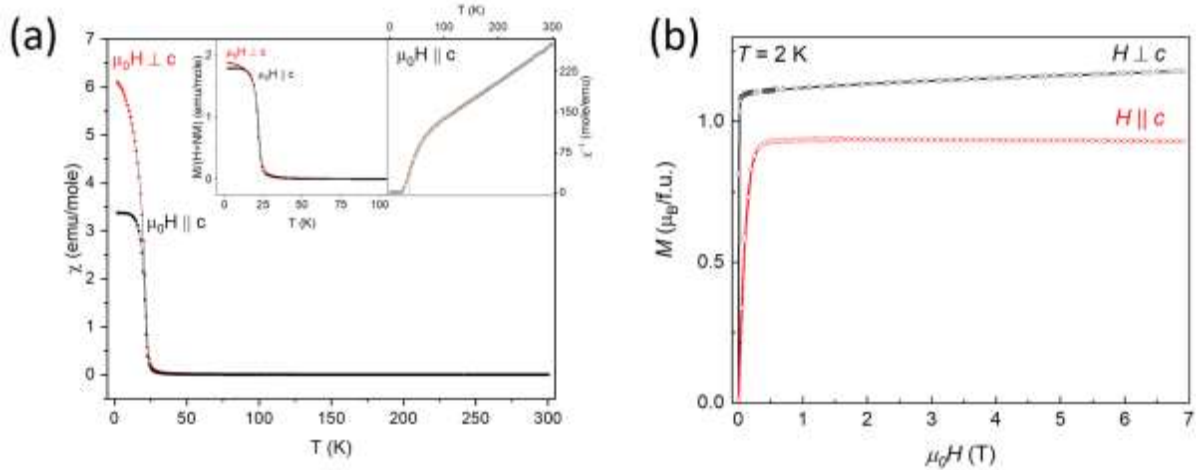

**Fig. 4.8.** (a) Temperature dependence of the magnetic susceptibility $\chi$ = $M/H$ in Cs$_2$Cu$_3$(SeO$_3$)$_4$·2H$_2$O for both $H\|c$ and $H\bot c$ taken at $\mu_0H$ = 0.1 T. Left inset: Temperature dependence of magnetic susceptibility corrected for demagnetization effects. Right inset: Temperature dependence of the inverse magnetic susceptibility for $H\|c$, where the solid line represents the Néel law. (b) Anisotropic 1/3 magnetization plateau in Cs$_2$Cu$_3$(SeO$_3$)$_4$(H$_2$O)$_2$.[58] (Reproduced with permission from reference 58.)

When measured for a single crystal sample of Cs$_2$Cu$_3$(SeO$_3$)$_4$(H$_2$O)$_2$ parallel ($\|$) and perpendicular ($\bot$) to the $c$-direction (**Fig. 4.8a,b**), the values of the magnetization at $\mu_0H$ = 7 T are quite different, namely, $M_\bot$ = 1.18 $\mu_B$, whereas $M_\|$ = 0.93 $\mu_B$.[58] There are three factors contributing to this highly anisotropic magnetization plateau; the nearly orthogonal arrangements of the Cu2O$_4$ square planes around each Cu1O$_4$ square plane (**Fig. 4.7b**), the strong nearest-neighbor antiferromagnetic exchange J$_1$, and the anisotropic g-factor of Cu$^{2+}$ ions in a square-planar coordination site. The magnetic anisotropy of a magnetic ion arises from SOC. In the spin-only description, in which orbital information is suppressed, the effect of SOC on magnetic anisotropy is discussed by introducing g-factor g different from 2.[2] That is, the magnetic moment $\mu$ of a spin site with spin S is given by $\mu = gS$, where g is the anisotropic g-factor of the magnetic ion. The g-factors of Cu$^{2+}$ at a square planar coordination site along the $c$-axis ($\|c$ for short) and perpendicular to the $c$-axis ($\bot c$ for short) are written as

$g_\| = 2 + \Delta g_\|$

$$g_\perp = 2 + \Delta g_\perp$$

where $\Delta g_\parallel > \Delta g_\perp$ (approximately, 0.25 vs. 0.05) (**Fig. 4.9a**). $\Delta g_\parallel$ and $\Delta g_\perp$ are proportional to the amounts of unquenched orbital angular momenta,[2] so the associated magnetic moments are also anisotropic, namely,

$$\mu_\parallel = g_\parallel S = (2 + \Delta g_\parallel)S$$
$$\mu_\perp = g_\perp S = (2 + \Delta g_\perp)S$$

Thus, the magnetic moment of the $Cu^{2+}$ ion is greater along the $\parallel z$ direction than along the $\perp z$ direction.

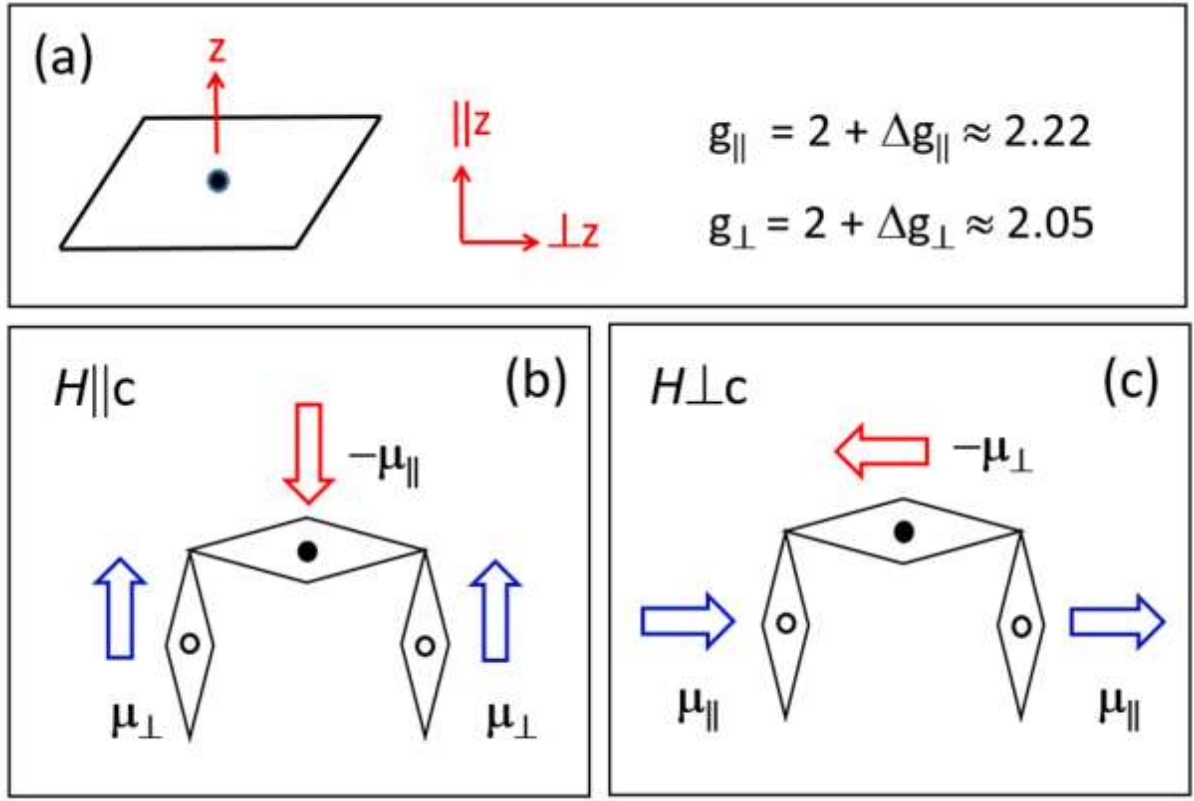

**Fig. 4.9.** (a) Anisotropic g-factors of the $Cu^{2+}$ ion at a square-planar coordination site. (b, c) The moments associated with the (↑↓↑) spin arrangement of a bent ferrimagnetic Cu2-Cu1-Cu2 fragment. The magnetic field is applied along the $\parallel c$ direction in (b), and along the $\perp c$ direction in (c). The filled and unfilled circles in (b) and (c) represent the $Cu1^{2+}$ and $Cu2^{2+}$ ions, respectively.[58]

To simplify our analysis of the observed magnetization anisotropy, we assume that each $Cu_2O_4$ unit is truly square planar in shape, and the planes of the $Cu1O_4$ units are truly orthogonal to the plane of the idealized $Cu2O_4$ unit (**Fig. 4.9b,c**). Then, all three $CuO_4$ square planes of a bent Cu2-Cu1-Cu2 ferrimagnetic fragment are identical except for their spatial arrangement. Since $J_1$ is AFM, the three $Cu^{2+}$ spins of a bent Cu2-Cu1-Cu2 fragment have a (↑↓↑) spin arrangement. For the magnetic field $H\parallel c$, the three spins are aligned along the crystallographic $\parallel c$ direction (**Fig. 4.7b**), so that the magnetic moments on the two up-spin sites are both $\mu_\perp$, while that on the down-spin site is $-\mu_\parallel$ (**Fig. 4.9b**). For the magnetic field $H\perp c$, however, the three spins are aligned along the $\perp c$ direction, so that the magnetic moments on the two up-spin sites are both $\mu_\parallel$, while that on the down-spin site is $-\mu_\perp$ (**Fig. 4.9c**). Therefore, the total moments of the ferrimagnetic fragment are given by

$$\mu_{tot}\,(\|c) = 2\mu_\perp - \mu_\| = (2g_\perp - g_\|)S \approx 1 + (\Delta g_\perp - \Delta g_\|/2) \approx 0.94$$
$$\mu_{tot}\,(\perp c) = 2\mu_\| - \mu_\perp = (2g_\| - g_\perp)S \approx 1 + (\Delta g_\| - \Delta g_\perp/2) \approx 1.195$$

This difference explains why the 1/3-magnetization plateau has the moment larger than 1 $\mu_B$ for $H\perp c$, but the moment smaller than 1 $\mu_B$ for $H\|c$, and why the magnetization plateau deviates more from 1 $\mu_B$ for $H\perp c$ than for $H\|c$.

### 4.1.5. Haldane chain of Cu$_6$ clusters and a 1/3-magnetization plateau in fedotovite K$_2$Cu$_3$O(SO$_4$)$_3$

Fedotovite, K$_2$Cu$_3$O(SO$_4$)$_3$, consists of Cu$_6$ clusters (**Fig. 4.10a**), which are made up of three different Cu atoms, Cu1, Cu2 and Cu3, in strongly distorted square planar coordination. Two Cu3O$_4$ planes are edge-shared to form a twisted Cu$_3$$_2$O$_6$ dimer, and one bridging O atom of this dimer is corner-shared with two Cu1O$_4$ square planes while the other bridging O atom is corner-shared with two Cu2O$_4$ square planes. Thus, the atoms of a Cu$_6$ cluster have the shape of an edge-sharing tetrahedra (**Fig. 4.10b**). Such Cu$_6$ clusters form chains along the $b$-direction (**Fig. 4.10c**).[59]

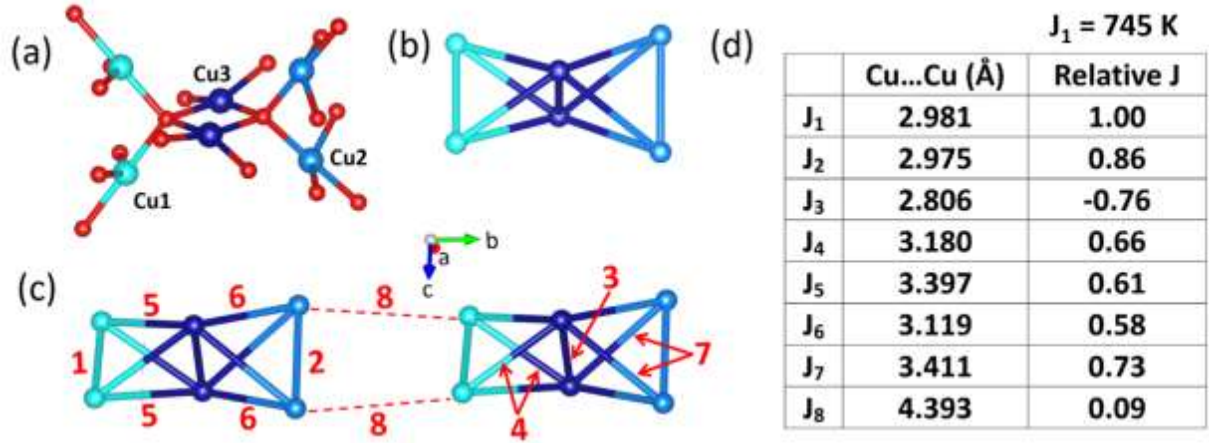

| | Cu...Cu (Å) | Relative J |
|---|---|---|
| $J_1$ | 2.981 | 1.00 |
| $J_2$ | 2.975 | 0.86 |
| $J_3$ | 2.806 | -0.76 |
| $J_4$ | 3.180 | 0.66 |
| $J_5$ | 3.397 | 0.61 |
| $J_6$ | 3.119 | 0.58 |
| $J_7$ | 3.411 | 0.73 |
| $J_8$ | 4.393 | 0.09 |

$J_1 = 745$ K

**Fig. 4.10**. (a) The structure of a Cu$_6$ cluster present in K$_2$Cu$_3$O(SO$_4$)$_3$, which is constructed from distorted Cu1O$_4$, Cu2O$_4$ and Cu3O$_4$ square planes. (b) View of a Cu$_6$ cluster resulting from two Cu$_4$ tetrahedra by edge-sharing. (c) Definitions of the eight spin exchanges $J_1 - J_8$. The labels 1 – 8 refer to $J_1 - J_8$, respectively. (d) Values of the $J_1 - J_8$ determined by DFT+U calculations.

K$_2$Cu$_3$O(SO$_4$)$_3$ undergoes a 3D AFM ordering at $T_N$ = 3.1 K. Above this temperature, K$_2$Cu$_3$O(SO$_4$)$_3$ behaves as an $S$ = 1 Haldane chain system (**Fig. 4.11a**), with each Cu$_6$ cluster acting as an $S$ = 1 species.[60] This implies that the spin exchange coupling between six Cu$^{2+}$ ions of the Cu$_6$ cluster is very strong so that thermal excitations within each Cu$_6$ cluster are weak. In addition, K$_2$Cu$_3$O(SO$_4$)$_3$ exhibits a 1/3-plateau above $T_N$ (**Fig. 4.11b**),[60] implying that each Cu$_6$ cluster forms a ferrimagnetic fragment with (4↑2↓) spin configuration. To confirm this interpretation, we examine the eight spin exchanges $J_1 - J_8$ defined in **Fig. 4.10c**. The values of these exchanges determined by DFT+U calculations are summarized in **Fig. 4.10d** (see Section S9 of the SI). The exchange $J_1$ between the Cu1$^{2+}$ ions is strongly AFM, and so is the exchange $J_2$ between the Cu2$^{2+}$ ions. In contrast, the exchange $J_3$ between the Cu3$^{2+}$ ions is strongly FM. There are four strong AFM exchanges between the Cu1$^{2+}$ and Cu3$^{2+}$ ions (namely, 2$J_4$ + 2$J_5$), and between the Cu2$^{2+}$

and Cu3$^{2+}$ ions (namely, 2J$_6$ + 2J$_7$). Since these AFM interactions dominate over J$_1$ and J$_2$, the energetically favorable spin arrangement for a Cu$_6$ cluster is either a (2↑2↓2↑) or a (2↓2↑2↓) configuration (**Fig. 4.12a**), which are both ferrimagnetic. Due to the AFM inter-cluster exchange J$_8$, the ferrimagnetic Cu$_6$ clusters prefer to couple antiferromagnetically (**Fig. 4.12b**). The gradual increase in the magnetization with magnetic field from 0 to about 20 T is explained by the field-induced breaking of the inter-cluster magnetic bonds J$_8$, one at a time, eventually reaching the ferrimagnetic state (**Fig. 4.12c**), in which all J$_8$ bonds are broken with of $M = M_{sat}/3$.

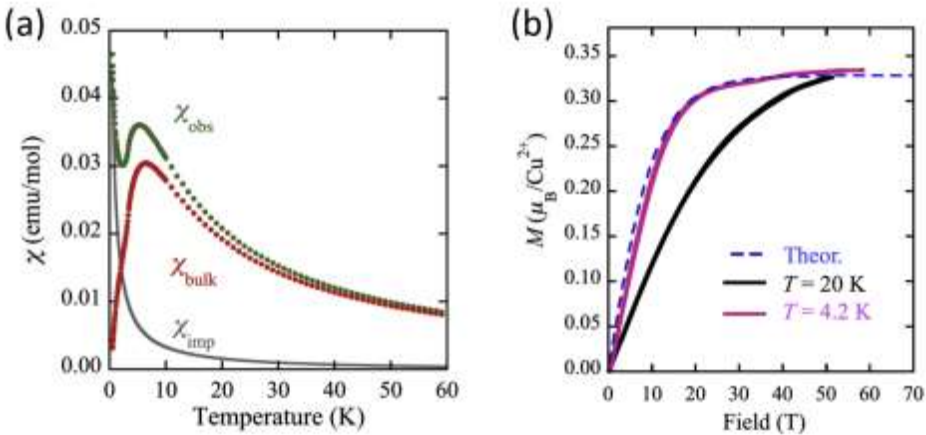

**Fig. 4.11**. (a) Temperature dependence of the magnetic susceptibility $\chi_{bulk}$ (filled red circles) of K$_2$Cu$_3$O(SO$_4$)$_3$ measured at 0.1 T, obtained by subtracting Pascal's diamagnetic contribution $\chi_{dia}$ and an estimated contribution of impurity $\chi_{imp}$ (gray solid line) from the experimental data $\chi_{obs}$ (filled green circles). (b) High-field magnetization at 4.2 K (pink solid line) and 20 K (black solid line). The blue dashed line denotes a theoretical magnetization curve.[60] (Reproduced with permission from reference 60.)

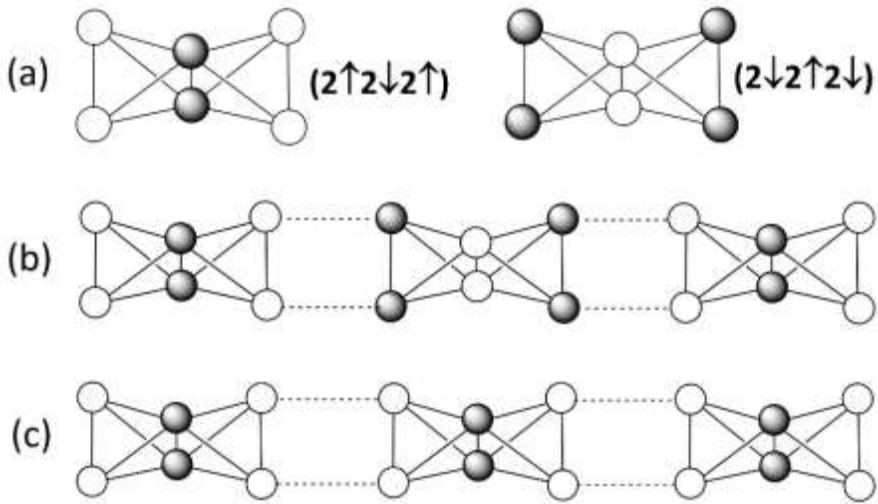

**Fig. 4.12**. (a) Ferrimagnetic state of a Cu$_6$ cluster in K$_2$Cu$_3$O(SO$_4$)$_3$. (b) AFM arrangement of ferrimagnetic Cu$_6$ clusters. (c) Ferrimagnetic arrangement of ferrimagnetic Cu$_6$ clusters.

It should be noted that the magnetic susceptibility of $K_2Cu_3O(SO_4)_3$ is rather weak (**Fig. 4.11a**). This is a direct consequence of the fact that the spin exchanges $J_1 - J_7$ leading to the ferrimagnetic fragment $Cu_6$ are rather strong. The latter is necessary for the effective $S = 1$ behavior of the $Cu_6$ clusters. The magnetic susceptibility of this Haldane chain system is weak despite the presence of six $Cu^{2+}$ cations in each cluster due to the (↑↓↑) arrangement of three FM dimers.

### 4.1.6. Trigonal arrangement of ferromagnetic chains in $Ca_3Co_2O_6$

$Ca_3Co_2O_6$ consists of $Co_2O_6$ chains in which $Co2O_6$ trigonal prisms alternate with $Co1O_6$ octahedra by sharing their triangular faces (**Fig. 4.13a**).[61] These chains running along the $c$-direction have a trigonal arrangement (**Fig. 4.13b**), with $Ca^{2+}$ cations occupying the positions in between these chains. Each $Co_2O_6$ chain is FM[62] so that the spin lattice of $Ca_3Co_2O_6$ can be described as a trigonal lattice by treating each FM chain a pseudo-magnetic ion with giant spin moment. Both Co1 and Co2 atoms of $Ca_3Co_2O_6$ are in the oxidation state of +3,[3,63] indicating that each $Co2O_6$ trigonal prism has six electrons to occupy its d-states, and so does each $Co1O_6$ octahedron. This made it difficult to explain why $Ca_3Co_2O_6$ exhibits a uniaxial magnetism[1,3,63] because the configuration $(3z^2-r^2)^1(xy, x^2-y^2)^0$ predicted for a $Co2O_6$ trigonal prism does not lead to uniaxial magnetism (**Fig. 4.14a**, Left). A systematic study[3] of $Ca_3Co_2O_6$ based on DFT+U and DFT+U+SOC calculations, including geometry relaxations allowing for Jahn-Teller distortions, showed that the uniaxial magnetism of $Ca_3Co_2O_6$ is a consequence of three effects: (a) the FM spin arrangement between the $Co^{3+}$ ions of adjacent $Co2O_6$ and $Co1O_6$ polyhedra, (b) a direct metal-metal interaction between adjacent $Co^{3+}$ ions mediated by their $3z^2-r^2$ orbitals (**Fig. 4.13c**), and (c) the SOC of the $Co^{3+}$ ion at the trigonal prism site (**Fig. 4.14a**, Right).

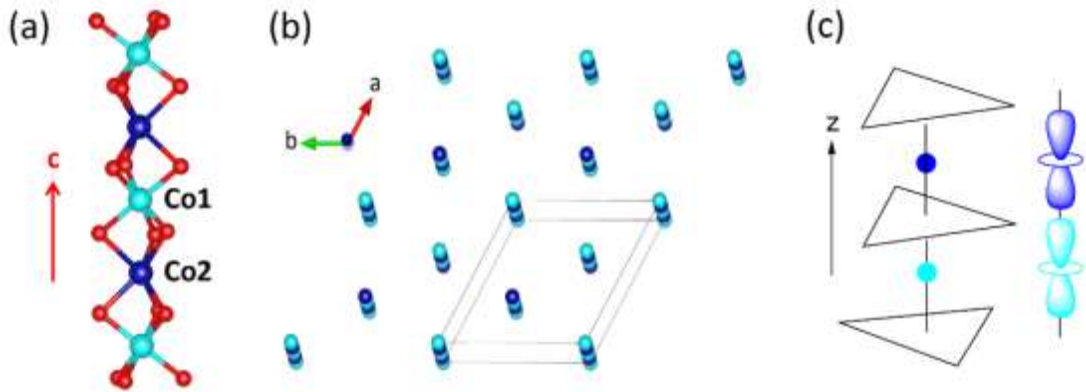

**Fig. 4.13.** (a) An isolated $Co_2O_6$ chain of $Ca_3Co_2O_6$, in which $Co1O_6$ octahedra alternate with $Co2O_6$ trigonal prisms by sharing their triangular faces. (b) Trigonal arrangement of the $Co_2O_6$ chains in $Ca_3Co_2O_6$, where each chain is represented by showing only the Co atoms. (c) The $3z^2-r^2$ orbitals of the Co1 and Co2 atoms in each $Co_2O_6$ chain.

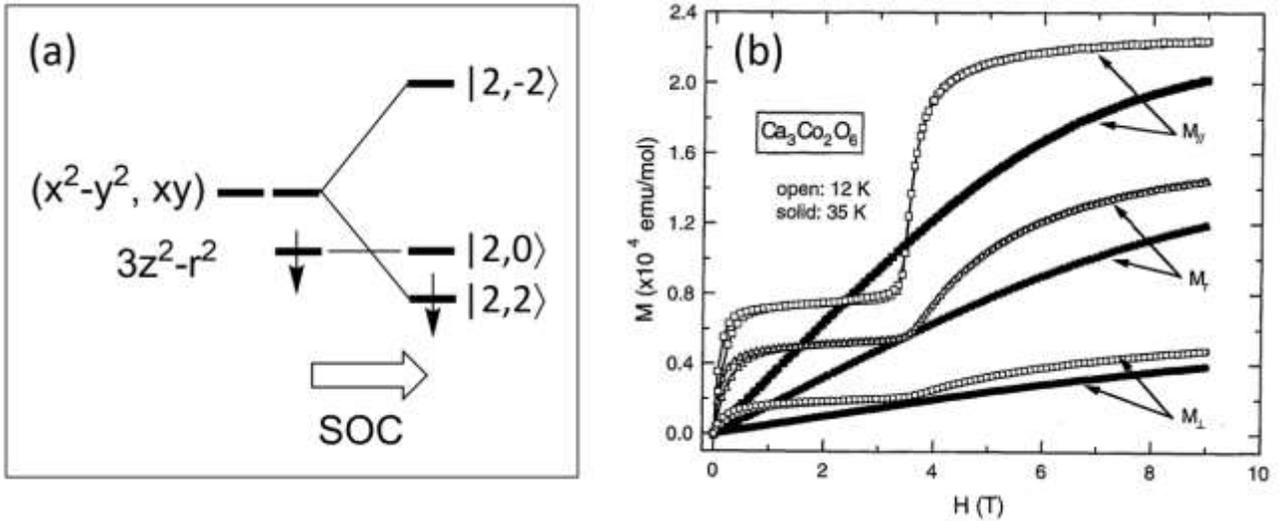

**Fig. 4.14.** (a) Left: Down-spin split d-states of a $Co^{2+}$ ion at the $Co2O_6$ trigonal prism with $(3z^2-r^2)^1(xy, x^2-y^2)^0$ configuration in the absence of SOC in $Ca_3Co_2O_6$. Right: Effect of SOC on the down-spin split d-states. In terms of the spherical harmonics, the angular parts of the $xy$ and $x^2-y^2$ states are given as linear combinations of $|2, 2\rangle$ and $|2, -2\rangle$, and that of $3z^2-r^2$ as $|2, 0\rangle$.[3] (b) Field dependence of the magnetization measured for a single crystal sample ($M_\parallel$ and $M_\perp$) and a powder sample ($M_\Gamma$) of $Ca_3Co_2O_6$ at 12 and 35 K.[62] (Reproduced with permission from reference 62.)

A single crystal sample of $Ca_3Co_2O_6$ exhibits a 1/3-magnetization plateau when the field is parallel to the chain direction, with the magnetization curve showing a step-like feature. When the field is perpendicular to the chain direction, there occurs no magnetization plateau.[62] (Experimentally, it is very difficult to align a single crystal sample of a uniaxial magnet precisely perpendicular to the field. A very slight misalignment can easily give rise to a nonzero magnetization.) As discussed for $CoGeO_3$ in the previous section, these observations are a direct consequence of the fact that $Ca_3Co_2O_6$ is a uniaxial magnet with spin moment along the chain direction. $Ca_3Co_2O_6$ can be described in terms of a regular trigonal spin lattice of simple magnetic ions once each FM $Co_2O_6$ chain is treated as a single magnetic ion (see Section 5). It is noteworthy that $Ca_3Co_2O_6$ reaches a full saturation magnetization at a rather low field (namely, at about 3.5 T). This reflects that the interchain magnetic bonds are weak.

### 4.2. Distorted triangular fragments
### 4.2.1. Diamond chains of $NaFe_3(HPO_3)_2(H_2PO_3)_6$

$NaFe_3(HPO_3)_2(H_2PO_3)_6$ has two nonequivalent Fe atoms, Fe1 and Fe2, forming $Fe1O_6$ and $Fe2O_6$ octahedra.[64] The $HPO_3$ unit occurs in two different forms, i.e., H-$PO_3$ and $PO_2(OH)$, but the $H_2PO_3$ unit only in the form H-$PO_2(OH)$. Consequently, both Fe1 and Fe2 atoms are present as $Fe^{3+}$ (S = 5/2) ions. These $Fe^{3+}$ ions are bridged by H-$PO_3$, $PO_2(OH)$, or H-$PO_2(OH)$, as illustrated by **Fig. 4.15a**. DFT+U calculations[65] showed that four AFM spin exchanges (i.e., $J_2$, $J_3$, $J_4$ and $J_6$ depicted in **Fig. 4.15a**) are relevant and comparable in magnitude (~2K). The three spin exchanges $J_2$, $J_3$ and $J_6$ couple the $Fe^{3+}$ cations to diamond chains, which are interlinked by $J_4$ to form 2D layers (**Fig. 4.15b**). Such layers are stacked to form the 3D spin lattice of $NaFe_3(HPO_3)_2(H_2PO_3)_6$. In addition, there are weak inter-layer AFM spin exchanges (see below for further discussion).

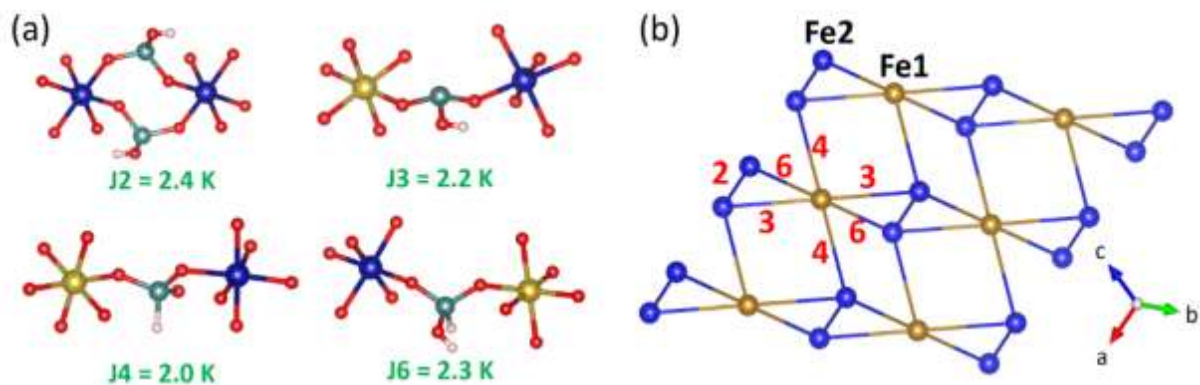

**Fig. 4.15**. (a) Geometrical arrangements associated with the spin exchange paths, $J_2$, $J_3$, $J_6$ and $J_4$ in $NaFe_3(HPO_3)_2(H_2PO_3)_6$. All these exchanges are of the Fe-O…O-Fe type with the O…O contact making a O…$P^{5+}$…O bridge. (b) 2D spin lattice of $NaFe_3(HPO_3)_2(H_2PO_3)_6$ made up of the spin exchanges $J_2$, $J_3$, $J_6$ and $J_4$, which are indicated by the labels 2, 3, 6 and 4, respectively.

As shown in **Fig. 4.16a** (inset), $NaFe_3(HPO_3)_2(H_2PO_3)_6$ undergoes a ferrimagnetic ordering below $T_C$ = 9.5 K and exhibits a 1/3-magnetization plateau in the magnetization.[65] The plateau extends to ~8 T. Above this field, the magnetization increases linearly with field until the saturation is reached at ~27 T. As discussed in Section 2.4, we suppose that the spin lattice of $NaFe_3(HPO_3)_2(H_2PO_3)_6$ undergoes field-induced partitioning into ferrimagnetic triangular clusters (**Fig. 4.16b**). Then, the spin arrangement, (↑↓↑), (↑↑↓) or (↓↑↑), of each cluster leads to one positive moment per cluster. Among these three, the (↑↓↑) arrangement at each triangular fragment is energetically most favorable because of the inter-diamond spin exchange $J_4$, thereby leading to the ferrimagnetic state with $M = M_{sat}/3$ (**Fig. 4.16c**). To increase the magnetization beyond $M_{sat}/3$, each ferrimagnetic triangle must break two magnetic bonds (**Fig. 4.16d**) within a cluster, which is accompanied by the breaking of two inter-diamond $J_4$ bonds. This needs high enough magnetic field, hence explaining the 1/3-plateau extending to ~8 T. When the field increases beyond 8 T toward the saturation magnetization, each ferrimagnetic triangle begins to break two magnetic bonds (**Fig. 4.16d**) within a cluster.

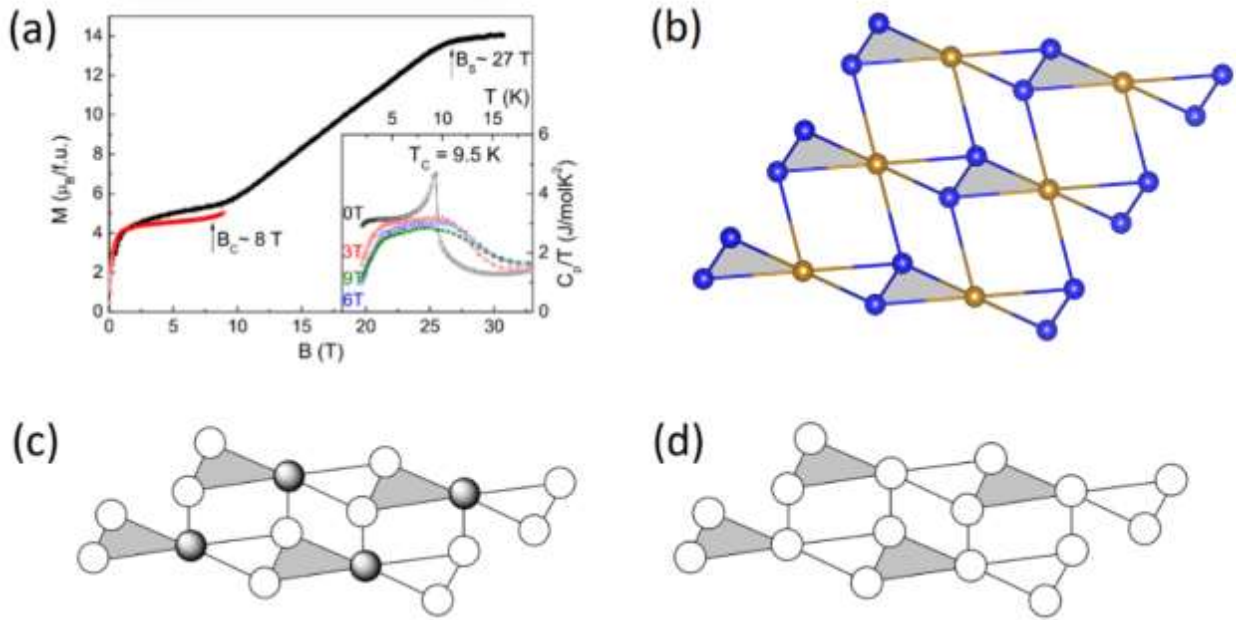

**Fig. 4.16**. (a) Magnetization of $NaFe_3(HPO_3)_2(H_2PO_3)_6$ in static field up to 9 T and pulsed field up to 32 T. The inset shows the temperature dependence of the specific heat $C_p/T$ taken at various magnetic fields.[65] (Reproduced with permission from reference 65.) (b) Spin lattice of $NaFe_3(HPO_3)_2(H_2PO_3)_6$ in terms of triangular ferrimagnetic fragments. (c) Ferrimagnetic ground state of a layer made up of the exchange paths $J_2$, $J_3$, $J_6$ and $J_4$ in $NaFe_3(HPO_3)_2(H_2PO_3)_6$, which has diamond chains (defined by $J_2$, $J_3$ and $J_6$) antiferromagnetically coupled (via $J_4$). (d) Spin arrangement in the FM state reached at magnetic saturation.

In general, the ground state of a magnet composed of ferrimagnetic layers is AFM because the weak high-spin orbital interactions between adjacent ferrimagnetic layers favor an AFM coupling rather than an FM coupling.[66] Indeed, DFT+U calculations found[65] that the interlayer spin exchanges $J_1$ and $J_5$, which are weakly AFM (~0.6 and ~0.4 K, respectively), and form spin-frustrated ($J_1$, $J_5$, $J_4$) triangles between adjacent ferrimagnetic layers as depicted in **Fig. 4.17a**. The interlayer FM coupling (**Fig. 4.17b**) leads to the ($J_1$, $J_5$, $J_4$) triangles, which have $J_5$ magnetic bonds and $J_1$ broken magnetic bonds. In contrast, the interlayer AFM coupling (**Fig. 4.17c**) leads to the ($J_1$, $J_5$, $J_4$) triangles, which have $J_5$ broken magnetic bonds and $J_1$ magnetic bonds. Since $J_1$ is slightly stronger than $J_5$, the interlayer AFM coupling is energetically favored over the interlayer FM coupling. This is consistent with the general observation that the magnetic ground state of a magnet composed of ferrimagnetic layers is AFM rather than ferrimagnetic. Furthermore, we note from **Fig. 4.16a** that, below ~2 T, the magnetization rises sharply with field toward $M_{sat}/3$. This observation can be related to the breaking of the weak interlayer magnetic bonds (i.e., $J_5$ and $J_1$) in the AFM ground state. It will be interesting to examine whether the magnetic ground state of $NaFe_3(HPO_3)_2(H_2PO_3)_6$ is AFM or ferrimagnetic.

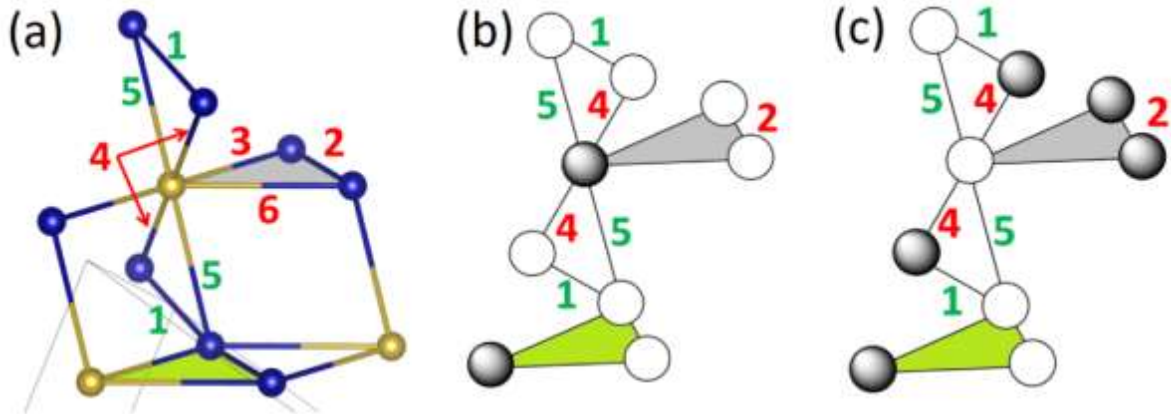

**Fig. 4.17**. (a) Spin exchanges $J_1$ and $J_5$ between adjacent layers of diamond chains linked by $J_4$ in $NaFe_3(HPO_3)_2(H_2PO_3)_6$. The ferrimagnetic triangular clusters belonging to two different layers are marked with different colors. The red labels 2, 3, 6 and 4 refer to the spin exchanges $J_2$, $J_3$, $J_6$ and $J_4$ of one layer, respectively. The green labels 1 and 5 refer to the interlayer spin exchanges $J_1$ and $J_5$, respectively. (b) Interlayer FM coupling between adjacent ferrimagnetic layers leading to $J_5$ magnetic bonds and $J_1$ broken magnetic bonds. (c) Interlayer AFM coupling between adjacent ferrimagnetic layers leading to $J_5$ broken magnetic bonds and $J_1$ magnetic bonds.

### 4.2.2. Three-dimensional spin lattice and anisotropic plateau width in azurite $Cu_3(CO_3)_2(OH)_2$

The important structural building blocks of azurite $Cu_3(CO_3)_2(OH)_2$[67] are the $Cu1O_4$ and $Cu2O_4$ square planes containing their $x^2-y^2$ magnetic orbitals. Each $Cu1^{2+}$ ion is surrounded by four $Cu2^{2+}$ ions to form a $Cu_5$ ribbon (**Fig. 4.18a**), where the four $Cu2O_4$ square planes of a $Cu_5$ ribbon are nearly perpendicular to the central $Cu1O_4$ square plane. This structural feature implies that the orientations of the $x^2-y^2$ magnetic orbitals are the key to understanding the magnetic properties and especially the magnetization plateau observed for azurite. By sharing their edges, such $Cu_5$ ribbons form 'diamond chains' along the $b$-direction (**Fig. 4.18b**). Each $Cu_5$ ribbon is described by three spin exchanges $J_1 – J_3$, as used early on by Rule et al. to discuss the temperature dependence of the magnetic susceptibility.[68] Kang et al. carried out DFT+U calculations[69] to find that adjacent diamond chains interact through the spin exchanges $J_4$, which take place through the bridging $CO_3^{2-}$ ions (**Fig. 4.18c**), to form a layer of interacting diamond chains, and that the dimer exchange $J_2$ (= 363 K) dominates with $J_1/J_2 \approx J_3/J_2 = 0.24$, and $J_4/J_2 = 0.13$. Jeschke et al. proposed a modified diamond chain model by including a spin exchange between the Cu1 cations.[70] Topologically, the 2D spin lattice of azurite is identical with that for $NaFe_3(HPO_3)_2(H_2PO_3)_6$ discussed in the previous section. So, one might expect that each layer of azurite is ferrimagnetic as shown in **Fig. 4.16c,** and such ferrimagnetic layers are antiferromagnetically coupled to form an ordered 3D AFM state, and that azurite exhibits a 1/3-plateau as $NaFe_3(HPO_3)_2(H_2PO_3)_6$ does. Indeed, azurite orders antiferromagnetically at $T_N \approx 1.9$ K[71] and exhibits a 1/3-magnetization plateau below this temperature, as shown in **Fig. 4.19**.

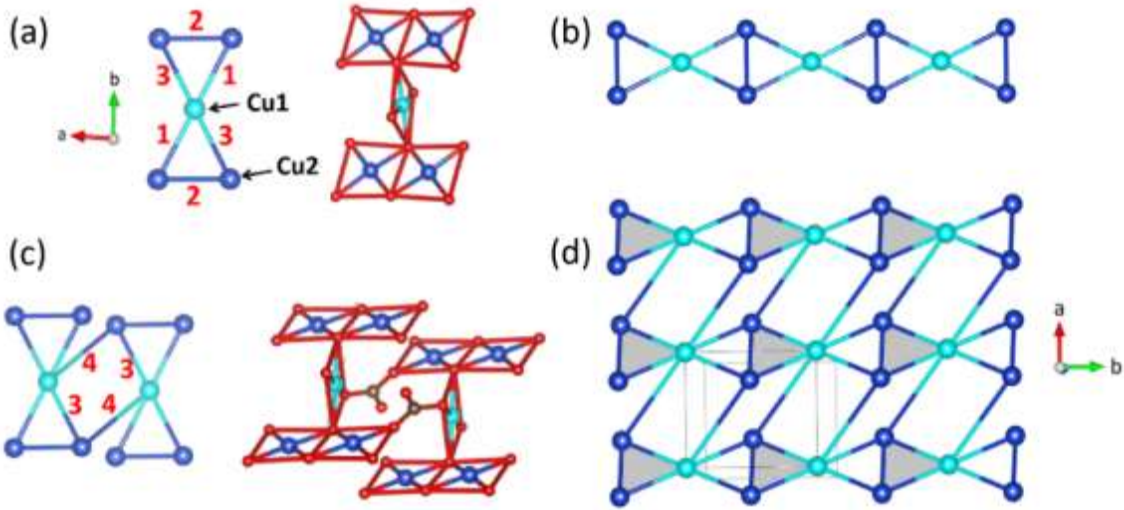

**Fig. 4.18.** (a) (left) A $Cu_5$ ribbon made up of one Cu1 and four Cu2 atoms in azurite $Cu_3(CO_3)_2(OH)_2$, where the labels 1 – 3 refer to the spin exchange paths $J_1 - J_3$, respectively. In this ribbon, the four $Cu2O_4$ square planes are nearly orthogonal to the $Cu1O_4$ square plane (right). (b) One diamond chain made up of edge-sharing ribbons. (c) Arrangement between two $Cu_5$ ribbons leading to the inter-ribbon exchanges $J_4$, which occur through a $CO_3$ bridge (right). (d) A layer of diamond chains parallel to the ab plane made up of edge-sharing $Cu_5$ ribbons.

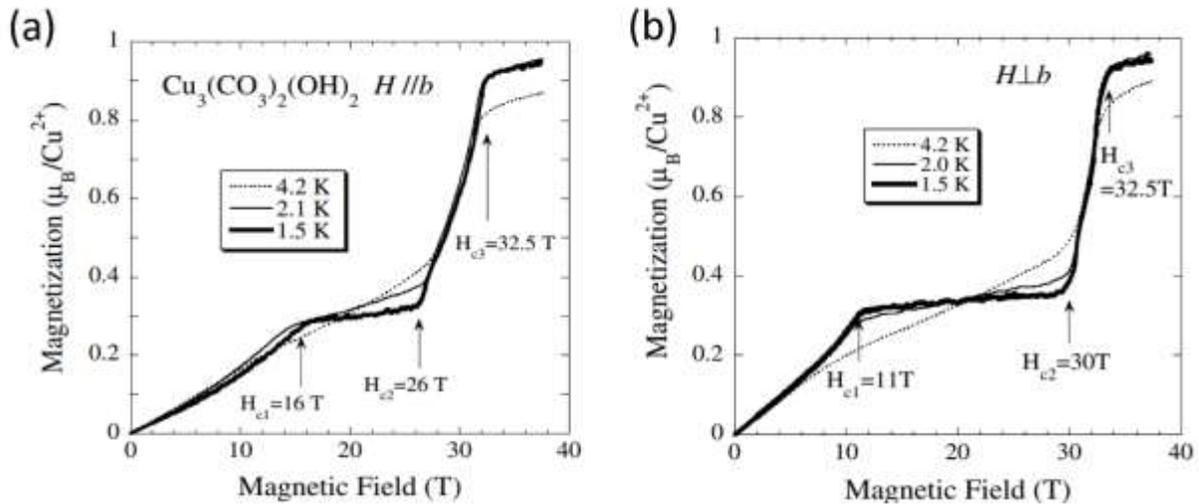

**Fig. 4.19.** (a) Field dependence of magnetization of $Cu_3(CO_3)_2(OH)_2$ for $H\|b$. (b) Field dependence of magnetization in $Cu_3(CO_3)_2(OH)_2$ for $H\perp b$.[71] (Reproduced with permission from reference 71.)

The 1/3-plateau of azurite presents two features remarkably different from that of $NaFe_3(HPO_3)_2(H_2PO_3)_6$ (**Fig. 4.16a**): (1) The field $H_{c1}$ where the $M = M_{sat}/3$ point starts on increasing the field from zero is much greater for azurite (over 10 T) than for $NaFe_3(HPO_3)_2(H_2PO_3)_6$ (~1 T). In $NaFe_3(HPO_3)_2(H_2PO_3)_6$, the gradual increase in $M$ with $\mu_0H$ in the region of $0 - H_{c1}$ is ascribed to the breaking of the inter-layer magnetic bonds. Since $H_{c1}$ is much higher for azurite, the interlayer AFM spin exchange of azurite must be substantial. (2) The width of the 1/3-plateau is much wider when the field is perpendicular to the b-axis ($\perp b$) than

parallel to the $b$-axis ($\|b$); the $H_{c1}$ is greater for $H\|b$ than for $H\perp b$ (16 vs. 11 T), whereas the $H_{c2}$ is smaller for $H\|b$ than for $H\perp b$ (26 vs. 30 T). Thus, the plateau width is significantly larger for $H\perp b$. However, in contrast to these differences in the plateau widths, the saturation fields $H_{c3}$ in both orientations are identical (32.5 T).[71] In the following we examine why these observations occur.

### A. Interlayer spin exchange in azurite

2D layers of interlinked diamond chains are stacked as depicted in **Fig. 4.20a**. There occur two Cu-O…O-Cu type spin exchange paths, $J_5$ and $J_6$, between adjacent layers (**Fig. 4.20b**). The $J_5$ paths take place between Cu1$^{2+}$ and Cu2$^{2+}$ ions (**Fig. 4.20c**), and the $J_6$ paths between two Cu2$^{2+}$ ions (**Fig. 4.20d**). The values of $J_5$ and $J_6$, determined by using the energy-mapping analysis (see Section S10 of the SI) are not negligible compared with the inter-diamond exchange $J_4$ within a layer; $J_5$ is AFM while $J_6$ is FM, and $J_5/J_4 = 0.7$ and $J_6/J_4 = -0.5$. The presence of the AFM interlayer exchange $J_5$, which is only slightly weaker than $J_4$, confirms our suggestion that the increase of magnetization with field in the $0 - H_{c1}$ region is related to the breaking of the inter-layer magnetic bonds, and azurite reaches the state consisting of ($\uparrow\downarrow\uparrow$) ferrimagnetic triangular fragments at $H_{c1}$.

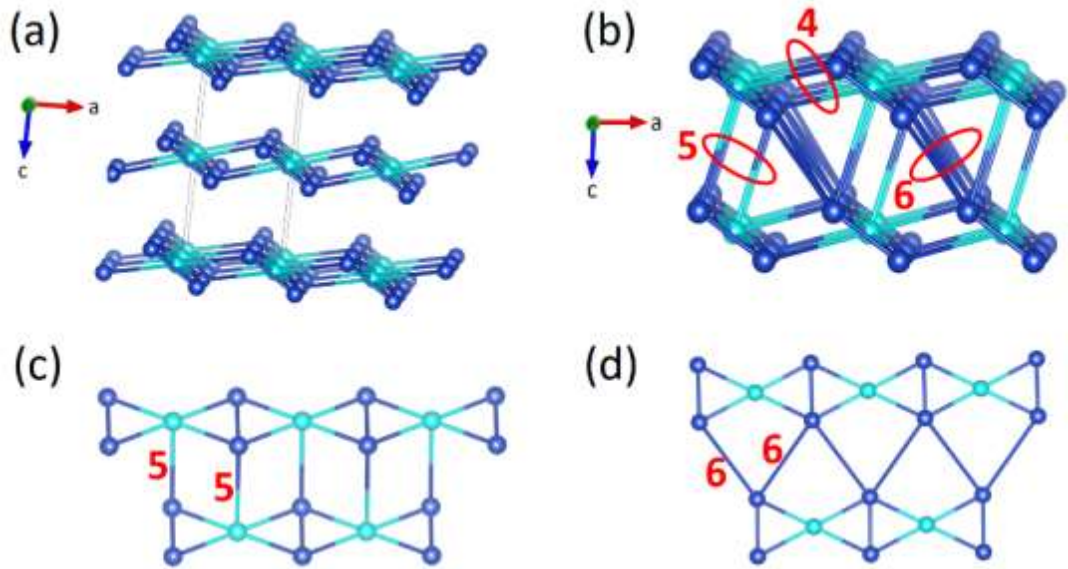

**Fig. 4.20**. (a) Stacking of 2D layers made up of interlinked diamond chains in Cu$_3$(CO$_3$)$_2$(OH)$_2$. (b) Two 2D layers with interlayer spin exchange paths $J_5$ and $J_6$. (c) Interlayer exchange paths $J_5$ between Cu1 and Cu2 atoms. (d) Interlayer exchange paths $J_6$ between two Cu2 atoms. The labels 4 – 6 refer to the spin exchange paths $J_4 – J_6$, respectively.

### B. Magnetic anisotropy affecting Dzyaloshinskii-Moriya (DM) interactions

The different widths of the 1/3 plateaus were explained by Kikuchi et al.[71] in terms of DM interactions by assuming a DM vector perpendicular to both the $J_2$ bond and the $b$-axis. So far, however, it is unknown why such a DM vector should exist in azurite, or why the 1/3-plateau starts at a higher field for $H\|b$ than for $H\perp b$. To resolve these questions, we examine how the Zeeman energies of the Cu$^{2+}$ ions in azurite are affected by the field direction based on the following three observations:

1) The g-factor for the $Cu^{2+}$ ion of a $CuO_4$ square plane is anisotropic; the g-factor along the four-fold rotational axis, $g_{\parallel} = 2 + \Delta g_{\parallel} \approx 2.25$, is substantially greater than that perpendicular to this axis, $g_{\perp} = 2 + \Delta g_{\perp} \approx 2.05$.[72]

2) In general, the g-factor of a magnetic ion measured with magnetic field $H$ in a certain direction can be written as $g = 2 + \Delta g$, where $\Delta g$ is related to the unquenched orbital moment $\delta L$ on the magnetic ion along that direction as[2]

$$\Delta g = \lambda \frac{\delta L}{\mu_B H} \propto \delta L, \qquad (4.1)$$

where $\lambda$ is the SOC constant of the magnetic ion, i.e., $\Delta g$ is a measure of $\delta L$.

3) In a DM interaction $\vec{D}_{ab} \cdot (\vec{S}_a \times \vec{S}_b)$ between two spins located at the sites $a$ and $b$ and coupled by spin exchange $J_{ab}$, the DM vector $\vec{D}_{ab}$ is related to the unquenched orbital moments $\delta\vec{L}_a$ and $\delta\vec{L}_b$ of the magnetic ions at the sites a and b, respectively, as[2,73]

$$\vec{D}_{ab} = \lambda J_{ab}(\delta\vec{L}_a - \delta\vec{L}_b) \propto (\Delta\vec{g}_a - \Delta\vec{g}_b) \qquad (4.2)$$

The essential key to understanding the observation of the different widths of the plateaus in azurite is that each $Cu1O_4$ square plane is nearly perpendicular to the four $Cu2O_4$ square planes within each diamond chain, and also nearly perpendicular to the two $Cu2O_4$ square planes between two adjacent diamond chains (**Fig. 4.21a**). To simplify our analysis, we assume that the $Cu1O_4$ and $Cu2O_4$ units have an ideal planar square shape, and that the arrangement of these square plane are ideally orthogonal such that the two edges of the $Cu1O_4$ plane are aligned along the y- and z-axes, but those of the $Cu2O_4$ planes along the x- and y-axes (**Fig. 4.21b**). Then, the four-fold rotational axis of the $Cu1O_4$ plane is parallel to the x-axis ($\parallel x$), but that of each $Cu2O_4$ plane is parallel to the z-axis ($\parallel z$). With this choice of the Cartesian coordinate system, the y-direction is approximately aligned along the b-direction, i.e., the diamond chain direction. Then, for the $Cu1O_4$, the g-factor of the $Cu^{2+}$ cations is $g_{\parallel}$ for $H\parallel x$, but $g_{\perp}$ for $H\perp x$ (**Fig. 4.21c**). For the $Cu2O_4$ planes, however, the g-factor of the $Cu^{2+}$ is $g_{\parallel}$ for $H\parallel z$, but $g_{\perp}$ for $H\perp z$ (**Fig. 4.21c**).

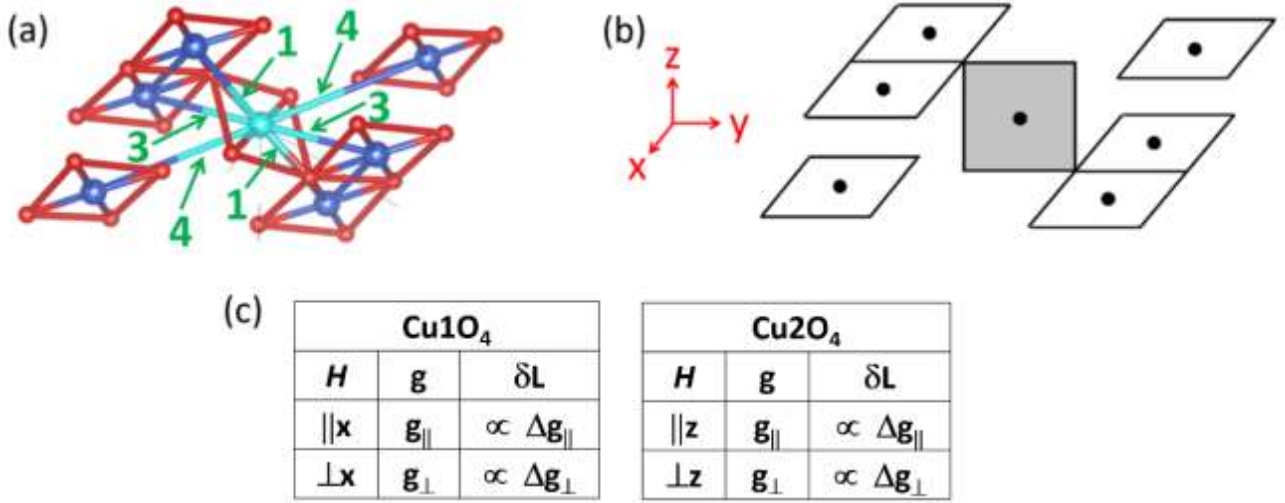

| $Cu1O_4$ | | | $Cu2O_4$ | | |
|---|---|---|---|---|---|
| H | g | $\delta L$ | H | g | $\delta L$ |
| $\parallel x$ | $g_{\parallel}$ | $\propto \Delta g_{\parallel}$ | $\parallel z$ | $g_{\parallel}$ | $\propto \Delta g_{\parallel}$ |
| $\perp x$ | $g_{\perp}$ | $\propto \Delta g_{\perp}$ | $\perp z$ | $g_{\perp}$ | $\propto \Delta g_{\perp}$ |

**Fig. 4.21**. (a) One Cu1$^{2+}$ ion making 2J$_1$ + 2J$_3$ + 2J$_4$ exchange bonds with six adjacent Cu2$^{2+}$ in azurite. (b) Idealized description associated with the 2J$_1$ + 2J$_3$ + 2J$_4$ exchange bonds. The idealized Cu1O$_4$ (shaded) and Cu2O$_4$ units (unshaded) units are treated as ideal square planes with four-fold rotation symmetry, with the edges of Cu1O$_4$ parallel to the x- and z-axes, and those of Cu2O$_4$ parallel to the x- and y-axes. (c) g-factors of and the amount of unquenched orbital moment on the Cu$^{2+}$ ion in the Cu1O$_4$ and Cu2O$_4$ square planes.

Using the results summarized in **Fig. 4.21c**, the Zeeman energy for the three Cu$^{2+}$ ions of each ferrimagnetic triangle (namely, one Cu1$^{2+}$ and two Cu2$^{2+}$ ions, **Fig. 4.18d**) is calculated as follows:

For $H\|x$: $\quad E_Z = (g_\| + 2g_\perp)\mu_0\mu_B HS$
For $H\|y$: $\quad E_Z = 3g_\perp\mu_0\mu_B HS$
For $H\|z$: $\quad E_Z = (2g_\| + g_\perp)\mu_0\mu_B HS$

For the Cu$^{2+}$ ion, $g_\perp < g_\|$ (i.e., ~2.05 vs. ~2.25). Thus, at a given magnetic field strength $\mu_0 H$, the Zeeman energy is lower for $H\|y$ than either for $H\|x$ or for $H\|z$. This implies that in reaching the energy required for breaking a certain magnetic bond, a higher magnetic field is necessary when the field is aligned along the y-direction. This explains why the $H_{c1}$ is higher for $H\|b$ than for $H\perp b$ (16 vs. 11 T) since the y-direction is approximately aligned along the b-direction of azurite. Interestingly, this identifies the magnetic bonds to break in this process are the inter-layer magnetic bonds, i.e., the different widths of the 1/3 plateaus in azurite is ultimately a consequence of interlayer exchange coupling.

We now examine why $H_{c1}$ is lower for $H\|b$ than for $H\perp b$ (i.e., 26 vs. 30 T) by noting that the $H_{c2}$ marks the point where each (2↑1↓) ferrimagnetic triangle of **Fig. 4.18d** begins to change into a (3↑0↓) ferromagnetic triangle. This change breaks six magnetic bonds (namely, 2J$_1$ + 2J$_3$ +2J$_4$) around a Cu1$^{2+}$ ion. As pointed earlier, J$_1$/J$_2$ ≈ J$_3$/J$_2$ ≈ 0.24 and J$_4$/J$_2$ ≈ 0.13, so (2J$_1$ + 2J$_3$ +2J$_4$) ≈ 0.74J$_2$. The DM interactions of the six magnetic bonds are identical except for the magnetic bond strengths. Therefore, we treat the six DM interactions involving one Cu1$^{2+}$ ion as one DM interaction of the Cu1$^{2+}$ ion with a hypothetical Cu2$^{2+}$ ion with effective bond J$_{eff}$ = 0.74J$_2$. Then, by considering that the Cu1$^{2+}$ and the hypothetical Cu2$^{2+}$ ions at sites $a$ and $b$, respectively, we obtain the following results,

For $\mu_0 H\|x$: $\quad D_{ab} \propto \lambda J_{eff}(\Delta g_\| - \Delta g_\perp) \approx 0.2\lambda J_{eff} < 0$
For $\mu_0 H\|y$: $\quad D_{ab} \propto \lambda J_{eff}(\Delta g_\perp - \Delta g_\perp) \approx 0$
For $\mu_0 H\|z$: $\quad D_{ab} \propto \lambda J_{eff}(\Delta g_\perp - \Delta g_\|) \approx -0.2\lambda J_{eff} > 0$,

where we used the fact that $\lambda < 0$ for Cu$^{2+}$ with more than half-filled d-shell. The above results show that the DM interaction vanishes for $H\|y$. The DM vector for $H\|x$ is opposite in sign to that for $H\|z$. For $H\|z$, the DM interaction raises the Zeeman energy, so the magnetic bond breaking occurs at a lower field (compared with the $H\|y$ case). For $H\|x$, however, the DM interaction lowers Zeeman energy, forcing the magnetic bond breaking to a higher field. What is observed for azurite can be understood if the $\perp b$ direction is close to the $\|x$ direction. The $\|z$ direction is also approximately the $\perp b$ direction, but the DM interaction for $H\|z$ raises the Zeeman energy while that for $H\|x$ lowers it. This leads to the prediction that the plateau widths increase in the order,

$H\|z < H\|y < H\|x$.

It would be interesting to verify this prediction experimentally.

## 5. Trigonal vs. kagomé magnets

The magnetization plateaus of magnets with triangular[74] and kagomé[75-80] spin lattices have received more attention in theoretical studies than in experimental studies. These plateaus are less prominent compared with those found for other magnets of lower symmetry.

### 5.1. Cause for the presence or absence of a clear-cut 1/3-magnetization plateau

Magnets of triangular and kagomé spin lattices show contrasting behaviors in their magnetization, especially, in the development of magnetization plateaus. The 1/3-magnetization plateaus observed for trigonal spin-lattice magnets are generally narrow in their widths.[81,82] In the case of kagomé spin lattice magnets, it is often difficult to detect magnetization plateaus in terms of their $M$ vs. $H$ curves. Therefore, sometimes d$M$/d$H$ vs. $H$ plots have been employed to discuss the plateaus.[83-85] However, 1/3-plateaus are clearly observed in their $M$ vs. $H$ plots for trigonal spin lattice magnets. In the following, we examine the probable cause of this difference by regarding the kagomé and trigonal spin lattices as made up of non-overlapping ferrimagnetic fragments, namely, ferrimagnetic triangles indicated by shading in **Fig. 5.1a** and **5.1b**, respectively. As discussed in Section 2.4, each ferrimagnetic triangle can assume three different spin arrangements (**Fig. 5.1c**). Then, all possible ordered and disordered spin configurations representing the 1/3-magnetization plateau are generated by how each ferrimagnetic triangle adopts one of the three spin arrangements. For example, **Fig. 5.2** shows three ordered spin arrangements creating the 1/3-plateau state for a kagomé spin lattice, and **Fig. 5.3** those for a trigonal spin lattice. To probe the question of whether kagomé and trigonal spin lattices have a 1/3-magnetization plateau, we note that the magnetization of the whole spin lattice remains at $M_{sat}/3$ regardless of whether there are more or fewer inter-fragment magnetic bonds. Thus, in the following, we examine the most and least stable arrangements that a given (↑↓↑) ferrimagnetic fragment can have with the surrounding ferrimagnetic fragments.

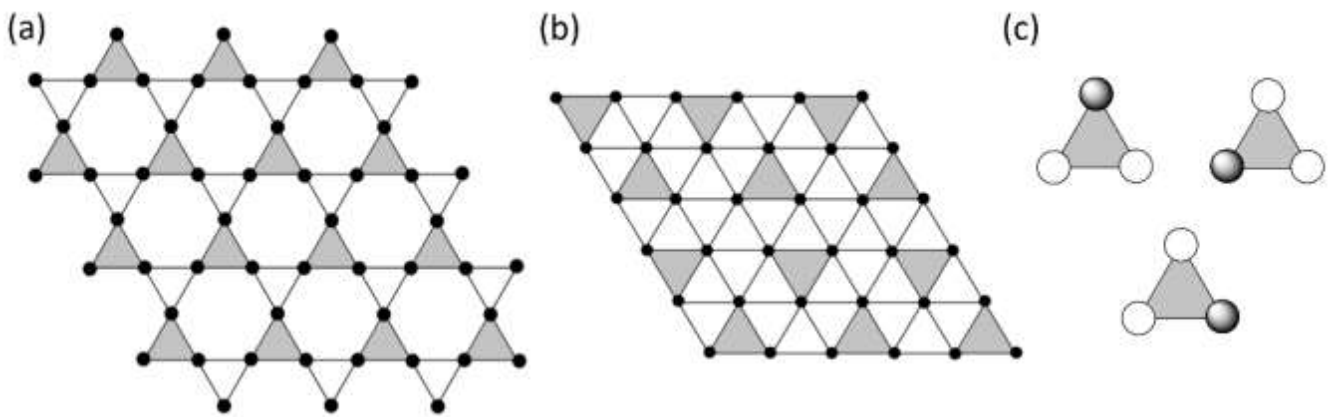

**Fig. 5.1.** Fragmentation of (a) a kagomé and (b) a trigonal spin lattice into non-overlapping ferrimagnetic triangles. (c) Three possible spin arrangements of a ferrimagnetic triangle, where the up-spin and down-spin sites are indicated by unshaded and shaded circles, respectively.

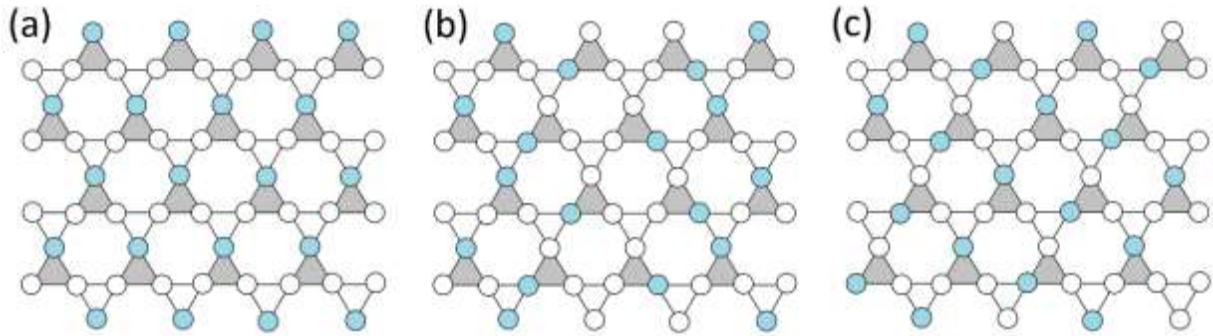

**Fig. 5.2.** Three ordered spin arrangements representing the 1/3-magnetization plateau state of a kagomé spin lattice.

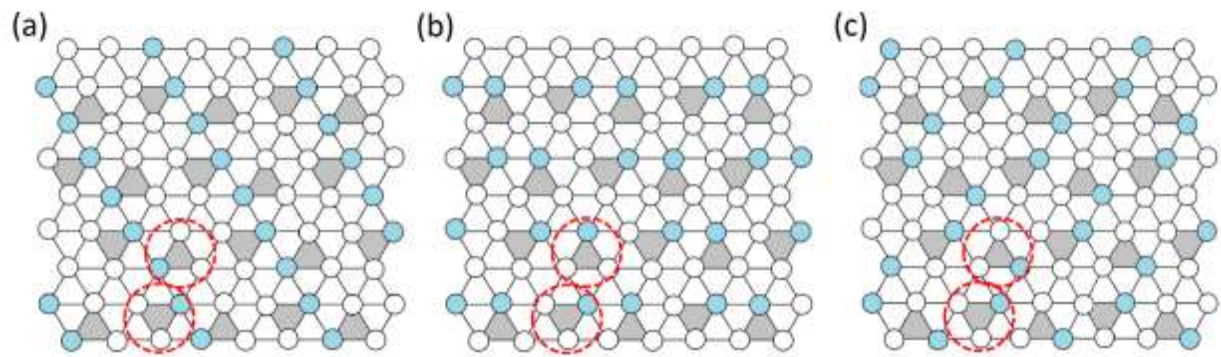

**Fig. 5.3.** Three ordered spin arrangements representing the 1/3-magnetization plateau state of a trigonal spin lattice. The spin arrangements of the two nonequivalent ferrimagnetic triangles are encircled for clarity.

Let us first examine possible spin arrangements around one ferrimagnetic fragment in a kagomé spin lattice. As depicted **Fig. 5.4a**, each shaded triangle, representing a ferrimagnetic fragment, is corner-shared with three unshaded triangles. The two sites on each edge of the unshaded triangle belong to two different ferrimagnetic fragments (see **Fig.5.2**) so that each ferrimagnetic fragment interacts with six different ferrimagnetic neighboring fragments. A chosen ferrimagnetic fragment makes the most stable inter-fragment spin arrangement by making six inter-fragment magnetic bonds (**Fig. 5.4b**), and the least stable spin arrangement by making six inter-fragment broken bonds (**Fig. 5.4c**). Obviously, it is not possible for every ferrimagnetic fragment to make six magnetic bonds with the six adjacent ferrimagnetic fragments. For, in making six bonds (broken bonds) with a chosen fragment, the six surrounding ferrimagnetic fragments should possess specific spin arrangements. These arrangements cannot be altered to make six bonds (broken bonds) for another ferrimagnetic fragment next to the chosen fragment. This means that there is a variation in the number of inter-fragment magnetic bonds each ferrimagnetic fragment can make, from six bonds to six broken bonds. The kagomé spin lattice as a whole is more (less) stable if it has more inter-fragment bonds (broken bonds) in average, implying that a good indicator for the width of the 1/3-plateau is the energy difference between the most and the least stable inter-fragment magnetic bonding. This energy difference corresponds to effectively 12 magnetic bonds, i.e., from six bonds (**Fig. 5.4b**) to six broken bonds (**Fig. 5.4c**) per ferrimagnetic fragment.

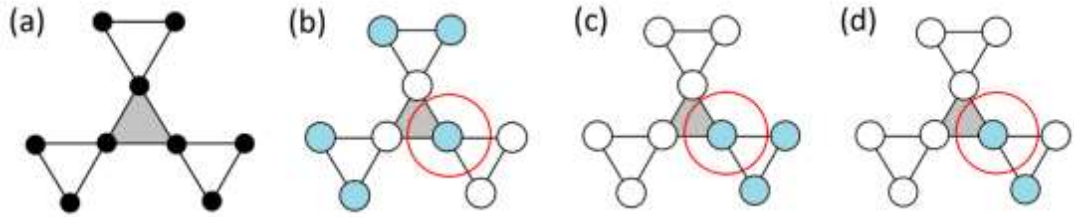

**Fig. 5.4.** (a) Arrangement of three unshaded triangles around a ferrimagnetic triangle, indicated by shading, in a kagomé spin lattice. (b) The most stable arrangement of six ferrimagnetic triangles around one ferrimagnetic fragment. (c) The least stable arrangement of six ferrimagnetic triangles around one ferrimagnetic fragment. (d) An arrangement that makes four broken bonds and requires the breaking of two bonds for the (↑↓↑) to (↑↑↑) spin flipping. The red circles indicate the down-spin site to go through (↑↓↑) to (↑↑↑) spin flip.

In a trigonal spin lattice, each ferrimagnetic fragment of a trigonal lattice is surrounded by 12 unshaded triangles and interacts with six adjacent ferrimagnetic fragments (**Fig. 5.5a**). Three of these six make interactions through a corner, and the remaining three through an edge (indicated by red rectangles in **Fig. 5.5a**). That is, in a trigonal spin lattice as well, a given ferrimagnetic fragment is surrounded by six ferrimagnetic fragments. In the interactions through a corner, the corner spin site can be either up-spin or down-spin. In the interactions through an edge, the two spins on the edge can be both up-spins or a combination of one up-spin and one down-spin, because this edge is a part of a (↑↓↑) triangle. Consequently, with six adjacent ferrimagnetic fragments, a ferrimagnetic fragment makes nine bonds and three broken bonds (i.e., effectively six bonds) in the most stable spin arrangement (**Fig. 5.5b**), but two bonds and 10 broken bonds (i.e., effectively, eight broken bonds) in the least stable arrangement (**Fig. 5.5c**). Thus, the energy difference between the most stable and the least stable arrangements is effectively 14 bonds.

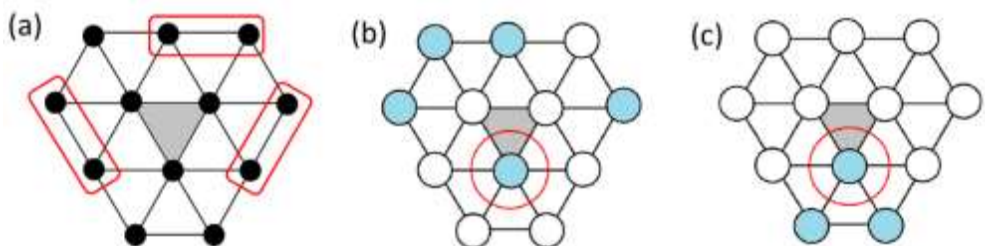

**Fig. 5.5.** (a) Arrangement of 12 unshaded triangles around a ferrimagnetic triangle, indicated by shading, in a trigonal spin lattice. (b) The most stable arrangement of six ferrimagnetic triangles around one ferrimagnetic fragment. (c) The least stable arrangement of six ferrimagnetic triangles around one ferrimagnetic fragment. The red circles indicate the down-spin site to go through (↑↓↑) to (↑↑↑) spin flip.

The above analysis indicates that, between the most stable and the least stable arrangements, the trigonal spin lattice has only a slightly greater energy difference than does the kagomé spin lattice (i.e., 14 vs 12 magnetic bonds). From this one might be led to speculate if trigonal and kagomé spin lattices have a similar 1/3-plateau properties. However, we note that the end point of the 1/3-plateau occurs when a ferrimagnetic fragment starts to have a configuration change from

(↑↓↑) to (↑↑↑). In a trigonal spin lattice, the (↑↓↑) to (↑↑↑) spin flip requires the breaking of six bonds in the most stable inter-fragment arrangement (**Fig. 5.5b**), and that of two bonds in the least stable inter-fragment arrangement (**Fig. 5.5c**). Namely, the spin flip requires energy in the most and least stable inter-fragment arrangements. This is not the case for a kagomé spin lattice. There, the (↑↓↑) to (↑↑↑) spin flip requires the breaking of four magnetic bonds at the site of the most stable inter-fragment arrangement (**Fig. 5.4b**), but no energy at the site of the least stable inter-fragment arrangement (**Fig. 5.4c**) because breaking two bonds within a ferrimagnetic fragment generates two bonds between the fragments. This implies that, during the field sweep from the most stable to the least stable distribution of the inter-fragment magnetic bonding, Zeeman energy causes (↑↓↑) to (↑↑↑) spin flips at certain down-spin sites with less favorable bonding connections with its neighboring fragments (e.g., **Fig. 5.4c,d**) because, in such a case, the spin flip requires less energy than does the breaking of the inter-fragment bonds. This reasoning predicts that a kagomé spin lattice has a narrower 1/3-magnetization plateau than does a trigonal spin lattice, and this might be the reason why the $M$ vs. $H$ curves of kagomé spin lattices show a steady increase in magnetization with field through the $M_{sat}/3$ point.

**5.2. Variation in the 1/3-plateau widths in RbFe(MoO$_4$)$_2$, Ba$_3$CoSb$_2$O$_9$ and Ba$_2$LaNiTe$_2$O$_{12}$**

The 2D antiferromagnets RbFe(MoO$_4$)$_2$,[86] Ba$_3$CoSb$_2$O$_9$[87] and Ba$_2$LaNiTe$_2$O$_{12}$[82] consist of trigonal layers made up of MO$_6$ (M = Fe, Co, Ni) octahedra (**Fig. 5.6a**). In RbFe(MoO$_4$)$_2$, The upper and lower surfaces of such a layer are condensed by corner-sharing with MoO$_4$ tetrahedra (**Fig. 5.6b,c**) in RbFe(MoO$_4$)$_2$, with Sb$_2$O$_9$ double octahedra in Ba$_3$CoSb$_2$O$_9$, and with TeO$_6$ octahedra in Ba$_2$LaNiTe$_2$O$_{12}$ (**Fig. 5.6d**). RbFe(MoO$_4$)$_2$ consists of trigonal layers of Fe$^{3+}$ (d$^5$, S = 5/2) ions, Ba$_3$CoSb$_2$O$_9$ those of Co$^{2+}$ (d$^7$, S = 3/2) ions, and Ba$_2$LaNiTe$_2$O$_{12}$ those of Ni$^{2+}$ (d$^8$, S = 1) ions. RbFe(MoO$_4$)$_2$ undergoes a phase transition at $T_N$ = 3.8 K into a 120° spin structure with all the spins confined in the basal plane. Application of an in-plane magnetic field induces a collinear spin state between 4.7 and 7.1 T, producing a 1/3-magnetization plateau (**Fig. 5.7a**).[88] Ba$_3$CoSb$_2$O$_9$ exhibits an AFM transition at $T_N$ = 3.8 K, and the powder neutron diffraction measurements show that it adopts a 120° spin structure in the $ab$-plane.[87] Under magnetic field applied in the $ab$-plane, Ba$_3$CoSb$_2$O$_9$ exhibits a 1/3-magnetization plateau between 10 – 15 T (**Fig. 5.7b**).[89] Ba$_2$LaNiTe$_2$O$_{12}$ undergoes successive magnetic phase transitions at $T_{N1}$ = 9.8 K and $T_{N2}$ = 8.9 K.[90] The ground state is accompanied by a weak ferromagnetic moment, suggesting that it adopts a slightly canted 120° spin structure. The magnetization curve exhibits a 1/3-magnetization plateau in the between 35 and 45 T (**Fig. 5.7c**).[82]

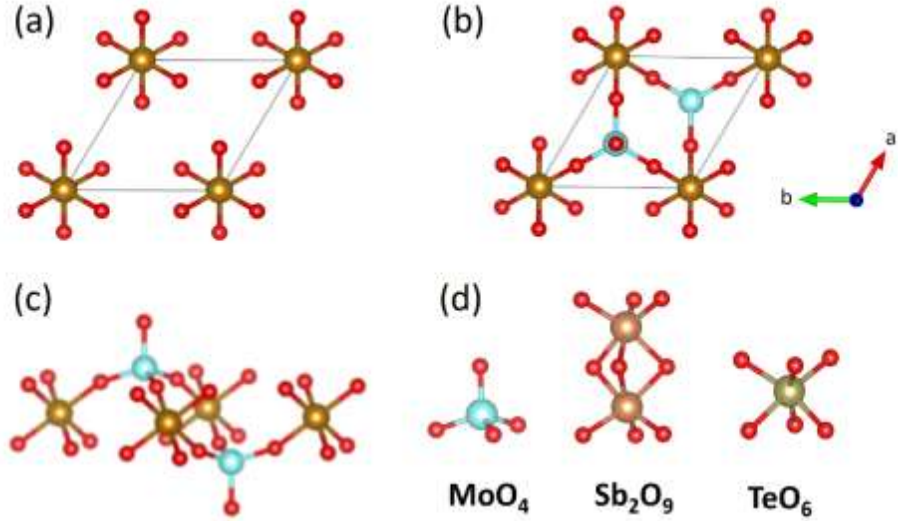

**Fig. 5.6.** (a) Trigonal layer of $MO_6$ (M = Fe, Co, Ni) octahedra found in $RbFe(MoO_4)_2$, $Ba_3CoSb_2O_9$ and $Ba_2LaNiTe_2O_{12}$. (b, c) Two views of how a trigonal layer of $FeO_6$ octahedra is capped by $MoO_4$ tetrahedra. (d) Views of a $MoO_4$ tetrahedron, a $Sb_2O_9$ double octahedron, and a $TeO_6$ octahedron capping the trigonal layers of $FeO_6$, $CoO_6$ and $NiO_6$ octahedra.

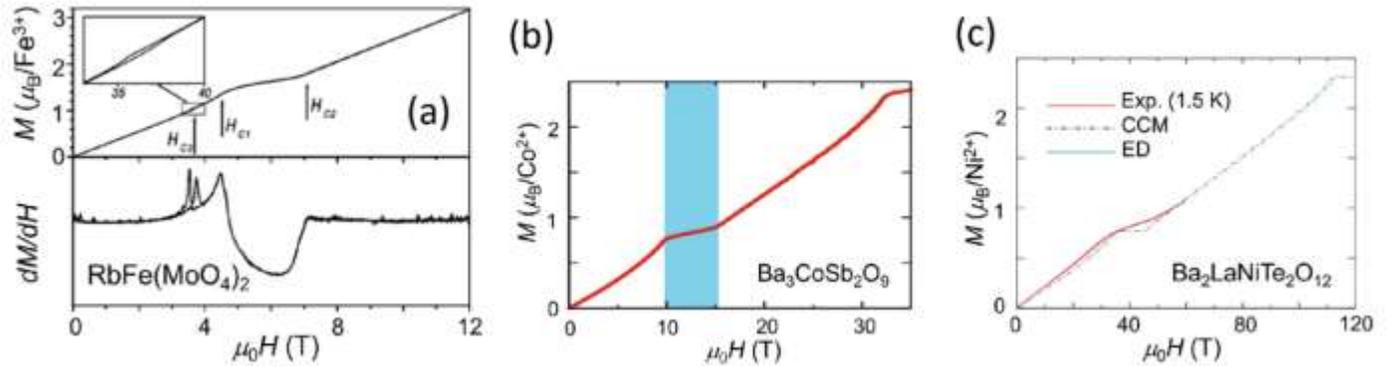

**Fig. 5.7.** Field dependence of magnetization observed for (a) $RbFe(MoO_4)_2$.[88] (Reproduced with permission from reference 88.) (b) $Ba_3CoSb_2O_9$ in the *ab*-plane at $T = 0.6$ K.[89] (c) $Ba_2LaNiTe_2O_{12}$ at 1.3 K.[82] (Reproduced with permission from reference 82.)

The observed widths $\Delta(\mu_0 H)$ of the 1/3-plateaus found for $RbFe(MoO_4)_2$, $Ba_3CoSb_2O_9$ and $Ba_2LaNiTe_2O_{12}$ are 2.4, 5.0 and ~10 T, respectively. In the previous section, we argued that the energy difference between the most stable and the least stable arrangements involving a (↑↓↑) ferrimagnetic triangle amounts to 14 nearest-neighbor magnetic bonds J. Thus, the $\Delta(\mu_0 H)$ values of these magnets should be related to their nearest-neighbor spin exchanges J as $\Delta(\mu_0 H) \propto J$. Precisely speaking, the spin exchange between two magnetic ions of spin S coupled by exchange constant J generates the energy $JS^2$ (Eq. 1.1). Since we compare the relative strengths of the magnetic bonds involving the ions of different spins, it is necessary to use the relationship $\Delta(\mu_0 H) \propto JS^2$. Then, according to the observed experimental $\Delta(\mu_0 H)$ values, the $JS^2$ values should increase in the order, $RbFe(MoO_4)_2 < Ba_3CoSb_2O_9 < Ba_2LaNiTe_2O_{12}$.

As depicted in **Fig. 5.8a**, the nearest-neighbor exchange J in the three magnets is of the M-O…O-M type exchange. In general, the strength of such an exchange becomes stronger as the O…O contact distance decreases.[2,91] As summarized in **Fig. 5.8b**, the O…O contact distances of the J exchange paths decrease in the order, $RbFe(MoO_4)_2$ > $Ba_3CoSb_2O_9$ > $Ba_2LaNiTe_2O_{12}$. The standard deviation of the O…O distance in $Ba_2LaNiTe_2O_{12}$ is rather large. However, even if the largest O…O distance of 2.74 Å allowed by the standard deviation is considered, the trend $RbFe(MoO_4)_2$ > $Ba_3CoSb_2O_9$ > $Ba_2LaNiTe_2O_{12}$ still remains valid. We carried out an energy mapping analysis based on DFT+U calculations (see Sections S11 – S13 of the SI) to find that J = 1.5 K for $RbFe(MoO_4)_2$, J = 6.2 K for $Ba_3CoSb_2O_9$ and J = 56 K for $Ba_2LaNiTe_2O_{12}$. (As expected, the J value of $Ba_2LaNiTe_2O_{12}$ is very large due to the unusually short O…O distance of 2.67 Å reported in the structure determination. A more accurate crystal structure would reduce the J value.) As summarized in **Fig. 5.8b**, the $JS^2$ values of the three antiferromagnets increase in the order, $RbFe(MoO_4)_2$ < $Ba_3CoSb_2O_9$ < $Ba_2LaNiTe_2O_{12}$. This provides experimental and theoretical support for our arguments presented in the previous section.

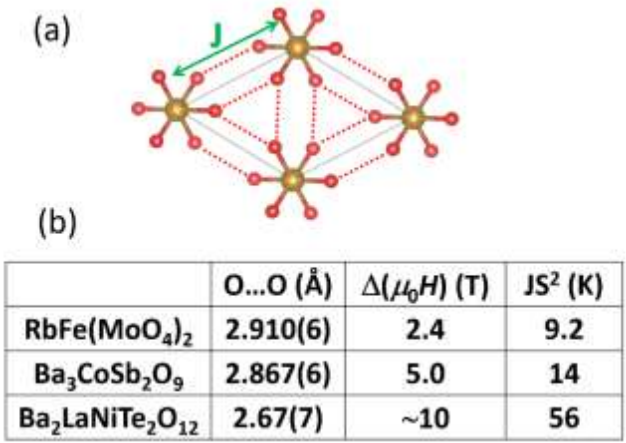

| | O…O (Å) | Δ($\mu_0H$) (T) | $JS^2$ (K) |
|---|---|---|---|
| $RbFe(MoO_4)_2$ | 2.910(6) | 2.4 | 9.2 |
| $Ba_3CoSb_2O_9$ | 2.867(6) | 5.0 | 14 |
| $Ba_2LaNiTe_2O_{12}$ | 2.67(7) | ~10 | 56 |

**Fig. 5.8.** (a) The M-O…O-M spin exchange paths J in a trigonal layer of $MO_6$ octahedra in $RbFe(MoO_4)_2$, $Ba_3CoSb_2O_9$, and $Ba_2LaNiTe_2O_{12}$. The dotted lines represent the O…O contacts, and each exchange path consists of two O…O contacts. (b) The O…O distance, the observed width Δ($\mu_0H$) of the 1/3-plateau, and the calculated energies $JS^2$ of the nearest-neighbor magnetic bonds in $RbFe(MoO_4)_2$, $Ba_3CoSb_2O_9$ and $Ba_2LaNiTe_2O_{12}$.

## 6. Complex clusters

### 6.1. Trimer-dimer zigzag chains for the 3/5-plateau in $Na_2Cu_5(Si_2O_7)_2$

Sodium copper pyrosilicate, $Na_2Cu_5(Si_2O_7)_2$, consists of ferrimagnetic zigzag chains in which trimer units alternate with dimer units (**Fig. 6.1a**).[92] Each trimer becomes ferrimagnetic due to the nearest-neighbor AFM exchange ($J_1$), the exchange ($J_2$) between adjacent trimer and dimer units is AFM, and the dimer exchange ($J_3$) is FM.[92] Thus, the ground state of the zigzag chain is ferrimagnetic (**Fig. 6.1b**). Since $Na_2Cu_5(Si_2O_7)_2$ undergoes a 3D AFM ordering below $T_N$ = 8K, the interchain interaction is weakly AFM. The fitting analysis of the magnetic susceptibility data using the ferrimagnetic chain model led to $J_1$ = 236 K, $J_2$ = 8 K and $J_3$ = -40 K.[92] The repeat unit of this ferrimagnetic state has the (3↑2↓) configuration with $M = M_{sat}/5$. Under a magnetic field, $Na_2Cu_5(Si_2O_7)_2$ exhibits a 3/5-magnetization plateau as shown in **Fig. 6.1c**.[93] It occurs because the weak $J_2$ bond is broken under field leading to the higher-energy ferrimagnetic state

(**Fig. 6.1d**) with the (4↑1↓) configuration and hence $M = 3M_{sat}/5$. To increase the magnetization beyond $3M_{sat}/5$, the two $J_1$ bonds in each trimer should be broken. Since $J_1$ is a strong bond, this does not occur unless the magnetic field is strong enough. Thus, the 3/5-magnetization plateau arises.

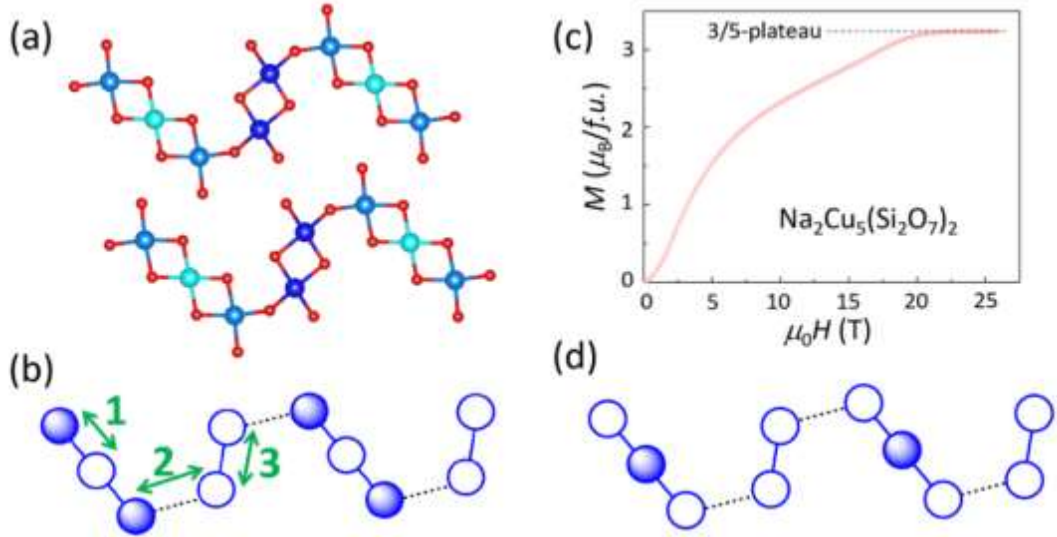

**Fig. 6.1**. (a) Zigzag chain of trimer and dimer unis in sodium copper pyrosilicate, $Na_2Cu_5(Si_2O_7)_2$. (b) Ferrimagnetic state of a chain obtained by an AFM coupling between ferrimagnetic trimers and FM dimers. (c) Field dependence of magnetization in $Na_2Cu_5(Si_2O_7)_2$ measured at 2 K.[93] (d) Ferrimagnetic state of a chain obtained by an FM coupling between ferrimagnetic trimers and FM dimers.

## 6.2. Linear heptamer of one trimer and two dimers for the 3/7-plateau in $Y_2Cu_7(TeO_3)_6Cl_6(OH)_2$

Viewed solely geometrically, the $Cu^{2+}$ ions of $Y_2Cu_7(TeO_3)_6Cl_6(OH)_2$ make chains of diamond-like tetramers which are interconnected by linear trimers.[94] In each trimer, a $CuO_2Cl_2$ plane corner-shares its oxygen atoms with two $CuO_3Cl$ planes such that the adjacent $CuO_2Cl_2$ and $CuO_3Cl$ planes are nearly perpendicular (**Fig. 6.2a**). In each diamond-like tetramer, the planes of the two $Cu_2O_3Cl$ dimers are separated and are nearly parallel to each other (**Fig. 6.2b**). When viewed from the point of the magnetic orbitals contained in square planar units containing $Cu^{2+}$ ions, a somewhat different picture emerges. In each diamond-like tetramer, the spin exchange between two $Cu_2O_3Cl$ dimer units cannot be strong, because their magnetic orbital planes are nearly parallel to each other. However, each $Cu_2O_3Cl$ dimer can interact with an adjacent trimer through the Cu-O…O-Cu type spin exchange because the O…O distance is short (2.659 Å) and the Cu-O bonds are nearly directed toward each other (∠Cu-O…O = 158.8, 165,7°). Thus, each trimer is connected to two adjacent $Cu_2O_3Cl$ dimers forming a linear heptamer (**Fig. 6.2c**), and such heptamers are expected to be important for $Y_2Cu_7(TeO_3)_6Cl_6(OH)_2$.

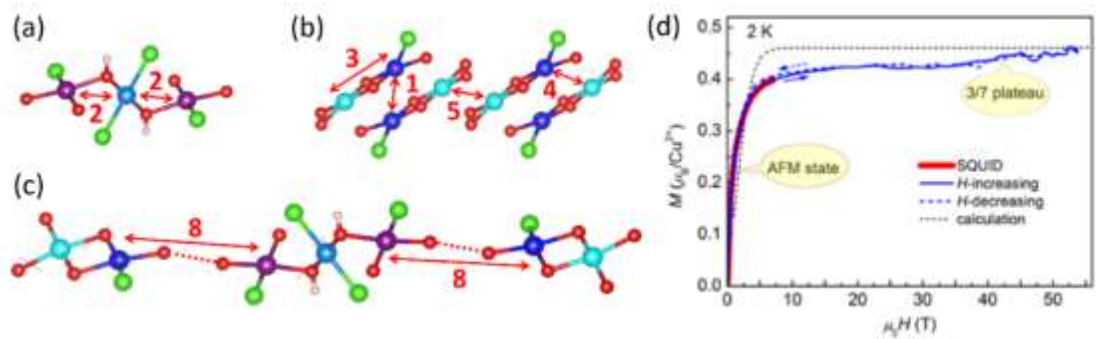

**Fig. 6.2.** (a – c) Building blocks of $Y_2Cu_7(TeO_3)_6Cl_6(OH)_2$. A trimer unit in (a), two diamond units in (b), and a heptamer formed by a trimer with two dimers using the Cu-O…O-Cu exchange paths in (c). (d) Field dependence of magnetization in $Y_2Cu_7(TeO_3)_6Cl_6(OH)_2$ showing a 3/7-magnetization plateau.[94] (Reproduced with permission from reference 94.)

The magnetization curve presents a field-induced metamagnetic transition at 0.2 T, which is followed by a magnetization plateau within a wide magnetic field range from 7 T to at least 55 T (**Fig. 6.2d**). To account for this observation, an energy-mapping analysis based on DFT+U calculations was carried out to find the spin exchanges (in K) summarized in **Fig. 6.3a** (see Section S14 of the SI). The latter shows that the dominating AFM spin exchanges are the inter-trimer-dimer exchange and the intra-dimer exchange $J_3$. The intra-trimer exchange $J_2$ is FM, and so are the exchanges between the $Cu_2O_3Cl$ units within each diamond-like tetramer but their magnitudes are weaker. Therefore, these considerations lead to the (5↑2↓) spin arrangement for each linear heptamer (**Fig. 6.3b**). One heptamer is coupled to two other heptamers through the AFM exchanges $J_5$, leading to an AFM chain of heptamers (**Fig. 6.3c,d**). (Since $Y_2Cu_7(TeO_3)_6Cl_6(OH)_2$ undergoes an AFM ordering below $T_N = 4.1$ K, the chains of heptamers have a very weak AFM inter-chain coupling.) Therefore, the breaking of all $J_5$ bonds is necessary to reach the $M = 3M_{sat}/7$ point. To increase the magnetization beyond $3M_{sat}/7$, it is necessary to break the intra-dimer $J_3$ bonds. These bonds are strong so that their breaking does not take place unless the magnetic field is strong enough. This explains the occurrence of the 3/7-plateau.

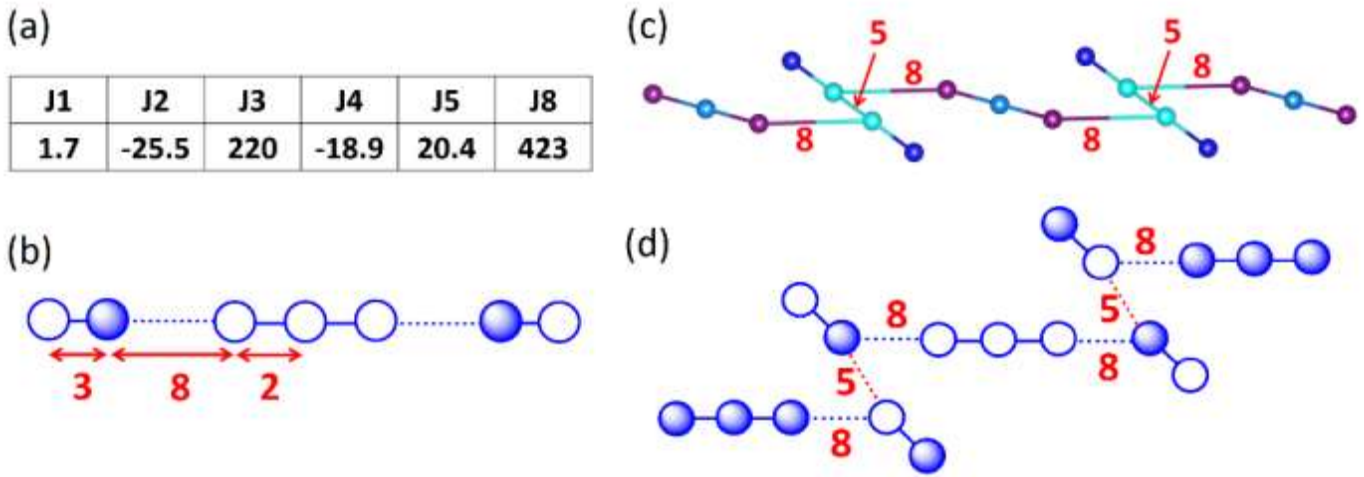

**Fig. 6.3.** (a) Values of the spin exchanges in K. (b) A (5↑2↓) spin arrangement of a linear heptamer in $Y_2Cu_7(TeO_3)_6Cl_6(OH)_2$. (c) Heptamers interacting through the $J_5$ spin exchanges. (d) AFM arrangements between adjacent heptamers leading to a heptamer chain.

### 6.3. Zigzag pentamer as an effective S = 1/2 unit in $Cu_5(VO_4)_2(OH)_4$

Turanite, $Cu_5(VO_4)_2(OH)_4$, has layers made up of three nonequivalent $CuO_6$ octahedra, which are interconnected by $VO_4$ groups. Within each layer, the $CuO_4$ square planes containing the $x^2-y^2$ orbitals are arranged as presented in **Fig. 6.4a**,[95] so the pattern of the $Cu^{2+}$ ion arrangement has interconnected chains of edge-sharing hexagons composed of six triangles (hereafter, the hexagon chains, for short) as depicted in **Fig. 6.4b**. This magnet undergoes a ferrimagnetic ordering at $T_C = 4.5$ K, and its magnetization evidences a rapid increase below about 0.01 T. The latter is followed by a much slower increase, eventually reaching a 1/5-magnetization plateau at 8 T (**Fig. 6.4c**).[96] There are two puzzling observations to note; the $M$ vs. $H$ curve is smooth and resembles that observed for a paramagnet of $S = 1/2$ ions, and only ~23.6 % of the spins participate in the ferrimagnetic ordering, which led to the suggestion that the remaining spins are still fluctuating.[96]

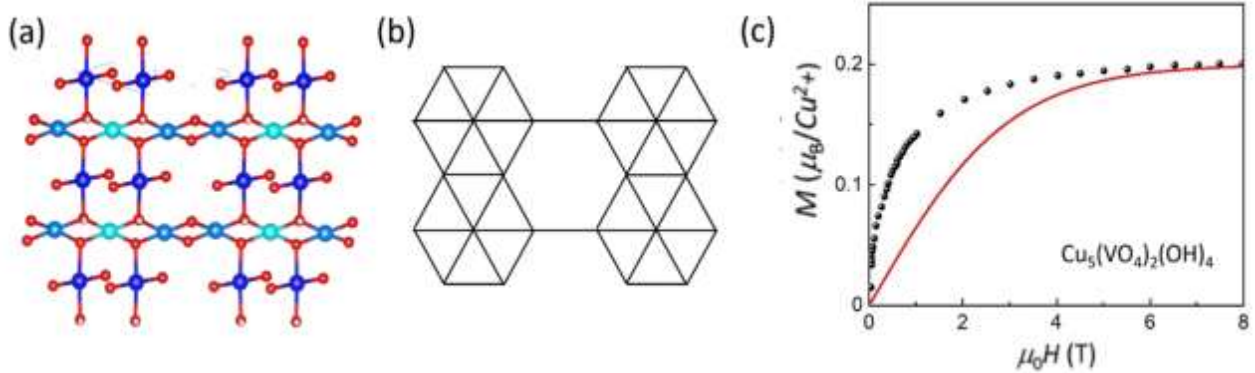

**Fig. 6. 4.**(a) A layer of $Cu_5(VO_4)_2(OH)_4$ made up of three nonequivalent $CuO_4$ planes by corner- and edge-sharing. (b) Pattern of the $Cu^{2+}$ ion arrangement showing chains of edge-sharing hexagons composed of six triangles. (c) Magnetization curve in $Cu_5(VO_4)_2(OH)_4$ at 2 K,[96] where the magnetization of a paramagnetic S = 1/2 ion (red curve) was added for comparison. (Reproduced with permission from reference 96.)

To probe the cause for these observations, it is necessary to know the spin exchanges in each layer of interlinked hexagon chains (**Fig. 6.4b**). What matters for spin exchanges is not the geometrical arrangement of magnetic ions but that of their magnetic orbitals. To see if the spin lattice of $Cu_5(VO_4)_2(OH)_4$ is spin frustrated, we examine the seven spin exchanges defined in **Fig. 6.5a**. Note that, due to the absence of a vertical mirror plane of symmetry in each hexagon chain, the exchanges $J_5$ and $J_6$ are treated as different (see **Fig. 6.5b**), and so are the spin exchanges $J_3$ and $J_4$. The spin exchanges of $J_1 - J_7$ determined by DFT+U calculations are summarized in **Fig. 6.5c** (see Section 15 of the SI), from which we observe the following:

(1) Within each hexagon chain, the AFM exchanges $J_1$ and $J_5$ dominate over the FM exchanges $J_2$, $J_4$ and $J_6$, so that there is effectively no spin frustration in all spin triangles of the hexagon chains.

(2) The exchanges $J_1$ and $J_5$ form zigzag pentamer ferrimagnetic fragments of (3↑2↓) spin configuration with $M = M_{sat}/5$ (**Fig. 6.5d**).

(3) Since $J_6$ is more strongly FM than $J_4$, each hexagon chain prefers to have adjacent (3↑2↓) ferrimagnetic fragments to have an AFM coupling than an FM coupling within each hexagon chain (**Fig. 6.6a,b**).

(4) Since the interchain exchange $J_7$ is AFM, an AFM coupling is preferred to an FM coupling between adjacent (3↑2↓) ferrimagnetic fragments between hexagon chains.

(5) Thus, the most stable arrangement between adjacent (3↑2↓) ferrimagnetic fragments is AFM in both within and between hexagon chains (**Fig. 6.6c**), and the least stable arrangement an FM in both within and between hexagon chains (**Fig. 6.6d**).

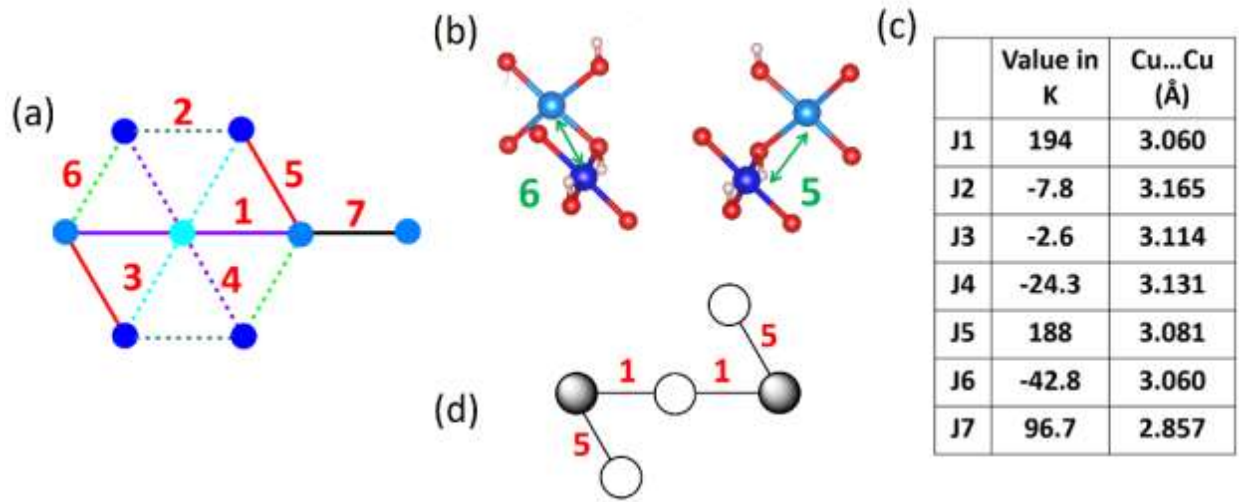

**Fig. 6.5.** (a) Seven spin exchange paths in $Cu_5(VO_4)_2(OH)_4$ defined with respect to the crystal structure given in Fig. 6.4a. (b) Different arrangements of the two $CuO_4$ planes associated with the $J_5$ and $J_6$ spin exchange paths. (c) Values of the calculated spin exchanges and the Cu…Cu distances associated with the spin exchange paths. (d) A zigzag ferrimagnetic fragment of $(3\uparrow 2\downarrow)$ spin configuration.

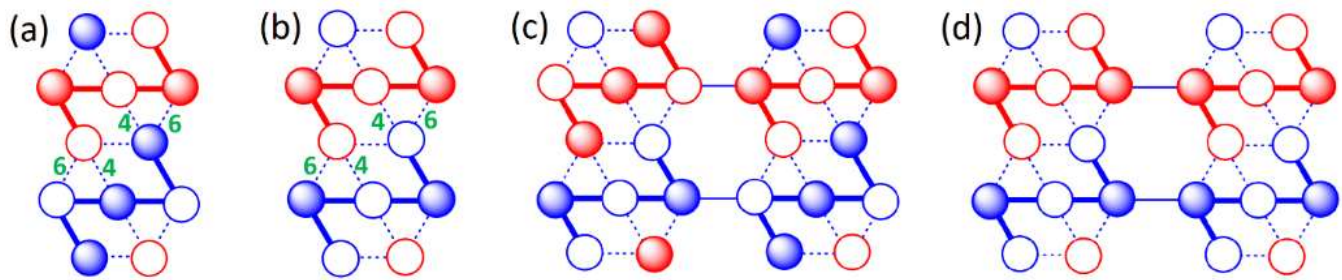

Fig. 6.6. (a) AFM arrangement of two adjacent ferrimagnetic fragments within a hexagon chain. (b) FM arrangement of two adjacent ferrimagnetic fragments within a hexagon chain. (c) AFM arrangement of adjacent ferrimagnetic fragments within and between hexagon chains. (d) FM arrangement of adjacent ferrimagnetic fragments within and between hexagon chains.

Though seemingly paradoxical, the above theoretical analysis is entirely consistent with the experimental observations for $Cu_5(VO_4)_2(OH)_4$. The key point to understand is that the AFM exchanges $J_5$ and $J_1$ leading to a $(3\uparrow 2\downarrow)$ ferrimagnetic fragment are very strong, so that each $(3\uparrow 2\downarrow)$ ferrimagnetic fragment acts as an effective $S = 1/2$ unit. Indeed, the observed magnetization curve is similar to the one found for a paramagnet of $S = 1/2$ ions (see the red curve in **Fig. 6.4c**). This realization explains the low-temperature magnetic properties of $Cu_5(VO_4)_2(OH)_4$, namely, why only one out of five spins appears to participate in the magnetic ordering below $T_C = 4.5$ K and why the magnetization behavior resembles that expected for a paramagnet of $S = 1/2$ ions. The magnetic susceptibility of $Cu_5(VO_4)_2(OH)_4$ reveals the presence of five spins per formula unit at high temperature, because thermal agitation would break the AFM coupling leading to the $(3\uparrow 2\downarrow)$ ferrimagnetic fragment. As already discussed in Section 4.1.3, such a phenomenon of reduced spin moments due to strong AFM coupling was also found for volborthite, which consists of two-leg spin ladders with rung made up of linear trimers of $Cu^{2+}$ ions. In this case, the AFM coupling between adjacent $Cu^{2+}$ ions is so strong that each linear trimer acts as an effective $S = 1/2$ unit.

### 6.4. Cu$_7$ cluster of corner-sharing tetrahedra for the 3/7-plateau in Pb$_2$Cu$_{10}$O$_4$(SeO$_3$)$_4$Cl$_7$ and Na$_2$Cu$_7$(SeO$_3$)$_4$O$_2$Cl$_4$

Pb$_2$Cu$_{10}$O$_4$(SeO$_3$)$_4$Cl$_7$ has a complex crystal structure consisting of one nonmagnetic $Cu^+$ ($S = 0$) ion and nine $Cu^{2+}$ ($S = 1/2$) ions per formula unit.[97] Of the nine $Cu^{2+}$ ions, two are found in a dimer unit and seven in a heptamer unit made up of two corner-sharing $(Cu^{2+})_4$ tetramers (**Fig. 6.7a**). Pb$_2$Cu$_{10}$O$_4$(SeO$_3$)$_4$Cl$_7$ orders antiferromagnetically at $T_N = 10.2$ K and below this temperature exhibits a sequence of spin-flop transition at 1.3 T and a 1/3-plateau at 4.4 T, which persists at least up to 53.5 T (**Fig. 6.7b**). The spin exchanges of Pb$_2$Cu$_{10}$O$_4$(SeO$_3$)$_4$Cl$_7$ evaluated by DFT+U calculations showed that the magnetic properties of Pb$_2$Cu$_{10}$O$_4$(SeO$_3$)$_4$Cl$_7$ are governed by the ferrimagnetic heptamer $(Cu^{2+})_7$ with (5↑2↓) spin configuration (**Fig. 6.7c**). The heptamers form chains (**Fig. 6.7d**) with interchain AFM coupling, so the magnetic ground state of the chain is an AFM state (**Fig. 6.7e**). Under magnetic field, the inter-cluster bonds become broken, eventually reaching the ferrimagnetic state in which the ferrimagnetic clusters are ferromagnetically coupled (**Fig. 6.7f**). A further increase in magnetization requires high magnetic field because it is necessary to break the magnetic bonds within a ferrimagnetic cluster, hence leading to the 3/7-plateau. Note that this discussion is based solely on the seven $Cu^{2+}$ ions of a heptamer $(Cu^{2+})_7$. The two $Cu^{2+}$ ions, strongly coupled antiferromagnetically in a dimer, are magnetically "silent". If we include these two magnetic ions in our analysis, the 3/7-plateau discussed above becomes equivalent to a 1/3-plateau. A similar 3/7-plateau was found for $Na_2Cu_7(SeO_3)_4O_2Cl_4$ (**Fig. 6.8a**),[98] which also consists of $(Cu^{2+})_7$ heptamers made up of two corner-sharing $(Cu^{2+})_4$ tetramers (**Fig. 6.8b**). $Na_2Cu_7(SeO_3)_4O_2Cl_4$ differs from $Pb_2Cu_{10}O_4(SeO_3)_4Cl_7$ in the bridging mode between adjacent heptamers (**Fig. 6.8c**), but the composition of the heptamers is identical, namely, it is composed of five square planar and two trigonal bipyramid units (**Fig. 6.9a**). Our DFT+U calculations summarized in **Fig. 6.8d** (see Section S16 of the SI) show that each $(Cu^{2+})_7$ heptamer has a (5↑2↓) spin configuration (**Fig. 6.7e**), thereby explain why $Na_2Cu_7(SeO_3)_4O_2Cl_4$ exhibits a 3/7-plateau as does $Pb_2Cu_{10}O_4(SeO_3)_4Cl_7$.

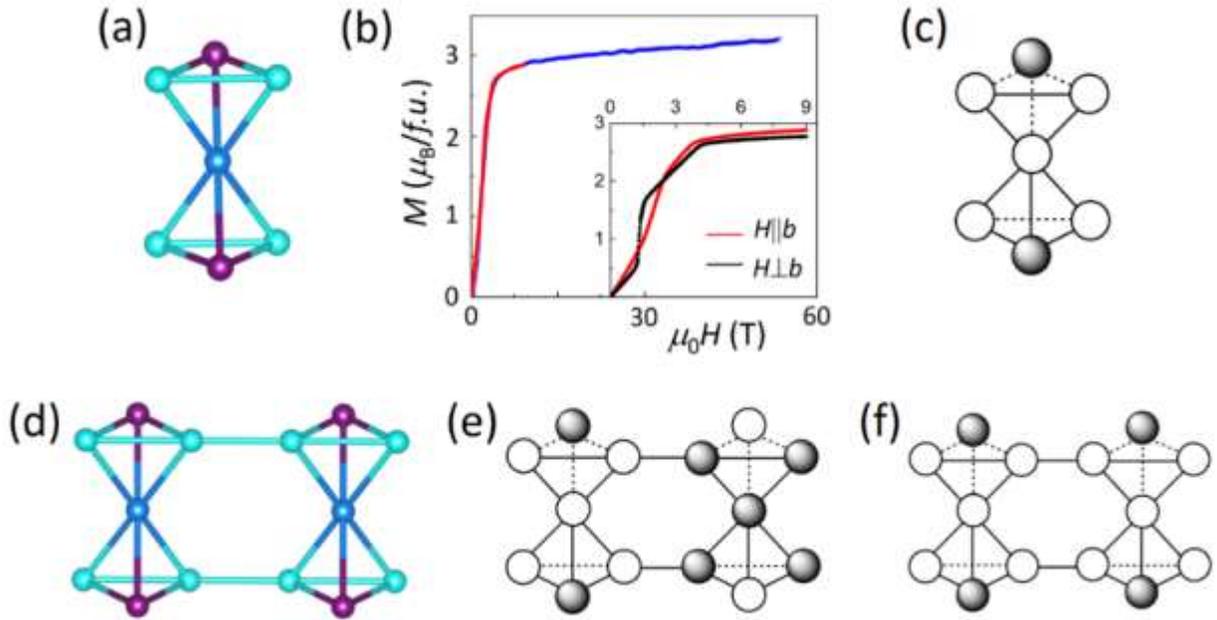

**Fig. 6.7.** (a) A $(Cu^{2+})_7$ heptamer made up of two corner-sharing tetrahedra in Pb$_2$Cu$_{10}$O$_4$(SeO$_3$)$_4$Cl$_7$. (b) Magnetization curve observed for Pb$_2$Cu$_{10}$O$_4$(SeO$_3$)$_4$Cl$_7$ at 2 K.[97] (Reproduced with permission from reference 97.) (c) (5↑2↓) spin configuration of a heptamer. (d) Bridging mode between heptamers to form a chain. (e) AFM coupling between heptamers (f) FM coupling between heptamers.

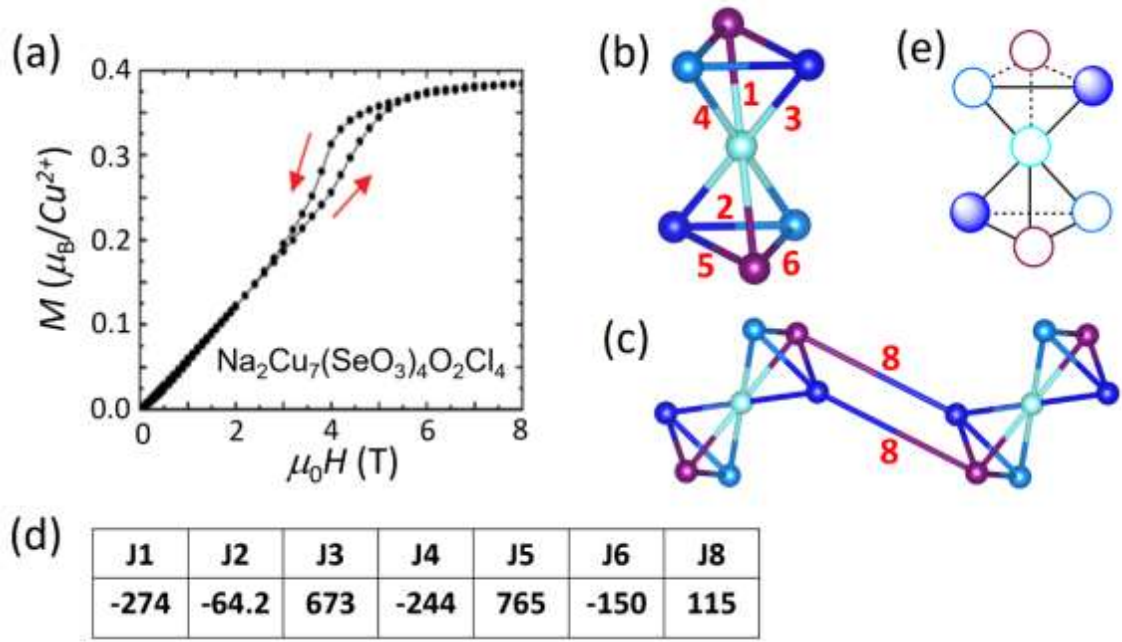

**Fig. 6.8**. (a) Magnetization curve in $Na_2Cu_7(SeO_3)_4O_2Cl_4$ at 2 K.[98] (Reproduced with permission from reference 98.) (b) A $(Cu^{2+})_7$ heptamer made up of two corner-sharing tetrahedra. (c) Bridging mode between heptamers to form a chain. (d) Spin exchanges (in K) determined by DFT+U calculations. (e) (5↑2↓) spin configuration of a heptamer.

Finally, we comment on why the $(Cu^{2+})_7$ heptamer adopts the (5↑2↓) spin configuration. From one trigonal bipyramid (TBP) to the central square plane (SP) to another trigonal bipyramid (TBP) in a heptamer, the three $Cu^{2+}$ ions form a linear path (**Fig. 6.9a**), and the two nearest-neighbor spin exchanges are of the Cu-O-Cu type. These two spin exchanges are strongly AFM in this linear path because the atoms associated with the Cu-O-Cu-O-Cu exchange paths are coplanar, so that the $a_1$ magnetic orbitals of the two TBPs[99] and the $x^2$-$y^2$ magnetic orbital of the central SP are coplanar (**Fig. 6.9b**). This makes the in-plane 2p orbitals of the two bridging O atoms intact efficiently with the $a_1$ and $x^2$-$y^2$ magnetic orbitals, leading to a strong ↑↓↑ coupling of the three $Cu^{2+}$ spins along the TBP-SP-TBP linear path. Note that the central SP is nearly orthogonal to every SP at both ends. This aligns their $x^2$-$y^2$ magnetic orbitals nearly orthogonal as well, so the associated spin exchange becomes FM. Consequently, the heptamer adopts the (5↑2↓) spin configuration shown in **Fig. 6.9**c.

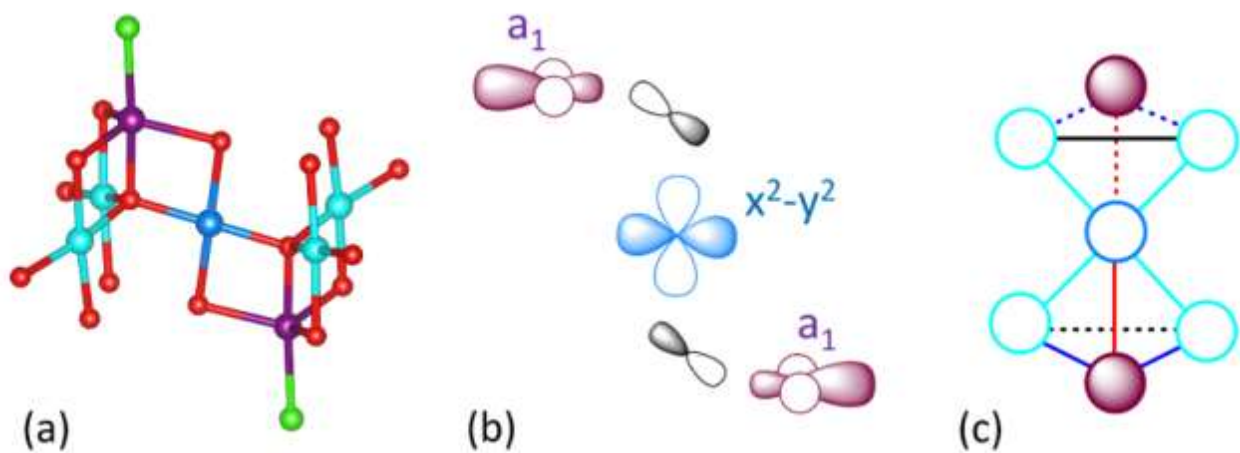

**Fig. 6.9**. (a) A $(Cu^{2+})_7$ heptamer of $Pb_2Cu_{10}O_4(SeO_3)_4Cl_7$ composed of five $CuO_4$ square planes and two $CuO_4Cl$ trigonal bipyramids. (b) The $x^2-y^2$ magnetic orbital of the central $CuO_4$ square plane (SP) interacting with the xy magnetic orbitals of the two trigonal bipyramids (TBPs), leading to a strong (↑↓↑) spin coupling of the three $Cu^{2+}$ ions of the linear TBP-SP-TBP paths. (c) (5↑2↓) spin configuration of a heptamer.

## 7. Concluding remarks

In an effort to find the conceptual picture describing the magnetization plateau phenomenon, we surveyed the crystal structures, the spin exchanges and the spin lattices of numerous magnets exhibiting magnetic plateaus. Our analyses show that an important key to understanding this phenomenon is the realization that a magnet under field absorbs Zeeman energy in accordance with Le Chartlier's principle, which occurs by breaking its magnetic bonds. For a magnet with spin lattice defined by several spin exchanges of different strengths, its weakest bonds are broken preferentially to partition the spin lattice into either antiferromagnetic or ferrimagnetic fragments, which fill the whole spin lattice without overlapping each other. For a magnet with spin-frustrated spin lattice defined by a few spin exchanges of comparable strengths, the weaker magnetic bonds are broken to partition the spin lattice into small ferrimagnetic fragments filling the whole spin lattice without overlapping each other. Such field-induced fragmentation is influenced by the spin-lattice interactions brought about by the fragmentation.

As illustrated in this survey, the conceptual aspects of the magnetization plateau phenomenon in any magnet can be readily explained once its crystal structure and its spin exchanges are known. It goes without saying that this approach is not designed to provide quantitative descriptions. The latter lie in the realm of quantitative calculations using model Hamiltonians with a minimal number of adjustable parameters (e.g., spin exchanges) to generate the magnetic energy spectrum of a magnet under investigation. Even with powerful computers currently available, such quantitative analyses cannot be carried out for most magnets because their spin lattices are complex and low in symmetry. The conceptual picture of the magnetization plateau phenomenon, based on the supposition of field-induced partitioning of a spin lattice into magnetic fragments, is valid for all magnets regardless of whether their spin lattices are complex or not.

The magnetization plateau phenomenon can be highly anisotropic as found for the Ising magnets $Ca_3Co_2O_6$ and $CoGeO_3$, in that their 1/3-magnetization plateaus observed with field along the easy axis do not occur if the field is perpendicular to the easy axis. A strong plateau anisotropy, though weaker than those found for the Ising magnets, is also observed for $Cs_2Cu_3(SeO_3)_4·2H_2O$ and azurite $Cu_3(CO_3)_2(OH)_2$. In $Cs_2Cu_3(SeO_3)_4·2H_2O$, the value of $M = M_{sat}/3$ depends on the field direction (i.e., $H\perp c$ vs. $H\|c$) because the $Cu^{2+}$ ion has a higher spin moment when the magnetic field is perpendicular than parallel to the $CuO_4$ square plane. In azurite, the width of the 1/3-magnetization plateau depends on the field direction ($H\|b$ vs. $H\perp b$) for two reasons; one is the Dzyaloshinskii-Moriya interactions between the $Cu^{2+}$ ions, which depend on the relative orientations of their $CuO_4$ square planes, and the other is the spin moment of a $Cu^{2+}$ ion, which depends on the field direction with respect to the $CuO_4$ square plane. In both $Cs_2Cu_3(SeO_3)_4·2H_2O$ and azurite, the strong anisotropy of their magnetization plateaus stems from the presence of near orthogonal arrangements of their $CuO_4$ square planes. This emphasizes once more the importance of analyzing the structural chemistry associated with magnetic ion arrangements.

Our supposition that the spin lattice of a magnet exhibiting one or more magnetic plateaus is partitioned into magnetic fragments is supported by the experimental observation that an Ising magnet can exhibit a magnetization plateau when the applied field is parallel to the easy axis of the magnet, but this magnetization plateau disappears when the field is perpendicular to the easy axis. This reflects that Zeeman energy, being a dot product between the magnetic field and the spin moment, is nonzero for the parallel field but zero for the perpendicular field. Another support comes from the observation that the highly anisotropic width of the 1/3-magnetization plateau ($H\|b$ vs. $H\perp b$) in azurite $Cu_3(CO_3)_2(OH)_2$ arises from the field-dependent Zeeman energy available for the magnet.

This survey reflects our efforts to comprehend the magnetization plateau phenomenon based on the relative strengths of magnetic bonds. Thus, our discussion focused on the arrangement of magnetic ions and their spin exchanges leading to the spin lattices responsible for magnetization plateaus. It is our hope that the conceptual picture of the magnetization plateau phenomenon presented in this survey will promote further developments in this and related research fields.

Near the completion stage of this survey, new magnets were reported to exhibit magnetization plateaus at low temperatures, which include $CsCo_2Br_4$,[100] $YCu_3(OD)_{6+x}Br_{3-x}$ ($x \approx 0.5$),[79] $Ni_2V_2O_7$,[101] $TbTi_3Bi_4$,[102] $TbRh_6Ge_4$,[103] $GdInO_3$,[104] $Cu_5(PO_4)_2(OH)_4$,[105] $Ba_2Cu_3(SeO_3)_4F_2$,[106] $Sr_2CoTeO_6$,[107] $Cu_3Bi(TeO_3)_2O_2Cl$,[108] and $Na_3Ni_2BiO_6$.[109] We expect that the magnetic plateau phenomena of all these new magnets can be readily explained using the concept of the field-induced partitioning of their spin lattices once their spin exchanges are determined.

In this survey, we focused on the "classical" magnetization plateaus that are readily described by field-induced partitioning of spin lattices into spin superstructures. It should be pointed that there may be cases that require a more sophisticated description beyond this classical picture [110-124]. For instance, as recently reviewed by Yoshida [125], quantum plateau phases may emerge from quantum spin liquids.

**Acknowledgements**

The work at KHU was supported by the Basic Science Research Program through the National Research Foundation of Korea (NRF) funded by the Ministry of Education (2020R1A6A1A03048004).

**References**
1. Dai, D.; Whangbo, M.-H. Analysis of the uniaxial magnetic properties of high-spin $d^6$ ions at trigonal prism and linear two-coordinate sites: Uniaxial magnetic properties of $Ca_3Co_2O_6$ and $Fe[C(SiMe_3)_3]_2$. *Inorg. Chem*. 2005, *44*, 4407-4414.
2. Dai, D.; Xiang, H. J.; Whangbo, M.-H. 2008 Effects of spin-orbit coupling on magnetic properties of discrete and extended magnetic systems. *J. Comput. Chem*. 2008, *29*, 2187-2209.
3. Zhang, Y.; Kan, E. J.; Xiang, H. J.; Villesuzanne, A.; Whangbo, M.-H. Density Functional Theory Analysis of the Interplay between Jahn-Teller Instability, Uniaxial Magnetism, Spin Arrangement, Metal-Metal Interaction, and Spin-Orbit Coupling in $Ca_3CoMO_6$ (M = Co, Rh, Ir). *Inorg. Chem*. 2011, *50*,1758-1766.
4. Honecker, A.; Schulenburg, J.; Richter, J. Magnetization plateaus in frustrated antiferromagnetic quantum spin models. *J. Phys.: Condens. Matter* 2004, 16, S749.
5. Takigawa, M.; Mila, F. Magnetization plateaus, in *Introduction to Frustrated Magnetism* (Lacroix C, Mendels P and Mila F, editors), Springer, New York, 2010, pp 241-268.
6. Xiang, H. J.; Lee, C.; Koo, H.-J.; Gong, X. G.; Whangbo, M.-H. Magnetic properties and energy-mapping analysis. *Dalton Trans.* 2013, *42*, 823-853.
7. Penc, K.; Shannon, N.; Shiba. H. Half-magnetization plateau stabilized by structural distortion in the antiferromagnetic Heisenberg model on a pyrochlore lattice. *Phys. Rev. Lett.* 2004, *93*, 197202.
8. Aoyama, K.; Gen, M.; Kawamura, H. Effects of spin-lattice coupling and a magnetic field in classical Heisenberg antiferromagnets on the breathing pyrochlore lattice. *Phys. Rev. B* 2021, *104*, 184411.
9. Gen, M.; Zhou, X.; Ikeda, A.; Aoyama, K.; Jeschke, H. O.; Nakamura, D.; Takeyama, S.; Ishii, Y.; Ishikawa, H.; Yajima, T.; Okamoto, Y.; Kindo, K.; Matsuda, Y. H.; Kohama, Y. Signatures of a magnetic superstructure phase induced by ultrahigh magnetic fields in a breathing pyrochlore antiferromagnet. *Proc. Nat. Acad. Sci. USA* 2023, *120*, e2302756120.
10. Onizuka, K.; Kageyama, H.; Narumi, Y.; Kindo, K.; Ueda, Y.; Goto, T. 1/3 magnetization plateau in $SrCu_2(BO_3)_2$ – stripe order of excited triplets. *J. Phys. Soc. Jpn.* 2000, *69*, 1016-1018.
11. M. Jaime, R. Daou, S. A. Crooker, B. D. Gaulin, Magnetostriction and magnetic texture to 100.75 Tesla in frustrated $SrCu_2(BO_3)_2$. *Proc. Natl. Acad. Sci. USA* 2012, *109*, 12404.
12. Y. H. Matsuda, N. Abe, S. Takeyama, H. Kageyama, P. Corboz, A. Honecker, S. R. Manmana, G. R. Foltin, K. P. Schmidt, and F. Mila, Magnetization of $SrCu_2(BO_3)_2$ in Ultrahigh Magnetic Fields up to 118 T. *Phys. Rev. Lett*. 2013, *111*, 137204.
13 T. Nomura, P. Corboz, A. Miyata, S. Zherlitsyn, Y. Ishii, Y. Kohama, Y. H. Matsuda, A. Ikeda, C. Zhong, H. Kageyama, and F. Mila, Unveiling new quantum phases in the Shastry-Sutherland compound $SrCu_2(BO_3)_2$ up to the saturation magnetic field. *Nat. Commun*. 2023, *14*, 3769.
14. Néel, L. 1957 Les métamagnétiques ou substances antiferromagnétiques à champ seuil. *Nuovo Cimento* 1957, *6*, 942-960.
15. Morosov, A. I.; Sigov, A. S. Surface spin-flop transition in an antiferromagnet. *Physics-Uspekhi* 2010, *53*, 677-689.
16. Strujewski, E. Giordano, N. Metamagnetism. *Adv. Phys.* 1977, *26*, 487-650.
17. Nawa, K.; Avdeev, M.; Perdonosov, P.; Sobolev, A.; Presniakov, I.; Aslandukova, A.; Kozlyakova, E.; Vasiliev, A.; Shchetinin, I.; Sato, T. Magnetic structure study of the sawtooth chain antiferromagnet $Fe_2Se_2O_7$. *Sci. Rep.* 2021, *11*, 24049.

18. Zakharov, K. V.; Zvereva, E. A.; Berdonosov, P. S.; Kuznetsova, E. S.; Dolgikh, V. A.; Clark, L.; Black, C.; Lightfoot, P.; Kockelmann, W.; Pchelkina, Z. V.; Streltsov, S. V.; Volkova, O. S.; Vasiliev, A. N. Thermodynamic properties, electron spin resonance, and underlying spin model in $Cu_3Y(SeO_3)_2O_2Cl$. *Phys. Rev. B* 2014, *90*, 214417.
19. Zakharov, K. V.; Zvereva, E. A.; Kuznetsova, E. S.; Kuznetsova, E. S.; Berdonosov, P. S.; Dolgikh V. A.; Markina, M. M.; Olenev, A. V.; Shakin, A. A.; Volkova, O. S.; Vasiliev, A. N. Two new lanthanide members of francisite family $Cu_3Ln(SeO_3)_2O_2Cl$ (Ln = Eu, Lu). *J. Alloys Compd.* 2016, *685*, 442-447.
20. Whangbo, M.-H.; Xiang, H. J. Magnetic Properties from the Perspectives of Electronic Hamiltonian: Spin Exchange Parameters, Spin Orientation and Spin-Half Misconception, in *Handbook of Solid State Chemistry, Volume 5: Theoretical Descriptions*; Dronskowski, R.; Kikkawa, S.; Stein, A., Eds.; Wiley, 2017, pp 285-343.
21. Whangbo, M.-H,; Koo, H.-J.; Kremer, R. K. Spin Exchanges between Transition Metal Ions Governed by the Ligand p-Orbitals in Their Magnetic Orbitals. *Molecules*. 2021, *26*, 531.
22. Kresse, G.; Joubert, D. From ultrasoft pseudopotentials to the projector augmented-wave method. *Phys. Rev. B* 1999, *59*, 1758-1775.
23. Kresse, G.; Furthmüller, J. Efficiency of ab-initio total energy calculations for metals and semiconductors using a plane-wave basis set. *Comput. Mater. Sci*. 1996, *6*, 15-50.
24. Perdew, J. P.; Burke, K.; Ernzerhof, M. Generalized Gradient Approximation Made Simple. *Phys. Rev. Lett.* 1996, *77*, 3865-3868.
25. Dudarev, S. L.; Botton, G. A.; Savrasov, S. Y.; Humphreys, C. J.; Sutton, A. P. Electron-energy-loss spectra and the structural stability of nickel oxide: An LSDA+U study. *Phys. Rev. B* 1998, *57*, 1505-1509.
26. Kuneš, K.; Novák, P.; Schmid, R.; Blaha, P.; Schwarz, K. Electronic structure of fcc Th: Spin-orbit calculation with $6p_{1/2}$ local orbital extension. *Phys. Rev. Lett.* 2001, *64*, 153102.
27. Mazurenko, V. V.; Skornyakov, S. L.; Anisimov, V. I.; Mila, F. First-principles investigation of symmetric and antisymmetric exchange interactions of $SrCu_2(BO_3)_2$. *Phys. Rev. B* 2008, *78*, 195110.
28. Miyahara, S.; Ueda, K. Theory of the orthogonal dimer Heisenberg spin model for $SrCu_2(BO_3)_2$. *J. Phys.: Condens. Matter* 2003, *15*, R237-R366.
29. Shastry, B. S.; Sutherland, B. Exact ground state of a quantum mechanical antiferromagnet. *Physica B&C* 1981, *108B*, 1069-1070.
30. Kageyama, H.; Yoshimura, K.; Stern, R.; Mushnikov, N.; Onizuka, K.; Kato, M.; Kosuge, K.; Slichter, C. P.; Goto, T.; Ueda, Y. Exact dimer ground state and quantized magnetization plateaus in the two-dimensional spin system $SrCu_2(BO_3)_2$. *Phys. Rev. Lett.* 1999, *82*, 3168-3171.
31. Ueda, H.; Katori, H. A.; Mitamura, H.; Goto, H.; Takagi, H. Magnetic-field induced transition to the 1/2 magnetization plateau state in the geometrically frustrated magnet $CdCr_2O_4$. *Phys. Rev. Lett.* 2005, *94*, 047202.
32. Chung, J. H.; Matsuda, M.; Lee, S.-H.; Kakurai, K.; Ueda, H.; Sato, T. J.; Takagi, H.; Hong, K.-P.; Park, S. Statics and dynamics of incommensurate spin order in a geometrically frustrated antiferromagnet $CdCr_2O_4$. *Phys. Rev. Lett.* 2005, *95*, 247204.
33. Peakor, D. R. The crystal structure of $CoGeO_3$. *Z. Krist.* 1968, *126*, 299-306.
34. Guo, H.; Zhao, M.; Baenitz, X.; Gukasov, A.; Melendez Sans, A.; Khomskii, D. I.; Tjeng, L. H.; Komareck, A. C. Emergent 1/3 magnetization plateaus in pyroxene $CoGeO_3$. *Phys. Rev. Res.* 2021, *3*, L032037.

35. Whangbo, M.-H.; Gordon, E. E.; Xiang, H. J.; Koo, H.-J.; Lee, C. Prediction of spin orientations in terms of HOMO-LUMO interactions using spin-orbit coupling as perturbation. *Acc. Chem. Res*. 2015, *48*, 3080-3087.
36. Zapf, V.: Jaime, M.; Batista, C. D. Bose-Einstein condensation in quantum magnets. *Rev. Mod. Phys*. 2014, *86*, 563; Erratum *Rev. Mod. Phys*. 2014, *86*, 1453.
37. Weller, M. T.; Skinner, S. J. $Ba_3Mn_2O_8$ determined from neutron powder diffraction. *Acta Crystallogr. C* 1999, *55*, 154-156.
38. Uchida M, Tanaka H, Mitamura H, Ishkawa F and Goto T 2002 High-field magnetization process in the S = 1 quantum spin system $Ba_3Mn_2O_8$. *Phys. Rev. B* **66** 054429.
39. Willett R D, Dwiggins Jr C, Kruh R F and Rundle R E 1963 Crystal structures of $KCuCl_3$ and $NH_4CuCl_3$. *J. Chem. Phys.* **38** 2429-2436.
40. Ryu, G.; Son, K. Surface defect free growth of a spin dimer $TlCuCl_3$ compound crystals and investigations on its optical and magnetic properties. *J. Solid State Chem.* 2016, *237*, 358-363.
41. O'Bannon, G.; Willet, R. D. A redetermination of the crystal structure of $NH_4CuCl_3$ and a magnetic study of $NH_4CuX_3$ (X = Cl, Br). *Inorg. Chim. Acta* 1981, *53*, L131-L132.
42. Rüegg, Ch.; Schefer, J.; Zaharko, O.; Furrer, A.; Tanaka, H.; Krämer, K. W.; Güdel, H.-U.; Vorderwisch, P.; Habicht, K.; Polinsli, T.; Meissner, M. Neutron scattering study of the field-dependent ground state and the spin dynamics in spin-one-half $NH_4CuCl_3$. *Phys. Rev. Lett.* 2004, *93*, 037207.
43. Matsumoto, M.; Normand, B.; Rice, T. M.; Sigrist, M. Field- and pressure-induced magnetic quantum phase transitions in $TlCuCl_3$. *Phys. Rev. B* 2004, *69*, 054423.
44. Oosawa, A.; Kato, T.; Tanaka, H.; Kakurai, K.; Müller, M.; Mikeska, H.-J. Magnetic excitations in the spin gap system $TlCuCl_3$. *Phys. Rev. B* 2002, *65*, 094426.
45. Tanaka, H.; Shiramura, W.; Takatsu, T.; Kurniawan, B.; Takahashi, M.; Kamishima, K.; Takizawa, K.; Mitamura, H.; Goto, T. High-field magnetization processes of quantum double spin chain systems $KCuCl_3$, $TlCuCl_3$ and $NH_4CuCl_3$. *Physica B* 1998, *246-247*, 230-233.
46. Rüegg, Ch., Diploma Thesis, 2001 and Ph. D. Thesis, 2005, Laboratory for Neutron Scattering at ETH Zurich and Paul Scherrer Institute.
47. Hay, P. J.; Thibeault, J. C.; Hoffmann, R. Orbital interactions in metal dimer complexes. *J. Am. Chem. Soc*. 1975, *97*, 4884-4899.
48. Volkova, O. S.; Shvanskaya, L. V.; Ovchenkov, E. A.; Zvereva, E. A.; Volkov, A. S.; Chareev, D. A.; Molla, K.; Rahaman, B.; Saha-Dasgupta, T.; Vasiliev, A. N. Structure-property relationships in α-, β'- and γ-modifications of $Mn_3(PO_4)_2$. *Inorg. Chem.* 2016, *55*, 10692-10700.
49. Baies, R.; Caignaert, V.; Pralong, V.; Raveau, B. Copper hydroxydiphosphate with a one-dimensional arrangement of copper polyhedra: $Cu_3[P_2O_6OH]_2$. *Inorg. Chem.* 2005, *44*, 2376-2380.
50. Hase, M.; Kohno, M.; Kitazawa, H.; Tsujii, N.; Suzuki, O.; Ozawa, K.; Kido, G.; Imai, M.; Hu, X. 1/3 magnetization plateau observed in the spin-1/2 trimer chain compound $Cu_3(P_2O_6OH)_2$. *Phys. Rev. B* 2006, *73*, 104419.
51. Koo, H.-J.; Whangbo, M.-H. Finite magnetization plateau from a two-dimensional antiferromagnet: density functional analysis of the magnetic structure of $Cu_3(P_2O_6OH)_2$. *Inorg. Chem.* 2010, *49*, 9253-9256.
52. Ishikawa, H.; Yamaura, J.; Okamoto Y.; Yoshida, H.; Nilsen, G. J.; Hiroi, Z. A novel crystal polymorph of volborthite, $Cu_3V_2O_7(OH)_2·2H_2O$. Acta Cryst. C 2012, 68, i41–i44.
53. Whangbo, M.-H.; Koo, H.-J.; Brücher, E.; Puphal, P.; Kremer, R. K. Absence of spin frustration in the kagomé layers of $Cu^{2+}$ ions in volborthite $Cu_3V_2O_7(OH)_2·2H_2O$ and observation of the suppression and re-entrance of specific heat anomalies in volborthite under an external magnetic field. *Condens. Matter* 2022, *7*, 24.
54. Yoshida, M.; Takigawa, M.; Yoshida, H.; Okamoto, Y. Hiroi, Z. Phase diagram and spin dynamics in volborthite with a distorted kagomé lattice. *Phys. Rev. Lett*. 2009, *103*, 077207.
55. Ishikawa, H.; Yoshida, M.; Nawa, K.; Jeong, M.; Krämer, S.; Horvatić, M.; Berthier, C.; Takigawa, M.; Akaki, M.; Miyake, A.; Tokunaga, M.; Kindo, K.; Yamaura, J.; Okamoto, Y.; Hiroi, Z. One-third magnetization plateau with a preceding novel phase in volborthite. *Phys. Rev. Lett.* 2015, *114*, 227202.
56. Janson, O.; Furukawa, S.; Momoi, T.; Sindzingre, P.; Richter, J.; Held, K. Magnetic behavior of volborthite $Cu_3V_2O_7(OH)_2 \cdot 2H_2O$ determined by coupled trimers rather than frustrated chains. Phys. Rev. Lett. 2016, 117, 037206.
57. Wiessner, R. M.; Fledderjohann, A.; Mütter, K.-H.; Karbach, M. Spontaneous magnetization in spin-ladder systems with competing interactions. Phys. Rev. B 1999, *60*, 6545.
58. Moskin, A. V.; Kozlyakova, E. S.; Shvanskaya, L. V.; Chareev, D. A.; Koo, H.-J.; Whangbo, M.-H.; Vasiliev, A. N. Highly anisotropic 1/3-magnetization plateau in a ferrimagnet $Cs_2Cu_3(SeO_3)_4 \cdot 2H_2O$: Topology of magnetic bonding necessary for magnetization plateau. *Dalton Trans.* 2023, *52*, 118-127.
59. Starova, L. P.; Filatov, S. K.; Fundamenskii, V. S.; Vergasova, L. P. The crystal structure of fedotovite, $K_2Cu_3O(SO_4)_3$. *Mineral. Mag*. 1991, *55*, 613-616.
60. Hase, M.; Rule, R. C.; Hester, J. R.; Fernandez-Baca, J. A.; Masuda, T.; Matsuo, Y. A possible magnetic structure of the cluster-based Haldane compound fedotovite $K_2Cu_3O(SO_4)_3$. *J. Phys. Soc. Jpn*. 2019, *88*, 094708.
61. Miyazaki, Y.; Yang, H. X.; Kajitani, T. Compounds and subsolidus phase relations in the CaO-$Co_3O_4$-CuO system. *J. Solid State Chem*. 2005, *178*, 2973-2979.
62. Kageyama, H.; Yoshimura, K.; Kosuge, K.; Azuma, M.; Takano, M.; Mitamura, H.; Goto, T. Magnetic anisotropy of $Ca_3Co_2O_6$ with ferromagnetic Ising chains. *J. Phys. Soc. Jpn.* 1997, *66*, 3996-4000.
63. Whangbo. M.-H.; Dai, D.; Koo, H.;-J.; Jobic, S. Investigations of the oxidation states and spin distributions in $Ca_3Co_2O_6$ and $Ca_3CoRhO_6$ by spin-polarized electronic band structure calculations. *Solid State Commun*. 2003, *125*, 413-417.
64. Munaò, I.; Zvereva, E. A.; Volkova, O. A.; Vasiliev, A. N.; Armstrong, A. R.; Lightfoot, P. $NaFe_3(HPO_3)_2((H,F)PO_2OH)_6$: A Potential Cathode Material and a Novel Ferrimagnet. *Inorg. Chem*. 2016, *55*, 2558−2564.
65. Vasiliev, A. N.; Volkova, O. S.; Zvereva, E. A.; Ovchenkov, E. A.; Munao, I.; Clark, L.; Lightfoot, P.; Vavilova, E. L.; Kamusella, S.; Klauss, H.-H.; Werner, J.; Koo, C.; Klingeler, R.; Tsirlin, A. A. 1/3 magnetization plateau and frustrated ferrimagnetism in a sodium iron phosphite. *Phys. Rev. B* 2016, *93*, 134401.
66. Koo, H.-J.; Kremer, R. K.; Whangbo, M.-H. High-Spin Orbital Interactions Across van der Waals Gaps Controlling the Interlayer Ferromagnetism in van der Waals Ferromagnets. *J. Am. Chem. Soc.* 2022, *144*, 16272-16275.
67. Zigan, F.; Schuster, H. D. 1972 Verfeinerung der Struktur von Azurit, $Cu_3(OH)_2(CO_3)_2$, durch Neutronenbeugung. *Z. Kristallogr*. 1972, *135*, 416-436.
68. Rule, K. C.; Reehuis, M.; Gibson, M. C. R.; Ouladdiaf, B.; Gutmann, M. J.; Hoffmann, J.-U.; Gerischer, S.; Tennant, D. A.; Süllow, S.; Lang, M. Magnetic and crystal structure of azurite $Cu_3(CO_3)_2(OH)_2$ as determined by neutron diffraction. *Phys. Rev. B* 2011, *83*, 104401.

69. Kang, J.; Lee, C.; Kremer, R. K.; Whangbo, M.-H. 2009 Consequences of the intrachain dimer–monomer spin frustration and the interchain dimer–monomer spin exchange in the diamond-chain compound azurite $Cu_3(CO_3)_2(OH)_2$. *J. Phys.: Condens. Matter* 2009, *21*, 392201.
70. Jeschke, H.; Opahle, I.; Kandpal, H.; Valentí, R.; Das, H.; Saha-Dasgupta, T.; Janson, O.; Rosner, H.; Brühl, A.; Wolf, B.; Lang, M.; Richter, J.; Hu, S.; Wang, X.; Peters, R.; Pruschke, Th.; Honecker, A. Multistep Approach to Microscopic Models for Frustrated Quantum Magnets: The Case of the Natural Mineral Azurite. *Phys. Rev. Lett*. 2011, *106*, 217201.
71. Kikuchi, H.; Fujii, Y.; Chiba, M.; Mitsudo, S.; Idehara, T.; Tonegawa, T.; Okamoto, K.; Sakai, T.; Kuwai, T.; Ohta, H. Experimental Observation of the 1/3 Magnetization Plateau in the Diamond-Chain Compound $Cu_3(CO_3)_2(OH)_2$. *Phys. Rev. Lett*. 2005, *94*, 227201.
72. Ohta, H.; Okubo, S.; Kamikawa, T.; Kunimoto, T.; Inagaki, Y.; Kikuchi, H.; Saito, T.; Azuma, M.; Takano, M. High Field ESR Study of the S = 1/2 Diamond-Chain Substance $Cu_3(CO_3)_2(OH)_2$ up to the Magnetization Plateau Region. *J. Phys. Soc. Jpn*. 2003, *72*, 2464.
73. Lee, C.; Kan, E. J.; Xiang, H. J.; Kremer, R. K.; Lee, S.-H.; Hiroi, Z.; Whangbo, M.-H. Spin Reorientation in the Square-Lattice Antiferromagnets RMnAsO (R = Ce, Nd): Density Functional Analysis of the Spin-Exchange Interactions between the Rare-Earth and Transition-Metal Ions. *Inorg. Chem*. 2012, *51*, 6890−6897.
74. Chubukov, A. V.; Golosov, D. I. Quantum theory of an antiferromagnet on a triangular lattice in a magnetic field. *J. Phys.: Condens. Matter* 1991, *3*, 69.
75. Nishimoto, S,; Shibata, N.; Hotta, C. Controlling frustrated liquids and solids with an applied field in a kagome Heisenberg antiferromagnet. *Nat. Commun*. 2013, *4*, 2287/.
76. Capponi, S.; Derzhko, O.; Honecker, A.; Läuchli, A. M.; Richter, J. Numerical study of magnetization plateaus in the spin-1/2 kagome Heisenberg antiferromagnet. *Phys. Rev. B* 2013, *88*, 144416.
77. Schnack, J.; Schulenburg, J.; Honecker, A.; Richter, J. Magnon Crystallization in the Kagome Lattice Antiferromagnet. *Phys. Rev. Lett*. 2020, *125*, 117207.
78. Suetsugu, S.; Asaba, T.; Kasahara, Y.; Kohsaka,Y.; Totsuka, K.; Li, B.; Zhao, Y.; Li, Y.; Tokunaga, M.; Matsuda, Y. Emergent spin-gapped magnetization plateaus in a spin-1/2 perfect kagome antiferromagnet. arXiv:2310.10069.
79. Jeon, S.; Wulferding, D.; Choi, Y.; Lee, S.; Nam, K.; Kim, K. H.; Lee, M.; Jang, T.-H.; Park, J.-H.; Lee, S.; Choi,S.; Lee, C.; Nojiri, H.; Choi, K.-Y. One-ninth magnetization plateau stabilized by spin entanglement in a kagome antiferromagnet. *Nat. Phys*. 2024, *20*, 435-441.
80. Cabra, D. C.; Grynberg, M. D.; Holdsworth, P. C. W.; Honecker, A.; Pujol, P.; Richter, J.; Schmalfuss, D.; Schulenburg, J. Quantum kagome antiferromagnet in a magnetic field: Low-lying non-magnetic excitations versus valence-bond crystal order. *Phys. Rev. B* 2005, *71*, 144420.
81. Petrenko, O. A.; Lees, M. R.; Balakrishnan, G.; de Brion, S.; Chouteau, G. Revised magnetic properties of $CuFeO_2$ – a case of mistaken identity. *J. Phys.: Condens. Matter* 2005, *17*, 2741−2747.
82. Saito, M.; Watanabe, M.; Kurita, N.; Matsuo, A.; Kindo, K.; Avdeev, M.; Jeschke, H. O.; Tanaka, H. Successive phase transitions and magnetization plateau in the spin-1 triangular-lattice antiferromagnet $Ba_2LaNiTe_2O_{12}$ with small easy-axis anisotropy. Phys. Rev. B 2019, 100, 064417.
83. Goto, M.; Ueda, H.; Michioka, C.; Matsuo, A.; Kindo, K.; Yoshimura, K. Various disordered ground states and 1/3 magnetization-plateau-like behavior in the S=1/2 $Ti^{3+}$ kagomé lattice antiferromagnets $Rb_2NaTi_3F_{12}$, $Cs_2NaTi_3F_{12}$, and $Cs_2KTi_3F_{12}$. *Phys. Rev B* 2016, *94*, 104432.

84. Okuta K, Hara S, Sato H, Narumi Y and Kindo K Observation of 1/3 magnetization-plateau-like anomaly in S=3/2 perfect kagomé lattice antiferromagnet KCr$_3$(OH)$_6$(SO$_4$)$_2$ (Cr-jarosite). *J. Phys. Soc. Jpn.* 2011, *80*, 063703.
85. Okuma, R.; Nakamura, D.; Takeyama, S. Magnetization plateau observed by ultrahigh-field Faraday rotation in the kagomé antiferromagnet herbertsmithite. *Phys. Rev. B* 2020, *102*, 104429.
86. Smirnov, A. I.; Soldatov, T. A.; Petrenko, O. A.; Takata, A.; Kida, T.; Hagiwara, M.; Shapiro, A. Ya.; Zhitomirsky, M. E. Order by quenched disorder in the model triangular antiferromagnet RbFe(MoO$_4$)$_2$. *Phys. Rev. Lett.* 2017, *119*, 047204.
87. Svistov, L. E.; Smirnov,. A. I.; Prozorova, L. A.; Petrenko, O. A.; Demianets, L. N.; Shapiro, A. Ya. Quasi-two-dimensional antiferromagnet on a triangular lattice RbFe(MoO$_4$)$_2$. *Phys. Rev. B* 2006, *67*, 139901.
88. Doi, Y.; Hinatsu, Y.; Ohoyama, K. Structural and magnetic properties of pseudo-two-dimensional triangular antiferromagnets Ba$_3$MSb$_2$O$_9$ (M = Mn, Co, and Ni). *J. Phys.: Condens. Matter* 2004, *16*, 8923-8935.
89. Kamiya, Y.; Ge, L.; Hong, T.; Qiu, Y.; Quintero-Castro, D. L.; Lu, Z.; Cao, H. B.; Matsuda, M.; Choi, E. S.; Batista, C. D.; Mourigal, M.; Zhou, H. D.; Ma, J. The nature of spin excitations in the one-third magnetization plateau phase of Ba$_3$CoSb$_2$O$_9$. *Nature Comm.* 2018, *9*, 2666.
90. Tanaka, H.; Ono, T.; Katori, H. Magnetic phase transition and magnetization plateau in Cs$_2$CuBr$_4$. *Progr. Theor. Phys. Suppl.* 2022, *145*, 101−106.
91. Whangbo, M.-H.; Koo, H.-J.; Dai, D. Spin exchange interactions and magnetic structures of extended magnetic solids with localized spins: theoretical descriptions on formal, quantitative and qualitative level. *J. Solid State Chem.* 2005, *176*, 417-481.
92. dos Santos A M, Brandao P, Fitch A, Reis M S, Amaral V S and Rocha J. Synthesis, crystal structure and magnetic characterization of Na$_2$Cu$_5$(Si$_2$O$_7$)$_2$: An inorganic ferrimagnetic chain. J. Solid State Chem. 2007, 180, 16-21.
93. Vasiliev A N, unpublished work.
94. Liu, X. C.; Ouyang, Z. W.; Yue, X. Y.; Jiang, X.; Wang, Z. X.; Wang, J. F.; Xia, Z. C. Structure and 3/7-like magnetization plateau of layered Y$_2$Cu$_7$(TeO$_3$)$_6$Cl$_6$(OH)$_2$ containing diamond chains and trimers. *Inorg. Chem.* 2019, *58*, 10680−10685.
95. Volkova, L. M.; Marinin, D. V. Antiferromagnetic spin-frustrated layers of corner-sharing Cu$_4$ tetrahedra on the kagomé lattice in volcanic minerals Cu$_5$O$_2$(VO$_4$)$_2$(CuCl), NaCu$_5$O$_2$(SeO$_3$)$_2$Cl$_3$, and K$_2$Cu$_5$Cl$_8$(OH)$_4$·2H$_2$O. *J. Phys.: Condens. Matter* 2018, *30*, 425801.
96. Zhang, S.-Y.; Guo, W.-B.; Yang, M.; Tang, Y.-Y.; Cui, M.-Y.; Wang, N.-N.; He, Z.-Z. A frustrated ferrimagnet Cu$_5$(VO$_4$)$_2$(OH)$_4$ with a 1/5 magnetization plateau on a new spin-lattice of alternating triangular and honeycomb strips. *Dalton Trans.* 2015, **44**, 20562−20567.
97. Vasiliev, A. N.; Berdonosov, P. S.; Kozlyakova, E. S.; Maximova, O. V.; Murtazoev, A. F.; Dolgikh, V. A,; Lyssenko, K. A.; Pchelkina, Z. V.; Gorbunov, D. I.; Chung, S. H.; Koo, H.-J.; Whangbo, M.-H. Observation of a 1/3 magnetization plateau in Pb$_2$Cu$_{10}$O$_4$(SeO$_3$)$_4$Cl$_7$ arising from (Cu$^{2+}$)$_7$ clusters of corner-sharing (Cu$^{2+}$)$_4$ tetrahedra. *Dalton Trans.* 2022, *51*, 15017−15021.
98. Tang, Y.; Guo, W.; Zhang, S.; Xiang, H.; Cui, M.; He, Z. Na$_2$Cu$_7$(SeO$_3$)$_4$O$_2$Cl$_4$: a selenite chloride compound with Cu$_7$ units showing spin-frustration and a magnetization plateau. *Dalton Trans*. 2016, *45*, 8324−8326.
99. Albright, T. A.; Burdett, J. K. Whangbo, M.-H. Orbital Interactions in Chemistry, Wiley, New York, 1985, Ch. 17.

100. Facheris, L.; Povarov, K. Yu.; Nabi, S. D.; Mazzone, D. G.; Lass, J.; Roessli, B.; Ressouche, E.; Yan, Z.; Gvasaliya, S.; Zheludev, A. Spin Density Wave versus Fractional Magnetization Plateau in a Triangular Antiferromagnet. *Phys. Rev. Lett.* 2022, *129*, 087201.
101. Cao, J. J.; Ouyang, Z. W.; Liu, X. C.; Xiao, T. T.; Song, Y. R.; Wang, J. F.; Ishii, Y.; Zhou, X. G.; Matsuda, Y. H. Unusual dimerization and magnetization plateaus in S = 1 skew chain $Ni_2V_2O_7$ observed at 120 T. *Phys. Rev. B* 2022, *106*, 184409.
102. Guo, K.; Ma, Z.; Liu, H.; Wu, Z.; Wang, J.; Shi, Y.; Li, Y.; Jia, S. 1/3 and other magnetization plateaus in the quasi-one-dimensional Ising magnet $TbTi_3Bi_4$ with zigzag spin chain. *Phys. Rev. B* 2024, *110*, 064416.
103. Chen, Y.; Zhang, Y.; Li, R.; Su, H.; Shan, Z.; Smidman, M.; Yuan, H. Multiple magnetic phases and magnetization plateaus in $TbRh_6Ge_4$. *Phys. Rev. B* 2023, *107*, 094414.
104. Yuan, N.; Elghandour, A.; Hergett, W.; Ohlendorf, R.; Gries, L.; Klingeler, R. 1/3 plateau and 3/5 discontinuity in the magnetization and the magnetic phase diagram of hexagonal $GdInO_3$. *Phys. Rev. B*, 2023, *108*, 224403.
105. Kikuchi, H.; Fujii, Y.; Ishikawa, Y.; Matsuo, A.; Kindo, K.; Widyaiswari, U.; Watanabe, I. Magnetic Plateaux of the Frustrated Magnet, Pseudomalachite. *JPS Conf. Proc.* 2023, *38*, 011127.
106. Yalikun, A.; Wang, Y.; Shi, N.; Chen, Y.; Huang, H.; Koo, H.-J.; Ouyang, Z.; Xia, Z.; Kremer, R.K.; Whangbo, M.-H.; Lu, H. Polar Layered Magnet $Ba_2Cu_3(SeO_3)_4F_2$ Composed of Bitriangular Chains: Observation of 1/3-Magnetization Plateau. *Chem. Mater.* 2024, *36*, 7887–7896.
107. Tanaka, H.; Matsuo, A.; Kindo, K. One-half magnetization plateau in the spin-1/2 fcc lattice antiferromagnet $Sr_2CoTeO_6$. *Phys. Rev. B* 2024, *109*, 094414.
108. Liu, X.; Ouyang, Z.; Jiang, D.; Cao, J.; Xiao, T.; Wang, Z.; Xia, Z.; Tong, W. Syntheses, structure and magnetization plateau of staircase kagome-lattice antiferromagnet $Cu_3Bi(TeO_3)_2O_2Cl$. *J. Magn. Magn. Mater.* 2023, *565*, 170228.
109. Shangguan, Y.; Bao, S.; Dong, Z,-Y.; Xi, N.; Gao, Y.-P.; Ma, Z.; Wang, W.; Qi, Z.; Zhang, S.; Huang, Z.; Liao, J.; Zhao, X.; Zhang, B.; Cheng, S.; Xu, H.; Yu, D.; Mole, R. A.; Murai, N.; Ohira-Kawamura, S.; He, L.; Hao, J.; Yan, Q.-B.; Song, F.; Li, W.; Yu, S.-L.; Li, J.-X.; Wen, J. A one-third magnetization plateau phase as evidence for the Kitaev interaction in a honeycomb-lattice antiferromagnet. *Nat. Phys.* 2023, *19*, 1883-1889.
110. Hida, K.; Affleck, I. Quantum vs classical magnetization plateaus of S=1/2 frustrated Heisenberg chains. *J. Phys. Soc. Jpn.* 2005, *74*, 1849-1857.
111. Sakai, T.; Okamoto, K. Quantum magnetization plateaux of an anisotropic ferrimagnetic spin chain. *Phys. Rev. B* 2002, *64*, 214403.
112. Momoi, T.; Totsuka, K. Magnetization plateaus as insulator-superfluid transitions in quantum spin systems. *Phys. Rev. B* 2000, *61*, 3231-3234.
113. Heidrich-Meissner, F.; Sergienko, I. A.; Feiguin, A. E.; Dagotto, E. R. Universal emergence of the one-third plateau in the magnetization process of frustrated quantum spin chains. *Phys. Rev. B* 2007, *75*, 064413.
114. Takayoshi, S.; Totsuka, K.; Tanaka, A. Symmetry-protected topological order in magnetization plateau states of quantum spin chains. *Phys. Rev. B* 2015, *91*, 155136.
115. Narumi, Y.; Katsumata, K.; Honda, Z.; Domenge, J.-C.; Sindzingre, P.; Lhuillier, C.; Shimaoka, Y.;. Kobayashi, T.C; Kindo, K. Observation of a transient magnetization plateau in a quantum antiferromagnet on the kagomé lattice. *Europhys. Lett.* 2004, *65*, 705–711.

116. Hu, H.; Cheng, C.; Xu, Z.; Luo, H.-G.; Chen, S. Topological nature of magnetization plateaus in periodically modulated quantum spin chains. *Phys. Rev. B* 2014, *90*, 035150.
117. Coletta, T.; Zhitomirsky, M. E.; Mila, F. Quantum stabilization of classically unstable plateau structures. *Phys. Rev. B* 2013, *87*, 060407(R).
118. Tandon, K.; Lal, S.; Pati, S.K.; Ramasesha, S., Sen, D. Magnetization properties of some quantum spin ladders. *Phys. Rev. B* 1999, *59*, 396-410.
119. Adhikary, M.; Ralko, A.; Kumar, B. Quantum paramagnetism and magnetization plateaus in a kagomé-honeycomb Heisenberg antiferromagnet. *Phys. Rev. B* 2021, *104*, 094416.
120. Ohanyan, V.; Rojas, O.; Strečka, J.; Bellucci, S. Absence of actual plateaus in zero-temperature magnetization curves of quantum spin clusters and chains. *Phys. Rev. B* 2015, *92*, 214423.
121. Ananikian, N.S.; Strečka, J.; Hovhannisyan, V. Magnetization plateaus of an exactly solvable spin-1 Ising–Heisenberg diamond chain. *Sol. St. Comm.* 2014, *194*, 48-53.
122. Totsuka, K. Magnetization process of random quantum spin chains. *Phys. Rev. B* 2001, *64*, 134420.
123. Pal, S.; Lal, S. Magnetization plateaus of the quantum pyrochlore Heisenberg antiferromagnet. *Phys. Rev. B* 2019, *100*, 104421.
124. Plat, X.; Alet, F.; Capponi, S.; Totsuka, K. Magnetization plateaus of an easy-axis kagomé antiferromagnet with extended interactions. *Phys. Rev. B* 2015, *92*, 174402.
125. Yoshida, H. K. Frustrated kagomé antiferromagnets under high magnetic fields *J. Phys. Soc. Jpn*. 2022, *91*, 101003.

Supporting Information

for

**Magnetization Plateaus by the Field-Induced Partitioning of Spin Lattices**

**Myung-Hwan Whangbo[1],*  Hyun-Joo Koo,[2] Reinhard K. Kremer,[3] and Alexander N. Vasiliev[4],***

[1] Department of Chemistry, North Carolina State University, Raleigh, NC 27695-8204, USA
[2] Department of Chemistry and Research Institute for Basic Sciences, Kyung Hee University, Seoul 02447, Republic of Korea
[3] Max Planck Institute for Solid State Research, Heisenbergstrasse 1, Stuttgart D-70569, Germany
[4] Department of Low Temperature Physics and Superconductivity, Lomonosov Moscow State University, Moscow 119991, Russia

whangbo@ncsu.edu
anvas2000@yahoo.com

## S1. CoGeO$_3$

### A. Spin exchange paths

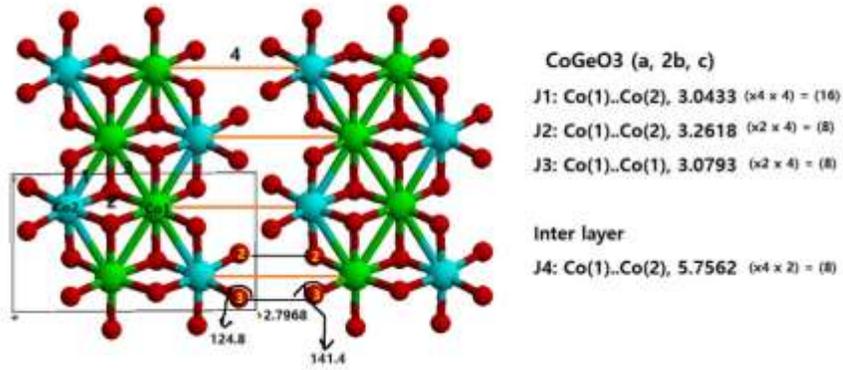

### B. Ordered spin states using a (a, 2b, c) superstructure

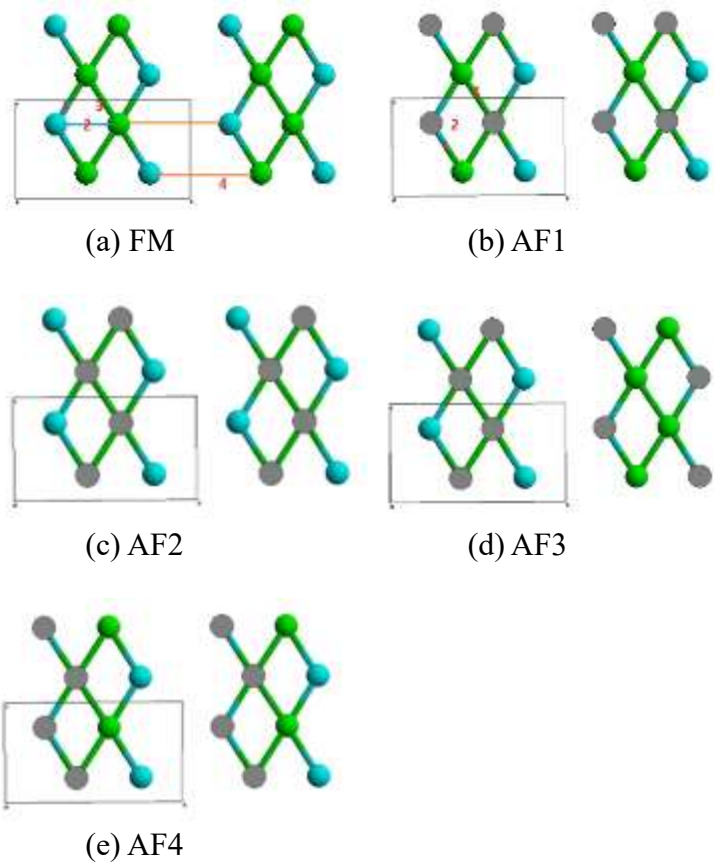

(a) FM            (b) AF1

(c) AF2           (d) AF3

(e) AF4

Figure 1. Ordered spin arrangements of (a) FM, (b) AF1, (c) AF2, (d) AF3 and (e) AF4 states.

### C. Energies of the ordered spin states in terms of the spin exchanges

$E_{FM} = (-16J_1 - 8J_2 - 8J_3 - 8J_4)(N^2/4)$
$E_{AF1} = (+16J_1 - 8J_2 + 8J_3 - 8J_4)(N^2/4)$
$E_{AF2} = (+16J_1 + 8J_2 - 8J_3 + 8J_4)(N^2/4)$
$E_{AF3} = (+16J_1 + 8J_2 - 8J_3 - 8J_4)(N^2/4)$
$E_{AF4} = (-16J_1 + 8J_2 + 8J_3 + 8J_4)(N^2/4)$

**D. Spin exchanges in terms of the ordered spin state energies**

$J_4 = (1/16)(E_{AF2} - E_{AF3})(4/N^2)$
$J_3 = (1/32)[(E_{AF4} - E_{FM}) - (E_{AF2} - E_{AF1})](4/N^2)$
$J_2 = (1/16)[\{(E_{AF4} - E_{FM})(4/N^2)\} - 16J_3 - 16J_4]$
$J_1 = (1/32)[(E_{AF3} - E_{AF4})(4/N^2) + 16J_3 + 16J_4]$

**E. Ordered spin state energies and spin exchanges from DFT+U calculations**

Table 1. Relative energies (in meV/FU) of the broken-symmetry states and the spin exchange parameters (in K) obtained from DFT+U calculations

|      | U = 3 eV | U = 4 eV |
|------|----------|----------|
| FM   | 61.13    | 65.07    |
| AF1  | 15.66    | 15.69    |
| AF2  | 3.22     | 2.61     |
| AF3  | 4.11     | 3.25     |
| AF4  | 0        | 0        |

(a, 2b, c) supercell
PBE functional for the exchange-correlation
SCF convergence criterion = $10^{-6}$ eV
Plane wave cutoff energy = 450 eV
kpoint set = (6x4x8)

|      | U = 3 eV | U = 4 eV |
|------|----------|----------|
| $J_1$ | 54.45   | 60.29    |
| $J_2$ | 185.03  | 198.14   |
| $J_3$ | 125.52  | 134.02   |
| $J_4$ | 4.60    | 3.33     |

**F. Ordered spin state energies and spin exchanges from DFT+U+SOC calculations**

Table 2. Relative energies (meV/Co) with respect to the spin orientation //c obtained from DFT+U(4eV)+SOC calculations.

|     | //a  | //b  | //c  |
| --- | ---- | ---- | ---- |
| Co1 | 0.27 | 0    | 1.19 |
| Co2 | 0.21 | 0.68 | 0    |

*The $Co^{2+}$ sites other than the one under investigation were replaced with $Zn^{2+}$ ions.

Table 3. Relative energies (in meV/FU) and spin exchange parameters (in K) obtained from DFT+U(4eV)+SOC calculations

|     | U = 4 eV |
| --- | -------- |
| FM  | 62.25    |
| AF1 | 4.84     |
| AF2 | 0        |
| AF3 | 7.35     |
| AF4 | 0.31     |

|       | U = 4 eV |
| ----- | -------- |
| $J_1$ | 74.39    |
| $J_2$ | 134.27   |
| $J_3$ | 147.19   |
| $J_4$ | 37.88    |

## S2. $Ba_3Mn_2O_8$

### A. Spin exchange paths using a (2a, b, 2c) supercell

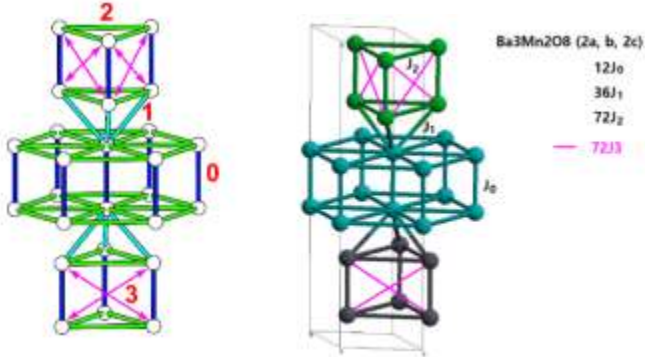

Figure 1. Spin exchange paths in $Ba_3Mn_2O_8$. The numbers 0 to 3 represent the spin exchange paths $J_0$ to $J_3$, respectively. The white circles indicate the $Mn^{2+}$ ions sites.

### B. Ordered spin states

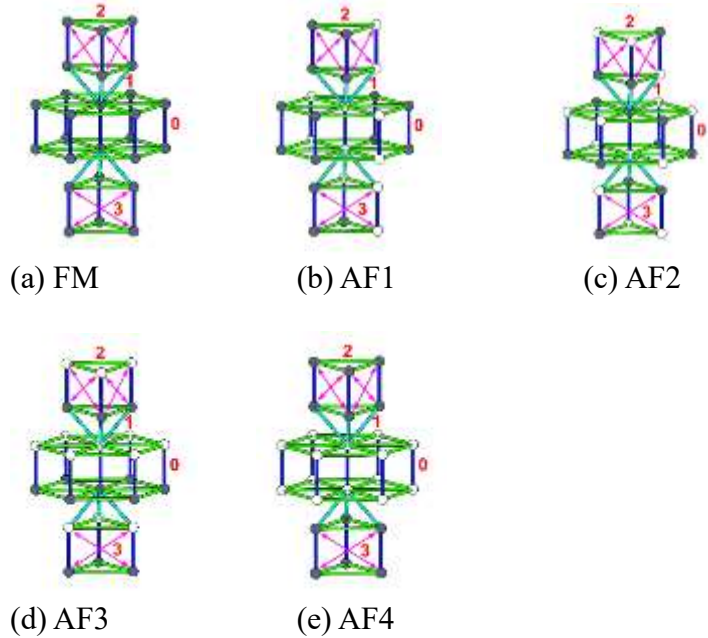

(a) FM  (b) AF1  (c) AF2

(d) AF3  (e) AF4

Figure 2. Ordered spin arrangements of (a) FM, (b) AF1, (c) AF2, (d) AF3 and (e) AF4 state. The gray and white circles indicate the up and down spin sites of $Mn^{2+}$ ions, respectively.

### C. Energies of the ordered spin states in terms of the spin exchanges

$E_{FM} = (-12J_0 - 36J_1 - 72J_2 - 72J_3)(N^2/4)$
$E_{AF1} = (-12J_0 - 4J_1 + 24J_2 + 24J_3)(N^2/4)$
$E_{AF2} = (+12J_0 + 4J_1 + 24J_2 - 24J_3)(N^2/4)$

$E_{AF3} = (+ 12J_0 + 36J_1 - 72J_2 + 72J_3)(N^2/4)$
$E_{AF4} = (- 12J_0 + 36J_1 - 72J_2 - 72J_3)(N^2/4)$

**D. Spin exchanges in terms of the ordered spin state energies**

$J_1 = (1/72)(E_{AF4} - E_{FM})(4/N^2)$
$J_3 = (1/192)[\{(E_{AF3} - E_{AF4}) - (E_{AF2} - E_{AF1})\}(4/N^2) + 8J_1]$
$J_0 = (1/24)[(E_{AF3} - E_{AF4})(4/N^2) - 144J_3]$
$J_2 = (1/96)[(E_{AF2} - E_{AF3})(4/N^2) + 32J_1 + 96J_3]$

**E. Ordered spin state energies and spin exchanges from DFT+U calculations**

Table 1. Relative energies (in meV/FU) and spin exchange parameters (in K) obtained from DFT+U calculations

|     | U = 2 eV | U = 3 eV | U = 4 eV |
| --- | --- | --- | --- |
| FM  | 13.55 | 9.77 | 7.23 |
| AF1 | 7.35 | 6.15 | 5.70 |
| AF2 | 0 | 0 | 0.47 |
| AF3 | 0.68 | 0.02 | 0 |
| AF4 | 10.10 | 7.45 | 5.78 |

(2a, b, 2c) super cell
PBE functional for the exchange-correlation
SCF convergence criterion = $10^{-6}$ eV
Plane wave cutoff energy = 450 eV
kpoint set = (6x6x3)

|     | U = 2 eV | U = 3 eV | U = 4 eV |
| --- | --- | --- | --- |
| $J_0$ | 21.98 | 18.21 | 15.23 |
| $J_1$ | 3.34 | 2.24 | 1.40 |
| $J_2$ | 2.50 | 1.32 | 0.39 |
| $J_3$ | 0.89 | 0.56 | 0.26 |

## S2. Supplementary figures

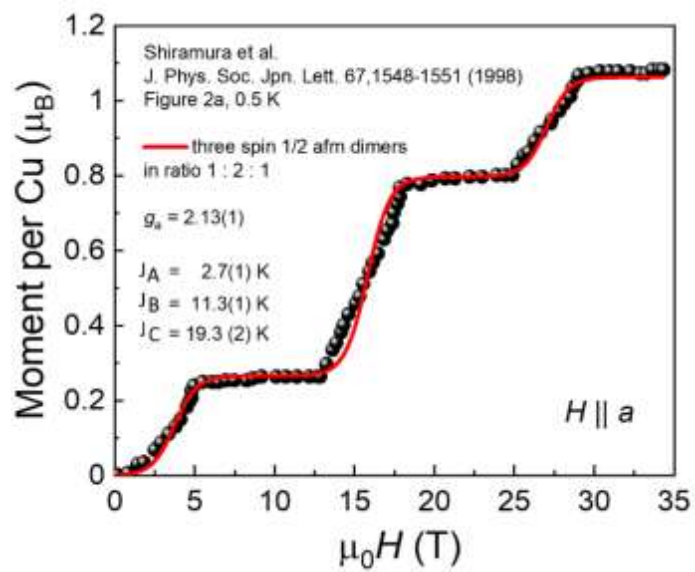

**Fig. S1**. Magnetization curve of $NH_4CuCl_3$ obtained by using $H\|a$ (black dots) simulated by assuming that dimers A, B and C are all singlet dimers (solid red curve).

(a) Interactions in the $J_2$ exchange path

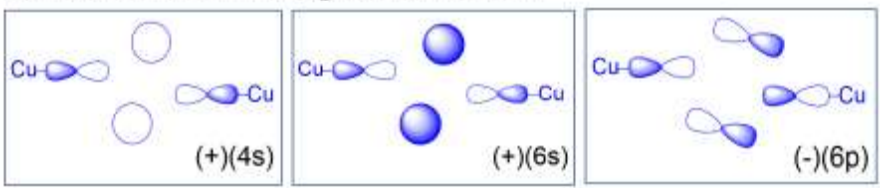

(b) Interactions in the $J_a'$ exchange path

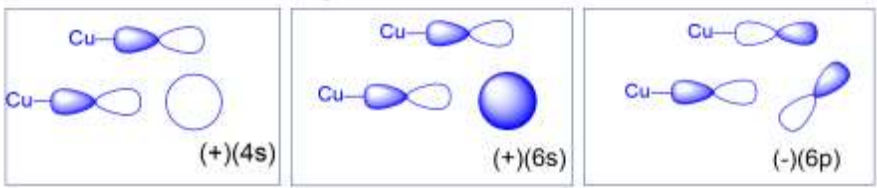

**Fig. S2**. Orbital interactions of the (+) and (-) d-states of (CuCl4)2 dimer with the frontier orbitals of the A⁺ cations making Cl…A+…Cl bridge in the (a) $J_2$ and (b) $J_a'$ exchange path. For simplicity, the (+) and (-) states are represented by showing only the Cl 3p-orbital of the Cu-Cl bond making the Cl…A+…Cl bridge.

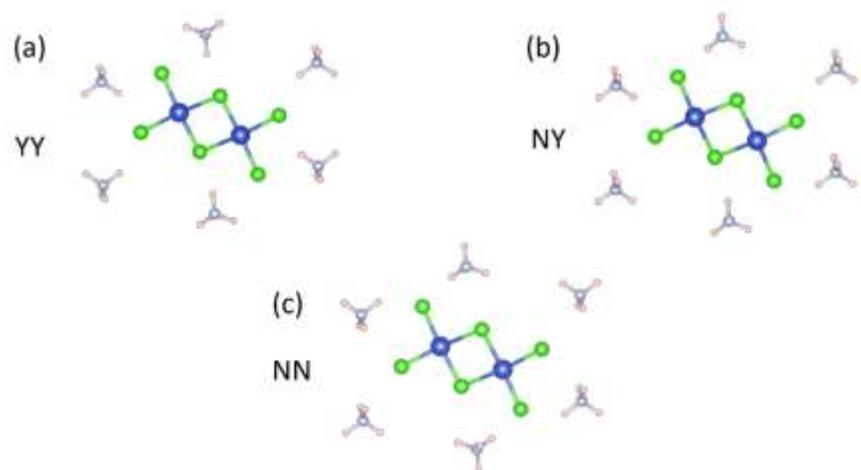

**Fig. S3**. Orientations of the six $NH_4^+$ cations surrounding each $Cu_2Cl_6^{2-}$ anion in $NH_4CuCl_3$ with the (a) YY, (b) NY and (c) NN arrangements of the $NH_4^+$ cations.

## S4. KCuCl₃

### A. Spin exchange paths

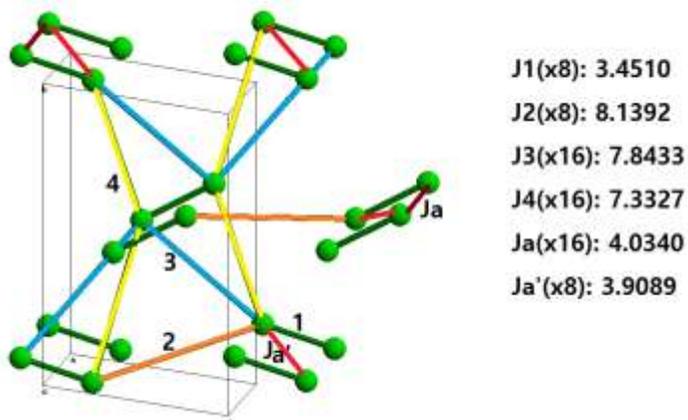

J1(x8): 3.4510
J2(x8): 8.1392
J3(x16): 7.8433
J4(x16): 7.3327
Ja(x16): 4.0340
Ja'(x8): 3.9089

### B. Ordered spin states using a (2a, b, 2c) superstructure

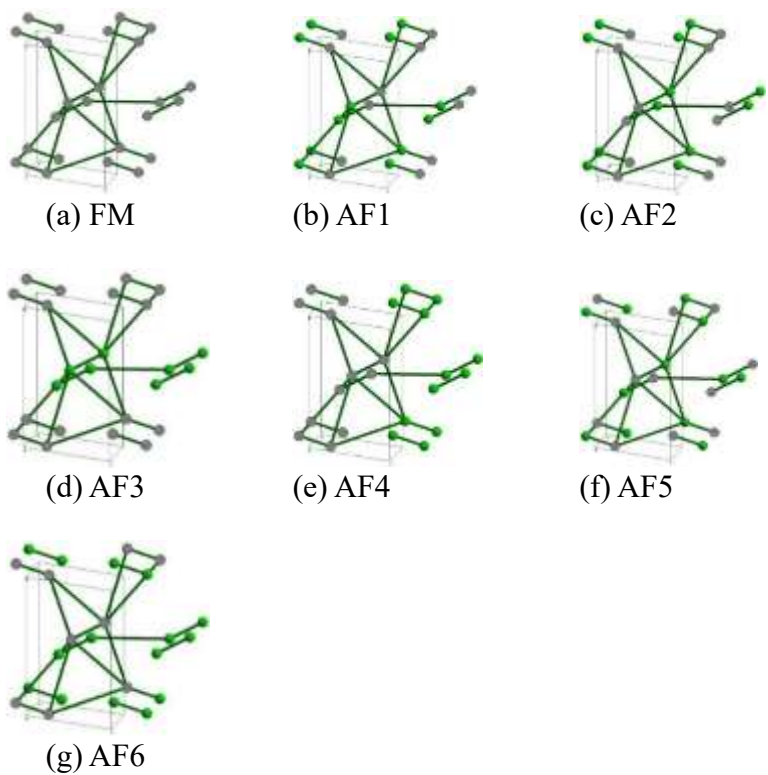

(a) FM    (b) AF1    (c) AF2

(d) AF3    (e) AF4    (f) AF5

(g) AF6

Figure 1. Ordered spin arrangements of (a) FM, (b) AF1, (c) AF2, (d) AF3, (e) AF4, (f) AF5 and (g) AF6 states.

### C. Energies of the ordered spin states in terms of the spin exchanges

$E_{FM} = (-8J_1 - 8J_2 - 16J_3 - 16J_4 - 16J_a - 8J_{a'})(N^2/4)$
$E_{AF1} = (+8J_1 + 8J_2 - 16J_3 + 16J_4 - 16J_a + 8J_{a'})(N^2/4)$
$E_{AF2} = (+8J_1 + 8J_2 + 16J_3 - 16J_4 - 16J_a + 8J_{a'})(N^2/4)$
$E_{AF3} = (-8J_1 - 8J_2 + 16J_3 + 16J_4 - 16J_a - 8J_{a'})(N^2/4)$
$E_{AF4} = (-8J_1 + 8J_2 \qquad\qquad - 16J_a - 8J_{a'})(N^2/4)$
$E_{AF5} = (+8J_1 + 8J_2 + 16J_3 - 16J_4 + 16J_a - 8J_{a'})(N^2/4)$
$E_{AF6} = (-8J_1 - 8J_2 - 16J_3 - 16J_4 + 16J_a + 8J_{a'})(N^2/4)$

## D. Spin exchanges in terms of the ordered spin state energies

$J_4 = (1/64)(4/N^2)[(E_{AF3} - E_{FM}) - (E_{AF2} - E_{AF1})]$
$J_3 = (1/32)[(E_{AF3} - E_{FM})(4/N^2) - 32J_4]$
$J_2 = (1/16)[\{(E_{AF4} - E_{FM})(4/N^2)\} - 16J_4 - 16J_3]$
$J_{a'} = (1/32)(4/N^2)[(E_{AF6} - E_{FM}) - (E_{AF5} - E_{AF2})]$
$J_a = (1/32)[(E_{AF5} - E_{AF2})(4/N^2) + 16J_{a'}]$
$J_1 = (1/16)[(E_{AF1} - E_{FM})(4/N^2) - 32J_4 - 16J_2 - 16J_{a'}]$

## E. Ordered spin state energies and spin exchanges from DFT+U calculations

Table 1. Relative energies (in meV/FU) of the broken-symmetry states and the spin exchange parameters (in K) obtained from DFT+U calculations

|     | U = 4 eV |
| --- | --- |
| FM  | 5.65 |
| AF1 | 0.89 |
| AF2 | 0 |
| AF3 | 4.72 |
| AF4 | 4.52 |
| AF5 | 1.20 |
| AF6 | 4.94 |

(2a, b, 2c) supercell
PBE functional for the exchange-correlation
SCF convergence criterion = $10^{-6}$ eV
Plane wave cutoff energy = 450 eV
kpoint set = (8x4x4)

|     | U = 4 eV |
| --- | --- |
| $J_1$ | 144.97 |
| $J_2$ | 30.65 |
| $J_3$ | 21.10 |
| $J_4$ | 0.38 |
| $J_a$ | -5.73 |
| $J_{a'}$ | 44.15 |

## S5. TlCuCl₃

### A. Spin exchange paths

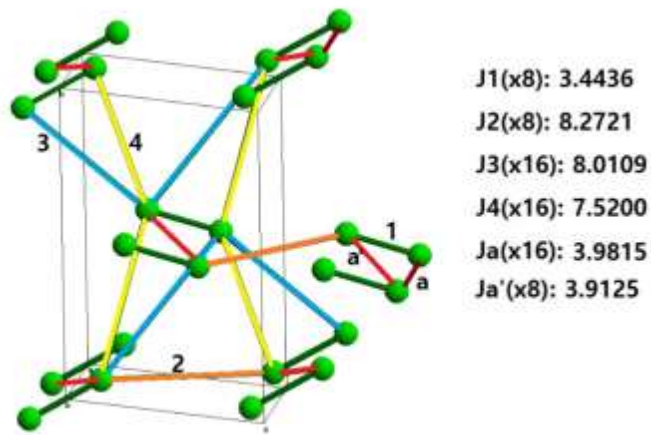

J1(x8): 3.4436
J2(x8): 8.2721
J3(x16): 8.0109
J4(x16): 7.5200
Ja(x16): 3.9815
Ja'(x8): 3.9125

### B. Ordered spin states using a (2a, b, 2c) superstructure

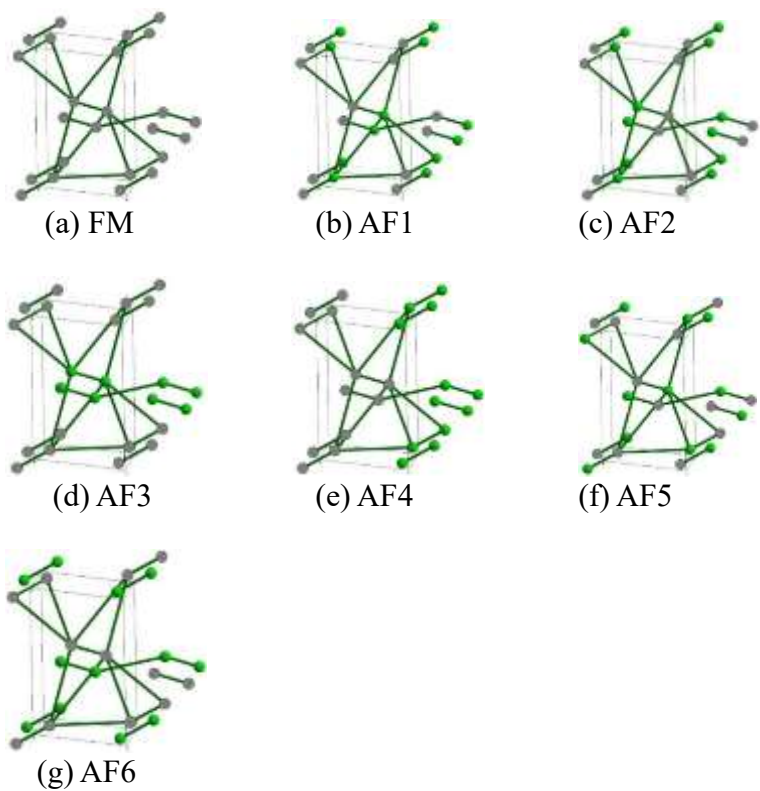

(a) FM   (b) AF1   (c) AF2

(d) AF3   (e) AF4   (f) AF5

(g) AF6

Figure 1. Ordered spin arrangements of (a) FM, (b) AF1, (c) AF2, (d) AF3, (e) AF4, (f) AF5 and (g) AF6 states.

## C. Energies of the ordered spin states in terms of the spin exchanges

$E_{FM} = (-8J_1 - 8J_2 - 16J_3 - 16J_4 - 16J_a - 8J_{a'})(N^2/4)$
$E_{AF1} = (+8J_1 + 8J_2 - 16J_3 + 16J_4 - 16J_a + 8J_{a'})(N^2/4)$
$E_{AF2} = (+8J_1 + 8J_2 + 16J_3 - 16J_4 - 16J_a + 8J_{a'})(N^2/4)$
$E_{AF3} = (-8J_1 - 8J_2 + 16J_3 + 16J_4 - 16J_a - 8J_{a'})(N^2/4)$
$E_{AF4} = (-8J_1 + 8J_2 \qquad - 16J_a - 8J_{a'})(N^2/4)$
$E_{AF5} = (+8J_1 + 8J_2 + 16J_3 - 16J_4 + 16J_a - 8J_{a'})(N^2/4)$
$E_{AF6} = (-8J_1 - 8J_2 - 16J_3 - 16J_4 + 16J_a + 8J_{a'})(N^2/4)$

## D. Spin exchanges in terms of the ordered spin state energies

$J_4 = (1/64)(4/N^2)[(E_{AF3} - E_{FM}) - (E_{AF2} - E_{AF1})]$
$J_3 = (1/32)[(E_{AF3} - E_{FM})(4/N^2) - 32J_4]$
$J_2 = (1/16)[\{(E_{AF4} - E_{FM})(4/N^2)\} - 16J_4 - 16J_3]$
$J_{a'} = (1/32)(4/N^2)[(E_{AF6} - E_{FM}) - (E_{AF5} - E_{AF2})]$
$J_a = (1/32)[(E_{AF5} - E_{AF2})(4/N^2) + 16J_{a'}]$
$J_1 = (1/16)[(E_{AF1} - E_{FM})(4/N^2) - 32J_4 - 16J_2 - 16J_{a'}]$

## E. Ordered spin state energies and spin exchanges from DFT+U calculations

Table 1. Relative energies (in meV/FU) of the broken-symmetry states and the spin exchange parameters (in K) obtained from DFT+U calculations

|     | U = 4 eV |
| --- | --- |
| FM  | 8.67 |
| AF1 | 1.85 |
| AF2 | 0 |
| AF3 | 6.56 |
| AF4 | 5.72 |
| AF5 | 2.47 |
| AF6 | 6.77 |

(2a, b, 2c) supercell
PBE functional for the exchange-correlation
SCF convergence criterion = $10^{-6}$ eV
Plane wave cutoff energy = 450 eV
kpoint set = (8x4x4)

|     | U = 4 eV |
| --- | --- |
| $J_1$ | 121.2 |
| $J_2$ | 87.7 |
| $J_3$ | 45.9 |
| $J_4$ | 3.1 |
| $J_a$ | -6.6 |
| $J_{a'}$ | 101.5 |

## S6. Unoptimized and optimized YY structures of NH₄CuCl₃

### A. Spin exchange paths

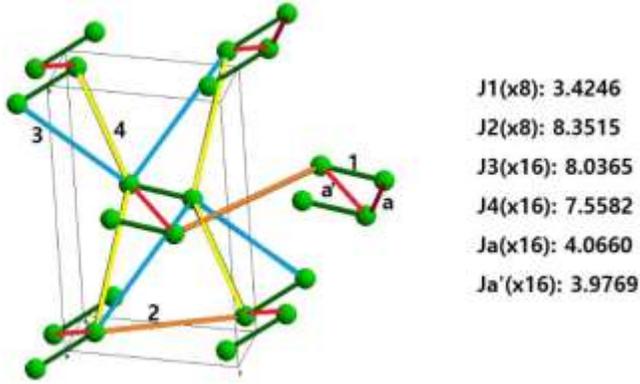

J1(x8): 3.4246
J2(x8): 8.3515
J3(x16): 8.0365
J4(x16): 7.5582
Ja(x16): 4.0660
Ja'(x16): 3.9769

### B. Ordered spin states using a (2a, b, 2c) superstructure

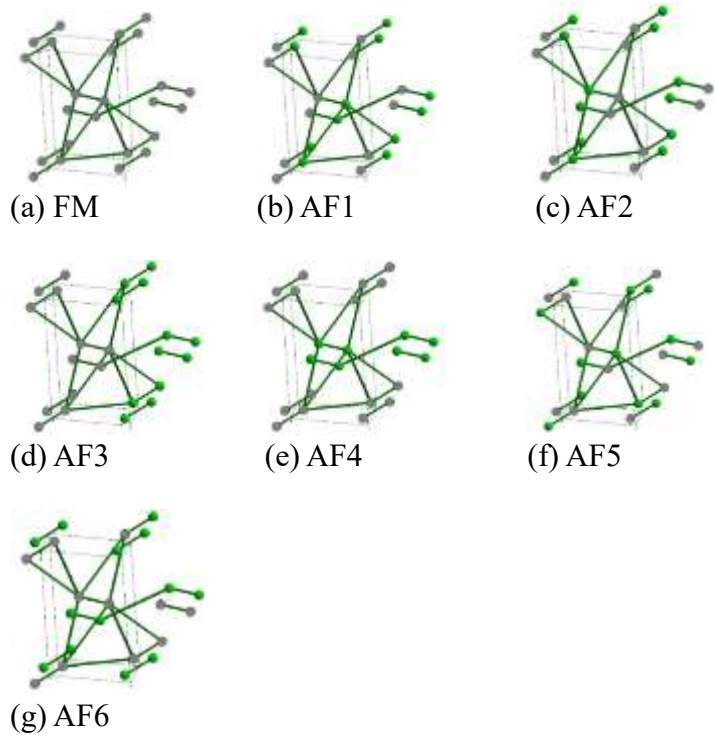

(a) FM    (b) AF1    (c) AF2

(d) AF3    (e) AF4    (f) AF5

(g) AF6

Figure 1. Ordered spin arrangements of (a) FM, (b) AF1, (c) AF2, (d) AF3, (e) AF4, (f) AF5 and (g) AF6 states.

### C. Energies of the ordered spin states in terms of the spin exchanges

$E_{FM} = (-8J_1 - 8J_2 - 16J_3 - 16J_4 - 16J_a - 8J_{a'})(N^2/4)$

$E_{AF1} = (+ 8J_1 + 8J_2 - 16J_3 + 16J_4 - 16J_a + 8J_{a'})(N^2/4)$
$E_{AF2} = (+ 8J_1 + 8J_2 + 16J_3 - 16J_4 - 16J_a + 8J_{a'})(N^2/4)$
$E_{AF3} = (- 8J_1 - 8J_2 + 16J_3 + 16J_4 - 16J_a - 8J_{a'})(N^2/4)$
$E_{AF4} = (- 8J_1 + 8J_2 \qquad\quad - 16J_a - 8J_{a'})(N^2/4)$
$E_{AF5} = (+ 8J_1 + 8J_2 + 16J_3 - 16J_4 + 16J_a - 8J_{a'})(N^2/4)$
$E_{AF6} = (- 8J_1 - 8J_2 - 16J_3 - 16J_4 + 16J_a + 8J_{a'})(N^2/4)$

**D. Spin exchanges in terms of the ordered spin state energies**

$J_4 = (1/64)(4/N^2)[(E_{AF3} - E_{FM}) - (E_{AF2} - E_{AF1})]$
$J_3 = (1/32)[(E_{AF3} - E_{FM})(4/N^2) - 32J_4]$
$J_2 = (1/16)[\{(E_{AF4} - E_{FM})(4/N^2)\} - 16J_4 - 16J_3]$
$J_{a'} = (1/32)(4/N^2)[(E_{AF6} - E_{FM}) - (E_{AF5} - E_{AF2})]$
$J_a = (1/32)[(E_{AF5} - E_{AF2})(4/N^2) + 16J_{a'}]$
$J_1 = (1/16)[(E_{AF1} - E_{FM})(4/N^2) - 32J_4 - 16J_2 - 16J_{a'}]$

**E. Ordered spin state energies and spin exchanges from DFT+U calculations**

Table 1. Relative energies (in meV/FU) of the broken-symmetry states and the spin exchange parameters (in K) obtained from DFT+U (4eV) calculations

|     | Unoptimized | Optimized |
| --- | --- | --- |
| FM  | 6.01 | 5.93 |
| AF1 | 0.55 | 0.65 |
| AF2 | 0 | 0 |
| AF3 | 5.47 | 5.26 |
| AF4 | 5.56 | 5.33 |
| AF5 | 0.09 | 0.11 |
| AF6 | 6.14 | 6.08 |

(2a, b, 2c) supercell
PBE functional for the exchange-correlation
SCF convergence criterion = $10^{-6}$ eV
Plane wave cutoff energy = 450 eV
kpoint set = (8x4x4)

|     | Unoptimized | Optimized |
| --- | --- | --- |
| $J_1$ | 246.20 | 233.57 |
| $J_2$ | 8.16 | 12.28 |
| $J_3$ | 12.70 | 15.20 |
| $J_4$ | -0.13 | 0.22 |
| $J_a$ | -2.60 | -3.10 |
| $J_{a'}$ | -0.90 | -1.01 |

## S7. Unoptimized and optimized NY structures of NH4CuCl3

### A. Spin exchange paths

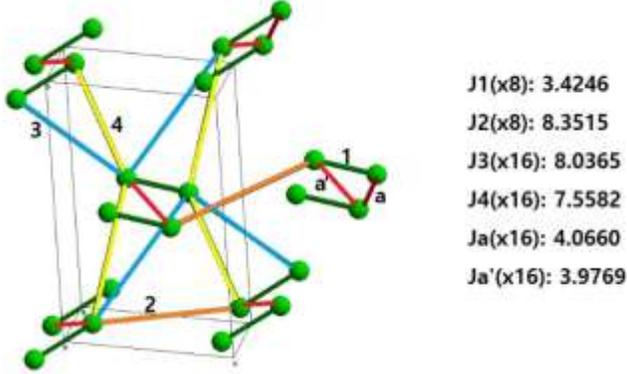

J1(x8): 3.4246
J2(x8): 8.3515
J3(x16): 8.0365
J4(x16): 7.5582
Ja(x16): 4.0660
Ja'(x16): 3.9769

### B. Ordered spin states using a (2a, b, 2c) superstructure

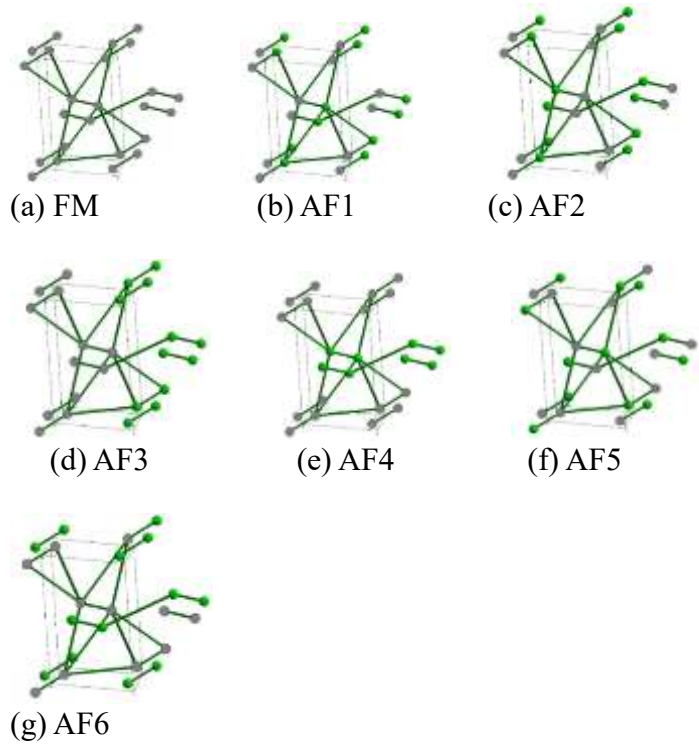

(a) FM     (b) AF1     (c) AF2

(d) AF3     (e) AF4     (f) AF5

(g) AF6

Figure 1. Ordered spin arrangements of (a) FM, (b) AF1, (c) AF2, (d) AF3, (e) AF4, (f) AF5 and (g) AF6 states.

### C. Energies of the ordered spin states in terms of the spin exchanges

$E_{FM} = (-8J_1 - 8J_2 - 16J_3 - 16J_4 - 16J_a - 8J_{a'})(N^2/4)$
$E_{AF1} = (+8J_1 + 8J_2 - 16J_3 + 16J_4 - 16J_a + 8J_{a'})(N^2/4)$

$E_{AF2} = (+ 8J_1 + 8J_2 + 16J_3 − 16J_4 − 16J_a + 8J_{a'})(N^2/4)$
$E_{AF3} = (− 8J_1 − 8J_2 + 16J_3 + 16J_4 − 16J_a − 8J_{a'})(N^2/4)$
$E_{AF4} = (− 8J_1 + 8J_2 \quad\quad − 16J_a − 8J_{a'})(N^2/4)$
$E_{AF5} = (+ 8J_1 + 8J_2 + 16J_3 − 16J_4 + 16J_a − 8J_{a'})(N^2/4)$
$E_{AF6} = (− 8J_1 − 8J_2 − 16J_3 − 16J_4 + 16J_a + 8J_{a'})(N^2/4)$

### D. Spin exchanges in terms of the ordered spin state energies

$J_4 = (1/64)(4/N^2)[(E_{AF3} − E_{FM}) − (E_{AF2} − E_{AF1})]$
$J_3 = (1/32)[(E_{AF3} − E_{FM})(4/N^2) − 32J_4]$
$J_2 = (1/16)[\{(E_{AF4} − E_{FM})(4/N^2)\} − 16J_4 − 16J_3]$
$J_{a'} = (1/32)(4/N^2)[(E_{AF6} − E_{FM}) − (E_{AF5} − E_{AF2})]$
$J_a = (1/32)[(E_{AF5} − E_{AF2})(4/N^2) + 16J_{a'}]$
$J_1 = (1/16)[(E_{AF1} − E_{FM})(4/N^2) − 32J_4 − 16J_2 − 16J_{a'}]$

### E. Ordered spin state energies and spin exchanges from DFT+U calculations

Table 1. Relative energies (in meV/FU) of the broken-symmetry states and the spin exchange parameters (in K) obtained from DFT+U (4eV) calculations

|     | Unoptimized | Optimized |
| --- | --- | --- |
| FM  | 5.12 | 4.48 |
| AF1 | 0.49 | 0.39 |
| AF2 | 0    | 0    |
| AF3 | 4.62 | 3.88 |
| AF4 | 4.69 | 4.06 |
| AF5 | 0.07 | 0.08 |
| AF6 | 5.26 | 4.59 |

(2a, b, 2c) supercell
PBE functional for the exchange-correlation
SCF convergence criterion = $10^{-6}$ eV
Plane wave cutoff energy = 450 eV
kpoint set = (8x4x4)

|     | Unoptimized | Optimized |
| --- | --- | --- |
| $J_1$ | 207.68 | 179.90 |
| $J_2$ | 8.60   | 5.18   |
| $J_3$ | 11.50  | 11.47  |
| $J_4$ | 0.20   | 2.51   |
| $J_a$ | -2.52 | -2.25 |
| $J_{a'}$ | -1.59 | -0.71 |

## S8. Unoptimized and optimized NN structures of NH$_4$CuCl$_3$

### A. Spin exchange paths

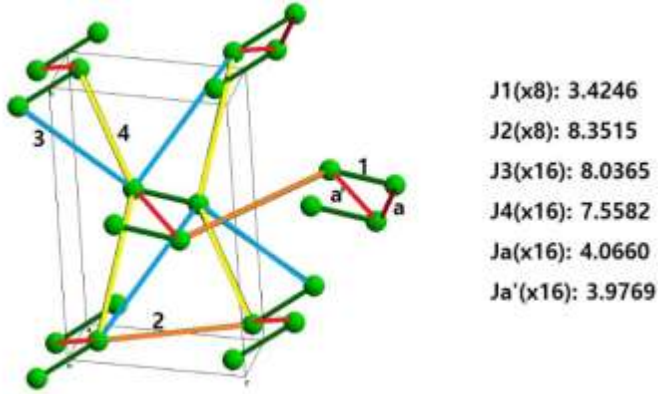

J1(x8): 3.4246
J2(x8): 8.3515
J3(x16): 8.0365
J4(x16): 7.5582
Ja(x16): 4.0660
Ja'(x16): 3.9769

### B. Ordered spin states using a (2a, b, 2c) superstructure

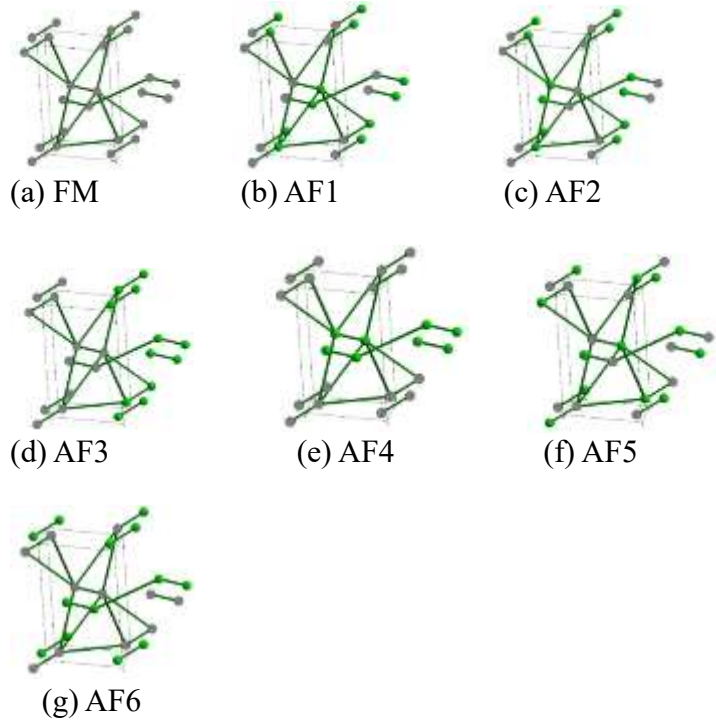

(a) FM  (b) AF1  (c) AF2

(d) AF3  (e) AF4  (f) AF5

(g) AF6

Figure 1. Ordered spin arrangements of (a) FM, (b) AF1, (c) AF2, (d) AF3, (e) AF4, (f) AF5 and (g) AF6 states.

### C. Energies of the ordered spin states in terms of the spin exchanges

$E_{FM} = (-8J_1 - 8J_2 - 16J_3 - 16J_4 - 16J_a - 8J_{a'})(N^2/4)$

$E_{AF1} = (+ 8J_1 + 8J_2 - 16J_3 + 16J_4 - 16J_a + 8J_{a'})(N^2/4)$
$E_{AF2} = (+ 8J_1 + 8J_2 + 16J_3 - 16J_4 - 16J_a + 8J_{a'})(N^2/4)$
$E_{AF3} = (- 8J_1 - 8J_2 + 16J_3 + 16J_4 - 16J_a - 8J_{a'})(N^2/4)$
$E_{AF4} = (- 8J_1 + 8J_2 \qquad\qquad - 16J_a - 8J_{a'})(N^2/4)$
$E_{AF5} = (+ 8J_1 + 8J_2 + 16J_3 - 16J_4 + 16J_a - 8J_{a'})(N^2/4)$
$E_{AF6} = (- 8J_1 - 8J_2 - 16J_3 - 16J_4 + 16J_a + 8J_{a'})(N^2/4)$

### D. Spin exchanges in terms of the ordered spin state energies

$J_4 = (1/64)(4/N^2)[(E_{AF3} - E_{FM}) - (E_{AF2} - E_{AF1})]$
$J_3 = (1/32)[(E_{AF3} - E_{FM})(4/N^2) - 32J_4]$
$J_2 = (1/16)[\{(E_{AF4} - E_{FM})(4/N^2)\} - 16J_4 - 16J_3]$
$J_{a'} = (1/32)(4/N^2)[(E_{AF6} - E_{FM}) - (E_{AF5} - E_{AF2})]$
$J_a = (1/32)[(E_{AF5} - E_{AF2})(4/N^2) + 16J_{a'}]$
$J_1 = (1/16)[(E_{AF1} - E_{FM})(4/N^2) - 32J_4 - 16J_2 - 16J_{a'}]$

### E. Ordered spin state energies and spin exchanges from DFT+U calculations

Table 1. Relative energies (in meV/FU) of the broken-symmetry states and the spin exchange parameters (in K) obtained from DFT+U (4eV) calculations

|     | Unoptimized | Optimized |
| --- | --- | --- |
| FM  | 4.03 | 2.35 |
| AF1 | 0.44 | 0.19 |
| AF2 | 0    | 0.01 |
| AF3 | 3.57 | 1.91 |
| AF4 | 3.63 | 2.06 |
| AF5 | 0.06 | 0    |
| AF6 | 4.19 | 2.39 |

(2a, b, 2c) supercell
PBE functional for the exchange-correlation
SCF convergence criterion = $10^{-6}$ eV
Plane wave cutoff energy = 450 eV
kpoint set = (8x4x4)

|     | Unoptimized | Optimized |
| --- | --- | --- |
| $J_1$ | 160.31 | 91.81 |
| $J_2$ | 8.03   | 3.58  |
| $J_3$ | 10.44  | 7.26  |
| $J_4$ | 0.26 | 2.90 |
| $J_a$ | -2.52 | -0.41 |
| $J_{a'}$ | -2.03 | -1.06 |

## S9. K₂Cu₃O(SO₄)₃

### A. Spin exchange paths

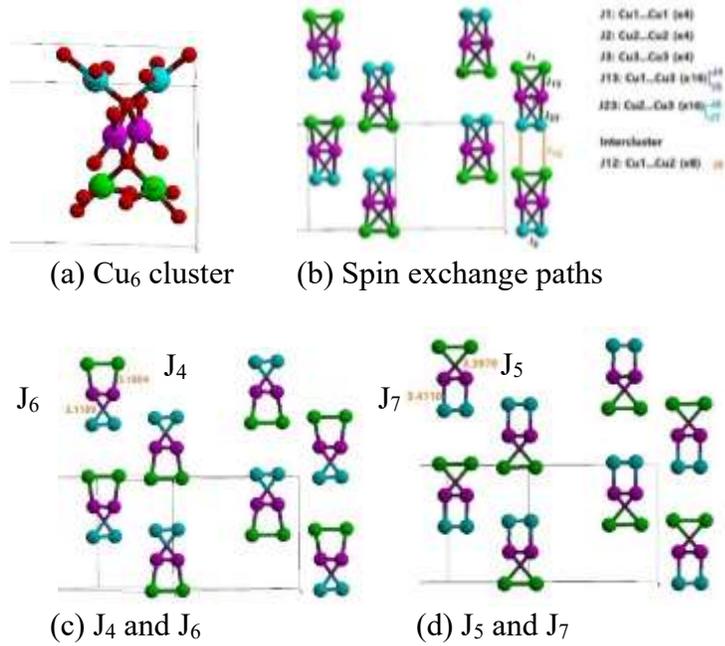

(a) Cu₆ cluster  (b) Spin exchange paths

(c) J₄ and J₆  (d) J₅ and J₇

Figure 1. (a) Cu₆ cluster. (b) Spin exchange paths between Cu₆ clusters. (c) J₄ and J₆. (d) J₅ and J₇

### B. Ordered spin states

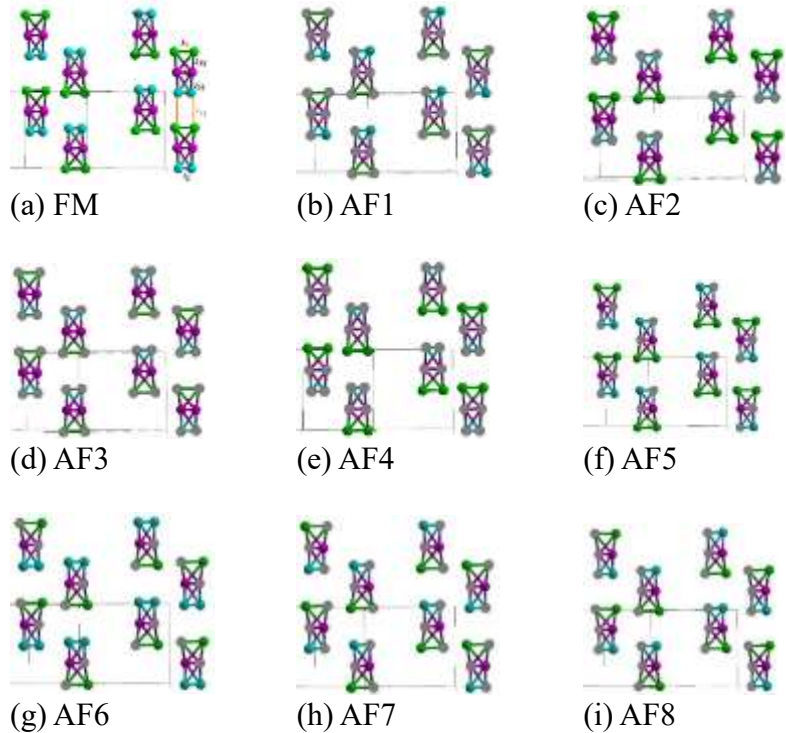

(a) FM  (b) AF1  (c) AF2

(d) AF3  (e) AF4  (f) AF5

(g) AF6  (h) AF7  (i) AF8

Figure 2. Ordered spin arrangements

**C. Energies of the ordered spin states in terms of the spin exchanges**

FM = $-4J_1 - 4J_2 - 4J_3 - 8J_4 - 8J_5 - 8J_6 - 8J_7 - 8J_8$
AF1 = $-4J_1 + 4J_2 - 4J_3 - 8J_4 - 8J_5$
AF2 = $-4J_1 - 4J_2 + 4J_3 - 8J_8$
AF3 = $+4J_1 - 4J_2 - 4J_3 - 8J_6 - 8J_7$
AF4 = $-4J_1 - 4J_2 - 4J_3 - 8J_4 - 8J_5 + 8J_6 + 8J_7 - 8J_8$
AF5 = $-4J_1 - 4J_2 - 4J_3 + 8J_4 + 8J_5 + 8J_6 + 8J_7 - 8J_8$
AF6 = $-4J_1 - 4J_2 - 4J_3 + 8J_4 + 8J_5 - 8J_6 - 8J_7 + 8J_8$
AF7 = $-4J_1 + 4J_2 + 4J_3 - 8J_6 + 8J_7$
AF8 = $+4J_1 - 4J_2 + 4J_3 + 8J_4 - 8J_5$

**D. Energy differences between ordered spin states in terms of the spin exchanges**

| Final | $J_1$ | $J_2$ | $J_3$ | $J_4$ | $J_5$ | $J_6$ | $J_7$ | $J_8$ |
|---|---|---|---|---|---|---|---|---|
| AF2 - FM | 0 | 0 | 0 | 0 | 0 | 16 | 16 | 16 |
| AF3 - AF4 | 0 | 0 | 0 | 0 | 0 | 16 | 16 | -16 |
| AF1 - AF2 | 0 | 8 | 0 | 0 | 0 | -8 | -8 | -8 |
| AF7 - AF6 | 0 | 8 | 0 | 0 | 0 | -8 | 8 | -8 |
| AF3 - FM | 0 | 0 | 0 | 16 | 16 | 16 | 16 | 0 |
| AF7 - AF8 | 0 | 0 | 0 | 16 | -16 | -16 | 16 | 0 |
| AF5 - AF4 | 0 | 8 | 8 | -8 | -8 | 0 | 16 | -8 |
| AF6 - AF5 | 8 | -8 | 0 | 8 | -8 | 8 | -8 | 0 |

**E. Spin exchanges in terms of the ordered spin state energies**

$J_8 = (1/32)(4/N^2)[(E_{AF2} - E_{FM}) - (E_{AF3} - E_{AF4})]$
$J_7 = (1/16)(4/N^2)[(E_{AF7} - E_{AF6}) - (E_{AF1} - E_{AF2})]$
$J_6 = (1/16)[(E_{AF3} - E_{AF4})(4/N^2) - 16J_7 + 16J_8]$
$J_2 = (1/8)[(E_{AF7} - E_{AF6})(4/N^2) + 16J_6 - 8J_7 + 8J_8]$
$J_5 = (1/32)[\{(E_{AF3} - E_{FM}) - (E_{AF7} - E_{AF8})\}(4/N^2) - 32J_6]$
$J_4 = (1/16)[\{(E_{AF7} - E_{AF8})(4/N^2)\} + 16J_5 + 16J_6 - 16J_7]$
$J_3 = (1/8)[\{(E_{AF5} - E_{AF4})(4/N^2)\} - 8J_2 + 8J_4 + 8J_5 - 16J_7 + 8J_8]$
$J_1 = (1/8)[\{(E_{AF6} - E_{AF5})(4/N^2)\} + 8J_2 - 8J_4 + 8J_5 - 8J_6 + 8J_7]$

**F. The energies of the ordered spin states and the spin exchanges from DFT+U calculations**

Table 1. Relative energies (in meV/FU) and spin exchange parameters (in K) obtained from DFT+U calculations

|     | ΔE (meV/FU) | |
| --- | --- | --- |
|     | U = 3 eV | U = 4 eV |
| FM  | 100.75 | 83.15 |
| AF1 | 64.62 | 54.45 |
| AF2 | 41.38 | 34.69 |
| AF3 | 0 | 0 |
| AF4 | 42.20 | 35.69 |
| AF5 | 37.84 | 33.43 |
| AF6 | 36.78 | 32.81 |
| AF7 | 32.61 | 29.02 |
| AF8 | 40.30 | 35.66 |

(a, b, c) unit cell
PBE functional for the exchange-correlation
SCF convergence criterion = $10^{-6}$ eV
Plane wave cutoff energy = 450 eV
kpoint set = (2x 8 x 4)

|     | U = 3 eV | U = 4 eV |
| --- | --- | --- |
| $J_1$ | 899.26 | 744.73 |
| $J_2$ | 841.58 | 637.49 |
| $J_3$ | -552.51 | -524.04 |
| $J_4$ | 622.14 | 495.29 |
| $J_5$ | 537.11 | 457.58 |
| $J_6$ | 542.33 | 429.91 |
| $J_7$ | 635.91 | 546.27 |
| $J_8$ | 199.23 | 148.09 |

## S10. Azurite $Cu_3(CO_3)_2(OH)_2$: Evaluation of the interlayer spin exchanges

### A. Spin exchange paths

In the main text, the diamond triangle is defined by $J_2$, $J_1$ and $J_3$. However, $J_1$ is very close to $J_3$. Thus, we simplify our analysis by an ideal diamond triangle defined by $J_2$, $J_1$ and $J_1$.

The intra-diamond exchanges $J_1$ and $J_2$ together with the inter-diamond exchange $J_3$ form layers. (In the main text, $J_3$ is referred to as $J_4$.)

There are two inter-layer exchanges $J_4$ and $J_5$. (In the main text, $J_4$ and $J_5$ are referred to as $J_5$ and $J_6$, respectively.)

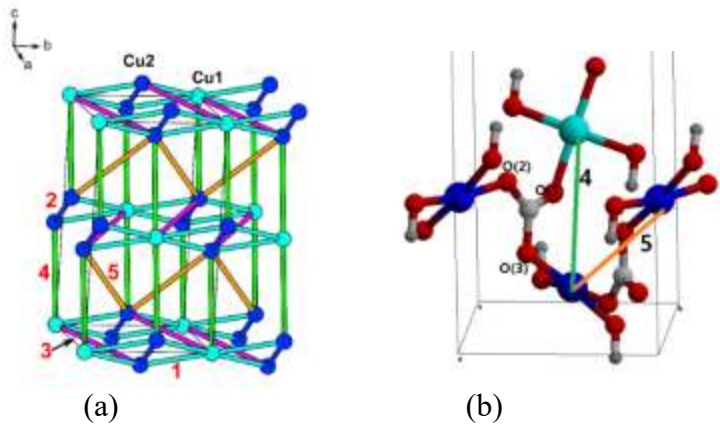

(a)          (b)

Figure 1. (a) Spin exchange paths, $J_1$ to $J_5$ and (b) Interlayer paths, $J_4$ and $J_5$.

Table 1. Geometrical parameters of interlayer paths $J_4$ and $J_5$

|  | Cu…Cu | O…O | ∠Cu-O…O, O…O-Cu |
|---|---|---|---|
| $J_4$ | 4.5391 | 2.2120 | 83.05, 147.38 |
| $J_5$ | 5.0959 | 2.2298 | 98.23, 141.53 |

### B. Ordered spin states using a (2a, 2b, c) super cell containing 8 Fus

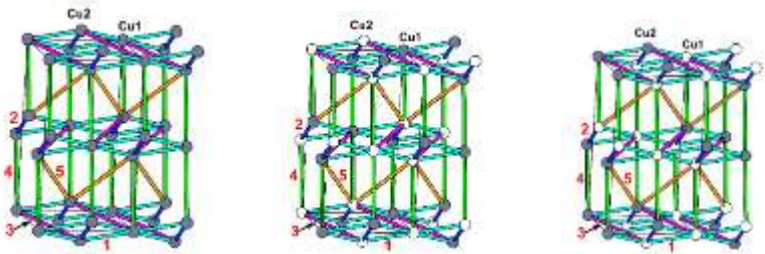

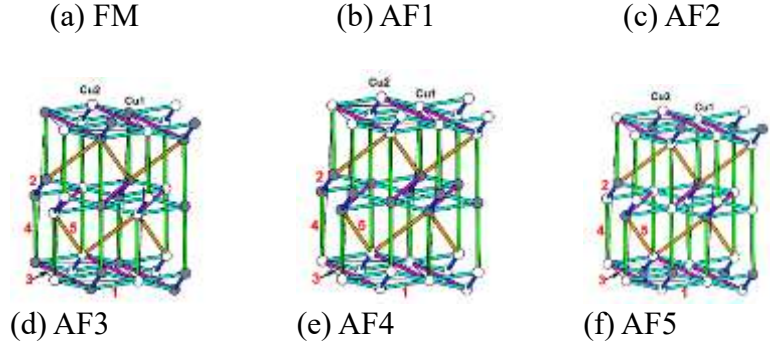

(a) FM  (b) AF1  (c) AF2
(d) AF3  (e) AF4  (f) AF5

Figure 2. Ordered spin arrangements of FM and AFi (i = 1 to 5).

## C. Energies of the ordered spin states in terms of the spin exchanges

Table 2. Values of $n_i$ in the energy expressions, $E_{spin} = \sum_{i=1}^{5} n_i J_i S^2$, for the ordered spin states FM and $AF_i$ (i = 1 – 5).

|  | $J_1$ | $J_2$ | $J_3$ | $J_4$ | $J_5$ |
|---|---|---|---|---|---|
| FM | -32 | -8 | -16 | -16 | -16 |
| AF1 | 0 | 8 | -16 | 0 | 0 |
| AF2 | 0 | 8 | 16 | 0 | 0 |
| AF3 | 32 | -8 | -16 | -16 | -16 |
| AF4 | -32 | -8 | -16 | 16 | 16 |
| AF5 | 0 | 8 | 0 | 0 | 16 |

## D. Spin exchanges in terms of the ordered spin state energies

$J_3 = (1/32)(4/N^2)(AF2 - AF1)$
$J_1 = (1/64)(4/N^2)(AF3 - FM)$
$J_2 = (1/32)[\{(AF1 - FM) - (AF4 - AF2)\}(4/N^2) - 64J_1 - 32J_3]$
$J_4 = (1/16)[\{(AF4 - AF5)(4/N^2)\} + 32J_1 + 16J_2 + 16J_3]$
$J_5 = (1/16)[\{(AF5 - AF2)(4/N^2)\} + 16J_3]$

## E. Ordered spin state energies and spin exchanges from DFT+U calculations

Table 3. Relative energies (in meV/FU) and spin exchange interactions (in K) obtained from DFT+U calculations

|  | U = 3 eV | U = 4 eV |
|---|---|---|
| FM | 29.20 | 22.62 |
| AF1 | 4.07 | 3.17 |
| AF2 | 0 | 0 |
| AF3 | 13.11 | 10.04 |
| AF4 | 28.01 | 21.98 |
| AF5 | 2.85 | 2.34 |

(2a, 2b, c) super cell  
PBE functional for the exchange-correlation  
SCF convergence criterion = $10^{-6}$ eV  
Plane wave cutoff energy = 450 eV  
kpoint set = (6x4x6)

| | U = 3 eV | U = 4 eV |
|---|---|---|
| $J_1$ | 93.29 | 73.00 |
| $J_2$ | 382.64 | 297.93 |
| $J_3$ | 47.20 | 36.74 |
| $J_4$ | 32.77 | 25.10 |
| $J_5$ | -18.95 | -17.65 |

## S11. RbFe(MoO$_4$)$_2$

### A. Spin exchange paths

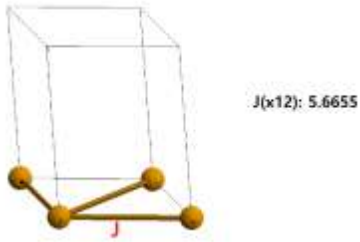

### B. Ordered spin states using a (2a, 2b, c) superstructure

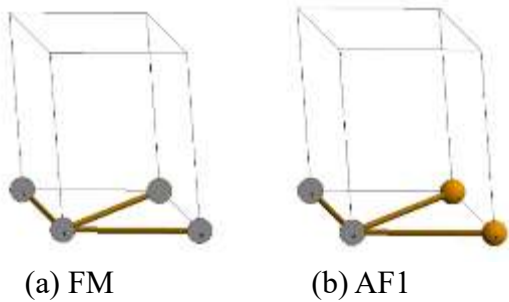

(a) FM          (b) AF1

Figure 1. Ordered spin arrangements of (a) FM and (b) AF1 states.

### C. Energies of the ordered spin states in terms of the spin exchanges

$E_{FM} = (-12J)(N^2/4)$
$E_{AF1} = (+4J)(N^2/4)$

### D. Spin exchanges in terms of the ordered spin state energies

$J = (1/16)(4/N^2)(E_{AF1} - E_{FM})$

### E. Ordered spin state energies and spin exchanges from DFT+U calculations

Table 1. Relative energies (in meV/FU) of the broken-symmetry states and the spin exchange parameters (in K) obtained from DFT+U calculations

|    | U = 4 eV |
|----|----------|
| FM | 3.18     |

| AF1 | 0 |

(2a, 2b, c) supercell
PBE functional for the exchange-correlation
SCF convergence criterion = $10^{-6}$ eV
Plane wave cutoff energy = 450 eV
kpoint set = (6x6x9)

|   | U = 4 eV |
|---|----------|
| J | 1.47     |

# S12. $Ba_3CoSb_2O_9$

## A. Spin exchange paths

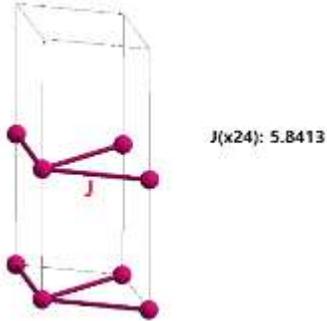

## B. Ordered spin states using a (2a, 2b, c) superstructure

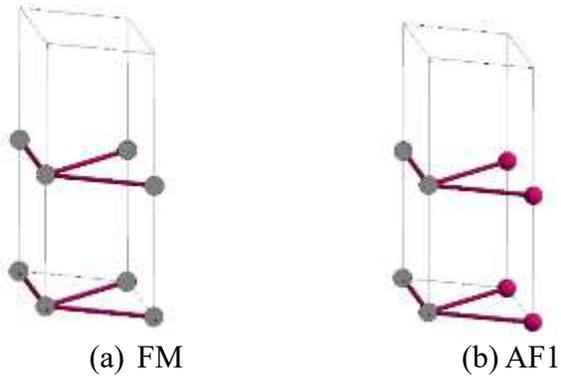

(a) FM  (b) AF1

Figure 1. Ordered spin arrangements of (a) FM and (b) AF1 states.

## C. Energies of the ordered spin states in terms of the spin exchanges

$E_{FM} = (-24J)(N^2/4)$
$E_{AF1} = (+8J)(N^2/4)$

## D. Spin exchanges in terms of the ordered spin state energies

$J = (1/32)(4/N^2)(E_{AF1} - E_{FM})$

## E. Ordered spin state energies and spin exchanges from DFT+U calculations

Table 1. Relative energies (in meV/FU) of the broken-symmetry states and the spin exchange parameters (in K) obtained from DFT+U calculations

|      | U = 4 eV |
|------|----------|
| FM   | 4.81     |
| AF1  | 0        |

(2a, 2b, c) supercell
PBE functional for the exchange-correlation
SCF convergence criterion = $10^{-6}$ eV
Plane wave cutoff energy = 450 eV
kpoint set = (6x6x4)

|   | U = 4 eV |
|---|----------|
| J | 6.20     |

## S13. Ba$_2$LaNiTe$_2$O$_{12}$

### A. Spin exchange paths

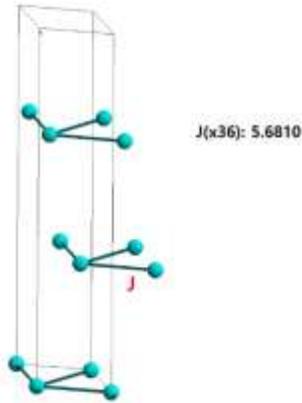

### B. Ordered spin states using a (2a, 2b, c) superstructure

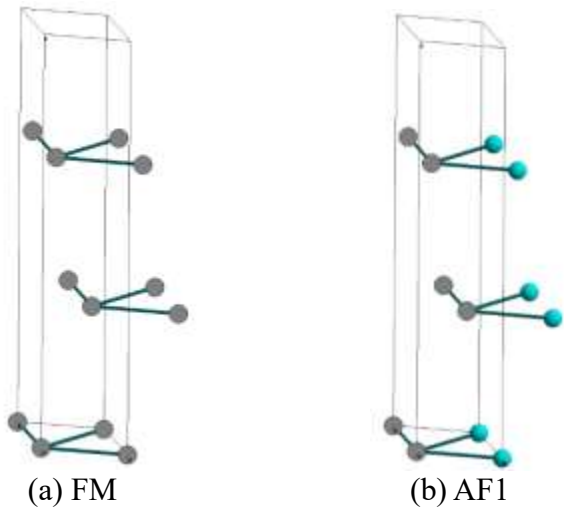

    (a) FM                (b) AF1

Figure 1. Ordered spin arrangements of (a) FM and (b) AF1 states.

### C. Energies of the ordered spin states in terms of the spin exchanges

$E_{FM} = (-36J)(N^2/4)$
$E_{AF1} = (+12J)(N^2/4)$

### D. Spin exchanges in terms of the ordered spin state energies

$J = (1/48)(4/N^2)(E_{AF1} - E_{FM})$

### E. Ordered spin state energies and spin exchanges from DFT+U calculations

Table 1. Relative energies (in meV/FU) of the broken-symmetry states and the spin exchange parameters (in K) obtained from DFT+U calculations

|     | U = 4 eV |
|-----|----------|
| FM  | 19.33    |
| AF1 | 0        |

(2a, 2b, c) supercell
PBE functional for the exchange-correlation
SCF convergence criterion = $10^{-6}$ eV
Plane wave cutoff energy = 450 eV
kpoint set = (6x6x2)

|   | U = 4 eV |
|---|----------|
| J | 56.04    |

## S14. $Y_2Cu_7(TeO_3)_6Cl_6(OH)_2$

### A. Spin exchange paths

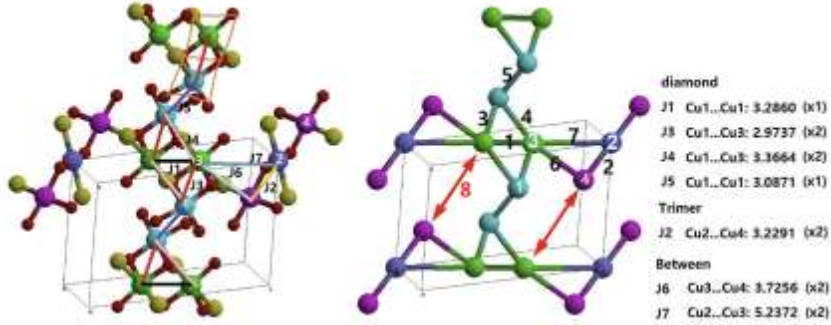

Figure 1. Spin exchange paths, $J_1$ to $J_8$. The cyan, purple, green and magenta circles represent the Cu1, Cu2, Cu3 and Cu4 ions, respectively. [$J_8$ Cu3…Cu4 = 6.2263 (x2)]

### B. Ordered spin states using a (2a, b, c) supercell

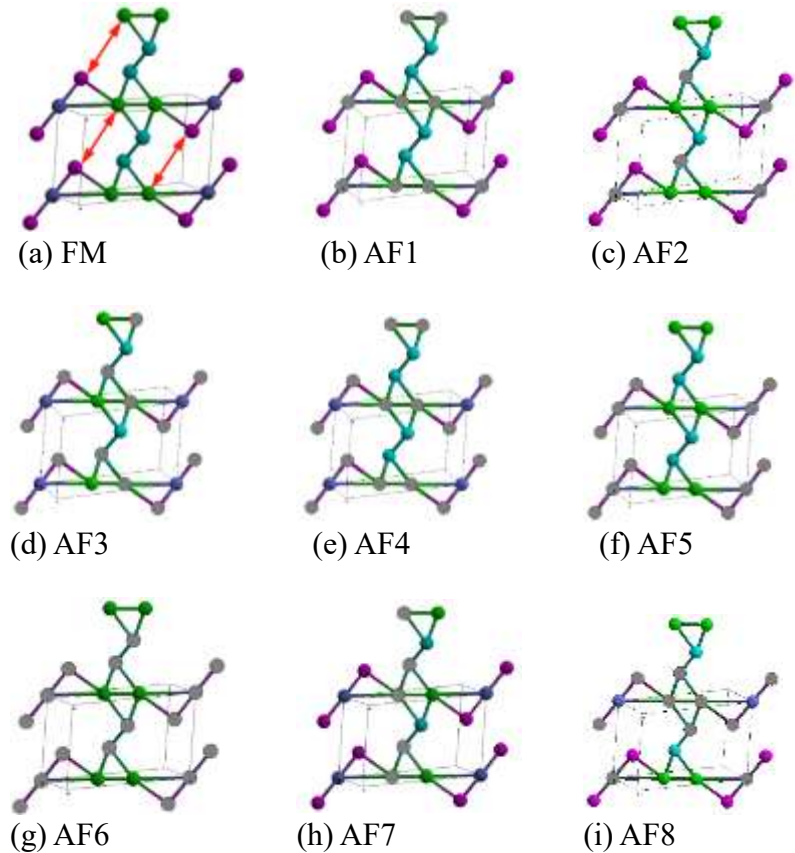

(a) FM  (b) AF1  (c) AF2
(d) AF3  (e) AF4  (f) AF5
(g) AF6  (h) AF7  (i) AF8

Figure 2. Ordered spin arrangements of FM, AF1 – AF8 states. The shaded and unshaded circles indicate the up and down spin sites, respectively.

## C. Energies of the ordered spin states in terms of the spin exchanges

Table 1. Coefficients of $n_i$ of $E = \sum_{i=1}^{7} n_i J_i S^2$ per (2a, b, c) supercell.

|     | $J_1$ | $J_2$ | $J_3$ | $J_4$ | $J_5$ | $J_6$ | $J_7$ | $J_8$ |
|-----|-----|-----|-----|-----|-----|-----|-----|-----|
| FM  | -2  | -4  | -4  | -4  | -2  | -4  | -4  | -4  |
| AF1 | -2  | 4   | 4   | 4   | -2  | 4   | -4  | 4   |
| AF2 | -2  | 4   | 0   | 0   | 2   | -4  | 4   | -4  |
| AF3 | 2   | 4   | 4   | -4  | 2   | 0   | 0   | 0   |
| AF4 | -2  | 4   | 4   | 4   | -2  | -4  | 4   | -4  |
| AF5 | -2  | -4  | -4  | -4  | -2  | 4   | 4   | 4   |
| AF6 | -2  | -4  | 4   | 4   | -2  | 4   | 4   | 4   |
| AF7 | 2   | -4  | -4  | 4   | 2   | 0   | 0   | 0   |
| AF8 | -2  | 4   | -4  | -4  | 2   | -4  | 4   | 4   |

## D. Spin exchanges in terms of the ordered spin state energies

$J_7 = (1/16)(4/N^2)[(AF5 - FM) - (AF1 - AF4)]$
$J_2 = (1/8)[\{(AF5 - FM) - (AF6 - AF4)\}(4/N^2) - 8J_7]$
$J_3 = (1/16)[\{(AF6 - AF7) - (AF5 - AF3)\}(4/N^2) - 8J_2]$
$J_4 = (1/8)[(AF6 - AF5)(4/N^2) - 8J_3]$
$J_8 = (1/8)[\{(AF7 - AF2) - (AF3 - AF8)\}(4/N^2) + 8J_2 + 12J_3 - 4J_4]$
$J_6 = (1/8)[(AF5 - FM)(4/N^2) - 8J_7 - 8J_8]$
$J_1 = (1/4)[(AF3 - AF8)(4/N^2) - 8J_3 - 4J_6 + 4J_7 + 4J_8]$
$J_5 = (1/4)[(AF2 - AF4)(4/N^2) + 4J_3 + 4J_4]$

## E. Relative energies of the ordered spin states and the spin exchanges from DFT+U calculations

Table 2. Relative energies (meV/FU) of the ordered spin states obtained from DFT+U calculations

|     | U = 3 eV | U = 4 eV |
|-----|----------|----------|
| FM  | 66.28    | 53.93    |
| AF1 | 1.62     | 1.95     |
| AF2 | 56.27    | 46.82    |
| AF3 | 20.76    | 17.90    |
| AF4 | 46.15    | 39.02    |
| AF5 | 21.81 | 17.36 |
| AF6 | 0 | 0 |
| AF7 | 44.98 | 36.32 |
| AF8 | 23.25 | 19.00 |

(2a, b, c) super cell
PBE functional for the exchange-correlation
SCF convergence criterion = $10^{-6}$ eV
Plane wave cutoff energy = 450 eV
kpoint set = (4x6x4)

Table 3. Spin exchange parameters (in K) from DFT+U calculations

| | U = 3 eV | U = 4 eV |
|---|---|---|
| $J_1$ | 7.08 | 1.73 |
| $J_2$ | -19.20 | -25.47 |
| $J_3$ | 276.63 | 220.24 |
| $J_4$ | -23.58 | -18.90 |
| $J_5$ | 18.39 | 20.38 |
| $J_6$ | 6.70 | 3.84 |
| $J_7$ | -0.35 | -2.85 |
| $J_8$ | 509.34 | 423.26 |

## S15. $Cu_5(VO_4)_2(OH)_4$

### A. Spin exchange paths

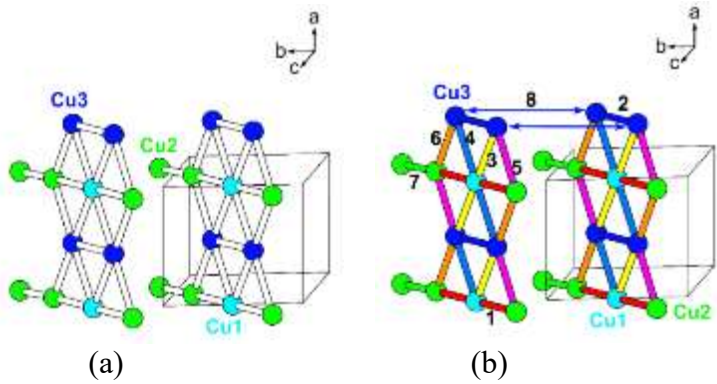

(a)         (b)

Figure 1. (a) Two Cu5-layers in (2a, 2b, c) super cell and (b) spin exchange paths. The numbers 1 to 8 indicate the spin exchange paths J1 to J8, respectively. The $J_8$ is interlayer spin exchange between Cu(2) cations.

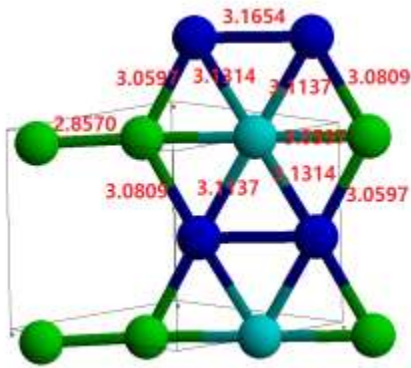

Figure 2. Cu-Cu bond distances associated with the intralayer spin exchange paths $J_1$ to $J_7$.

### B. Ordered spin states using a (2a, 2b, c) super cell

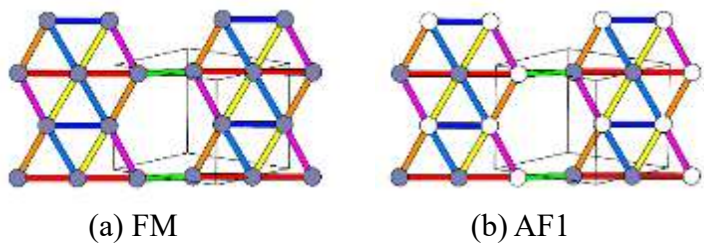

(a) FM         (b) AF1

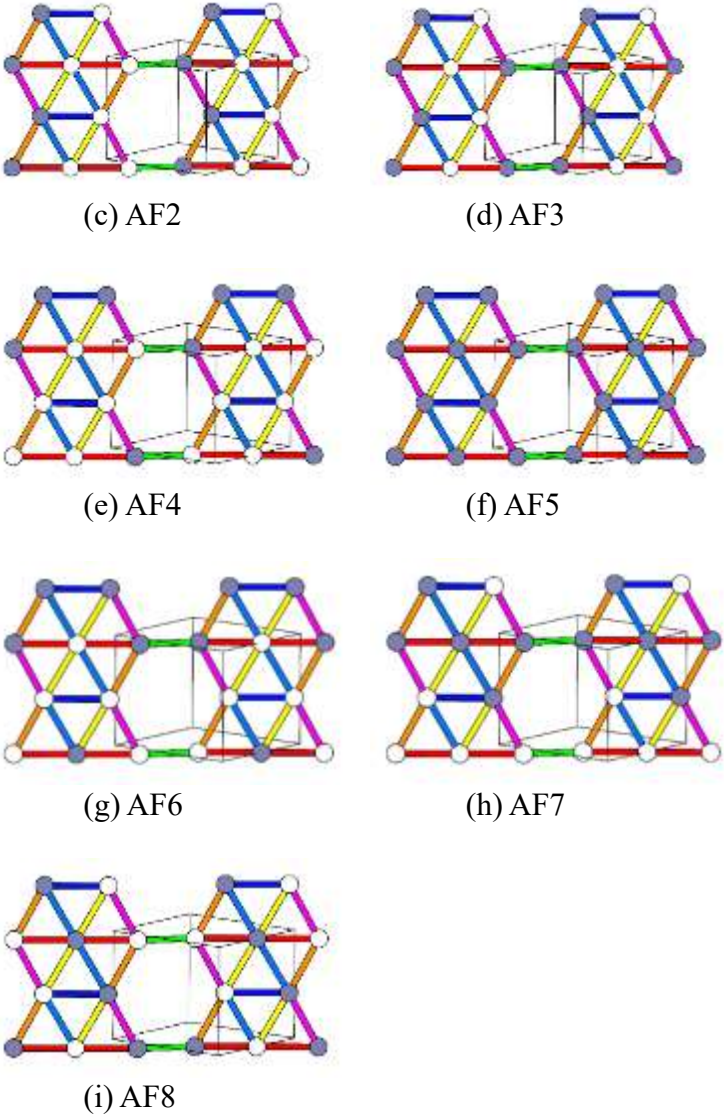

(c) AF2　　　　　　　　(d) AF3

(e) AF4　　　　　　　　(f) AF5

(g) AF6　　　　　　　　(h) AF7

(i) AF8

Figure 3. Ordered spin arrangements of FM and AF(i) (i = 1 to 8). The spin arrangements of the $J_8$ are AFM except for the FM state.

## C. Energies of the ordered spin states in terms of the spin exchanges

Table 1. Coefficients $n_i$ of $E_i = \sum_{i=1}^{8} n_i J_i S^2$

|     | $J_1$ | $J_2$ | $J_3$ | $J_4$ | $J_5$ | $J_6$ | $J_7$ | $J_8$ |
|-----|-------|-------|-------|-------|-------|-------|-------|-------|
| FM  | -16   | -8    | -16   | -16   | -16   | -16   | -8    | -8    |
| AF1 | 0     | -8    | 16    | 16    | 0     | 0     | 8     | 8     |
| AF2 | 0     | 8     | 0     | 0     | -16   | -16   | 8     | 8     |
| AF3 | 16    | 8     | 0     | 0     | 0     | 0     | -8    | 8     |
| AF4 | 0 | -8 | 0 | 0 | 16 | -16 | 8 | 8 |
| AF5 | -16 | -8 | -16 | -16 | -16 | -16 | -8 | 8 |
| AF6 | 16 | -8 | 0 | 0 | 0 | 0 | -8 | 8 |
| AF7 | -16 | 8 | 16 | -16 | 16 | -16 | -8 | 8 |
| AF8 | 16 | 8 | 16 | -16 | -16 | 16 | -8 | 8 |

**D. Spin exchanges in terms of the ordered spin state energies**

$J_8 = (1/16)(4/N^2)(E_{AF5} - E_{FM})$
$J_2 = (1/16)\{(4/N^2)[(E_{AF3} - E_{FM}) - (E_{AF6} - E_{AF5})] - 16J_8\}$
$J_5 = (1/32)[(4/N^2)(E_{AF4} - E_{AF2}) + 16J_2]$
$J_1 = (1/32)\{(4/N^2)[(E_{AF4} - E_{AF5}) - (E_{AF1} - E_{AF3})] - 16J_2 - 32J_5\}$
$J_6 = (1/32)[(4/N^2)(E_{AF8} - E_{AF7}) - 32J_1 + 32J_5]$
$J_7 = (1/16)[(4/N^2)(E_{AF4} - E_{AF3}) + 16J_1 + 16J_2 - 16J_5 + 16J_6]$
$J_3 = (1/32)[(4/N^2)(E_{AF7} - E_{AF5}) - 16J_2 - 32J_5]$
$J_4 = (1/32)[(4/N^2)(E_{AF1} - E_{AF7}) - 16J_1 + 16J_2 + 16J_5 - 16J_6 - 16J_7]$

**E. Relative energies of the ordered spin states and spin exchanges from DFT+U calculations**

Table 2. Relative energies (in meV/FU) obtained from DFT+U calculations

| | U = 3 eV | U = 4 eV | U = 5 eV |
|---|---|---|---|
| FM | 68.15 | 55.06 | 44.29 |
| AF1 | 26.52 | 22.25 | 18.60 |
| AF2 | 40.85 | 33.16 | 26.84 |
| AF3 | 13.60 | 12.24 | 11.03 |
| AF4 | 0 | 0 | 0 |
| AF5 | 67.57 | 55.20 | 44.99 |
| AF6 | 12.65 | 11.56 | 10.55 |
| AF7 | 27.96 | 23.83 | 20.14 |
| AF8 | 35.45 | 30.29 | 25.94 |

(2a, 2b, c) super cell
PBE functional for the exchange-correlation
SCF convergence criterion = $10^{-6}$ eV
Plane wave cutoff energy = 450 eV
kpoint set = (6x4x8)

Table 3. Spin exchange parameters (in K) obtained from DFT+U calculations

|       | U = 3 eV | U = 4 eV | U = 5 eV |
|-------|----------|----------|----------|
| $J_1$ | 240.93   | 193.76   | 154.66   |
| $J_2$ | -11.02   | -7.83    | -5.51    |
| $J_3$ | 3.80     | -2.55    | -6.09    |
| $J_4$ | -27.06   | -24.32   | -21.36   |
| $J_5$ | 231.42   | 188.40   | 152.93   |
| $J_6$ | -52.92   | -42.80   | -35.35   |
| $J_7$ | 103.30   | 96.66    | 88.75    |
| $J_8$ | 6.76     | -1.62    | -8.12    |

## S16. Na$_2$Cu$_7$(SeO$_3$)$_4$O$_2$Cl$_4$

### A. Spin exchanges

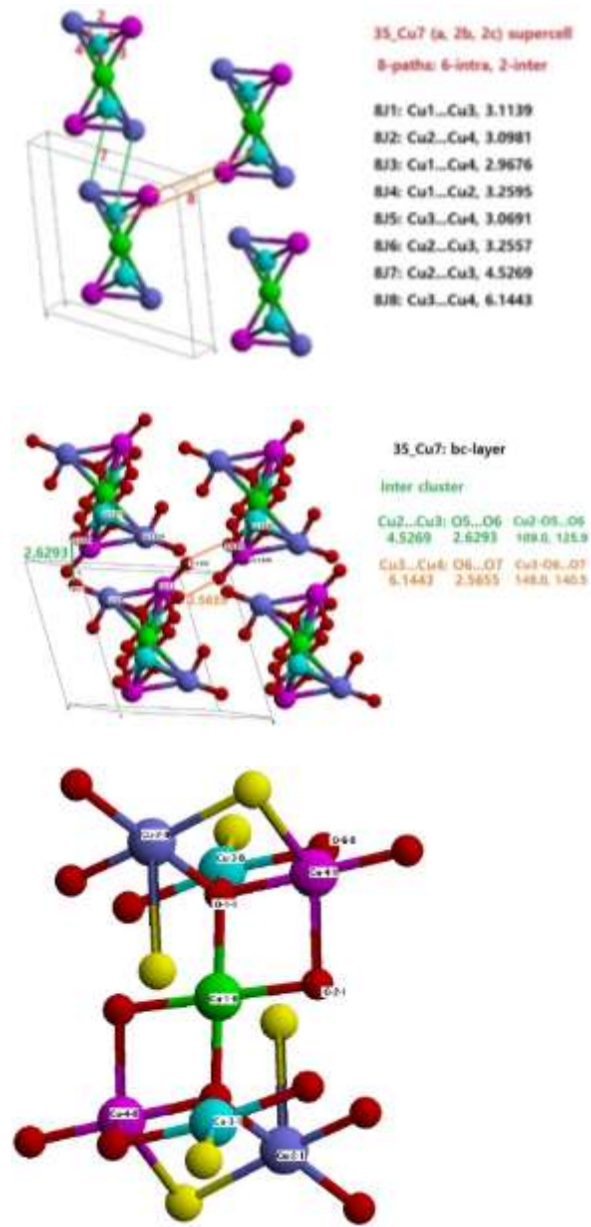

Figure 1. Spin exchange paths

|  | Cu…Cu | ∠Cu-O-Cu | O…O | ∠Cu-O…O |
|---|---|---|---|---|
| J$_1$ | 3.1139 | 107.7 |  |  |
| J$_2$ | 3.0981 | 108.3 |  |  |
| J$_3$ | 2.9676 | 94.7, 101.7 |  |  |
| J$_4$ | 3.2595 | 116.8 | | |
| J$_5$ | 3.0691 | 95.5, 105.6 | | |
| J$_6$ | 3.2557 | 115.3 | | |
| J$_7$ | 4.5269 | | 2.6293 | 109.0, 125.9 |
| J$_8$ | 6.1443 | | 2.5655 | 148.0, 140.5 |

## B. Ordered spin states using a (a, 2b, 2c) supercell

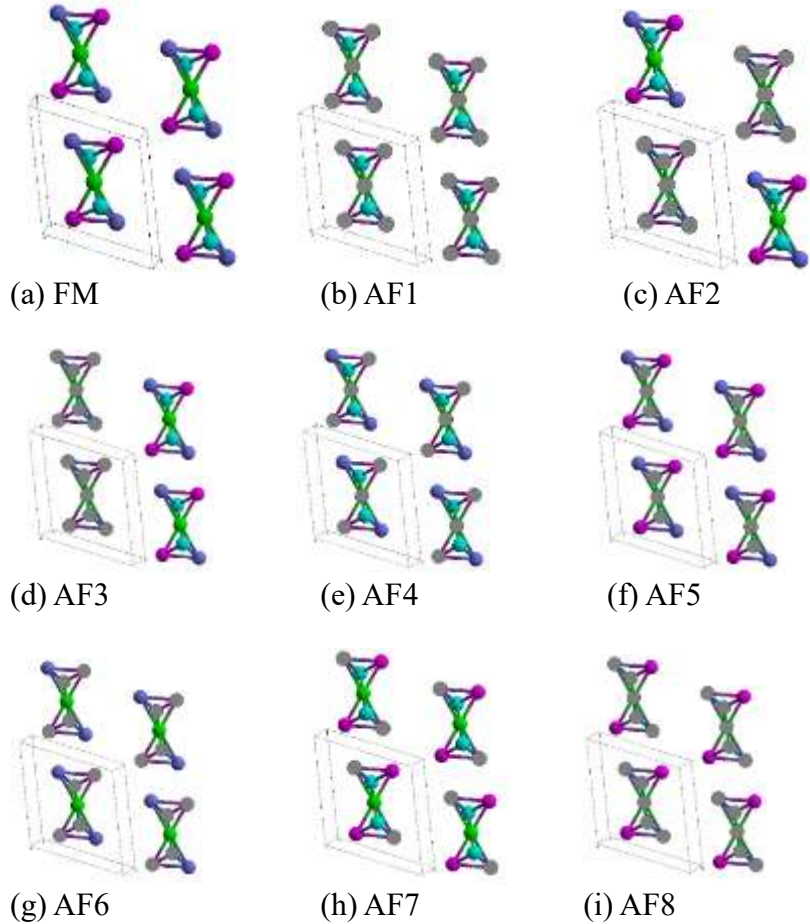

(a) FM   (b) AF1   (c) AF2

(d) AF3   (e) AF4   (f) AF5

(g) AF6   (h) AF7   (i) AF8

Figure 2. Ordered spin arrangements of FM and AF(i) (i = 1 to 8). The green, purple, cyan and magenta indicate the Cu1, Cu2, Cu3 and Cu4 ions. The shaded and unshaded circles represent the up and down spin sites of Cu ions, respectively.

## C. Energies of the ordered spin states in terms of the spin exchanges

Table 1. Coefficients $n_j$ of $E_i = \sum_{j=1}^{8} n_j J_j S^2$  The $E_i$(i= 1 to 9) = $E_{FM}$ and $E_{AF1}$ to $E_{AF8}$

| | J$_1$ | J$_2$ | J$_3$ | J$_4$ | J$_5$ | J$_6$ | J$_7$ | J$_8$ |
|---|---|---|---|---|---|---|---|---|
| $E_{FM}$ | -8 | -8 | -8 | -8 | -8 | -8 | -8 | -8 |
| $E_{AF1}$ | 8 | -8 | -8 | -8 | 8 | 8 | -8 | -8 |
| $E_{AF2}$ | -8 | -8 | -8 | -8 | -8 | -8 | 8 | -8 |
| $E_{AF3}$ | -8 | -8 | -8 | -8 | -8 | -8 | -8 | 8 |
| $E_{AF4}$ | 8 | 8 | -8 | 8 | 8 | -8 | -8 | 8 |
| $E_{AF5}$ | -8 | -8 | 8 | 8 | 8 | 8 | 8 | 8 |
| $E_{AF6}$ | 8 | 8 | 8 | -8 | -8 | 8 | 8 | -8 |
| $E_{AF7}$ | -8 | 8 | -8 | 8 | -8 | 8 | 8 | -8 |
| $E_{AF8}$ | -8 | 8 | 8 | -8 | 8 | -8 | -8 | 8 |

**D. Spin exchanges in terms of the ordered spin state energies**

$J_7 = (1/16)(4/N^2)(E_{AF2} - E_{FM})$
$J_8 = (1/16)(4/N^2)(E_{AF3} - E_{FM})$
$J_6 = (1/32)[\{(E_{AF1} - E_{AF4}) - (E_{AF3} - E_{AF7})\}(4/N^2) - 16J_7 + 32J_8]$
$J_4 = (1/32)[\{(E_{AF5} - E_{AF8}) - (E_{AF3} - E_{AF7})\}(4/N^2) - 32J_6 - 32J_7 + 16J_8]$
$J_3 = (1/32)[\{(E_{AF5} - E_{AF7}) - (E_{AF1} - E_{AF6})\}(4/N^2) - 16J_7 - 16J_8]$
$J_5 = (1/16)[\{(E_{AF1} - E_{AF6}) - (E_{AF5} - E_{AF8})\}(4/N^2) + 16J_3 + 16J_4 + 16J_6 + 32J_7]$
$J_1 = (1/16)[(E_{AF6} - E_{AF8})(4/N^2) + 16J_5 - 16J_6 - 16J_7 + 16J_8]$
$J_2 = (1/16)[(E_{AF8} - E_{AF5})(4/N^2) + 16J_4 + 16J_6 + 16J_7]$

**E. Relative energies of the ordered spin states and the spin exchanges from DFT+U calculations**

Table 2. Relative energies (meV/FU) of the ordered spin states obtained from DFT+U calculations

| | U = 3 eV | U = 4 eV |
|---|---|---|
| $E_{FM}$ | 94.86 | 73.60 |
| $E_{AF1}$ | 103.05 | 86.09 |
| $E_{AF2}$ | 134.24 | 107.45 |
| $E_{AF3}$ | 119.29 | 95.43 |
| $E_{AF4}$ | 105.88 | 88.92 |
| $E_{AF5}$ | 34.34 | 29.79 |
| $E_{AF6}$ | 120.70 | 102.63 |
| $E_{AF7}$ | 172.85 | 144.25 |
| $E_{AF8}$ | 0 | 0 |

(a, 2b, 2c) super cell
PBE functional for the exchange-correlation

SCF convergence criterion = $10^{-6}$ eV
Plane wave cutoff energy = 450 eV
kpoint set = (6x4x2)

Table 3. Spin exchange parameters (in K) obtained from DFT+U calculations

|  | U = 3 eV | U = 4 eV |
| --- | --- | --- |
| $J_1$ | -311.62 | -274.35 |
| $J_2$ | -54.00 | -64.19 |
| $J_3$ | 672.71 | 545.29 |
| $J_4$ | -243.88 | -211.77 |
| $J_5$ | 765.01 | 625.92 |
| $J_6$ | -150.03 | -150.89 |
| $J_7$ | -58.46 | -47.10 |
| $J_8$ | 115.01 | 92.28 |

**F. Additional calculations using ordered spin states not considered above**

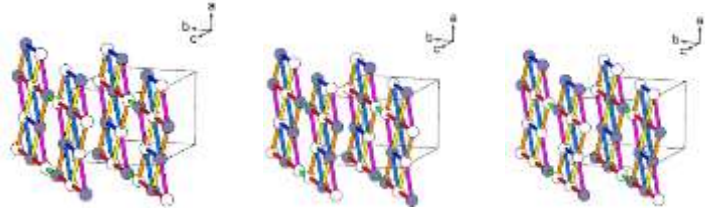

(a) Model_1    (b) Model_2    (c) Model_3 (AF4)

Table 4. Coefficients $n_i$ of $E_i = \sum_{i=1}^{8} n_i J_i S^2$ of Model_1, Model_2 and Model_3 (AF4) state

| (2a, 2b, 2c) | $J_1$ | $J_2$ | $J_3$ | $J_4$ | $J_5$ | $J_6$ | $J_7$ | $J_8$ |
| --- | --- | --- | --- | --- | --- | --- | --- | --- |
| Model_1 | 32 | 16 | -32 | 32 | 32 | -32 | 16 | 16 |
| Model_2 | 32 | 16 | -32 | 32 | 32 | -32 | -16 | 16 |
| Model_3 (AF4) | 0 | -16 | 0 | 0 | 32 | -32 | 16 | 16 |

Table 5. Relative energies of (meV/FUs) associated with the J-values obtained from DFT+U calculations for the Model_1 and Model_2 (AF4) state

| ΔE(mV/FUs) | U = 3 eV | U = 4 eV | U = 5 eV |
|---|---|---|---|
| Model_1 | 0 | 0 | 0 |
| Model_2 | 8.91 | 8.33 | 7.65 |
| Model_3 (AF4) | 17.16 | 14.15 | 11.54 |

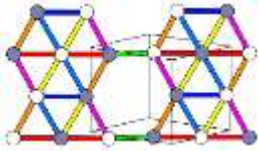 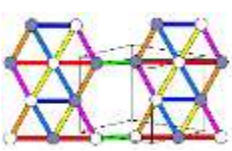 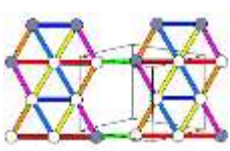

(a) Model_1     (b) Model_2     (c) Model_3

## S15. Na$_2$Cu$_7$(SeO$_3$)$_4$O$_2$Cl$_4$

### A. Spin exchanges

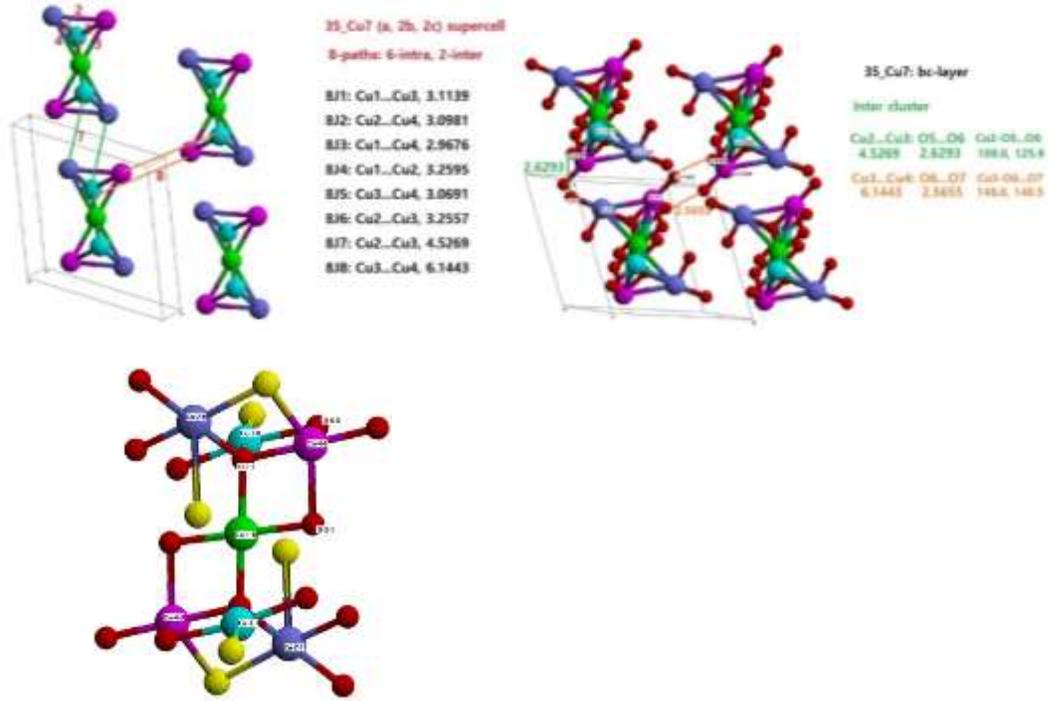

Figure 1. Spin exchange paths

|       | Cu…Cu  | ∠Cu-O-Cu    | O…O    | ∠Cu-O…O       |
|-------|--------|-------------|--------|---------------|
| J$_1$ | 3.1139 | 107.7       |        |               |
| J$_2$ | 3.0981 | 108.3       |        |               |
| J$_3$ | 2.9676 | 94.7, 101.7 |        |               |
| J$_4$ | 3.2595 | 116.8       |        |               |
| J$_5$ | 3.0691 | 95.5, 105.6 |        |               |
| J$_6$ | 3.2557 | 115.3       |        |               |
| J$_7$ | 4.5269 |             | 2.6293 | 109.0, 125.9  |
| J$_8$ | 6.1443 |             | 2.5655 | 148.0, 140.5  |

### B. Ordered spin states using a (a, 2b, 2c) supercell

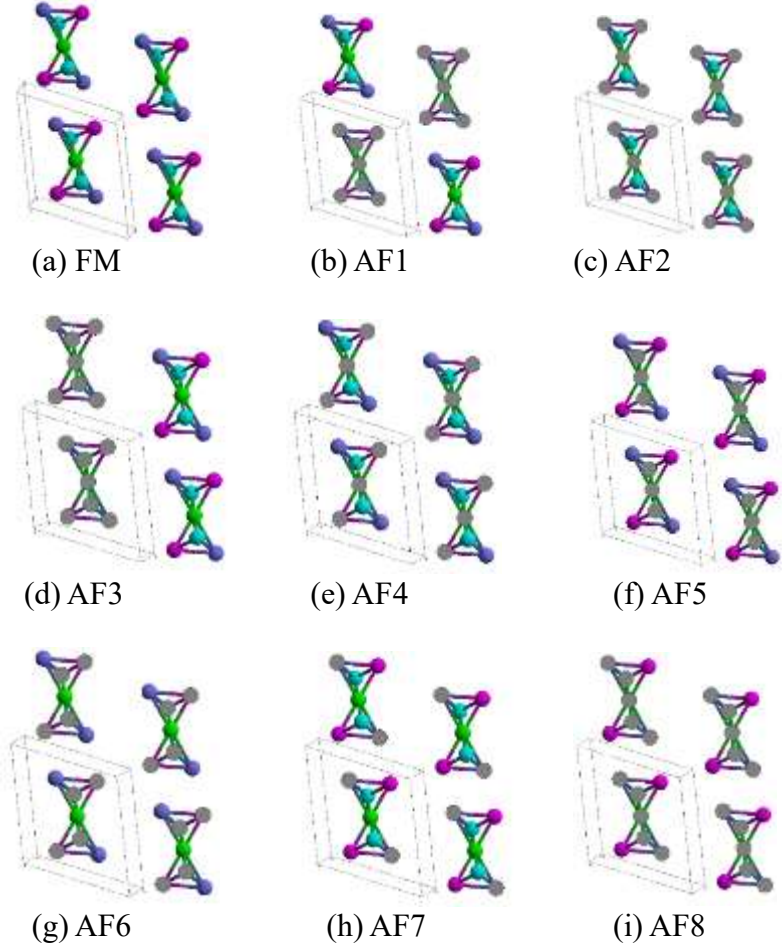

| | (a) FM | (b) AF1 | (c) AF2 |
| | (d) AF3 | (e) AF4 | (f) AF5 |
| | (g) AF6 | (h) AF7 | (i) AF8 |

Figure 2. Ordered spin arrangements of FM and AF(i) (i = 1 to 8). The green, purple, cyan and magenta indicate the Cu1, Cu2, Cu3 and Cu4 ions. The shaded and unshaded circles represent the up and down spin sites of Cu ions, respectively.

## C. Energies of the ordered spin states in terms of the spin exchanges

Table 1. Coefficients $n_j$ of $E_i = \sum_{j=1}^{8} n_j J_j S^2$, where $E_i$ (i = 1 – 9) = $E_{FM}$, $E_{AF1}$ - $E_{AF8}$.

|  | $J_1$ | $J_2$ | $J_3$ | $J_4$ | $J_5$ | $J_6$ | $J_7$ | $J_8$ |
|---|---|---|---|---|---|---|---|---|
| $E_{FM}$ | -8 | -8 | -8 | -8 | -8 | -8 | -8 | -8 |
| $E_{AF1}$ | 8 | -8 | -8 | -8 | 8 | 8 | -8 | -8 |
| $E_{AF2}$ | -8 | -8 | -8 | -8 | -8 | -8 | 8 | -8 |
| $E_{AF3}$ | -8 | -8 | -8 | -8 | -8 | -8 | -8 | 8 |
| $E_{AF4}$ | 8 | 8 | -8 | 8 | 8 | -8 | -8 | 8 |
| $E_{AF5}$ | -8 | -8 | 8 | 8 | 8 | 8 | 8 | 8 |
| $E_{AF6}$ | 8 | 8 | 8 | -8 | -8 | 8 | 8 | -8 |
| $E_{AF7}$ | -8 | 8 | -8 | 8 | -8 | 8 | 8 | -8 |
| $E_{AF8}$ | -8 | 8 | 8 | -8 | 8 | -8 | -8 | 8 |

**D. Spin exchanges in terms of the ordered spin state energies**

$J_7 = (1/16)(4/N^2)(E_{AF2} - E_{FM})$
$J_8 = (1/16)(4/N^2)(E_{AF3} - E_{FM})$
$J_6 = (1/32)[\{(E_{AF1} - E_{AF4}) - (E_{AF3} - E_{AF7})\}(4/N^2) - 16J_7 + 32J_8]$
$J_4 = (1/32)[\{(E_{AF5} - E_{AF8}) - (E_{AF3} - E_{AF7})\}(4/N^2) - 32J_6 - 32J_7 + 16J_8]$
$J_3 = (1/32)[\{(E_{AF5} - E_{AF7}) - (E_{AF1} - E_{AF6})\}(4/N^2) - 16J_7 - 16J_8]$
$J_5 = (1/16)[\{(E_{AF1} - E_{AF6}) - (E_{AF5} - E_{AF8})\}(4/N^2) + 16J_3 + 16J_4 + 16J_6 + 32J_7]$
$J_1 = (1/16)[(E_{AF6} - E_{AF8})(4/N^2) + 16J_5 - 16J_6 - 16J_7 + 16J_8]$
$J_2 = (1/16)[(E_{AF8} - E_{AF5})(4/N^2) + 16J_4 + 16J_6 + 16J_7]$

**E. Relative energies of the ordered spin states and the spin exchanges from DFT+U calculations**

Table 2. Relative energies (meV/FU) of the ordered spin states obtained from DFT+U calculations

| | U = 3 eV | U = 4 eV |
|---|---|---|
| $E_{FM}$ | 94.86 | 73.60 |
| $E_{AF1}$ | 103.05 | 86.09 |
| $E_{AF2}$ | 134.24 | 107.45 |
| $E_{AF3}$ | 119.29 | 95.43 |
| $E_{AF4}$ | 105.88 | 88.92 |
| $E_{AF5}$ | 34.34 | 29.79 |
| $E_{AF6}$ | 120.70 | 102.63 |
| $E_{AF7}$ | 172.85 | 144.25 |
| $E_{AF8}$ | 0 | 0 |

(a, 2b, 2c) super cell
PBE functional for the exchange-correlation
SCF convergence criterion = $10^{-6}$ eV
Plane wave cutoff energy = 450 eV
kpoint set = (6x4x2)

Table 3. Spin exchange parameters (in K) obtained from DFT+U calculations

| | U = 3 eV | U = 4 eV |
|---|---|---|
| $J_1$ | -311.62 | -274.35 |
| $J_2$ | -54.00 | -64.19 |
| $J_3$ | 672.71 | 545.29 |
| $J_4$ | -243.88 | -211.77 |
| $J_5$ | 765.01 | 625.92 |
| $J_6$ | -150.03 | -150.89 |
| $J_7$ | -58.46 | -47.10 |
| $J_8$ | 115.01 | 92.28 |